%% file: Madaiah_Chandrashekar.tex
\documentclass[a4paper, 12pt, openright, twoside]{Thesis} 
\graphicspath{{Figures/}}  

\usepackage[square, numbers, comma, sort&compress]{natbib}  
\usepackage{verbatim}  
\usepackage{vector} 
\hypersetup{urlcolor=blue, colorlinks=false}  
\hypersetup{urlcolor=blue, colorlinks=true}  

\usepackage{amsmath}
\usepackage{epsfig}
\usepackage{hyperref}
\usepackage{bm}
\usepackage{bbm}
\usepackage{tabularx}
\newcommand{\be}{\begin{equation}}
\newcommand{\ee}{\end{equation}}
\newcommand{\bc}{\begin{center}}
\newcommand{\ec}{\end{center}}
\newcommand{\ba}{\begin{array}}
\newcommand{\ea}{\end{array}}
\newcommand{\bea}{\begin{eqnarray}}
\newcommand{\eea}{\end{eqnarray}}
\newtheorem{thm}{Theorem}

\newcommand{\ra}{\rangle}

\begin{document}
\frontmatter	  

\title  {Discrete-Time Quantum Walk - Dynamics and Applications}
\authors  {\texorpdfstring
          {Chandrashekar Madaiah\\
            (C. M. Chandrashekar)}
          {Chandrashekar Madaiah}
            }
\addresses  {\groupname\\\deptname\\\univname}  
\date       {\today}
\subject    {}
\keywords   {}

\maketitle

\setstretch{1.3} 

\fancyhead{}  
\cfoot{\thepage}  
\lhead{}  

\addtotoc{Abstract}
\abstract{
\addtocontents{toc}{\vspace{1em}} 
This dissertation presents investigations on dynamics of discrete-time quantum walk and some of its applications. Quantum walks has been exploited as an useful tool for quantum algorithms in quantum computing. Beyond quantum computational purposes, it has been used to explain and control the dynamics in various physical systems.  In order to use the quantum walk to its fullest potential, it is important to know and optimize the properties purely due to quantum dynamics and in presence of noise. Various studies of its dynamics in the absence and presence of noise have been reported. We propose new approaches to optimize the dynamics, discuss symmetries and effect of noise on the quantum walk. Making use of its properties, we propose the use of quantum walk as an efficient new tool for various applications in physical systems and quantum information processing.

In the first and second part of this dissertation, we discuss evolution process of the quantum walks, propose and demonstrate the optimization of discrete-time quantum walk using quantum coin operation from SU(2) group and discuss some of its properties. 
We investigate symmetry operations and environmental effects on dynamics of the walk on a line and an $n-$cycle highlighting the interplay between noise and topology.

Using the properties and behavior of quantum walk discussed in part two, in part three we propose the application of quantum walk to realize quantum phase transition in optical lattice, that is to efficiently control and redistribute ultracold atoms in optical lattice. We also discuss the implementation scheme. Another application we consider is creation of spatial entanglement using quantum walk on a quantum many body system. }

\clearpage  

\cleardoublepage

\setstretch{1.3}  

\acknowledgements{
\addtocontents{toc}{\vspace{1em}}  
I have been fortunate in having the benefit of interactions and collaboration with numerous people  to whom I feel grateful and would like to thank at this occasion.  I would particularly like to thank Prof. Raymond Laflamme for giving me support, guidance and many opportunities at Waterloo. 
It has been a great opportunity to be around so many talented people at Waterloo, the experience will have a lasting impact on my academic career.  I am grateful to the Mike and Ophelia Lazaridis fellowship and University of Waterloo for their financial support.

I had number of great teachers back in India, during my stay at Oxford, and Waterloo. I take this opportunity to thank them all for the insights they have shared. Particularly, I thank Prof. R. Simon at The Institute of Mathematical Sciences, Chennai, India for his continuous encouragement and support and Prof. Keith Burnett, who was at Oxford during my stay for his support and guidance. At Waterloo, I thank  Frank Wilhelm, Joseph Emerson, Andrew Childs, Andris Ambainis and Ashwin Nayak for their help in refining my ideas.  

Working with my collaborators has greatly benefited me. I thank R. Srikanth, Subhashish Banerjee, Sandeep Goyal for all that I have gained from our discussions during some of the work done in collaboration.  

I thank all my colleagues at the Institute for Quantum Computing including Jonathan Baugh, Jeremy Chamilliard, Mike Ditty, Osama Moussa, Gina Passante, Colm Ryan,  Marcus da Silva, Urbasi Sinha and Jingfu Zhang for making my stay simulating and enjoyable.  I also thank Wendy Reibel, Judy McDonnell and all other staff members at Institute for Quantum Computing, University of Waterloo and Perimeter Institute for Theoretical Physics for their support during my stay at Waterloo. 
I thank  Srinath Reddy,  R. Srikanth, and Sarvagya Upadhyay for their comments on the draft.

This page will not be complete without thanking my parents, brother, sister, relatives and friends for understanding and supporting me all these years.  I also thank Indu for her love.}

\clearpage  
\cleardoublepage

\setstretch{1.3}  

\pagestyle{fancy}  
\dedicatory{

\addtocontents{toc}{\vspace{1em}}  

To my parents\ldots

}
\clearpage

\lhead{\emph{Contents}}  
\tableofcontents  
\addtocontents{toc}{\vspace{1em}}  
\clearpage
\lhead{\emph{List of Figures}}  
\listoffigures  
\addtocontents{toc}{\vspace{1em}}  
\clearpage

\pagestyle{empty}
\preface{
\addtocontents{toc}{\vspace{1em}}  
This dissertation which is divided into four parts discusses {\it the dynamics of discrete-time quantum walk and some of its applications}. Part I, {\it Introduction} consists of Chapter \ref{Chapter1} which introduces the reader to an overview of the subject followed by brief discussion of results presented in the latter chapters.  Part II, {\it Quantum walks and its dynamics}  consists of Chapters \ref{Chapter2} and \ref{Chapter3}. In Chapter \ref{Chapter2}, we review the two standard forms of quantum walks (continuous- and discrete-time quantum walk) and study the dynamics of discrete-time quantum walk in detail. In Chapter \ref{Chapter3}, we study symmetries and effect of noise on the dynamics of the discrete-time quantum walk.  Part III, {\it Applications} consists of Chapters \ref{Chapter4} and \ref{Chapter5}. In Chapter \ref{Chapter4}, we demonstrate the use of discrete-time quantum walk to control the dynamics of ultracold in optical lattice and in Chapter \ref{Chapter5}, we create spatial entanglement using quantum walk on distinguishable multiparticle system. Part IV, {\it Conclusion} consists of Chapter \ref{Chapter6} which concludes with summery and future perspective.}
\cleardoublepage  

\pagestyle{fancy}
\prefacea{
\addtocontents{toc}{\vspace{1em}}  
\lhead{\emph{Glossary of Notations}}  
\bc
{\bf NOTATIONS INTRODUCED IN CHAPTER \ref{Chapter1}}
\ec
\begin{tabular}{p{7.5cm}p{6.5cm}}
 $\mathcal H$    &   Hilbert space \\
 \\
$\mathcal H_{p}$  &  {\em Position} Hilbert space \\
 \\
 $\mathcal H_{c}$  &    {\em Coin} Hilbert space \\
 \\
$H = \frac{1}{\sqrt 2} \left( \begin{array}{clcr}
 1  & &  ~1   \\
1  & &  -1 
\end{array} \right)$    &  Hadamard operation\\
\\
$B_{\theta} \equiv \left( \begin{array}{clcr}
 \cos(\theta)  &   {\mbox ~~}\sin(\theta)   \\
\sin(\theta)  &  -\cos(\theta)
\end{array} \right)$ & U(2) operator with a single-variable parameter  \\
\\
$|0\rangle = \left(   \begin{array}{clcr}   1 \\  0
\end{array} \right)$ & Single qubit notation\\
\\
$|1\rangle =   \left(   \begin{array}{clcr}   0 \\  1
\end{array} \right)$  & 
Single qubit notation\\
\\
$\frac{1}{\sqrt 2} \left ( |0\rangle + i |1\rangle \right )$  & Symmetric superposition state of the particle \\
\\
$|\psi_{0}\rangle$  & State of the position\\
\\
$|\Psi_{ins}\rangle \equiv \frac{1}{\sqrt 2}\left ( |0\rangle + i |1\rangle \right ) \otimes |\psi_{0}\rangle$ 
&  Symmetric superposition state of particle in position space as initial state \\
\\
$B_{\xi,\theta,\zeta} \equiv \left( \begin{array}{clcr}  \mbox{~}e^{i\xi}\cos(\theta)  & &   e^{i\zeta}\sin(\theta)   \\
-e^{-i\zeta} \sin(\theta)  & &  e^{-i\xi}\cos(\theta)
\end{array} \right)$ &  SU(2) operator : generalized three parameter quantum coin operation
\end{tabular}
\newpage
\bc
{\bf NOTATIONS INTRODUCED IN CHAPTER \ref{Chapter2}}
\ec
\underline{In Section \ref{ctqwm}} \\
\begin{tabular}{p{7.0cm}p{7cm}}
$G=(V, E)$  &  Graph where $V$ and $E$ are vertex and edges set \\
\\
$(j,k)$  & The vertex set, form the edge when $j$ and $k$ are connected\\
\\
$A_{j,k} =  \begin{cases}
1  &   ~~ (j,k) \in E \\
0  &   ~~ (j,k) \notin E 
\end{cases}$  & Adjacency matrix\\
\\
${\bf H}$  &  Hamiltonian \\
\\
${\bf H}_{j,k} =  \begin{cases}
d_j \gamma   &   ~~ j =  k \\
-\gamma  &   ~~ (j,k) \in E \\
0  &  ~~ {\rm otherwise}
\end{cases}$  & Generator matrix
\end{tabular}\\
\\
The Hamiltonian ${\bf H}$ of the walk process acts as the generator matrix ${\bf H}_{j,k}$ which will transform the probability amplitude at the rate of $\gamma$ to the neighboring sites. Therefore we represent both Hamiltonian and generator matrix by ${\bf H}$.  $d_{j}$ is the degree of the vertex $j$.\\
\\
\begin{tabular}{p{6.5cm}p{7.5cm}}
$P_{j}(t)$  &   Probability of being at vertex $j$ at time $t$
\end{tabular}

\underline{In Section \ref{dtqw}} \\
\begin{tabular}{p{7cm}p{7cm}}
$j$  &  Position (vertex)\\
\\
$t$  &  Time ($t$ is also used for number of steps when unit time is required to implement each step of quantum walk)\\
\\
$|\Psi_{in}\rangle = \left ( \cos(\delta)|0\rangle + e^{i\eta}\sin(\delta)|1\rangle \right )\otimes |\psi_{0}\rangle$ & General form of initial state : coin (particle) and position space
\end{tabular}
\begin{tabular}{p{6.3cm}p{7cm}}
$|\Psi_{t}\rangle = |\Psi(t)\rangle$ & Coin and position state after time $t$ over entire position space\\
\\
$|\Psi_{j, t}\rangle = |\Psi(j, t)\rangle$  & Coin and position state after time $t$ at position $j$ \\
\\
${\mathbbm  1} =  \left(   \begin{array}{clcr} 1 & &  0 \\ 0 & & 1
\end{array} \right)$  & Identity operation \\
\\
$\sigma_x = X  
=    \left(   \begin{array}{clcr}   0 & & 1 \\  1 & & 0
\end{array} \right)$ & Pauli $x$ operation\\
\\
$\sigma_{y}  = Y =   \left(   \begin{array}{clcr} 0 & & -i \\ i  & & ~0
\end{array} \right)$ & Pauli $y$ operation\\
\\
$\sigma_{z} = Z =   \left( \begin{array}{clcr}   1 & & ~0 \\  0 & & -1
\end{array} \right)$ & Pauli $z$ operation\\
\\
$B_{\zeta, \alpha, \beta, \gamma} = e^{i \zeta} e^{i\alpha \sigma_{x}}e^{i\beta \sigma_{y}}e^{i\gamma \sigma_{z}}$ & Quantum coin operation $B \in U(2)$ \\
\\
$S  =  |0\rangle  \langle 0|\otimes  \sum_{j  \in
\mathbb{Z}}|\psi_{j-1}\rangle  \langle \psi_{j} |+|1\rangle  \langle 1
|\otimes \sum_{j \in \mathbb{Z}} |\psi_{j+1}\rangle \langle \psi_{j}|$ & Conditional unitary shift operator on a line \\
\\
$W_{\zeta, \alpha, \beta, \gamma} =
S(B_{\zeta, \alpha, \beta, \gamma}  \otimes   {\mathbbm  1})$&  Single step quantum walk operation \\
\\
$P(j,\tau)$  & Probability at position $j$ after time $t$
\end{tabular}

\underline{In Section \ref{gen-qw}}\\
\begin{tabular}{p{6.3cm}p{7cm}}
$\mathcal{A}_{m,t}$ & Amplitudes of state $|0\rangle$ at position $m$ after $t$ steps of quantum walk \\
\\
$\mathcal{B}_{m,t}$ & Amplitudes of state $|1\rangle$ at position $m$ after $t$ steps of quantum walk
\end{tabular}

\begin{tabular}{p{7.5cm}p{7cm}}
$B^{\prime}_{\xi, \theta, \zeta} \equiv \left( \begin{array}{clcr}  e^{i\xi}\cos(\theta)  & &   e^{i\zeta}\sin(\theta)   \\
e^{-i\zeta} \sin(\theta)  & &  -e^{-i\xi}\cos(\theta)
\end{array} \right)$ & Three parameter quantum coin operation which is not in SU(2) group
 \\
 \\
$P(j) = P_{j}$  & probability at position $j$\\
\\
$P(j,t)$  &  Probability at position $j$ after time $t$\\
\\
$f(\phi) = t \cos(\theta)\sin(\phi)$ & For any given $\theta$, the position $j$ can be parametrized by function $f(\phi)$ \\
\\
$H(j)= -\sum_j P_j \log P_j$ & Shannon entropy\\
\\
$S^{c}  =  |0\rangle  \langle 0|\otimes  \sum_{j=0}^{n-1}|\psi_{j-1~{\rm mod}~n}\rangle  \langle \psi_{j} | +|1\rangle  \langle 1
|\otimes \sum_{j=0}^{n-1} |\psi_{j+1~{\rm mod}~n}\rangle \langle \psi_{j}|$ & Conditional unitary shift operator on an $n-$cycle 

\end{tabular}

\underline{In Section \ref{recurrQW}}\\
\begin{tabular}{p{7.0cm}p{7cm}}

$A_{o}$ &  Observable  \\
\\
$X_{p}$ &  Position operator \\
\\
${\mathcal P}_{crw} \equiv 1 - \frac{1}{\sum_{t=0}^\infty P_{0}(t)}$ &  Classical P\'olya number \\
\\
$P_{0}(t) = P(0, t)$ & The probability of  periodicity of dynamics that the particle returns to origin during the $t$ steps of walk (recurrence probability)\\
\\
$|\Psi_{j, T_{M}+1}\rangle$  & state at position $j$ after evolving up to time $(T+1)$ with intermediate measurement at time $T$\\
\\
${\mathcal P}_{qw} = 1 - \prod_{t=1}^{\infty} \left [ 1- P_{0}(t) \right ]$ & Quantum P\'olya number 
\end{tabular}

\newpage

\bc
{\bf NOTATIONS INTRODUCED IN CHAPTER \ref{Chapter3}}
\ec

\underline{In Section \ref{sec:symmline}}\\
\\
\begin{tabular}{p{6.5cm}p{7.5cm}}

$\hat{a}$ and $\hat{a}^{\dag}$  &  Unitary operators that are
notationally reminiscent of annihilation and creation operations \\
\\
$X$ & Bit flip\\
\\
$Z$ & Phase flip\\
\\
$G$ &  Operation represent interchanging of coin operation which leave quantum walk symmetric\\
\\
$P$ &  Parity ($\hat{a}  \leftrightarrow \hat{a}^{\dag}$)   \\
\\
$R$ & Angular  reflection ($\theta \rightarrow \pi/2-\theta$, i.e.,    $\sin\theta    \leftrightarrow    \cos\theta$,    and    $\xi \leftrightarrow  -\zeta$)\\
\\
{\bf X}, {\bf Z}, {\bf G}, {\bf P} and {\bf R} &   Refers to application of $X, Z, G, P$ and $R$ operations at each step of quantum walk. \\
 \\
$\Phi(\phi)   \equiv |0\rangle\langle0| +  e^{i\phi}|1\rangle\langle1|$ & Phase shift operation\\
\\
${\cal E}(\rho) = (1-p)\rho + pZ\rho Z$ & Phase flip channel \\
\\
${\cal E}(\rho) = (1-p)\rho + pX \rho X$ & Bit flip channel\\

\end{tabular}

\begin{tabular}{p{7.0cm}p{7cm}}

${\bf H} = {\bf H}_{S} + {\bf H}_{R} + {\bf H}_{SR}$ &  Interaction Hamiltonian\\
\\
${\bf H}_S$ &  Hamiltonian of the system (S)\\
\\
${\bf H}_R$ & Hamiltonian of the reservoir (R)\\ 
\\
${\bf H}_{SR}$ & Hamiltonian of the system-reservoir (SR) interaction\\
\\
$\hat{b}$ and $\hat{b}^{\dag}$  & Annihilation and creation operators\\
\\
$\sigma_{+} = |1\rangle \langle 0| = \frac{1}{2} (\sigma_{x} + i \sigma_{y})$ & Raising operator\\
\\
 $\sigma_{-} = |0\rangle \langle 1| = \frac{1}{2} (\sigma_{x} - i \sigma_{y})$ & Lowering operator

\end{tabular}
\underline{In Section \ref{qwpg}}\\
\\
\begin{tabular}{p{7.0cm}p{7cm}}
$G(\beta)=  \left(\begin{array}{ll}
1 & 0 \\ 0 & e^{i\beta}\end{array}\right)$ & Generalized phase gate\\
\\
$d(t)  = (1/2)\sum_j|P(j,t)-P^{\prime}(j,t)|$ & Kolmogorov distance (trace  distance) : distance between the particle position distributions obtained without and  with the symmetry operation given by $P(j,t)$ and $P^{\prime}(j,t)$\\
\\
$\tau$ & Number of turns on an n-cycle\\
\\
$D(\tau)$ &  Normalized Kolmogorov distance\\
\\
${\bm C}$  &  Coherence\\
\\
$C(m)$ and  $c(m)$ &  Coherence function and  normalized coherence function
\end{tabular}
\newpage

\bc
{\bf NOTATIONS INTRODUCED IN CHAPTER \ref{Chapter4}}
\ec

\underline{In Section \ref{pt}}\\
\\
\begin{tabular}{p{6.0cm}p{8cm}}
${\bf H}_{B}$  &  Bose-Hubbard Hamiltonian \\
\\
$\hat{b}$ and $\hat{b}^{\dag}$  & Boson annihilation and creation operators\\
\\
$\hat{b}_{j}$ and $\hat{b}_{k}^{\dag}$  & Boson annihilation and creation operators at position $j$ and $k$ respectively\\
\\
$\hat{n}_{j} = \hat{b}_{j}^{\dag} \hat{b}_{j}$  &  Boson number operator, counts the number of bosons at position $j$ \\
\\
$\mu$ & Chemical potential of the bosons \\
\\
$U$ &  Repulsive interaction between atoms in single lattice site\\
\\
$J$ & Hopping element which allow hopping of bosons between the nearest neighbor sites\\
\\
$\hat{N}_{b}$  &  Boson number operator, counts the total number of atoms in the entire system\\
\\
${\bf H}_{MF}$ & Mean-field Hamiltonian \\
\\
$E_{MF}(\Psi_{B})$ & Ground state energy of $H_{MF}$\\
\\
$M$ & Number of lattice site\\
\\
\end{tabular}

\newpage
\underline{In Section \ref{impQW}}\\
\\
\begin{tabular}{p{7.0cm}p{7cm}}
$|j\rangle$ &  For notational convenience, position state  $|\psi_{j}\rangle$ is written as $|j\rangle$\\
\\
${\bf H}_{BH_{2}}$  &  Bose-Hubbard Hamiltonian for two-state atoms\\
\\
$\hat{b}_{j \uparrow}$ and $\hat{b}_{k \downarrow}^{\dag}$  & Boson annihilation and creation operators at position $j$ and $k$ respectively. $\uparrow$ and $\downarrow$ represent the terms for atoms in state $|0\rangle$ and $|1\rangle$ respectively
\end{tabular}
\\
\underline{In Section \ref{impl}}\\
\\
\begin{tabular}{p{7.0cm}p{7cm}}

$|0_{BEC}\rangle$ and $|1_{BEC}\rangle$   &  Macroscopic state of $N$ condensed atoms in state $|0\rangle$ and $|1\rangle$\\
\\
$z_{R}$ &  Rayleigh range : distance at which the diameter of the laser beam size increases by a factor of $\sqrt{2}$\\
\\
$\omega_{R}$ &  Rabi frequency\\
\\
$\Delta$ & Laser detuning from radio frequency (rf) resonance\\
\\
\end{tabular}

\newpage
\bc
{\bf NOTATIONS INTRODUCED IN CHAPTER \ref{Chapter5}}
\ec
\underline{In Section \ref{cpent}}\\
\\
\begin{tabular}{p{6.0cm}p{8.0cm}}
$E_{c}(t)$  & Entropy of the reduced density matrix to quantify the position - particle entanglement\\
\\
$\lambda_{j}$  &   Eigenvalues of the reduced density matrix of the coin\\
\end{tabular}
\underline{In Section \ref{spentqw}}\\
\\
\begin{tabular}{p{7.0cm}p{7cm}}
$|1_{j}\rangle$ &  $j-$th position is occupied\\
\\
$|0_{j}\rangle$ &  $j-$th position is unoccupied\\
\\
$|\Psi_{lat}\rangle$ &  State of the lattice\\
\\
\end{tabular}
\underline{In Section \ref{mbqw}}\\
\\
\begin{tabular}{p{7.0cm}p{7.5cm}}
$|\Psi_{0}^{M}\rangle = \bigotimes_{j=-\frac{M-1}{2}}^{j=\frac{M-1}{2}}
  \left( \frac{|0\rangle + i|1\rangle}{\sqrt{2}} \right) \otimes
  |\psi_{j}\rangle$  &  Initial state of $M$ particle system\\
\\
${\rm A}$, ${\rm B}$ and ${\rm C}$ & label for 3 particles\\
\\
\end{tabular}
\underline{In Section \ref{mpent}}\\
\\
\begin{tabular}{p{6.0cm}p{8cm}}
$|\psi \rangle$, $|\psi_{1}\rangle \cdots |\psi_{k}\rangle$  &   General form of state\\
\\
$E$ and $E(|\Psi_{lat}\rangle)$ &  Meyer-Wallach measure for entanglement\\
\\
$\rho_{i}$ & Reduced density matrix of each subsystem $i$\\
\\
$\rho_{j}$ & Reduced density matrix of the $j$ lattice point
\end{tabular} 
}

\clearpage  
\cleardoublepage

\mainmatter	  
\pagestyle{fancy}  

\part{Introduction}

\input{./Chapters/Chapter1}

\part{Quantum Walks and Its Dynamics}

\input{./Chapters/Chapter2}

\input{./Chapters/Chapter3}

\part{Applications}

\input{./Chapters/Chapter4}

\input{./Chapters/Chapter5}

\part{Conclusion}

\input{./Chapters/Chapter6}



\addtocontents{toc}{\vspace{2em}} 


\newpage
\addtotoc{Appendix}  
\appendix{
\addtocontents{toc}{\vspace{1em}} 

\input{./Appendices/Appendix0}

\input{./Appendices/AppendixA}

\input{./Appendices/AppendixC}

\addtocontents{toc}{\vspace{2em}}  
\backmatter

\label{Bibliography}
\lhead{\emph{Bibliography}}  
\bibliographystyle{unsrtnat}  
\bibliography{Bibliography}  

\end{document}

%% file: Chapters/Chapter1.tex

\chapter{Introduction} 
\label{Chapter1}
\lhead{Chapter 1. \emph{Introduction}} 

\section{Introduction}

Theoretical studies and experimental evidences from the early and mid twentieth century led us to consider the physical world is governed by laws of quantum mechanics (see any standard text book on quantum mechanics. For example, Feynman (1970) \cite{fey70} and R. Shenkar (1994) \cite{sha94}).  This, in early 1980's led Yuri Manin (1980) \cite{Man80} and Feynman (1982) \cite{Fey82} to independently observe and suggest that the physical world can be ideally simulated using quantum computers. A decade later, theoretical studies and the experimental possibility to construct quantum computers based on the laws of quantum mechanics such as superposition and interference between quantum amplitudes became one of the active areas of research~\cite{NC00, KLM07}. Deutsch algorithm (1985) \cite{Deu85} and Deutsch-Jozsa algorithm (1992) \cite{DJ92} were among the first algorithms to show that quantum computers are capable of solving certain computational problems much more efficiently than classical deterministic computers.  Simon's algorithm (1994) was one of the first example to show that a quantum algorithm can solve the problem of computing an unknown function with a polynomial number of queries, where a classical algorithm required  an exponential number of queries \cite{Sim94, Sim97}. In 1994, Shor devised a quantum algorithm to factor arbitrary integers exponentially faster than the best known classical counterpart ~\cite{Sho94, Sho97}. In 1996, Grover devised an algorithm which can in principle search an unsorted database quadratically faster than any known classical algorithm \cite{Gro97}.  Shor's algorithm unleashed active research in the area of quantum information and quantum computation across a broad range of disciplines: quantum physics, computers science, mathematics and engineering. This research has unveiled many new effects that are strikingly different from their classical counterparts. Design and analysis of quantum algorithms to solve various problems more efficiently than classical algorithms has become a vibrant research area (see Mosca (2008) \cite{Mos08} and Childs and van Dam (2008) \cite{CD08} for recent reviews). 
\par
Among classical algorithms, many are based on classical random walk. Markov chain simulation, which has emerged as a powerful algorithmic  tool  \cite{MT05} is one such example. Like classical random walk, the quantum version of it has also become an important  constituent  of quantum algorithms and computation.
\par
\begin{figure}[h]
\begin{center}
\epsfig{figure=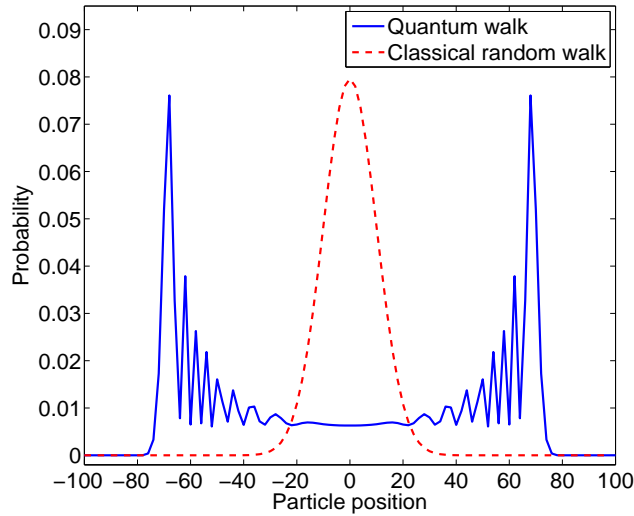, width=9.0cm}
\caption[Probability distribution of classical random walk and its quantum counterpart, quantum walk. Difference in the variance can been seen in the distribution. The distribution is after 100 steps of walk.]{Probability distribution of classical random walk and its quantum counterpart, quantum walk. Difference in the variance can been seen in the distribution. The distribution is after 100 steps of walk.}
\label{quantumwalk} 
\end{center}
\end{figure}

The quantum walk as it is known today is a generalization of the classical random walk developed by using the aspects of quantum mechanics such as superposition and interference of quantum amplitudes. In the classical random walk the particle moves in the position space with a certain probability, whereas  the quantum walk, which involves a superposition of states, moves by exploring multiple possible paths simultaneously with the amplitudes corresponding to different  paths interfering. 
This makes the variance of the quantum walk on a line to grow quadratically with the number of steps (time), compared to the linear growth for the classical random walk \cite{ABN01, NV01}. A probabilistic  result is obtained in quantum walk upon measurement.  Figure \ref{quantumwalk} shows the comparative difference in the probability distribution of classical random walk and quantum walk. 
\par
Like its classical counterpart, which has found applications in almost all branches of sciences \cite{BN70, Cha43}, quantum walk has also proven to be a useful tool for quantum algorithms, to study, control and explain the dynamics in various physical systems.  Below, we will briefly summarize the history of quantum walk and the progress made in employing it for quantum algorithms and the study of the dynamics of physical systems. 
\par
Though quantum random walk\footnote{Unlike its classical counterpart, the evolution of the quantum version is unitary, reversible and has no randomness associated with it during the evolution. Therefore, keeping away the term `random',  quantum walk has been the preferred usage.} was first introduced by Aharonov \emph{et~al.} (1993) \cite{ADZ93}, the idea of exploring multiple paths simultaneously using path integrals dates back to works by Riazanov (1958) \cite{Ria58} and Feynman and Hibbs (1965) \cite{FH65}. Meyer (1996) studied the dynamics of quantum particle in a quantum lattice gas formulation. The dynamics involved the interference of the left and the right moving components of the amplitude at given nodes \cite{Mey96, Mey97}. Building on these ideas, the concept of quantum walk was developed and now is studied in two standard forms: continuous-time quantum walk and discrete-time quantum walk. Continuous-time quantum walk was introduced by Farhi and Gutmann (1998) \cite{FG98} and discrete-time quantum walk was introduced by Watrous (2001) \cite{Wat01}. A specific form of discrete-time quantum walk know as Hadamard walk was introduced by Ambainis \emph{et~al.} (2001) \cite{ABN01}. An 
early review by Kempe (2003) \cite{Kem03} discusses the two variants in detail with some algorithmic applications. 
\par
Both continuous- and discrete-time quantum walk have been widely used in algorithms for a variety of problems. Several quantum search algorithms have been proposed (see for example, Shenvi \emph{et~al.} (2003) \cite{SKW03}; Childs and Goldstone (2004) \cite{CG04a, CG04b}; Ambainis \emph{et~al.} (2005) \cite{AKR05}; Aaronson and Ambainis (2005) \cite{AA05}; Magniez \emph{et~al.} (2007) \cite{MNR07}). The continuous-time quantum walk has been used by Childs \emph{et~al.} (2003)  \cite{CCD03} to demonstrate the exponential speedup over classical computation for a hitting time problem on a glued tree.
Ambainis (2004) \cite{Amb04} review discusses some of these algorithms in detail. Ambainis (2007) \cite{Amb07} applied quantum walk to give an optimal quantum algorithm for the element distinctness problem. It has been applied to various other problems in query model like triangle finding by Magniez \emph{et~al.} (2005) \cite{MSS05}, checking matrix  multiplication by Buhrman and {\v S}palek (2006) \cite{BS06}, testing group commutativity by Maganiez and Nayak (2007) \cite{MN07}, for evaluating balanced binary game trees by Farhi \emph{et~al.}  (2007)\cite{FGG07}, Boolean formulas by Ambainis \emph{et~al.} (2007) \cite{ACR07} and Reichardt and {\v S}palek \cite{RS08} and to obtain general adversary bound to characterize the quantum query complexity by Reichardt \cite{Rei09}.
\par
Beyond applications in quantum algorithms, quantum walk is emerging as a potential tool to understand various phenomenon in physical systems and has been employed to demonstrate coherent control over quantum many body systems. Oka \emph{et~al.} (2005)  \cite{OKA05} mapped the breakdown of electric-field driven system to quantum walk, Engel \emph{et~al.} (2007) \cite{ECR07} and Mohseni \emph{et~al.} (2008) \cite{MRL08} explained wavelike energy transfer within photosynthetic systems, Somma \emph{et~al.} (2008) \cite{SBB08} described quantum simulation of classical annealing processes, and Chandrashekar and Laflamme (2008) \cite{CL08} demonstrated coherent quantum control over redistribution of atoms in optical lattice. 
\par
Some experimental progress on the implementation of quantum walk has been reported. Du \emph{et~al.} (2003) \cite{DLX03} and Ryan \emph{et~al.} (2005) \cite{RLB05} implemented continuous-time quantum walk on 2- qubit and discrete-time quantum walk on a 3- qubit Nuclear Magnetic Resonance (NMR) system respectively. Grossman \emph{et~al.} (2004) \cite{GCD04} with sodium Bose-Einstein condensates and recently, Perets \emph{et~al.} (2008) \cite{PLF08} using propagating photons in waveguide lattices have implemented the discrete-time quantum walk. Other implementation schemes have also been proposed: Travaglione and Milburn (2002) \cite{TM02} in an ion trap, D\"ur \emph{et~al.} (2002) \cite{DRK02} and Eckert \emph{et~al.} (2005) \cite{EMB05} on neutral cold atoms in an optical traps, Chandrashekar (2006) \cite{Cha06} using Bose-Einstein condensates, Ma \emph{et~al.} (2006) \cite{MBD06} using quantum accelerator mode, Manouchehri and Wang (2008) \cite{MW08} in an array of quantum dots and Sanders \emph{et~al.} (2003) \cite{SBT03} introduced quantum quincunx to physically demonstrate quantum walk in cavity quantum electrodynamics capabilities.
\par
It is important to understand the dynamics and behavior of quantum walk, both in the presence and absence of environmental effects to effectively use it for developing algorithms, to simulate any natural physical process or to experimentally implement in any physical system.  Considerable study has been done in this direction. Since this thesis involves results of some of these investigations, we will first list the main work that constitutes this thesis and briefly elaborate on them in the following sections.
\par
{\bf The main constituents of this thesis are :} 
\begin{itemize}
\item Structure and dynamic properties of discrete-time quantum walk. Construction of a generalized model using quantum coin operation from SU(2) group to demonstrate control over the dynamics of the walk \cite{CSL08, Cha09}.
\item Using a coin-embedded shift operator, a generic quantum walk model which can be used to conveniently retrieve  discrete- and continuous-time quantum walk  under different conditions  is proposed.  The coin degree of freedom is retained while the standard variants are retrieved and the generic model simplifies the physical resources required to implement quantum walk compared to the resource required for the discrete-time quantum walk \cite{Cha08}.
\item Investigations on symmetries in discrete-time quantum walk on a line and its behavior in presence of environmental effects, breakdown of these symmetries for a walk on an $n-$cycle and restoration with noise. Relevance of these investigations in physical systems \cite{CSB07, BSC08}. 
\item Demonstration of coherent control over the redistribution of atoms in optical lattice, use of noisy channels- bit flip, phase damping, and amplitude damping to act as an additional toolbox to control the dynamics of atoms and an experimental implementation scheme \cite{CL08, Cha06}.  
\item Spatial entanglement in many body system using quantum walk and its physical relevance \cite{GC09}.
\end{itemize}

\section{Dynamics of quantum walk}
\label{dyna}

In the continuous-time quantum walk, one  can directly define the walk on the {\em position} Hilbert space $\mathcal H_{p}$ \cite{FG98}. In the discrete-time quantum walk, in addition to $\mathcal H_{p}$ it is necessary to introduce a quantum coin operation, a {\em coin} Hilbert space $\mathcal H_{c}$ to define the direction in which the particle has to move \cite{ABN01}. Due to the coin degree of freedom, the discrete-time variant is shown to be more powerful than the other  in  some contexts (see Ambainis \emph{et~al.} (2005) \cite{AKR05}). To match the performance of a spatial search using the discrete-time quantum walk, the coin degree of freedom has been introduced in the continuous-time quantum walk  model \cite{CG04b}.  In this thesis, we mainly consider the discrete-time quantum walk for our study.
\par
The dynamics of discrete-time quantum walk using the Hadamard operation 
\be
H \equiv \frac{1}{\sqrt 2} \left( \begin{array}{clcr}
1  &   {\mbox ~~}1   \\
1  &  -1
\end{array} \right)
\ee
as quantum coin operation was analyzed and defined as Hadamard walk \cite{ABN01}. Most of the work thereafter on discrete-time quantum walk considered using Hadamard operation as quantum coin operation. Nayak and Vishwanath (2001) \cite{NV01} analyzed using the single-variable parameter  U(2) operator 
\be
B_{\theta} \equiv \left( \begin{array}{clcr}
 \cos(\theta)  &   {\mbox ~~}\sin(\theta)   \\
\sin(\theta)  &  -\cos(\theta)
\end{array} \right) 
\ee
as a quantum coin\footnote{Meyer has used the U(2) operator as a scattering operator in lattice quantum gas automata model \cite{Mey96}}. The variance $\sigma^{2}$ was shown to be dependent on the parameter $\theta$. The discrete-time quantum walk on a particle initially in a symmetric superposition state ($\frac{1}{\sqrt 2} \left ( |0\rangle + i |1\rangle \right )$) at position space ($|\psi_{0}\rangle$), $|\Psi_{ins}\rangle = \frac{1}{\sqrt 2} \left ( |0\rangle + i |1\rangle \right ) \otimes |\psi_{0}\rangle$ using  $B_{\theta}$ as the quantum coin was shown to return the symmetric probability distribution of the walk in the position space. Asymmetry in the initial state of the particle introduced an asymmetric probability distribution. This led to their conclusion that obtaining a symmetric distribution depends  largely on the initial state of the particle \cite{NV01, ABN01, TFM03}.
\par
In Chapter \ref{Chapter2}, we will first define the continuous-time and discrete-time quantum walk, equivalence of their form to the Schr\"odinger and  Dirac equations respectively.
In Section \ref{gen-qw}, we present the discrete-time quantum walk using operation from the SU(2) group with three Caley-Klein  parameters $\xi$, $\theta$ and $\zeta$  (Euler angles) as  the  quantum  coin 
\be
B_{\xi, \theta, \zeta} \equiv \left( \begin{array}{clcr}
 {\mbox ~~} e^{ i \xi}\cos(\theta)  &   e^{i\zeta}\sin(\theta)   \\
- e^{-i\zeta}\sin(\theta)  &  e^{-i \xi} \cos(\theta)
\end{array} \right). 
\ee
We  show  that  the  variance is dependent on the parameter $\theta$ and takes the form $\sigma^{2} \approx (1-\sin(\theta))t^{2}$, where $t$ is number of steps of the quantum walk.
For a particle with symmetric superposition as the initial state parameters  $\xi$ and $\zeta$ introduce asymmetry in the probability  distribution and their effect on the variance is very small. For a particle with an asymmetric superposition as the initial state, the parameters $\xi$ and $\zeta$ can be configured to obtain a symmetric probability distribution. We  discuss the variation  of measurement  entropy in  position space  with the coin parameters and optimization of the quantum walk for the maximum variance, improving  mixing time in an $n$-cycle and controlling the probability distribution using coin parameters \cite{CSL08}. 
\par
Recurrence in the dynamics of physical systems is an important phenomenon. For a classical conservative system, whether discrete or continuous in time, the Poincar\'e recurrence theorem states that any phase-space  configuration of a system enclosed in a finite volume will be repeated as accurately as one wishes after a finite interval of time (with no restriction on the interval) \cite{Bar06, Kre85}. A similar recurrence theorem is shown to hold in quantum theory as well \cite{BL57, Sch78}. In a system with a discrete energy eigenvalue spectrum $\{ E_{n} \}$; if $\Psi(t_0)$  is its state vector at the time $t_0$ and $\epsilon$ is any positive number, there exists a finite time $T$ such that,
\be
|\Psi(T)-\Psi(t_0)| < \epsilon.
\ee 
Robinett's (2004) \cite{Rob04} review article discusses recurrence in quantum systems. In Section \ref{recurrQW}, we show the discrete-time quantum walk which evolves with interference of quantum amplitudes fails to recur completely in position space but fractional recurrence is seen \cite{Cha09}.
\par
Chapter \ref{Chapter3} focuses on the symmetries and effects of noise on the dynamics of discrete-time quantum walk. The transition of walk from the quantum to the classical regime has been studied earlier by introducing decoherence to Hadamard walk \cite{BCA03, KT03, RSA05}(see review by Kendon (2006) \cite{Ken06}). In a scheme proposed by Chandrashekar \cite{Cha06} to implement the discrete-time quantum walk on Bose-Einstein  condensates, stimulated Raman kick was used as shift operator. The stimulated Raman kick, in addition to shifting different states of atoms to different position space flips (bit-flip) the internal state of the atoms. However in spite of bit-flip on state of atoms along with shift in position space, the probability distribution remained invariant in position space. Motivated by this, the shift operator was augmented by other form of operations and the operation for which the distribution remains invariant have been studied and are called symmetries of quantum walk. These symmetries have been studied using generalized SU(2) coin operation $B_{0,\theta, 0}$ as quantum coin. We further generalize the observations of these symmetries in the presence of environmental effects, modeled by  various noise channels such as bit flip, phase-flip and generalized amplitude damping channels \cite{CSB07}. Interestingly,  we find that the symmetry operations are sensitive to the  walk topology. For example, a symmetry that holds for quantum walk on an one-dimensional line does not hold, in general, for a quantum  walk on an $n-$cycle but leads to other interesting behavior. Noise on an $n-$cycle tends  to  restore these symmetry, both  by ``classicalizing" the  walk and also desensitizing the
symmetry operation as a topology probe for the quantum walk \cite{BSC08}. From Chapter \ref{Chapter2}, we can conclude that the symmetries and addition of small amount of experimentally engineered noise or environmental effect can be used as an additional tool to control the dynamics and hence the probability distribution of the quantum walk. These observations can  have important implications  for a better insight into quantum walk, and for simplifying certain implementations and are also of relevance in the study of dynamics in environment-assisted quantum systems. 
\par
\section{Applications}
\label{app}

By employing the properties of quantum walks due to purely quantum dynamics and small amount of engineered noise  presented in Chapters \ref{Chapter2} and \ref{Chapter3},  we demonstrate coherent quantum control over atoms in optical lattice in Chapter \ref{Chapter4}.  Traditionally, theoretical studies of the dynamics of atoms in an optical lattice are done using mean field approaches \cite{OSS01} and quantum Monte Carlo methods \cite{BRS02, KPS02, WAT04}.  Use of quantum walk can serve as an alternate method.  In particular, we will consider the quantum phase transition from the Mott insulator (MI)  to the superfluid (SF) state \cite{JBC98, GME02} and vice versa. The simulation of the quantum phase transition using the quantum walk occurs quadratically faster in one dimension (1D) compared to varying the optical lattice depth and letting the atom-atom interaction follow the classical random walk behavior.  In our study,  we use the noise models discussed in Chapter \ref{Chapter3} to act as an enhanced toolbox to control the redistribution of atoms in an optical lattice \cite{CL08}. We will also discuss the experimental implementation of quantum walk on ultra cold atoms in an optical lattice \cite{Cha06}. 
\par
Entanglement in many body system has not only been a computational 
resource, it has also been used as a signature of quantum phase transition \cite{ON02, OAF02, ORO06}. With increase in dimension of the Hilbert space, the number of invariants and the measures of entanglement grows exponentially so that scaling becomes impractical.  In 2002, Meyer and Wallach proposed a global entanglement measure - a polynomial measure of multiparticle entanglement to address the scalability problem \cite{MW02}.  In Chapter \ref{Chapter5} we investigate the evolution of spatial entanglement, particle-number entanglement between regions of space in a many particle system subjected to  quantum walk evolution using Meyer-Wallach measure. We use the coin degree of freedom to demonstrate the dependency of entanglement with the coin parameter, number of particles in the system and number of steps of quantum walk.

%% file: Chapters/Chapter2.tex

\chapter{Quantum walks} 
\label{Chapter2}
\lhead{Chapter 2. \emph{Quantum walks}}

\section{Introduction}
Quantum walks are the quantum analog of the classical random walks \cite{Ria58, FH65, ADZ93}  developed using the aspects of quantum mechanics such as interference and superposition. Like their classical counterpart, the quantum walks are also widely studied in two forms: continuous-time quantum walk and discrete-time quantum walk. 
In the continuous-time quantum walk, one  can directly define the walk on the position space \cite{FG98}, whereas in  the discrete-time quantum walk, it is necessary to introduce a quantum coin operation to define the direction in which the particle has to move \cite{ABN01}. The results from the continuous-time quantum walk  and the discrete-time quantum walk  are often similar, but due to the coin degree of freedom, the discrete-time variant has been shown to be more powerful than the other  in  some context  \cite{AKR05}. To match the performance of the discrete-time quantum walk, the coin degree of freedom can be introduced in the continuous-time quantum walk  \cite{CG04b}. We will only consider the discrete-time quantum walk for our study.
\par
The main focus of this chapter is to present in detail the dynamics of the discrete-time quantum walk and demonstrate methods to control it. This chapter is organized as follows. For the completeness, we will first review the continuous-time quantum walk in Section \ref{ctqwm}. In Section \ref{dtqw}, we will review the discrete-time quantum walk before demonstrating the optimization and randomization of the dynamics using quantum coin operations from the SU(2) group.
In Section \ref{recurrQW}, we discuss the fractional recurrence nature of the quantum walk and conclude with a summary in Section \ref{summary2}.   
\section{Continuous-time quantum walk}
\label{ctqwm}
To define the continuous-time quantum walk, it is easier to first define the continuous-time classical random walk and quantize it by introducing quantum amplitudes in place of classical probabilities.  
\par
The continuous-time classical random walk takes place entirely in the {\it position} space. To illustrate, let us define continuous-time classical random walk on the position space $\mathcal  H_{p}$ spanned by a vertex set $V$ of a graph $G$ with edges set $E$,  $G=(V, E)$. A step of the random walk can be described by a adjacency matrix $A$ which transform the probability distribution over $V$, i.e.,
\bea
A_{j,k} =  \begin{cases}
1  &   ~~ (j,k) \in E \\
0  &   ~~ (j,k) \notin E  
\end{cases}
\eea
for every pair $j, k \in V$.  The other important matrix associated with the graph $G$ is the generator matrix ${\bf H}$ given by
\bea
{\bf H}_{j,k} =  \begin{cases}
d_j \gamma   &   ~~ j =  k \\
-\gamma  &   ~~ (j,k) \in E \\
0  &  ~~ {\rm otherwise}
\end{cases}, 
\eea
where $d_{j}$ is the degree of the vertex $j$ and $\gamma$ is the probability of transition between  neighboring nodes per unit time.
\par
If  $P_{j}(t)$ denotes the probability of being at vertex $j$ at time $t$ then the transition on graph $G$ is defined as the solution of differential equation
\be
\label{ctcrw}
\frac{d}{dt} P_{j}(t) = - \sum_{k \in V}  {\bf H}_{j,k} P_{k}(t).
\ee
The solution of the differential equation is given by
\be
P(t) = e^{-{\bf H}t} P(0).
\ee
By replacing the probabilities $P_{j}$ by quantum amplitudes $a_{j}(t) = \langle j | \psi(t) \rangle$ where $|j\rangle$ is spanned by the orthogonal basis of the position Hilbert space $\mathcal  H_{p}$ and introducing a factor of $i$ we obtain
\be
\label{ctqw}
i\frac{d}{dt} a_{j} (t) = \sum_{k \in V} {\bf H}_{j,k} a_{k}(t).
\ee
We can see that (\ref{ctqw}) is the Schr\"odinger equation
\be
i \frac{d}{dt} |\psi\rangle = {\bf H} |\psi \rangle.
\ee
Since generator matrix is an Hermitian operator, the normalization is preserved during the dynamics. 
The solution of the differential equation can be written in the form
\be
|\psi(t) \rangle = e^{-i{\bf H}t} |\psi(0)\rangle.
\ee
Therefore, the continuous-time quantum walk is of the form of Schr\"odinger equation, a non-relativistic quantum evolution.
\par
To implement the continuous-time quantum walk  on a line, the position Hilbert space $\mathcal H_{p}$ can be written as a state spanned by the basis states $|\psi_{j} \rangle$, where $j \in \mathbb{Z}$. The Hamiltonian ${\bf H}$ is defined such that,
\be
\label{ctqw1b}     {\bf H}|\psi_{j}\rangle     =     -\gamma |\psi_{j-1}\rangle     +
2 \gamma |\psi_{j}\rangle  - \gamma |\psi_{j+1}\rangle  
\ee 
and  is made  to evolve with time $t$ by applying the transformation
\be
\label{ctqw1a}
 U(t) =\exp(-i{\bf H}t).
 \ee
The Hamiltonian ${\bf H}$ of the process acts as the generator matrix which will transform the probability amplitude at the rate of $\gamma$ to the neighboring sites, where $\gamma$ is time-independent constant.

\section{Discrete-time quantum walk}
\label{dtqw}

We will first define the structure of the discrete-time classical random walk. The discrete-time classical random walk takes place on the position Hilbert space $\mathcal H_{p}$  with instruction from the coin operation.  A coin flip defines the direction in which the particle moves and a subsequent position shift operation moves the particle in position space. For a walk on a line,  a two sided coin with {\it head} and {\it tail} defines the movements to the {\it left} and {\it right} respectively. 
\par
The discrete-time quantum walk also has a very similar structure to that of its classical counterpart. The coin flip is replaced by the quantum coin operation which defines the superposition of direction in which the particle moves simultaneously. The quantum coin operation followed by the unitary shift operation is iterated without resorting to intermediate measurement to implement  a large number of steps.  During the walk on a line, interference between the left and the right propagating amplitude results in the quadratic growth of variance with the number of steps.
\par
The discrete-time quantum walk on a line is defined on a Hilbert space
\be
\mathcal  H=  \mathcal H_{c}  \otimes \mathcal H_{p},
\ee
where $\mathcal H_{c}$ is the {\it coin} Hilbert space and $\mathcal H_{p}$  
is the {\it position} Hilbert space. For a  discrete-time quantum walk  in one dimension, $\mathcal H_{c}$  is spanned  by the basis state (internal state) of the  particle $|0\rangle$ and  $|1\rangle$ and $\mathcal H_{p}$  is spanned by the basis state of the position $|\psi_{j}\rangle$, where $j  \in \mathbb{Z}$. 
To implement the discrete-time quantum walk on a particle at origin in state 
\be
\label{qw:ins}
|\Psi_{in}\rangle= \left ( \cos(\delta)|0\rangle + e^{i\eta}\sin(\delta)|1\rangle \right )\otimes |\psi_{0}\rangle,
\ee
the  quantum coin toss operation $B \in U(2)$ which in general can be written as  
\be 
\label{U2coin}
B_{\zeta, \alpha, \beta, \gamma} = e^{i \zeta} e^{i\alpha \sigma_{x}}e^{i\beta \sigma_{y}}e^{i\gamma \sigma_{z}}
\ee
is applied. Where $\sigma_{x}, \sigma_{y}$ and $\sigma_{z}$ are the Pauli spin operators. Parameters of the coin operations $\zeta, \alpha, \beta, \gamma$ can be varied to  get different superposition state of the particle. That is, quantum coin operation $B_{\zeta, \alpha, \beta, \gamma}$ is used to evolve the particle to superposition of its basis states such that it can serve as an instruction to simultaneously evolve the particle to the $left$ and $right$ of its initial position.
The quantum coin operation is followed by  the conditional unitary shift operation $S$ given by
\be
\label{eq:condshifta}
S = e^{-i(|0\rangle \langle 0|
- |1\rangle  \langle 1|)\otimes  Pl} = \left ( |0\rangle \langle
0|\otimes e^{-iPl} \right )+ \left ( |1\rangle \langle 1|\otimes e^{iPl} \right )  
\ee
where $P$ is  the  momentum operator, $l$ is the step length and $|0\rangle$ and $|1\rangle$ are the basis states of the particle. 
Therefore the operator $S$ which delocalizes the wave packet in different basis states $|0\rangle$ and $|1\rangle$ over the position $(j-1)$ and $(j+1)$ when when step length $l =1$ can also  be written as
\be
\label{eq:condshift}
S  =  |0\rangle  \langle 0|\otimes  \sum_{j  \in
\mathbb{Z}}|\psi_{j-1}\rangle  \langle \psi_{j} |+|1\rangle  \langle 1
|\otimes \sum_{j \in \mathbb{Z}} |\psi_{j+1}\rangle \langle \psi_{j}|.
\ee
The states in the new position is again evolved into the superposition of its basis state and the process of quantum coin toss operation $B_{\zeta, \alpha, \beta, \gamma}$ followed by the conditional unitary shift operation $S$,
\be
\label{dtqwev}
 W_{\zeta, \alpha, \beta, \gamma} =
S(B_{\zeta, \alpha, \beta, \gamma}  \otimes   {\mathbbm  1})
\ee
is iterated without resorting to intermediate
measurement, to realize a large number of steps of the  discrete-time quantum walk. The four variable parameters of  the quantum  coin, $\zeta, \alpha, \beta$ and $\gamma$ (\ref{U2coin}) can be varied  to  change and control the  probability  amplitude  distribution  in  the position space.  \par
Most widely studied form of the discrete-time quantum walk is the Hadamard walk, using Hadamard operation $H$ as quantum coin operation and the role of coin operation and initial state to control the probability amplitude distribution has been discussed in earlier studies \cite{ABN01, BCG04}. We will discuss the Hadamard walk in detail in Section \ref{hw}. In Section \ref{gen-qw}, we demonstrate that a three parameter SU(2) quantum coin operation is sufficient to describe the most general form of the discrete-time quantum walk. Before that, we will further analyze the structure of the discrete-time quantum walk.
\par
\vskip 1.2cm
\subsection{Discrete-time quantum walk and Klein-Gordon equation} 
\par
The standard symmetric discrete-time classical random walk leads to
\be
\label{crwDE}
P(j, t+1) = \frac{1}{2}\left [ P(j-1,t) + P(j+1,t) \right ],
\ee
when unit time is required each step of classical random walk, $P(j, t)$ denotes the probability of finding the particle at position $j$ at discrete time $t$. Subtracting $P(j, t)$ 
from both sides of (\ref{crwDE}) leads to the difference equation which corresponds to differential equation
\be
\label{eq:cla-dif}
\frac{\partial}{\partial t} P(j, t) = \frac{\partial^2}{\partial j^{2}} P(j,t).
\ee
The above equation is irreversible because the coin is effectively
thrown away after each toss. It is also non-relativistic in the
sense that it is not symmetric in time and space, and leads to
a dispersion relation that is essentially non-relativistic \cite{Str07}. In the continuum limit, (\ref{eq:cla-dif}) leads to the standard classical diffusion. On the contrary, in the discrete-time quantum walk the information of the state of the quantum coin in the previous step is retained and carried over to the next step. This makes the quantum walk reversible. 
\par
To illustrate this, we consider the wavefunction describing the position of a particle and analyze how it evolves with time $t$. Let $t$ be the time required to implement $t$ steps of quantum walk. The two component vector of amplitudes of the particle, being at position $j$, at time $t$, with left (L) and right (R) moving component is given by 
\begin{eqnarray}
\label{compa}
\Psi(j, t) = \left (  \begin{array}{cl} \Psi_L(j,t)  \\
\Psi_R(j,t)  \end{array} \right ). 
\end{eqnarray}
The dynamics for $\Psi$ driven by single parameter quantum coin operation is :
\be
B_{0, \theta, \frac{\pi}{2}} = \left( \begin{array}{ccc}
\mbox{~~~}\cos(\theta) &~& -i\sin(\theta) \\
-i\sin(\theta) &~& \mbox{~~~}\cos(\theta) 
\end{array}\right)
\ee
and followed by the conditional shift operator $S$, $[S(B \otimes {\mathbbm 1})]$ in terms of left (L) moving and right (R) moving component at a given position $j$ and time $t +1$ is given by
\begin{eqnarray}
\label{eq:compa}
\left (  \begin{array}{cl} \Psi_L(j, t +1)  \\
\Psi_R(j, t +1)  \end{array} \right ) = 
\left( \begin{array}{ccc}
\mbox{~~~}\cos(\theta)a &~& -i\sin(\theta)a^{\dagger} \\
-i\sin(\theta)a^{\dagger} &~& \mbox{~~~}\cos(\theta) a
\end{array}\right) \left ( \begin{array}{cl}  \Psi_L(j,t)  \\
\Psi_R(j, t)  \end{array}\right ), 
\end{eqnarray}
where action of operator $a$ and $a^{\dagger}$  on $\Psi(j, t)$ is given by
\begin{eqnarray}
\label{eq:compa1}
a \Psi(j, t) = \Psi(j+1, t) \nonumber \\ 
a^{\dagger} \Psi(j, t) = \Psi(j-1, t).
\end{eqnarray}
Therefore,
\begin{subequations}
\begin{eqnarray}
\label{eq:comp}
\Psi_L(j,t+1) &=& \cos(\theta)\Psi_L(j+1,t) - 
	i \sin(\theta)\Psi_R(j-1,t)  \\
\label{eq:compb}
\Psi_R(j,t+1) &=& \cos(\theta)\Psi_R(j-1,t) - 
	i \sin(\theta)\Psi_L(j+1,t).
\end{eqnarray}
\end{subequations}
We thus find that the coin degree of freedom is carried over during the dynamics of the discrete-time quantum walk making it reversible. 
\par
Further, the components $\Psi_L$ and $\Psi_R$ are decoupled to obtain
\begin{equation}
\Psi_R(j, t +1) + \Psi_R(j, t -1) = 
	\cos(\theta)[\Psi_R(j+1,t) + \Psi_R(j-1,t)].
\label{eq:dec}
\end{equation}
Subtracting $2\Psi_R(j,t)$ from both sides, we obtain a difference equation which corresponds to 
differential equation
\begin{equation}
\label{eq:k-g}
\left [ \cos(\theta) \frac{\partial^2}{\partial j^2} -  \frac{\partial^2}{\partial  t^2} \right ]\Psi_R(j,t) 
= 2(1-\cos(\theta))\Psi_R(j,t),
\end{equation}
and similar expression can be obtained for $\Psi_L (j, t)$, see Appendix \ref{Appendix0} for intermediate steps. This shows that each component
follows a Klein-Gordon equation of the form
\begin{equation}
\left(\nabla^{2} - \frac{1}{c^{2}} \frac{\partial^{2}}{\partial t ^2}\right ) \phi - \mu^{2}\phi = 0
\end{equation}
showing the essentially relativistic character of the discrete-time quantum walk. Equations (\ref{eq:comp}) and (\ref{eq:compb}) can also be written as a compact first-order
difference equations, which in the continuum limit would be
analogous to the one-dimensional Dirac equation for a spinor
(of spin $1/2$).
\par
Setting the time-step and spatial-step to 1, we obtain from (\ref{eq:k-g}), the equivalent of light speed $c$ and mass $m$ in the discrete-time quantum walk dynamics : 
\begin{eqnarray}
c &\equiv& \sqrt{\cos(\theta)} \nonumber \\
\mu= \frac{mc}{\hbar}&\equiv& \sqrt{2[\sec (\theta)-1]}.
\end{eqnarray}
Considering $\hbar =1$, we can write 
\begin{eqnarray}
 m = \sqrt {\frac{2[\sec (\theta) -1]}{\cos (\theta)}}~.
\end{eqnarray}
Thus, from Section \ref{ctqwm} and the above analysis we conclude that a continuous-time quantum walk takes the form of Schr\"odinger equation and a discrete-time quantum walk takes the form of the relativistic quantum equation of a free spin-0 particle. 

\subsection{Hadamard  walk}
\label{hw}

\begin{figure}[h]
\begin{center}
\epsfig{figure=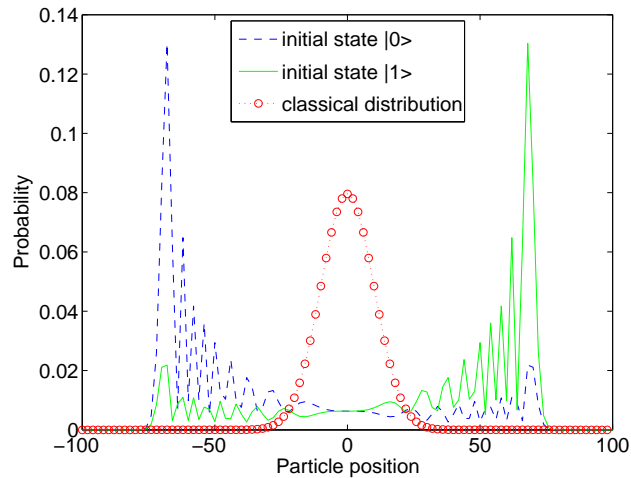, width=9.0cm}
\caption[Spread of probability distribution of the Hadamard walk with the initial state $|0\rangle$ and initial state $|1\rangle$ on the position space along with distribution of the classical random walk. ]{Spread of probability distribution of the Hadamard walk with the initial state $|0\rangle$ (dashed line)  and initial state $|1\rangle$ (solid line) on the position space along with distribution of the classical random walk (dashed-circle). The distribution is for 100 steps.}
\label{fig:qw1}
\end{center}
\end{figure}

The simplest version of the discrete-time quantum walk is the walk using Hadamard operation as quantum coin operation and is known as the Hadamard walk \cite{ABN01}. A particle at origin  in one of the basis state $|0\rangle$ or $|1\rangle$ of $\mathcal H_{c}$ (internal state of the particle) is evolved into the superposition of $\mathcal H_{c}$ with equal probability, by applying the Hadamard operation
\be
H = \frac{1}{\sqrt 2} 
\left( \begin{array}{clcr}
 1  & &   ~1   \\
1  & &  -1 
\end{array} \right),
\ee
such that
\begin{subequations}
\begin{eqnarray}
\label{eq:shift}
(H\otimes \mathbbm{1})  (|0\rangle\otimes|\psi_{0}\rangle) =\frac{1}
{\sqrt 2}\left ( |0\rangle+|1\rangle \right ) \otimes|\psi_{0}\rangle  \\
(H\otimes \mathbbm{1}) (|1\rangle\otimes|\psi_{0}\rangle) =\frac{1}
{\sqrt 2}\left (|0\rangle - |1\rangle \right ) \otimes|\psi_{0}\rangle.
\end{eqnarray}
\end{subequations}
The operation $H$ is then followed by the conditional shift operation $S$ (\ref{eq:condshift}), $W = S(H\otimes \mathbbm{1})$. The process is iterated without resorting to intermediate measurements to evolve the particle in superposition of position space and realize a  large number of steps  of the Hadamard walk. After the first  two iterations of $W$, the left and right evolving components of the amplitude begin to interfere, deviating from the classical evolution, resulting in a quadratic speedup in the growth of the variance. The probability  amplitude distribution arising from the iterated  application  of  $W$  is significantly  different  from  the probability distribution of the classical random walk, Figure \ref{fig:qw1}. The particle initially in the state $|0\rangle$ drifts to the left and the particle with an initial state $|1\rangle$ drifts to the right. This asymmetry arises from the fact that the Hadamard operation treats the two states $|0\rangle$ and $|1\rangle$ differently,  phase difference of $-1$ in case of state $|1\rangle$.  This phase difference, depending on the initial state of the particle contributes to the constructive interference on one side and to the destructive interference on the other side of the position space.
\par
To obtain left-right symmetry in the probability distribution, 
Figure \ref{fig:qw1a}(b), one needs to start the walk with the particle initially in a symmetric superposition state
\be
\label{inista}
|\Psi_{ins}\rangle \equiv \frac{1}{\sqrt 2}\left ( |0\rangle + i |1\rangle \right ) \otimes |\psi_{0}\rangle.
\ee 
For Hadamard walk on the line it is shown  that after $t$ steps, the  probability distribution is spread over the interval $[\frac{-t}{\sqrt  2 }, \frac{t}{\sqrt 2 }]$ and reduce quickly outside this  region  \cite{ABN01}. The  moments have been calculated for asymptotically large number of steps $t$ and the variance is shown to vary as 
\be
\sigma^{2}(t)    =    \left( 1-\frac{1}{\sqrt 2} \right ) t^2.
\ee
Hadamard walk has been extensively studied and the symmetric  distribution  depends on  the initial state of  the particle (see \cite{ABN01, NV01, TFM03, KNS04}). 
\par
In the following section we will present a quantum walk using an operation from SU(2) group as a generalized quantum coin.  The parameters in the SU(2) group will give control over the dynamics and amplitude distribution of the quantum walk. This eliminates the role of initial state of the particle to obtain a symmetric amplitude distribution. 

\subsection{Optimization using SU(2) coin}
\label{gen-qw}

In the introduction to this section we defined a discrete-time quantum walk using $B \in$U(2) (\ref{U2coin}) as quantum coin operation. To control and optimize the dynamics of the discrete-time quantum walk we consider the three parameter operator from the group SU(2) rather than the complete U(2) operator.  Apart from the global phase which does not affect the amplitude distribution, the effect of U(2) operator on the quantum walk can be completely reproduced using SU(2) operator. 
\par
We consider the group SU(2) which has for its elements, matrices of the form
\be
\label{2paraSU2}
B= \left( \begin{array}{clcr}  \mbox{~~}\alpha\mbox{~~}  & &   \beta   \\
-\beta^{*}  & &  \alpha^{*}
\end{array} \right) = \left( \begin{array}{clcr}  \mbox{~~}\alpha_{1} + i\alpha_{2}  & & \beta_{1} + i \beta_{2} \\
-\beta_{1}+i\beta_{2} & &  \alpha_{1}-i\alpha_{2}
\end{array} \right) , \mbox{~~~~}  \alpha_{1}^{2}+\alpha_{2}^{2}+\beta_{1}^{2}+\beta_{2}^{2} =1.
\ee
as the generalized quantum coin operation. $\alpha$'s and $\beta$'s are real values. For our present purpose the parametrization of SU(2) in terms of Euler angles turns out to be the convenient  one,
\be
\label{3paraU2}
B_{\xi,\theta,\zeta} \equiv \left( \begin{array}{clcr}  \mbox{~}e^{i\xi}\cos(\theta)  & &   e^{i\zeta}\sin(\theta)   \\
-e^{-i\zeta} \sin(\theta)  & &  e^{-i\xi}\cos(\theta)
\end{array} \right).
\ee
We can get the Hadamard operator up to global phase by choosing $\zeta = \xi = \pi/2$, $\theta =\pi/4$ or by choosing $\zeta = \xi = 0$, $\theta =\pi/4$ and multiplying the operator by $\sigma_{z}$ from left.
\par
\begin{figure}[h]
\begin{center}
\epsfig{figure=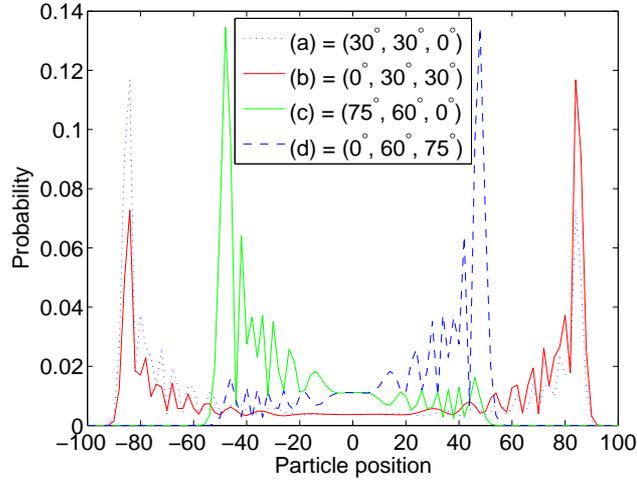, width=9.0cm}
\caption[Spread of probability distribution for different values of $\xi$, $\theta$, $\zeta$ using quantum coin operator $B_{\xi, \theta, \zeta}$. Parameter $\xi$ shifts the distribution to the left: (a) = $( \frac{\pi}{6}, \frac{\pi}{6}, 0)$  and (c) = $(\frac{5 \pi}{12}, \frac{\pi}{3}, 0 )$. Parameter $\zeta$  shifts it to the right: (b) = $(0, \frac{ \pi}{6}, \frac{\pi}{6})$ and  (d) = $(0, \frac{\pi}{3},  \frac{5 \pi}{12})$.]{Spread of probability distribution for different values of $\xi$, $\theta$, $\zeta$ using quantum coin operator $B_{\xi, \theta, \zeta}$. Parameter $\xi$ shifts the distribution to the left: (a) = $( \frac{\pi}{6}, \frac{\pi}{6}, 0)$  and (c) = $(\frac{5 \pi}{12}, \frac{\pi}{3}, 0 )$. Parameter $\zeta$  shifts it to the right: (b) = $(0, \frac{ \pi}{6}, \frac{\pi}{6})$ and  (d) = $(0, \frac{\pi}{3},  \frac{5 \pi}{12})$. The initial state of the particle $|\Psi_{ins}\rangle = \frac{1}{\sqrt{2}}(|0\rangle + i |1\rangle) \otimes |\psi_{0}\rangle$ and the distribution is for 100 steps.}
\label{fig:qw2}
\end{center}
\end{figure}

For the analysis of generalized discrete-time quantum walk we consider the initial state of a particle to be in the symmetric superposition of basis states $|\Psi_{ins}\rangle$ (\ref{inista}). 
Implementing $W_{\xi, \theta,  \zeta} = S(B_{\xi, \theta,  \zeta} \otimes {\mathbbm
1})$ on $|\Psi_{ins}\rangle$ 
evolves the particle to, 
\begin{eqnarray}
\label{eq:condshift2}
W_{\xi, \theta, \zeta}|\Psi_{ins}\rangle =  \frac{1}{\sqrt 2}
 [ \left(e^{i\xi}  \cos(\theta)+ i e^{i\zeta} \sin(\theta)\right )
|0\rangle\otimes |\psi_{-1}\rangle  \nonumber \\
+ \left( -e^{-i\zeta}\sin (\theta) + i e^{-i\xi} 
\cos(\theta)\right) |1\rangle\otimes |\psi_{+1}\rangle ]. 
\end{eqnarray}
The position probability distribution in (\ref{eq:condshift2}) after first step corresponding to
the left and right positions are $\frac{1}{2}[1 \pm \sin(2\theta)\sin(\xi - \zeta)]$.
 These probability distribution would be equal and lead to a left-right symmetry in position, if and only if $\xi = \zeta$. That is, the parameters $\xi \neq \zeta$ introduce asymmetry in the position space probability distribution from the first step itself. The effect of different values of coin parameters after 100 steps of walk is shown in Figure \ref{fig:qw2} plotted using numerically obtained values. We thus find that the generalized operator $B_{\xi, \theta, \zeta}$ as a quantum coin can bias the probability distribution of the quantum walk in spite of the symmetry of initial state of the particle. This is not true in case of the Hadamard walk, where obtaining  symmetric  distribution  depends largely on  the initial state of  the particle \cite{ABN01, NV01, TFM03, KNS04}. This can be verified by substituting $\xi = \zeta = 0, \theta=\pi/4$ and multiplying the components of state $|1\rangle$ in (\ref{eq:condshift2}) by $-1$.
\par
The walk beyond the first step involves interference and hence, it is appropriate to analyze the evolution after $t$ steps using $B_{\xi, \theta,  \zeta}$ as coin operator. The analysis shows how non-vanishing $\xi$ and $\zeta$ introduces bias.  Therefore, the state after $t$ steps can be written as 
\begin{equation}
[W_{\xi, \theta,  \zeta}]^t |\Psi_{ins}\rangle =
|\Psi(t)\rangle = \sum_{m=-t}^t 
(\mathcal{A}_{m,t}|0\rangle|\psi_m\rangle +\mathcal{B}_{m,t}|1\rangle|\psi_m\rangle)
\label{eq:lr}
\end{equation}
and proceeds according to the iterative relations,
\begin{subequations}
\label{eq:iter}
\begin{eqnarray}
\mathcal{A}_{m,t} = e^{i\xi}\cos(\theta) \mathcal{A}_{m+1,t-1} +
e^{i\zeta}\sin(\theta) \mathcal{B}_{m+1,t-1} \\
\mathcal{B}_{m,t} =  e^{-i\xi}\cos(\theta) \mathcal{B}_{m-1,t-1} - e^{-i\zeta}\sin(\theta) \mathcal{A}_{m-1,t-1}.
\end{eqnarray}
\end{subequations}
After $(t+1)^{\mbox{th}}$ step, $\mathcal{A}_{m,t}$ and $\mathcal{B}_{m,t}$  shifts to position $(m-1)$ and $(m+1)$ respectively.
A little algebra reveals that the solutions to $\mathcal{A}_{m,n}$ and $\mathcal{B}_{m,n}$ (\ref{eq:iter}) can be decoupled and shown to satisfy the following equations :
\begin{subequations}
\label{eq:iter0}
\begin{eqnarray}
\mathcal{A}_{m,t}  =  \cos(\theta)\left ( e^{i\xi} \mathcal{A}_{m+1,t-1} + e^{-i\xi} \mathcal{A}_{m-1,t-1} \right ) - \mathcal{A}_{m, t-2}\\
\mathcal{B}_{m,t} = 
\cos(\theta)\left ( e^{i\xi}\mathcal{B}_{m+1,t-1} + e^{-i\xi}\mathcal{B}_{m-1,t-1} \right ) - \mathcal{B}_{m,t-2}.
\end{eqnarray}
\end{subequations}
The decoupled expression (\ref{eq:iter0}) shows dependence of amplitude on the values from previous two steps. By repeatedly substituting the dependence relation for amplitudes on the right hand side of (\ref{eq:iter0}) until amplitude term at $t=0$ appears, reveals the dependency of 
$\mathcal{A}_{m,t}$ and $\mathcal{B}_{m,t}$ on amplitudes of the initial state and coin operation (\ref{eq:condshift2}).
We know that to obtain a spatial symmetry of the probability distribution from a particle initially in symmetric superposition state, the walk should  be invariant under an exchange of $|0\rangle \leftrightarrow |1\rangle$, and hence should evolve $\mathcal{A}_{m,t}$ and $\mathcal{B}_{m,t}$ alike (as in the Hadamard walk \cite{KRS03}).  
From the above analysis we see that $\mathcal{A}_{m,t}$ and $\mathcal{B}_{m,t}$ are symmetric to each other and evolve alike for all value of $\theta$ only when $\xi = \zeta$. Positive $\zeta$  contributes to constructive  interference towards right and destructive interference  to the left, and vice versa in case of $\xi$, as shown in Figure \ref{fig:qw2}. The inverse effect can be noticed when the $\xi$ and $\zeta$ are negative. 
\par
One can show that the coin operator 
\be
B_{\xi, \theta,  \zeta} = \left( \begin{array}{clcr}  e^{i\zeta}  & &   0   \\
0  & &  e^{-i\zeta}
\end{array} \right)B_{\xi - \zeta, \theta, 0} =
\left( \begin{array}{clcr}  e^{i \xi}  & &  0 \\
0  & &  e^{-i\xi}
\end{array} \right) B_{0, \theta,  \zeta  - \xi}.
\ee  
Setting $\xi \neq \zeta$ introduce asymmetry, biasing the walk. Therefore, a quantum coin operation $B_{\xi, \theta, \zeta}$ when $\xi \neq \zeta$ can be called as a {\it biased quantum coin operation} on a particle initially in symmetric superposition state\footnote{It should be noted that for a particle  initially in one of its basis state $|0\rangle$ or $|1\rangle$, the coin operation $B_{0,\theta, 0}$ itself biases the walk, state $|0\rangle$ to the left and $|1\rangle$ to the right. When the initial state is a symmetric superposition of the basis state, the biasing on state $|0\rangle$ is compensated by biasing on state $|1\rangle$  leaving the effective walk distribution symmetric.}. In  Figure \ref{fig:qw2} we show the biasing  effect for 
$(\xi,  \theta, \zeta)  = (  0^\circ ,  60^\circ ,
75^\circ)$ and for $(75^\circ,  60^\circ , 0^\circ)$.   
\par
\begin{figure}[h]
\begin{center}
\epsfig{figure=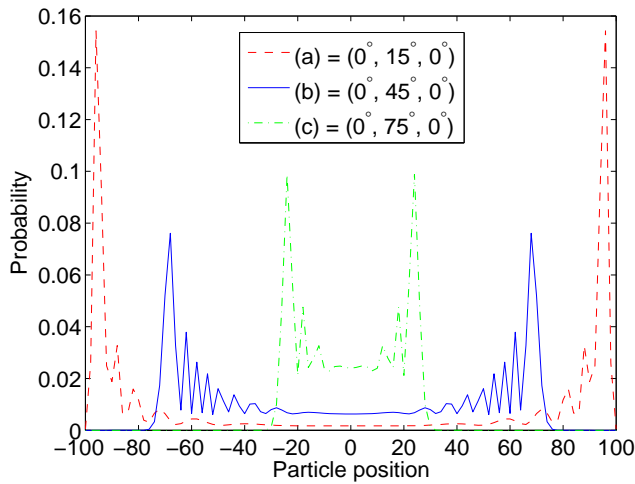, width=9.0cm}
\caption[Spread of probability distribution for different values of $\theta$ using the quantum coin operator $B_{0, \theta, 0}$. The distribution is wider for (a) = $(0,  \frac{\pi}{12},  0)$ than for (b) = $(0,  \frac{\pi}{4}, 0)$ and (c) = $(0, \frac{5 \pi}{12}, 0)$, showing the decrease in spread with increase in $\theta$.]{Spread of probability distribution for different values of $\theta$ using the quantum coin operator $B_{0, \theta, 0}$. The distribution is wider for (a) = $(0,  \frac{\pi}{12},  0)$ than for (b) = $(0,  \frac{\pi}{4}, 0)$ and (c) = $(0, \frac{5 \pi}{12}, 0)$, showing the decrease in spread with increase in $\theta$. The initial state of the particle $|\Psi_{ins}\rangle = \frac{1}{\sqrt 2}\left ( |0\rangle + i |1\rangle \right ) \otimes |\psi_{0}\rangle$ and the distribution is for 100 steps.}
\label{fig:qw1a}
\end{center}
\end{figure}

From (\ref{eq:iter0}) we know that only when $\xi = \zeta$ the quantum walk on a particle initially in symmetric superposition state evolve symmetrically in position space. From Section \ref{hw} we also know that the Hadamard walk on particle initially in basis state $|0\rangle$ biases the distribution to the left and in state $|1\rangle$ biases the distribution to the right. 
Therefore, for a particle initially in symmetric superposition state, by varying the parameter $\xi$ and $\zeta$ the results obtained for walk starting with one of the basis $|0\rangle$ or $|1\rangle$ (or any other nonsymmetric superposition) state can be reproduced. Similarly, a particle initially in one of its basis states $|0\rangle$ or $|1\rangle$ can also be evolved to obtain symmetric probability distribution by making an appropriate choice of the parameter $\xi$ and $\zeta$.
For example, quantum walk on a particle initially in state $|0\rangle$ with coin parameters $\theta$ being any value, $\xi =0$ and $\zeta = \pi/2$ and for a particle initially in state $|1\rangle$, $\xi = \pi/2$ and $\zeta = 0$ return a symmetric probability distribution in the position space (symmetric distribution similar for Figure \ref{fig:qw1a}).
Therefore quantum coin operation $B_{\xi, \theta, \zeta}$ and initial state of the particle complement each other.
The variance can be varied using the parameter $\theta$. Compared to the Hadamard walk or a walk using single variable parameter unitary coin \cite{NV01}, a walk using a SU(2) operator as a quantum coin gives better access to control the dynamics of discrete-time quantum walk irrespective of the initial state of particle.
\par
We note that using of three parameter operator U(2)$~\neq~$SU(2) of the form
\be
\label{U2}
B^{\prime}_{\xi, \theta, \zeta} \equiv \left( \begin{array}{clcr}  e^{i\xi}\cos(\theta)  & &   e^{i\zeta}\sin(\theta)   \\
e^{-i\zeta} \sin(\theta)  & &  -e^{-i\xi}\cos(\theta)
\end{array} \right)
\ee
in place of $B_{\xi, \theta, \zeta}$ will also produce the same effect on amplitude distribution during the quantum walk evolution. That is,
\be
S(B_{\xi, \theta, \zeta}\otimes {\mathbbm
1}) |\Psi_{ins}\rangle \equiv  S(B^{\prime}_{\xi, \theta, \zeta}\otimes {\mathbbm
1})|\Psi_{ins}\rangle
\ee 
when $|\Psi_{ins}\rangle$ is (\ref{inista}). The operator $B^{\prime}_{\xi, \theta, \zeta}$ reproduces the Hadamard operation for $\xi=\zeta = 0$ and $\theta = \pi/4$ (whereas $B_{\xi,\theta,\zeta}$ needs an additional Pauli $z$ operation, $\sigma_{z}$). For the convenience of reproducibility of properties of Hadamard walk, in this thesis we will switch between these two operators since it does not alter any of the results. We justify this claim as the existence of symmetries in quantum walk in Chapter \ref{Chapter3}.
\par
\par
\vskip 0.0cm
{\bf Approximating the variance:}
\par
\begin{figure}
\begin{center}
\epsfig{figure=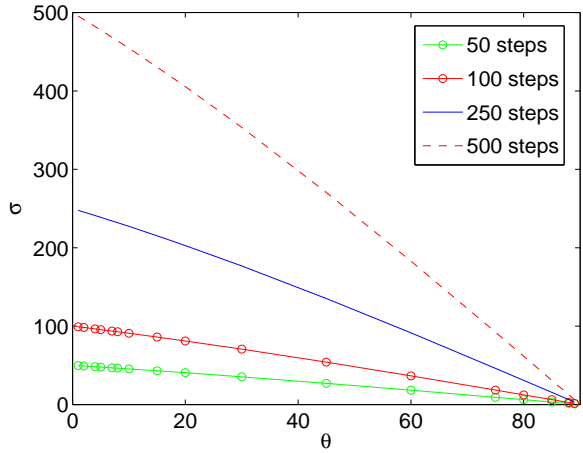, width=8.6cm}
\caption[A comparison of variation of $\sigma$  with  $\theta$ for different number of steps of walk using operator $B_{0,\theta, 0}$ using numerical integration.]{A comparison of variation of $\sigma$  with  $\theta$ for different number of steps of walk using operator $B_{0,\theta, 0}$ using numerical integration.}
\label{fig:qwsdtheta}
\end{center}
\end{figure}

The effect of parameter $\theta$ in coin operation $B_{\xi,\theta, \zeta}$ on the distribution  when $\xi=\zeta=0$ is shown in Figure \ref{fig:qw1a}, obtained by numerically evolving the density  matrix. The numerical results in Figure \ref{fig:qwsdtheta} shows decrease in variance with increase in the value of $\theta$ for different number of steps of quantum walk. 
\begin{figure}
\begin{center}
\epsfig{figure=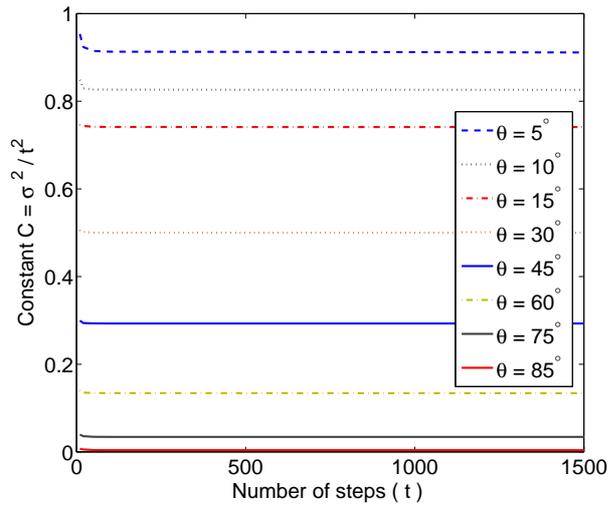,width=8.6cm}
\caption[Value of $C$ for a given $\theta$ with increase in number of steps.]{Value of $C$ for a given $\theta$ with increase in number of steps.  $C$ remains constant with increase in number of steps and decreases with increase in $\theta$.}
\label{fig:CwithTheta1a} 
\end{center}
\end{figure}
\par
\begin{figure}
\begin{center}
\epsfig{figure=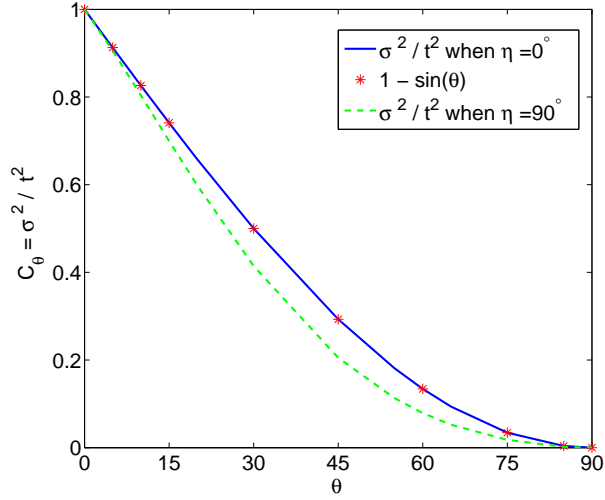,width=8.6cm}
\caption[Plot of the variation of $C_{\theta}$ when $\eta = \lvert \xi - \zeta \rvert = 0^\circ$ from numerical data and the function $(1-\sin(\theta))$ to which it fits. The effect of maximum biasing, $\eta = 90^\circ$ on $C_{\theta}$ is also shown and it is very small.]{Plot of the variation of $C_{\theta}$ when $\eta = \lvert \xi - \zeta \rvert = 0^\circ$ from numerical data and the function $(1-\sin(\theta))$ to which it fits. The effect of maximum biasing, $\eta = 90^\circ$ on $C_{\theta}$ is also shown and it is very small.}
\label{fig:CwithTheta}
\end{center}
\end{figure}

The change in variance for  different values of $\theta$ is attributed to change in the value of $C_{\theta}$, a constant for a given $\theta$, Figure \ref{fig:CwithTheta1a}. The dependence can be written in the form 
\be
\sigma^{2} = C_{\theta}t^{2}.
\ee
Therefore, starting from Hadamard walk ($\theta=\frac{\pi}{4}; \xi=\zeta=0$), the variance can be increased ($\theta<\frac{\pi}{4}$) or decreased ($\theta > \frac{\pi}{4}$) using $B_{\xi, \theta, \zeta}$. Figure \ref{fig:CwithTheta}  shows the dependence of $C$ on $\theta$. 
To analyze the effect of $\theta$ on variance further, it is instructive to first consider  the extreme values of parameters in $B_{\xi, \theta, \zeta}$. If $\xi=\theta=\zeta=0$, $B_{0, 0,0}=\sigma_{z}$, the Pauli $z$ operation, then $W_{\xi, \theta, \zeta} \equiv S$ and the two superposition states, $|0\rangle$ and $|1\rangle$, move away from  each other without any diffusion and interference having high $\sigma^{2}= t^{2}$. On the other hand, if $\theta=\frac{\pi}{2}$,  then $B_{0, \frac{\pi}{2}, 0}=\sigma_{x}$, the Pauli $x$ operation, and the two states $|0\rangle$ and $|1\rangle$ cross each other going back and forth, thereby remaining close to position $j=0$ and hence
giving very low $\sigma^{2} \approx 0$. These two extreme cases are not dominated by interference effect, but they define  the limits of the variance. Intermediate values of $\theta$ between these two extremes show intermediate drifts and quantum interference and after $t$ steps, the probability distribution is spread over the interval $(-t\cos(\theta), t\cos(\theta))$ \cite{NV01}. Probability distribution is almost $0$ beyond $|t\cos(\theta)|$ and for all practical purpose it can be considered to be equal to $0$. This is also verified by analyzing the distribution obtained using numerical integration technique. 
\par
\begin{figure}
\begin{center}
\epsfig{figure=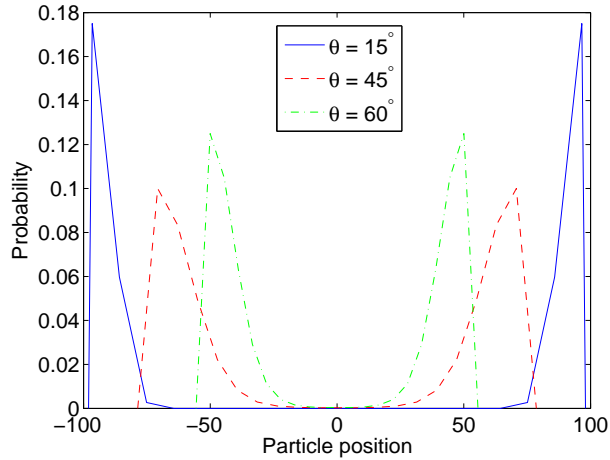, width=8.6cm}
\caption[The approximate probability distribution that envelop the numerically obtained probability distribution.]{The approximate probability distribution that envelop the numerically obtained probability distribution is obtained using (\ref{eq:proba1a}) for different value of $\theta$. The distribution is for 100 steps.}
\label{fig:qwfit}
\end{center}
\end{figure}

 
Numerically obtained data of variance for different $\theta$ overlaps with $(1-\sin(\theta))t^2$.
That is, the best fit function of $\theta$ for the numerical data can be written in the form
\be
\sigma^{2} = C_{\theta} t^{2} \approx [1-\sin(\theta)]t^{2}.
\ee
To arrive at an expression for variance as a function of $\theta$, a function of $\theta$ for the probability distribution of quantum walk can be used. A function of $\theta$ for probability distribution can be approximated by fitting a function that envelop the probability distribution obtained for different values of $\theta$ from the numerical integration technique, Figure \ref{fig:qwfit} \footnote{Arriving at $\theta$ dependent function from first principles, that is from the analytics of the dynamics of quantum walk is an ideal method. For this thesis, I have not considered that method due to constrains from the complexity of the dynamics of quantum walk.}. 
\par
One of the $\theta$ dependent function that envelop the probability distribution is,
\begin{eqnarray}
\int P(j) d j \approx \int_{-t\cos(\theta)}^{t\cos(\theta)} \frac{[1+\cos^{2}(2\theta)]e^{K(\theta)\left ( \frac{j^2}{t^2 \cos^2(\theta)} - 1 \right )}}{\sqrt t}  d j \approx 1,
\label{eq:proba1a}
\end{eqnarray}
\noindent where, $K(\theta) = \frac{\sqrt t}{2} \cos(\theta)[1+\cos^{2}(2\theta)] [1 + \sin(\theta)]$\footnote{Since $K(\theta)$ is an approximate function, other forms of $K(\theta)$ which closely fit the envelop can also be considered.}. The above probability distribution $P(j)$ as a function of $\theta$ was obtained by trying out different function of $\theta$ until a reasonable fit for the numerically obtained distribution for quantum walk was obtained. 
Figure \ref{fig:qwfit} shows the probability distribution obtained by using (\ref{eq:proba1a}).  
\par
The probability distribution for any $\theta$ is spread over the position $j$ in the interval $(-t\cos(\theta), t\cos(\theta))$. Therefore, for any $\theta$ the position $j$ can be parametrized by introducing a function of $\phi$, $f(\phi)$. That is, 
\be
j  \approx f(\phi) = t \cos(\theta)\sin(\phi)
\ee
where $\phi$ range from $-\frac{\pi}{2}$ to $\frac{\pi}{2}$. For a walk with coin $B_{0, \theta, 0}$, the mean of the distribution is zero and hence variance can be analytically obtained by evaluating
\begin{eqnarray}
\sigma^{2} = \int  P(j) j^2 d j \approx \int_{-t\cos(\theta)}^{t\cos(\theta)} P(j) j^2 d j = \int_{-\frac{\pi}{2}}^{\frac{\pi}{2}} P(f(\phi))(f(\phi))^2 f^{\prime}(\phi) d\phi.
\label{eq:varia}
\end{eqnarray}
Substituting appropriate values and simplifying we get 
\begin{eqnarray}
\sigma^{2} \approx \int_{-\frac{\pi}{2}}^{\frac{\pi}{2}} \frac{(1+\cos^{2}(2\theta))}{\sqrt t}{e^{K(\theta)\left (\sin^2(\phi)- 1 \right )}}\left( t\cos(\theta)\sin(\phi)\right )^2 \nonumber \\
\times \left(t\cos(\theta)\cos(\phi) \right) d \phi = t^2 (1-\sin(\theta)),
\label{eq:vari1}
\end{eqnarray}
That is,
\begin{eqnarray}
\sigma^{2} = C_{\theta}t^{2} \approx (1-\sin(\theta)) t^2.
\label{eq:vari2}
\end{eqnarray}
The function matches the result $C_{\theta} = (1  - \sin(\theta))$ obtained from the numerical data, as shown in Figure \ref{fig:CwithTheta}. Intermediate steps are presented in the Appendix \ref{AppendixA}. 
\par
We note that biasing the walk by setting $\xi \neq \zeta$ in $B_{\xi,\theta,\zeta}$ does
not alter the width of the distribution in the position space but the probability 
decreases as a function of $\cos(\eta)$ on one side and increases as a function  
of $\sin(\eta)$ on the other side, where $\eta = \lvert \xi-\zeta \rvert$. The mean value $\bar{j}$ of the distribution, which is zero for $B_{0, \theta, 0}$, attains finite value with non-vanishing $\eta$, that contributes for an additional term in (\ref{eq:varia}),
\begin{eqnarray}
\sigma^{2} \approx \int_{-t\cos(\theta)}^{t\cos(\theta)} P(j) (j-\bar{j})^2 d j. 
\label{eq:varimean}
\end{eqnarray}
The contribution of $\eta$ is a very small decrease in the variance of biased quantum walk, see Figure \ref{fig:CwithTheta}.

\subsubsection{Entropy  of measurement}
\label{entropy}
As  an alternative measure  of fluctuation in position, we consider the Shannon entropy of the quantum walk. Shannon entropy of the walk is obtained using position probability distribution  $P_j$ obtained  by tracing
over the coin basis,
\be
H(j)= -\sum_j P_j \log P_j.
\ee
\begin{figure}
\begin{center}
\epsfig{figure=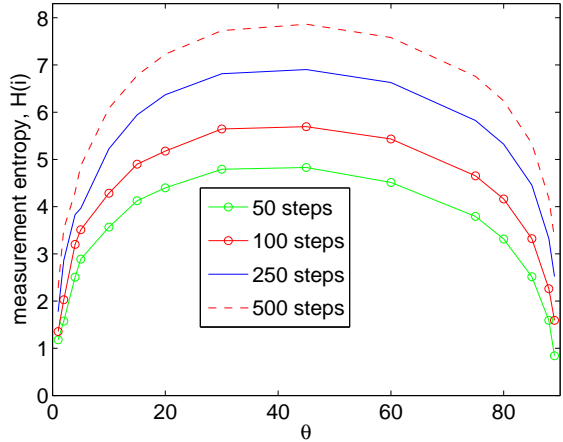, width=8.6cm}
\caption[Variation of entropy of measurement $H(j)$ with $\theta$ for different number of steps of quantum walk. The decrease in $H(j)$ is not drastic till $\theta$ is close to $0$ or $\frac{\pi}{2}$.]{Variation of entropy of measurement $H(j)$ with $\theta$ for different number of steps of quantum walk. The decrease in $H(j)$ is not drastic till $\theta$ is close to $0$ or $\frac{\pi}{2}$.}
\label{fig:qwEntropy}
\end{center}
\end{figure}
The quantum walk with a Hadamard coin toss, $B_{0,  \frac{\pi}{4}, 0}$, has maximum
uncertainty associated with the probability distribution and hence measurement  entropy  is  maximum.   For $\xi = \zeta =0$
and low  $\theta$, operator $B_{0,  \theta, 0}$ is almost  a Pauli $z$ operation$\sigma_{z}$, leading to localization of walker at $\pm t$.   
At $\theta$ close to  $\frac{\pi}{2}$, with $\xi=\zeta=0$, 
$S$  approaches the Pauli $x$ operation $\sigma_{x}$, leading to localization close to
the origin, and again, low entropy. However,
as $\theta$ approaches $\frac{\pi}{4}$, 
the splitting of amplitude in the position space increases
towards the maximum. The  resulting enhanced diffusion is reflected in
the relatively large entropy at $\frac{\pi}{4}$, 
as seen in Figure \ref{fig:qwEntropy}.  
Figure \ref{fig:qwEntropy}  is the measurement  entropy with variation
of  $\theta$ in the  coin $B_{0,  \theta, 0}$  for different  number of
steps of  quantum walk.  The decrease  in entropy from  the maximum by
changing $\theta$ on either side of $\frac{\pi}{4}$ is not drastic
until $\theta$  is close to $0$ or  $\frac{\pi}{2}$.  Therefore for
all practical purposes, small entropy can be  compensated for by
the relatively  large $C_{\theta}$, and hence  $\sigma^{2}$. The effect  of $\xi$ and $\zeta$ on  the  measurement  entropy   is  of  very  small  magnitude.  These parameters  do not  affect  the  spread of  distribution and variation in height reduces the entropy by a very small fraction. In other cases, such as mixing of quantum walk on an $n-$cycle briefly discuss in Section \ref{cycle}, it is ideal to use a  lower value of $\theta$. 
\subsubsection{Mixing time on an $n$-cycle}
\label{cycle}

The simplest finite Cayley graph is an $n$-cycle with $n$ vertices in closed path \cite{AAK01}. 
Due to closed structure of the $n-$cycle graph, with $n$ positions in  $\mathcal H_{p}$, 
the conditional shift operation $S$ (\ref{eq:condshift}) for a  discrete-time quantum walk on a line takes the form  
\begin{eqnarray}
\label{eq:condshift1}  S^{c}  =  |0\rangle  \langle 0|\otimes  \sum_{j=0}^{n-1}|\psi_{j-1~{\rm mod}~n}\rangle  \langle \psi_{j} | +|1\rangle  \langle 1
|\otimes \sum_{j=0}^{n-1} |\psi_{j+1~{\rm mod}~n}\rangle \langle \psi_{j}|.
\end{eqnarray}
The quantum state after $t$ steps of  discrete-time quantum walk  on particle initially in state $|\Psi_{in}\rangle$ (\ref{qw:ins}) is written as
\be
|\Psi_{t} \rangle = W_{\xi, \theta, \zeta}^t | \Psi_{in}\rangle= \sum_{j = 0}^{n-1} |\Psi_{j,t} \rangle,
\ee
where $|\Psi_{j,t}\rangle$ is the state at position $j$ after $t$ steps of quantum walk.
\begin{figure}[h]
\begin{center}
\epsfig{figure=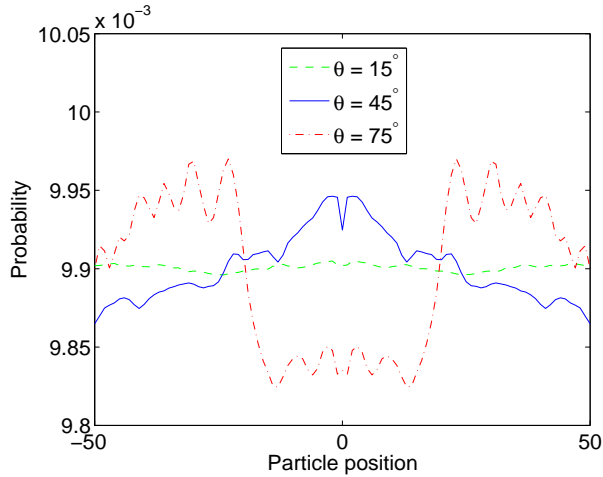, width=8.6cm}
\caption[A comparison of mixing time $M$ of the time-averaged probability distribution of a quantum walk on an $n$-cycle for different value of $\theta$ using coin operation $B_{0,\theta,0}$.]{A comparison of mixing time $M$ of the time-averaged probability distribution of a quantum walk on an $n$-cycle for different value of $\theta$ using coin operation $B_{0,\theta,0}$, where $n$, the number of position, is $101$. Mixing is faster for lower value of $\theta$. The distribution is for 200 cycles.} 
\label{fig:mixing}
\end{center}
\end{figure}

{\bf Mixing time}: Mixing time is the time it takes for the probability distribution on a graph (position space) to converge to a stationary distribution.
The  classical  random   walk  approaches  a  stationary  distribution independent of  its initial  state on a  finite graph. Unitary (i.e., non-noisy) quantum walk,
does not converge to any stationary distribution.
But by defining a  time-averaged probability distribution, 
\be
\overline{P(j,T)} =
\frac{1}{T}  \sum_{\tau=0}^{T-1}  P(j,  \tau), 
\ee
\noindent obtained  by  uniformly picking a random time $\tau$ between $0$ and $(T-1)$, evolving for $t$ time steps and  measure to find a particle at a given vertex (position $j$), a convergence in the  probability distribution can be  seen even in the quantum  case. It has been shown that for the quantum walk on an $n$-cycle, the mixing time is bounded above by $M = {\it O}(n \log n)$, almost quadratically faster than the classical case which is ${\it  O}(n^2)$ \cite{AAK01}. From previous section, we know that a quantum walk can be optimized for maximum variance and wide spread in position  space, between $\left (-t \cos(\theta),~ t \cos(\theta) \right )$ after $t$ steps. For a  walk on an $n$-cycle, choosing $\theta$  slightly above $0$
would give the maximum spread over $n$ vertices in the cycle for $n/2$ steps of the quantum walk. Maximum spread during each cycle distributes the probability over the cycle faster and this
would optimize the mixing time. 
For optimal mixing time, it turns out to be ideal to fix $\xi= \zeta$ in $B_{\xi, \theta,  \zeta}$, since  biasing  impairs a proper mixing.    Figure \ref{fig:mixing}  is the  time  averaged probability distribution of a  quantum walk on an $n$-cycle graph  after time $t= n \log n$, where $n$ is $101$. It can be seen that the variation of probability distribution over the position space is least for $\theta = 15^{\circ}$ compared to $\theta = 45^{\circ}$ and $\theta = 75^{\circ}$. 

\subsection{Randomizing coin operations from SU(2) group}
\label{rqw}
In Section \ref{hw} and \ref{gen-qw} we discussed quantum walk using Hadamard operation and SU(2) operation as quantum coins.  Though we used a generalized SU(2) coin $B_{\xi, \theta, \zeta}$ (\ref{3paraU2}), identical coin parameters were used during each step of the walk and we also saw the dependence of variance on the coin parameter $\theta$.
\par
\begin{figure}[h]
\begin{center}
\epsfig{figure=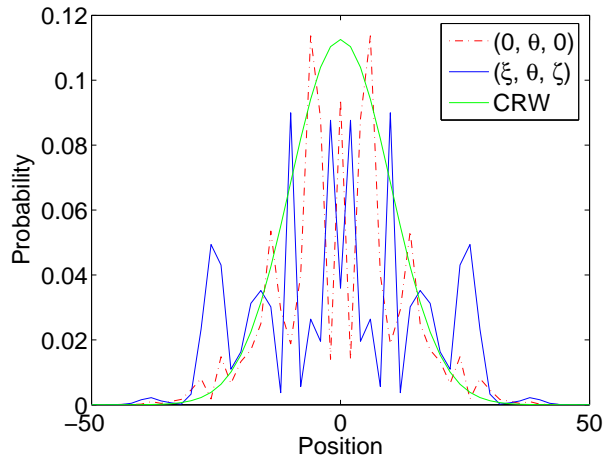, width=8.6cm}
\caption[Quantum walk using randomly picked coin operation for each step of walk  from SU(2) group with randomly picked $\xi, \theta, \zeta \in \{0, \pi/2 \}$ for $B_{\xi, \theta, \zeta}$ and $B_{0, \theta, 0}$ respectively.]{Quantum walk using randomly picked coin operation for each step of walk  from SU(2) group with randomly picked $\xi, \theta, \zeta \in \{0, \pi/2 \}$ for $B_{\xi, \theta, \zeta}$ and $B_{0, \theta, 0}$ respectively. We see that the variance of the distribution is very much close to the variance of the classical random walk (CRW) distribution.} 
\label{fig:randomqw} 
\end{center}
\end{figure}
\begin{figure}[h]
\begin{center}
\epsfig{figure=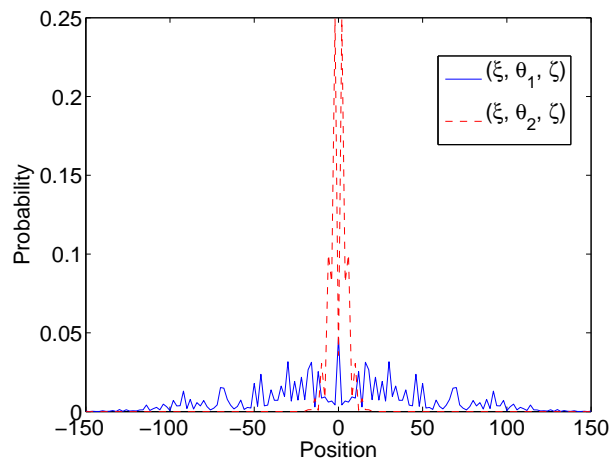, width=8.6cm}
\caption[Quantum walk using two different set of randomly picked parameters 
$\xi, \zeta \in \{0, \pi/2\},  \theta_{1} \in \{0, \pi/4\}$ and $\theta_{2} \in \{\pi/4, \pi/2\}$   for $B_{\xi, \theta_{2}, \zeta}$ and $B_{\xi, \theta_{2}, \zeta}$ respectively. ]{Quantum walk using two different set of randomly picked parameters 
$\xi, \zeta \in \{0, \pi/2\},  \theta_{1} \in \{0, \pi/4\}$ and $\theta_{2} \in \{\pi/4, \pi/2\}$   for $B_{\xi, \theta_{2}, \zeta}$ and $B_{\xi, \theta_{2}, \zeta}$ respectively.  The distribution is for 200 steps, walk using $\theta_{1} \in \{0, \pi/4\}$ spreads wider in position space, whereas the walk using $\theta_{2} \in \{\pi/4, \pi/2\}$  localizes the distribution.} 
\label{fig:randomqw1} 
\end{center}
\end{figure}

Evolution of quantum walk using randomized coin operation can be constructed by randomly chosen quantum coin operator for each step from a  set of operators in SU(2) group. That is,
\be
S(B_{\xi_{t}, \theta_{t}, \zeta_{t}}\otimes {\mathbbm
1})    \cdots S(B_{\xi_{x}, \theta_{x}, \zeta_{x}}\otimes {\mathbbm
1}) \cdots  S(B_{\xi_{0}, \theta_{0}, \zeta_{0}}\otimes {\mathbbm
1}) |\Psi_{in}\rangle
\ee
with randomly chosen parameters $\xi, \theta, \zeta \in \{0, \pi/2\}$ for each step. Though the coin parameters are randomly chosen for each step, the evolution is unitary and involves interference of amplitudes, and the effect is seen in probability distribution, see Figure \ref{fig:randomqw}. From the numerical evolution we also see that variance of the distribution is much closer to variance of the classical random walk distribution. However by restricting the range of the coin parameters that can be used for the walk, the probability distribution can be localized or made to spread wide in position space. One simple example we can consider is, by randomly picking different $\theta$ for each step from a subset of complete range of $\theta$, subset with $\theta_{1} \in \{0, \pi/4\}$ and $\theta_{2} \in \{\pi/4, \pi/2\}$ for $B_{\xi, \theta_{1}, \zeta}$ and $B_{\xi, \theta_{2}, \zeta}$ respectively, where other parameters are still picked from the complete range, $\xi, \zeta \in \{0, \pi/2\}$. Probability distribution obtained is shown in Figure \ref{fig:randomqw1}. Walk using $\theta_{1} \in \{0, \pi/4\}$ spreads wider in position space, whereas the walk using $\theta_{2} \in \{\pi/4, \pi/2\}$  localizes the distribution. 

\section{Recurrence of quantum walk} 
\label{recurrQW}

In the dynamics of physical system, from a single free particle to stellar dynamics, understanding of recurrence phenomenon have significantly contributed to a better understanding of the system dynamics \cite{Cha43}. For a classical conservative system, whether discrete or continuous in time, the Poincar\'e recurrence theorem states that any phase-space  configuration of a system enclosed in a finite volume will be repeated as accurately as one wishes after a finite interval of time (with no restriction on the interval) \cite{Bar06, Kre85}. A similar recurrence theorem is shown to hold in quantum theory as well \cite{BL57, Sch78}. 
In a system with a discrete energy eigenvalue spectrum $\{ E_{n} \}$; if $\Psi(t_0)$  is its state vector at the time $t_0$ and $\epsilon$ is any positive number, there exists a finite time $T$ such that,
\be
|\Psi(T)-\Psi(t_0)| < \epsilon.
\ee 
\par
For classical random walk we can consider the recurrence probability $P_{0}(t)$, i.e., the probability of  periodicity of dynamics that the particle returns to origin during the time evolution ($t$ steps). It is characterized by the P\'olya number
\be
{\mathcal P}_{crw} \equiv 1 - \frac{1}{\sum_{t=0}^\infty P_{0}(t)}.
\ee
If the P\'olya number equals one, the classical random walk is recurrent, otherwise the walk is transient, i.e., with nonzero probability the particle never returns to the origin. For a classical random walk to be transient the series $\sum_{t=0}^\infty P_{0}(t)$ must converge \cite{Rev90}. P\'olya proved that the one- and two- dimensional classical random walk are recurrent \cite{Pol21}, while in higher dimension for each dimension a unique P\'olya number is associated and the classical random walk is transient.
\par
In standard quantum mechanics, initially localized wave packet in state $|\Psi(t_0)\rangle$ which can spread significantly in a closed system can also reform later in the form of a quantum revival, i.e., the spreading reverses itself and the wave packet relocalizes \cite{Rob04}. The relocalized wave packet can again spread and the periodicity in the dynamics can be seen validating the quantum recurrence theorem. The time evolution of a state $|\Psi(t)\rangle$ over entire position $j$, or a wavefunction $\Psi(j,t)$ at a position $j$ after time $t$ is given by a deterministic unitary transformation associated with the Hamiltonian. During  quantum evolution we deal with amplitudes and probability density $P(j,t) = |\Psi(j,t)|^2$ at position $j$ after time $t$ appears only when we collapse the wave packet to perform measurement. 
\par
Unlike standard wave packet spreading in free space and harmonic potential\footnote{During standard wave packet spreading in free space and harmonic potential, a wave packet which is initially Gaussian, spreads retaining the Gaussian shape causing increase in the full-width at half maxima (FWHM).}, the quantum walk spreads the wave packet in multiple possible paths with amplitudes corresponding to different paths interfering. In this section we show that due to particle-position entanglement and interference effect during the evolution of the quantum walk, the wave packet delocalizes over the position space as a small copies of the initial wave packet.  These delocalized copies of fractional wave packet fails to satisfy the complete quantum recurrence theorem during quantum walk evolution.  That is for quantum walk, the recurrence we considered is return of the unit probability amplitude at the origin during the dynamics. However, due to revival of fractional wave packets, a fractional recurrence can be seen in the quantum walk. 
The probabilistic characterization of the recurrence in quantum walk has been done in \cite{ABN01} and the characterization using quantum P\'olya number defined in \cite{SJK08a} can be used to show the fractional recurrence nature of quantum walk. In Section \ref{qwrecurcycle}, we also show the exceptional cases of quantum walk that can be constructed by suppressing or minimizing the interference effect and get closer to  complete recurrence.

\subsection{Quantum recurrence theorem}
\label{qrecur}
Quantum recurrence theorem in the dynamics of a closed system states that there exist a time $T$  when 
\be
|\Psi(T) - \Psi(t_0) | <  \epsilon ,
\ee
where $\Psi(T) = |\Psi_T\rangle$ is the state of the system after time $T$,  $\Psi(t_0) = |\Psi_0\rangle$ is the initial state of the system and $\epsilon$ is any positive number \cite{BL57}.
\par
 The recurrence of the complete state of the system or exact revival happens when all the expectation values of observables $A_{o}$ of the two states $|\Psi_T\rangle$ and $|\Psi_0\rangle$ are equal to one another,  that is, 
\be
\langle \Psi_T| A_{o} | \Psi_T \rangle = \langle \Psi_0| A_{o} | \Psi_0 \rangle.
\ee
\par
In classical dynamics, the characterization of the nature of recurrence can be conveniently done using probabilistic measures. Measurements on a quantum system leaves the state in one of its basis states with certain probability. Therefore, recurrence in quantum systems can be analyzed using comparative evolution of the two identically prepared quantum system with the initial states $|\Psi_{0,0}\rangle$ at position $j =0$ and time $t = 0$. We will consider two cases of comparative analysis.
\par
{\it Case 1:} Consider two identically prepared particle wave packets which revive completely in the position space at time $T$. One of the two particle wave packets at position $j= 0$ and time $t = 0$ is first evolved to spread in position space and then reverse the spreading till it relocalizes completely at position $j=0$ at time $T$. The measurement performed on this particle will collapse the wave packet at the relocalized position with an expectation value
\be
 \langle \Psi_{0,T}| X_{p} | \Psi_{0,T}\rangle =  \langle \Psi_{0,0}| X_{p}| \Psi_{0,0} \rangle = 1,
 \ee 
where $X_{p}$ is the position operator. After the measurement at time $T$, the system is further evolved for an additional unit time and the corresponding state can be given by $|\Psi_{j, T_{M}+1}\rangle$. The subscripts $j$ and  $(T_{M}+1)$ stand for position and time $(T+1)$ with  intermediate measurement being performed at time $T$. The second  particle wave packet at position $j=0$ and time $t=0$ is evolved up to time $(T+1)$ directly without any measurement being performed at time $T$ and the state can be written as $|\Psi_{j, T+1}\rangle$. Since both the wave packets completely relocalize at position $j=0$ after the evolution for time period $T$, irrespective of the measurement being performed, expectation value for both the particles after time $(T+1)$ would be identical, 
\be 
\langle \Psi_{j, T_{M}+1} |X_{p}|\Psi_{j, T_{M}+1} \rangle = \langle \Psi_{j, T+1} |X_{p} |\Psi_{j, T+1} \rangle,
\label{rec}
\ee
with $j$ spanning over all position space.\\
\par
\noindent
{\it Case 2:} Consider two identically prepared particle wave packets which does not relocalize completely at position $j=0$ at time $T$, i.e., revive fractionally or does not revive at all.  
The measurement will collapse the wave packet and return the expectation value 
\be
\label{qwrev}
\langle \Psi_{0,T}| X_{p} | \Psi_{0,T} \rangle = \delta   ~~;  ~~0 < \delta < 1.
\ee
In this case the two identically prepared particle evolved to time $(T+1)$, one with a measurement being performed at time $T$ and an other without any measurement being performed. With very high probability the measurement will not return the same expectation value, that is 
 \be
 \label{revqw1}
 \langle \Psi_{j, T_{M}+1} |X_{p} |\Psi_{j, T_{M}+1} \rangle \neq \langle \Psi_{j, T+1} |X_{p} |\Psi_{j,T+1} ,\rangle
 \ee
with $j$  spanning over all position space. Non-zero values in both expectation values in the inequality can act as a signature of fractional recurrence of the quantum state at time $T$ and a zero expectation value on the left hand side shows transient dynamics.  
 \par
From the analysis of the above two cases we can conclude that, if a system is completely recurrent at time $T$ the two states  $|\Psi_{j, T_M+1}\rangle$ and $|\Psi_{j, T+1}\rangle$, one with intermediate measurement at time $T$ and other  without any measurement at time $T$ will be equal to one another. This will be helpful for us in understanding the fractional recurrence nature of the quantum walk.

\subsection{Fractional recurrence of quantum walk}
\label{qwrecur}
\subsubsection{On a line}
\label{qwrecurline}
 The state of the particle wave packet after implementing the discrete-time quantum walk of $t$ steps on a line with unit time required to implement each step can be written as 
\be
|\Psi_{t} \rangle = W_{\xi, \theta, \zeta}^t | \Psi_{0, 0}\rangle= \sum_{j} |\Psi_{j,t} \rangle.
\ee
Where $|\Psi_{t}\rangle$ is the state after time $t$ over entire position space, $|\Psi_{j,t} \rangle$ is the state of the delocalized wave packet at each position $j$ in $\mathcal H_{p}$ and $|\Psi_{0, 0}\rangle$ is the coin and position state of the wave packet before implementing the quantum walk. In the quantum walk process that involves a deterministic unitary evolution, the particle wave packet delocalize and spread over the position space forming a mini wave packet. During this delocalization process the mini wave packets interfere and entangle the position and coin Hilbert space, $\mathcal H_{p}$ and $\mathcal H_{c}$. The interference and the entanglement between the $\mathcal H_{p}$ and $\mathcal H_{c}$ during the standard quantum walk evolution does not permit complete relocalization of the wave packet at initial position after any given number of steps $t$.  Therefore the argument leading to (\ref{qwrev})  and (\ref{revqw1}) holds to show that the complete recurrence of the quantum state does not occur during the evolution of the quantum walk process on a line. That is,\\
\par
{\it In a discrete-time quantum walk evolution on a line and on an $n-$cycle dominated by the interference of quantum amplitude \footnote{By choosing extreme value of $\theta$ ($0$ or $\pi/2$) in the quantum coin operation, quantum walk on a line can be evolved without constructive or destructive interference taking place.}, there exists no time $T$ when the quantum state of the system revive completely and repeat the delocalization and revival at regular interval of time}. \\
 \begin{figure}
\begin{center}
\epsfig{figure=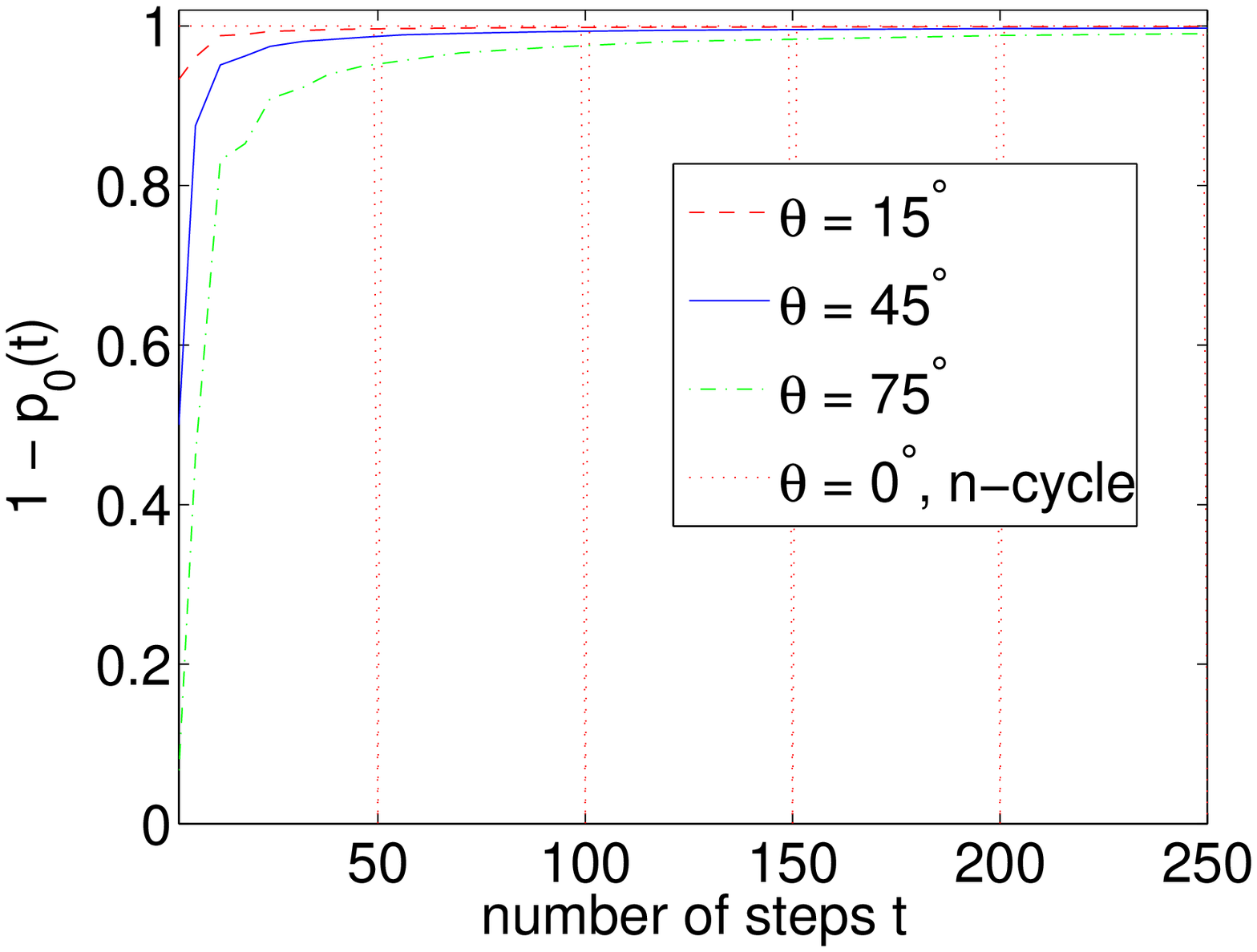, width=7.2cm}
\epsfig{figure=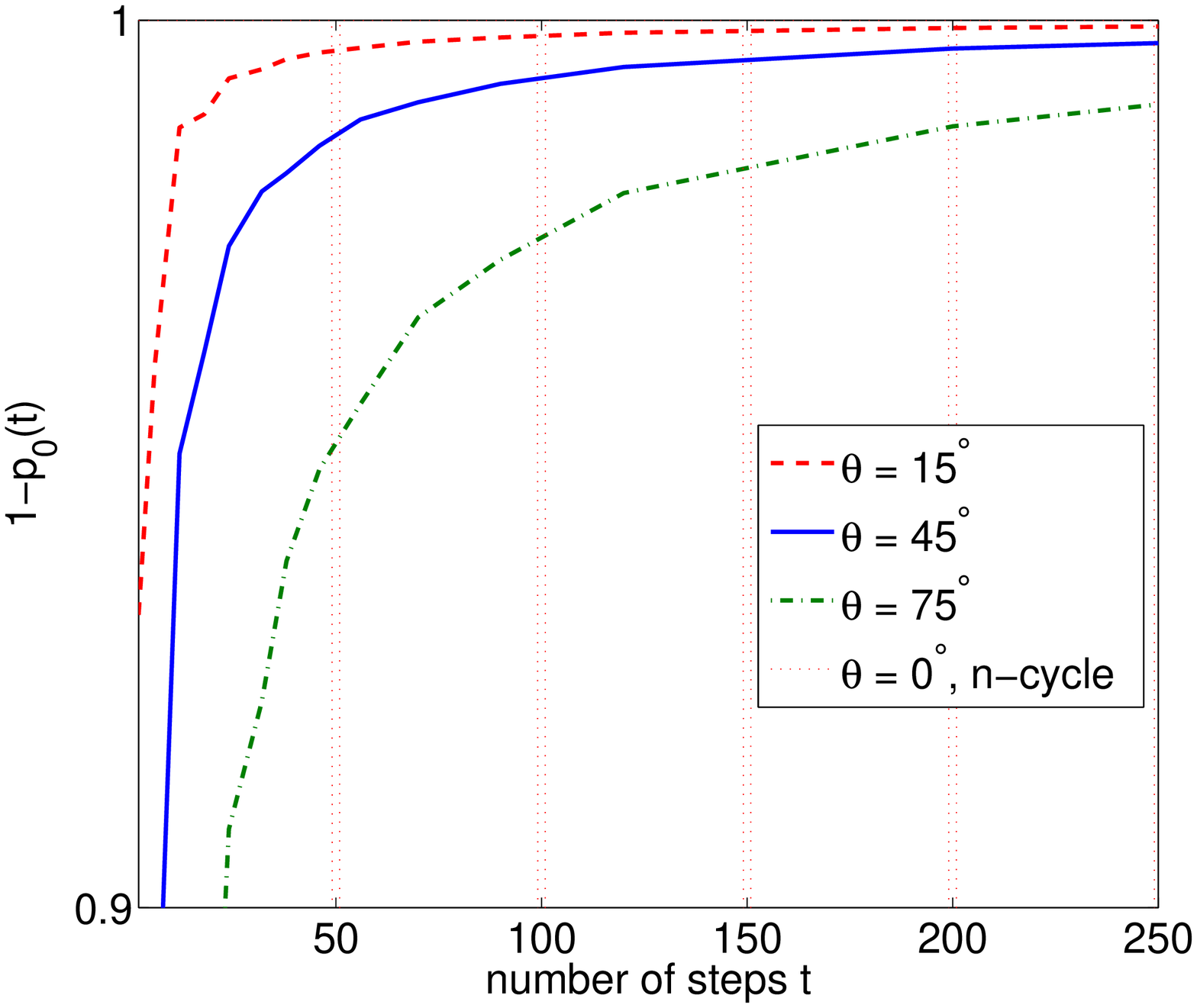, width=7.2cm, height=5.4cm}\\
~~~~~~(a)~~~~~~~~~~~~~~~~~~~~~~~~~~~~~~~~~~~~~~~~~~~~~~~~(b)
\caption[The plot of $(1-P_{0}(t))$,  where $P_{0}(t)$ is the probability of particle at the origin with $t$ being the number of steps in the discrete-time quantum walk evolution.  The plot for quantum walk on a line using different coin operation parameter $\theta = 15^{\circ}, 45^{\circ}$, and $75^{\circ}$ is shown.]{The plot of $[1-P_{0}(t)]$,  where $P_{0}(t)$ is the probability of particle at the origin with $t$ being the number of steps in the discrete-time quantum walk evolution.  The plot for quantum walk on a line using different coin operation parameter $\theta = 15^{\circ}, 45^{\circ}$, and $75^{\circ}$ is shown. With an increase in $\theta$, the quantum P\'olya number, which characterizes the fractional recurrence nature of the quantum walk, also increases. The plot with $\theta = 0^{\circ}$ is for a walk on an $n-$cycle with a completely suppressed interference effect, where $n=51$.  For a walk to be completely recurrent there should exist a $t = T$ where $[1-P_{0}(T)] = 0$. Any non-zero value can be used to characterize the fractional recurrence nature of the quantum walk and an absolute $1$ for all $t$ would show that the quantum walk is completely transient. (b) Close up of the plot.}
\label{polyaqw}
\end{center}
\end{figure}
\par
The above statement can also be quantified in the following way:  
After implementing a quantum walk, the wavefunction describing the particle at position $j$ and  time $t$ can be written as a two component vector of amplitudes of particle being at position $j$ at time $t$ with left and right propagating component
\be
\label{qstate}
|\Psi_{j,t}\rangle  =   \left(   \begin{array}{clr}   \Psi_{L} (j, t)   \\     \Psi_{R} (j, t)
\end{array} \right).
\ee
Lets analyze the dynamics of discrete-time quantum walk on wave packet $\Psi(j, t)$ driven by single parameter quantum coin 
\be
B_{\theta} = \left( \begin{array}{ccc}
\cos (\theta) &~~~& \sin(\theta) \\
\sin(\theta) &~~~& - \cos(\theta) 
\end{array}\right)
\ee
and shift operator $S$ ($W = S(B_{\theta}\otimes {\mathbbm
1})$).  In terms of left and right propagating component it is given by
\begin{eqnarray}
\label{eq:compa1}
\Psi_{L}(j, t+1) &=& \cos(\theta)\Psi_{R}(j-1,t) +
	 \sin(\theta)\Psi_{L}(j-1,t) \nonumber \\
\Psi_{R}(j, t+1) &=&  \sin(\theta)\Psi_{R}(j+1, t) -\cos(\theta)\Psi_{L}(j+1, t). 
	\end{eqnarray}
	
Then the probability of being at position $j$ and $t$ is
\be
\label{proba}
P(j,t) = |  \Psi_{L} (j, t)|^{2} + |\Psi_{R} (j, t)|^{2} 
\ee
and sum of probability over the entire position space is 
\be
\label{probsum}
\sum_{j} P(j, t) = 1.
\ee
After time $t$ with unit time required to implement each step of the quantum walk  on a line, the wavefunction will be spread between $j=-t$ to $+t$  and can be written over position space as 
\be
\sum_{j=-t}^{t} \Psi (j, t)  = \sum_{j=-t}^{t} \left [  \Psi_{L}(j, t) + \Psi_{R}(j, t) \right ].
\ee 
It should be noted that for even number of steps the amplitude at odd labeled positions is $0$ and for odd number of steps amplitude at even labeled position is $0$.  
\par
 For classical random walk, each step of walk is associated with the {\it randomness} and the {\it probability} of the entire particle therefore, however small the probability is at the origin, it is attributed to the recurrence  of the entire particle. That is, $0 < P_{crw}(0,T) < 1$ at some time $T$ shows the recurrence of complete particle at the origin with some finite probability.  
For the discrete-time quantum walk evolution described by (\ref{eq:compa1}) we found that the information of the coin degree of freedom is carried over to the later steps during the dynamics of the walk making it reversible (\ref{eq:iter0}). The {\it randomness} and the {\it probability} in quantum walk comes into consideration only when the wave packet is collapsed to discard the signature of the quantum coin degree of freedom from earlier steps. Therefore, for a discrete-time quantum walk to be completely recurrent,  the condition
\be
\label{prob1}
P(0,T) = |  \Psi_{L} (0, T)|^{2} + |\Psi_{R} (0, T)|^{2} = 1
\ee
has to be satisfied for some time $T$.  Any $P(0, t) < 1$ shows that the particle is present in superposition of position space (origin and other positions), which we will call as the fractional recurrence nature of the quantum walk. From (\ref{eq:compa1}) and (\ref{proba}), we can conclude that (\ref{prob1}) is satisfied only when $\theta = \pi/2$ and $t$ is even, that is when there is no interference of the quantum amplitudes.  For $\theta = 0$ the two left and right component move in opposite directions without returning and for any $0< \theta < \pi/2$ we get $P(0, t) < 1$ showing the fractional recurrence of the quantum walk.
\par
Alternatively, using quantum Fourier analysis to study the evolution of the discrete-time quantum walk on a line,  it is shown that the amplitude at the origin and to a very good approximation for any position $j$ between two dominating peaks (for example see Figure \ref{fig:qw1a}) decreases by  $O(1/\sqrt{t})$\cite{ABN01}. 
Therefore, 
\be
\label{qwexp}
\langle \Psi_{j,t} | \Psi_{j,t} \rangle =  O\left (\frac{1}{t} \right ) < 1.
\ee
and their exists no time $T = t$ where the walk is completely recurrent showing the fractional recurrence nature of the quantum walk.   
\par
A probability based characterization of the recurrence nature of the quantum walk, quantum  P\'olya number was defined for an ensemble of identically prepared quantum walk systems by the expression 
\be
\label{qpn}
{\mathcal P}_{qw} = 1 - \prod_{t=1}^{\infty} \left [ 1- P_{0}(t) \right ],
\ee
where $P_{0}(t)$ is the recurrence probability of the particle.
Each identically prepared particle is subjected to different number of steps of quantum walk from $1$ to $\infty$ and the probability of the particle at the origin is measured and discarded   \cite{SJK08a}.
The probability that the particle is found at the origin in a single series of such measurement records is the quantum P\'olya number. The quantum P\'olya number was calculated for various coined quantum walks and it is shown that in the higher dimension it depends both on the initial state and the parameters of the coin operator whereas, for the classical random walk the P\'olya number is uniquely determined by its dimensionality \cite{SJK08b}. 
\par
To show the quantum walk to be completely recurrent adopting a probability based characterization given by (\ref{qpn}) needs to have at least one of the many particles, each evolved to different steps $t$ from $1$ to $\infty$, to return $P_{0} (t)=1$.
 If $0< P_{0}(t) <1$ for any $t$, then only with certain probability $P_0(t)$ the wave packet  collapses at $j=0$, that is, prior to measurement the particle existed in superposition of position space. If  $0< P_{0}(t) <1$ for all the particles evolved to different steps from $1$ to $\infty$ then the fractional recurrence of the quantum walk is characterized by the ${\mathcal P}_{qw}$.  
\par
In Figure \ref{polyaqw} the plot of $[1-P_{0}(t)]$ is shown for a discrete-time quantum walk on a line where the different coin operation parameter $\theta = 15^{\circ}, 45^{\circ}, 75^{\circ}$.   With increase in $\theta$ the quantum P\'olya number, which can also be called as the {\it fractional recurrence number} ${\mathcal P}_{qw}$, also increases.

\subsubsection{On an $n-$cycle}
\label{qwrecurcycle}

We have discussed the dynamics of discrete-time quantum walk on an $n-$cycle and the mixing time using coin degree of freedom in Section \ref{cycle}. In this section we will discuss the recurrence nature of quantum walk on an $n-$cycle.
\par
As discussed earlier, in the case of classical random walk  walk on a line a non-zero probability is sufficient to show the recurrence of the entire particle and a uniform  distribution returns a non-zero probability at the origin. Whereas uniform distribution in the quantum case does not reveal the complete recurrence nature of the particle as it does in the classical case. 
\par
\begin{figure}
\begin{center}
\epsfig{figure=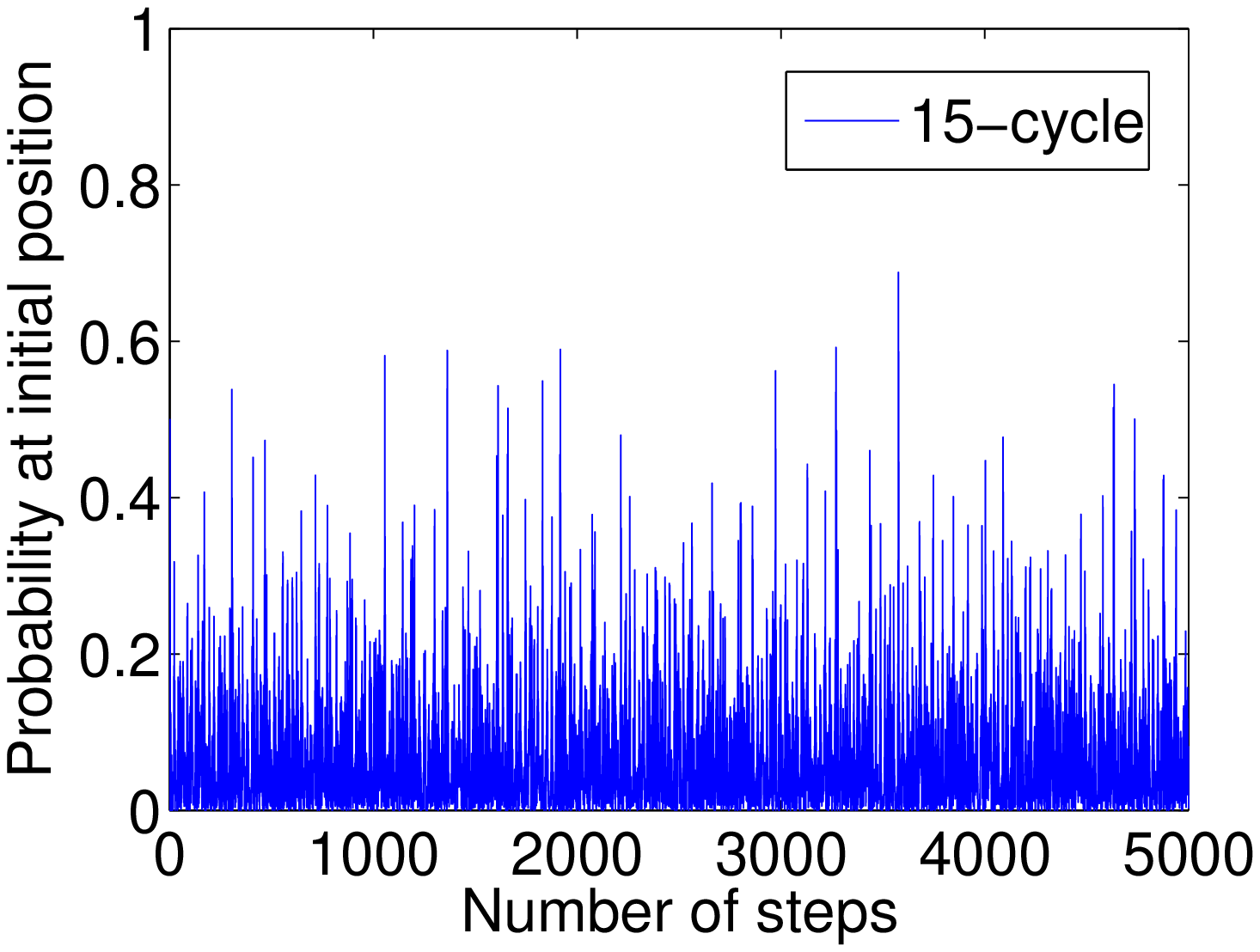, width=7.2cm}
\epsfig{figure=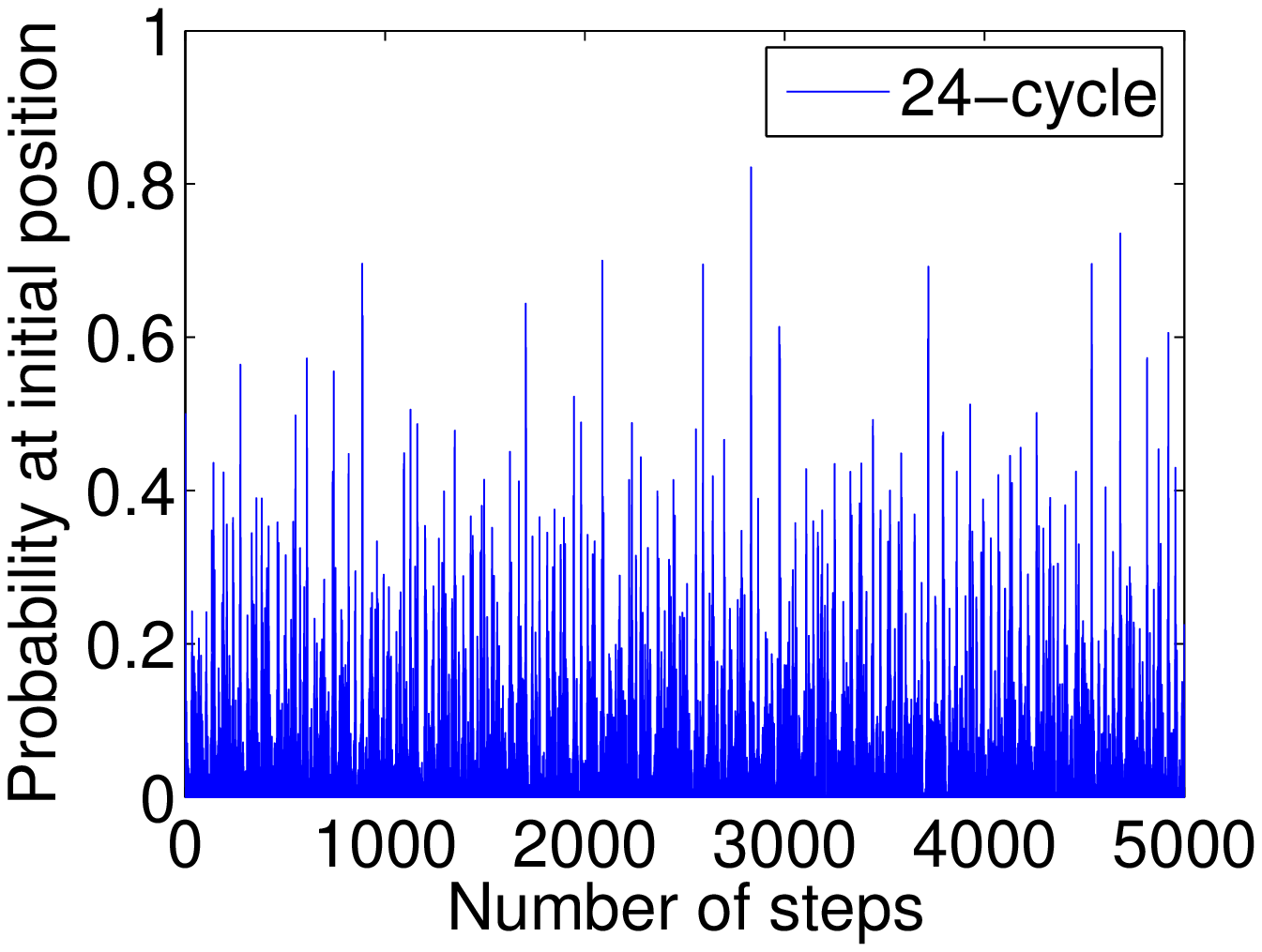, width=7.2cm}\\
~~~~~~~(a)~~~~~~~~~~~~~~~~~~~~~~~~~~~~~~~~~~~~~~~~~~~~~~~~(b)
\caption[Probability at the initial position after different number of steps of quantum walk on $15-$ and $24-$ cycle.]{Probability at the initial position after different number of steps of quantum walk on $15-$ and $24-$ cycle.  This clearly shows that even for steps as large as 5000, their is no signature of complete recurrence. The distribution is obtained using Hadamard operation, $B_{0, \pi/4, 0}$ as quantum coin operation.}
\label{fig:fracrev1}
\end{center}
\end{figure}

In Figure \ref{fig:fracrev1} we show numerically that the probability of finding the particle at the initial position $j=0$ after different number of steps of quantum walk on $15-$ and $24-$ cycle. The distribution is obtained for up to  as large as 5000 steps.  At no time $t=\mbox{number of steps}$, a unit probability value is returned at the initial position showing that the wave packet evolved using quantum walk on an $n-$cycle fails to revive completely at the initial position. 

\par
The failure of the wave packet to completely revive and recur at initial position can be attributed to the interference effect caused by the mixing of the left and right propagating components of the amplitude. By suppressing the interference effect during the evolution in a closed path one can get closer to the complete relocalization, revival at initial position $j=0$.  For example, the wave packet can be completely relocalized at $j=0$ on an $n-$cycle at $t=n$ and make the quantum walk recurrent  by choosing  an extreme value of coin parameters $(\xi, \theta, \zeta) =(0^\circ, 0^\circ, 0^\circ)$. By choosing $\theta = \delta$, $\delta$ being very small and close to $0^\circ$ during the evolution, the interference effect is minimized and will return a near complete recurrence on an $n-$cycle. In Figure \ref{polyaqw},  the plot of $[1-P_{0}(t)]$ is shown for 
$\theta = 0^{\circ}$ and $n=50$. The interference effect is completely suppressed and the quantum walk recurs  after every 50 steps. 
\par
\begin{figure}
\begin{center}
\epsfig{figure=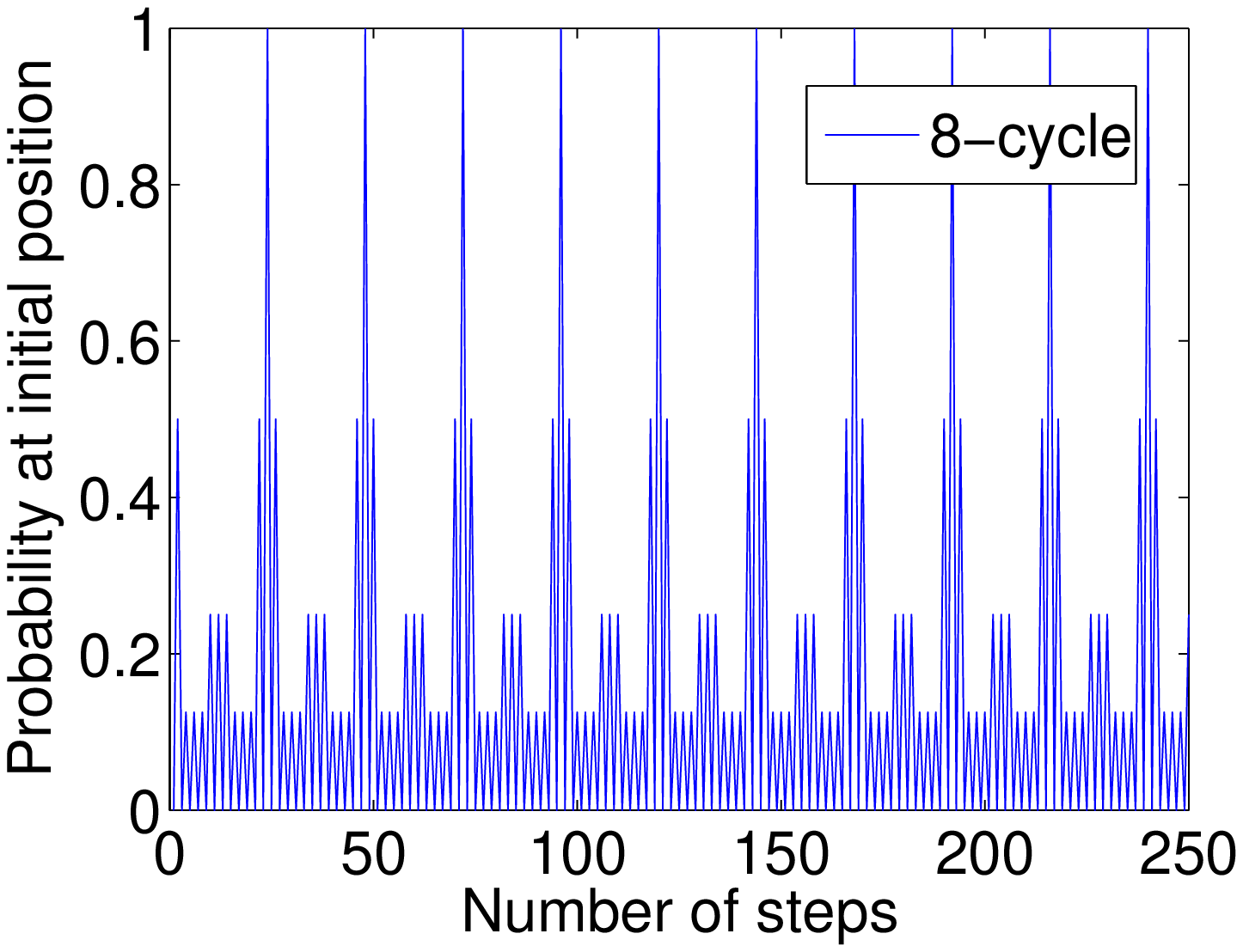, width=7.2cm}
\epsfig{figure=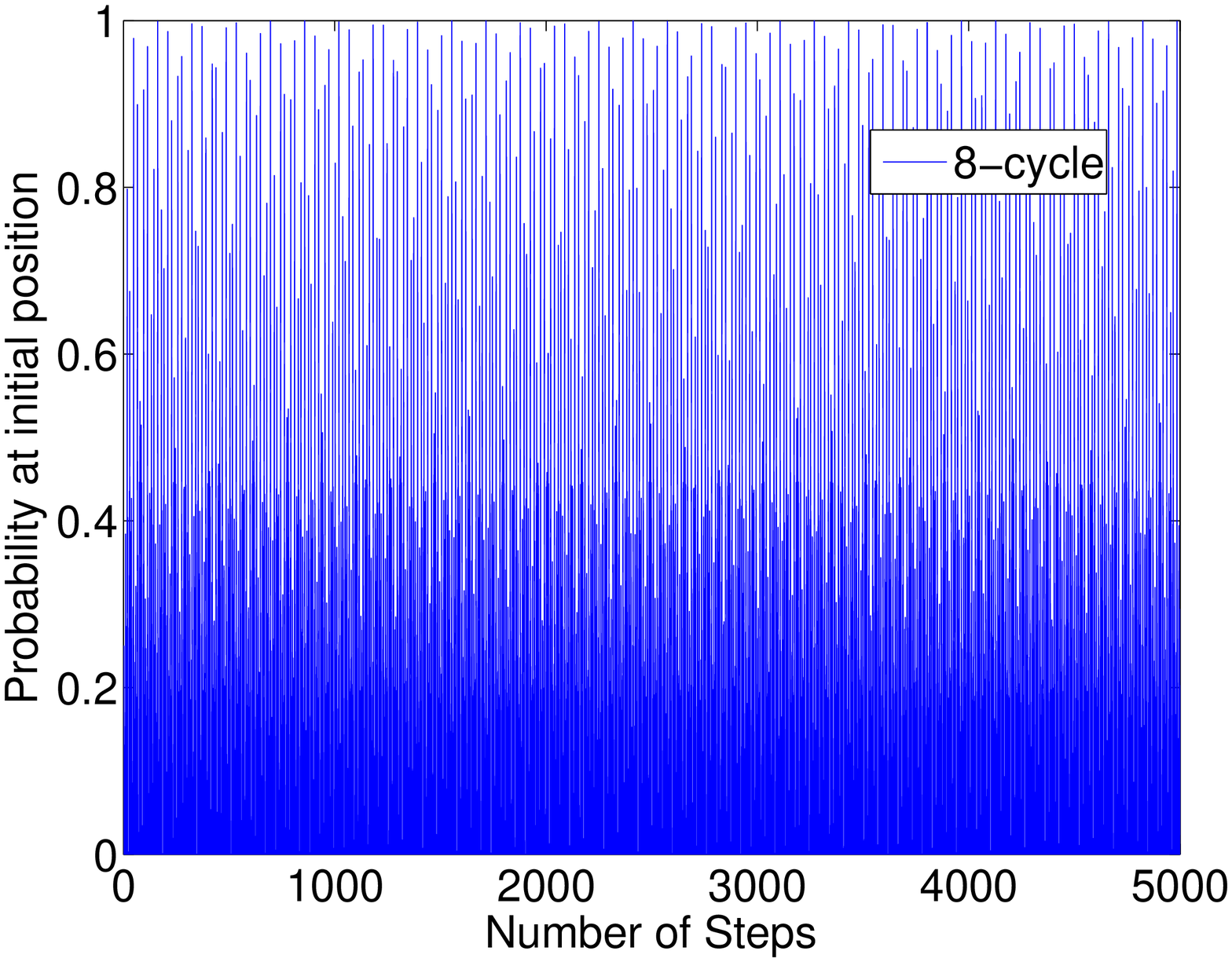, width=7.2cm}\\
~~~~(a)~~~~~~~~~~~~~~~~~~~~~~~~~~~~~~~~~~~~~~~~~~~~~~~~~(b)
\caption[Probability at the initial position on $8-$cycle using different coin operation.]{Probability at the initial position on $8-$cycle. (a)
The distribution is obtained using Hadamard operation $B_{0, \pi/4, 0}$, due to return of amplitudes to initial position (constructive interference at the origin) before interference of amplitude dominates uniformly over the entire vertices, recurrence is seen.  (b) The use of quantum coin, $B_{0, \pi/6, 0}$ during the evolution does not lead to complete constructive interference at origin and hence complete recurrence is not seen.} 
\label{fig:rev8cycle} 
\end{center}
\end{figure}
\begin{figure}
\begin{center}
\epsfig{figure=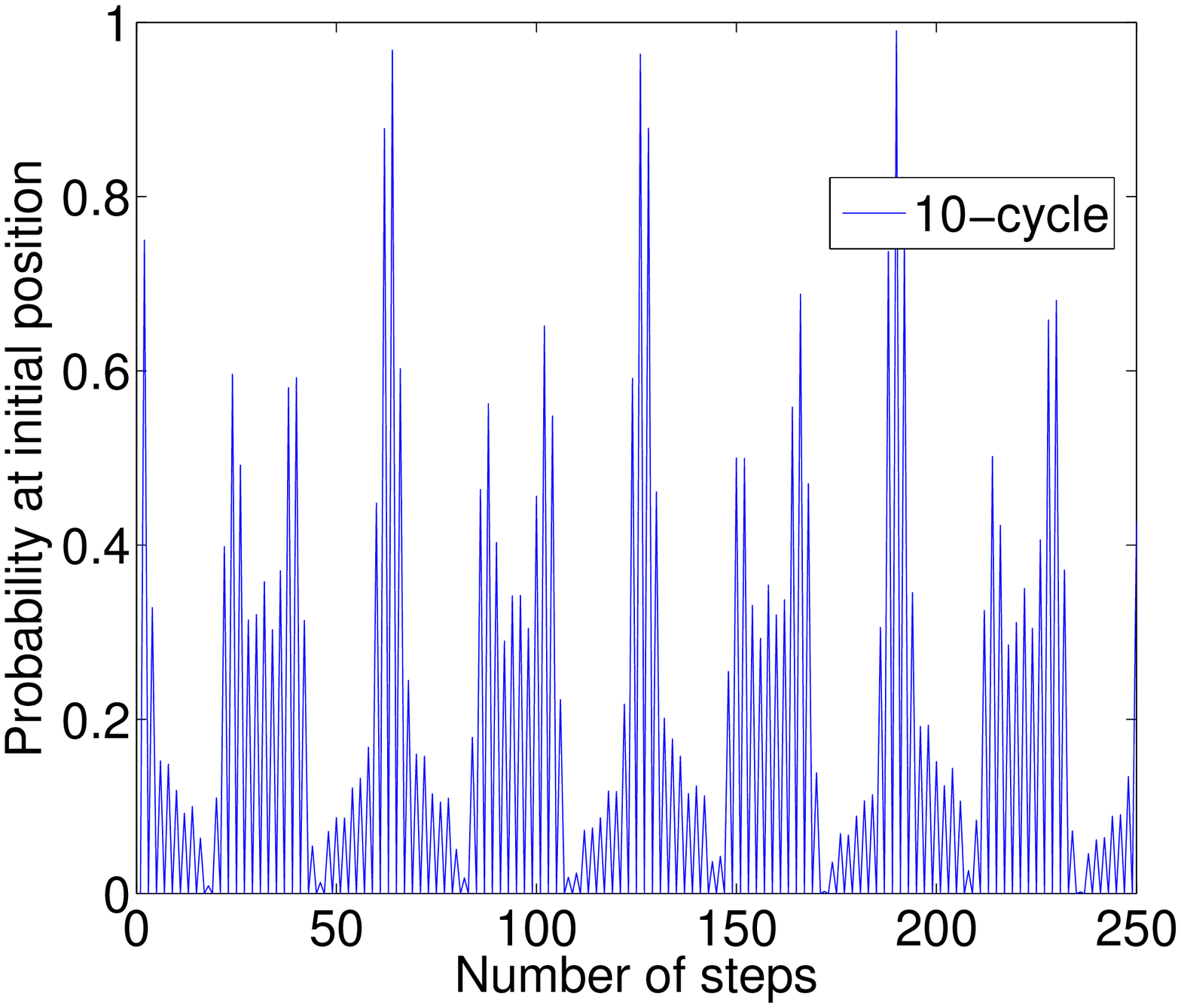, width=7.1cm}
\epsfig{figure=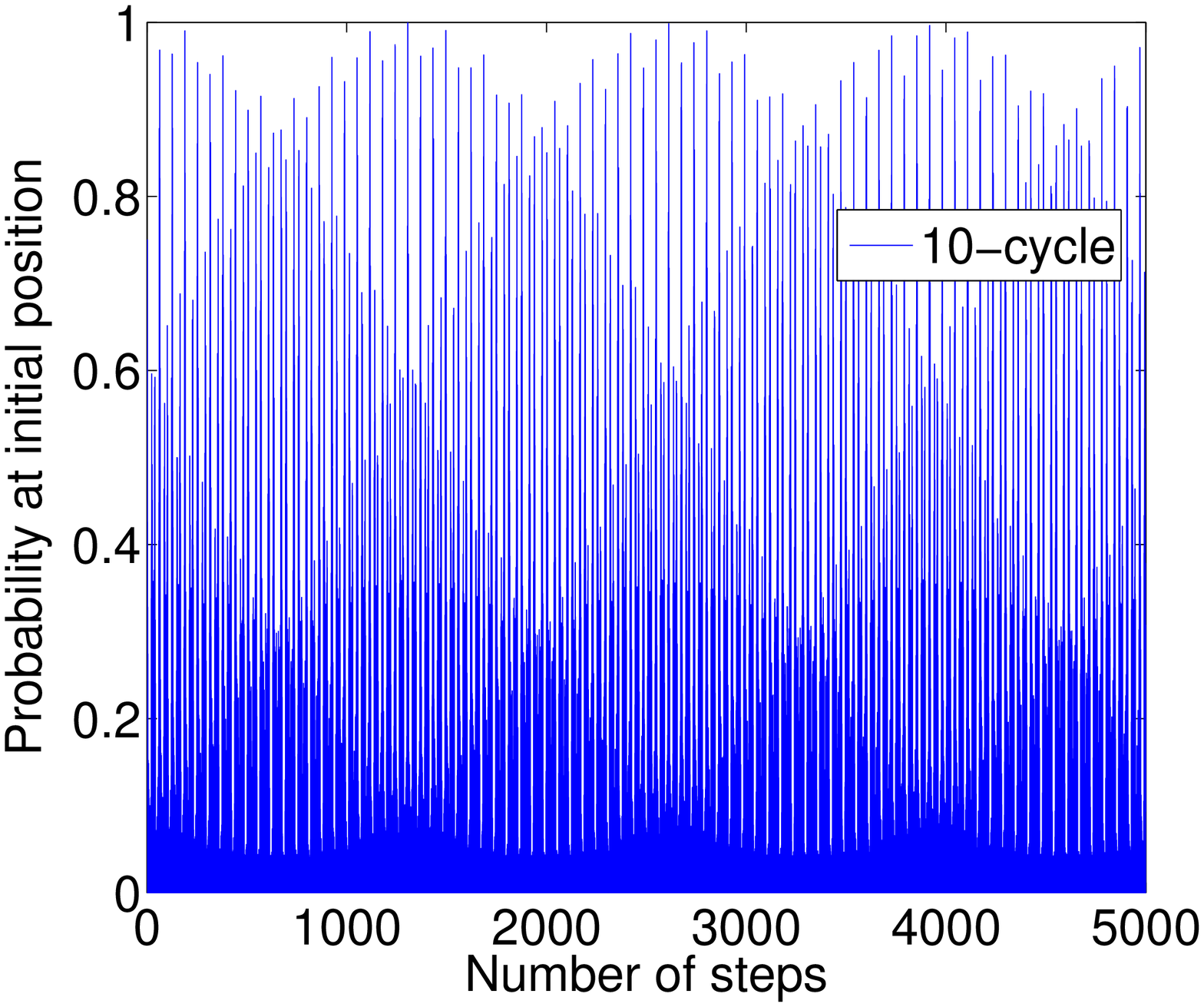, width=7.1cm}\\
~~~~(a)~~~~~~~~~~~~~~~~~~~~~~~~~~~~~~~~~~~~~~~~~~~~~~~~~(b)\\
\epsfig{figure=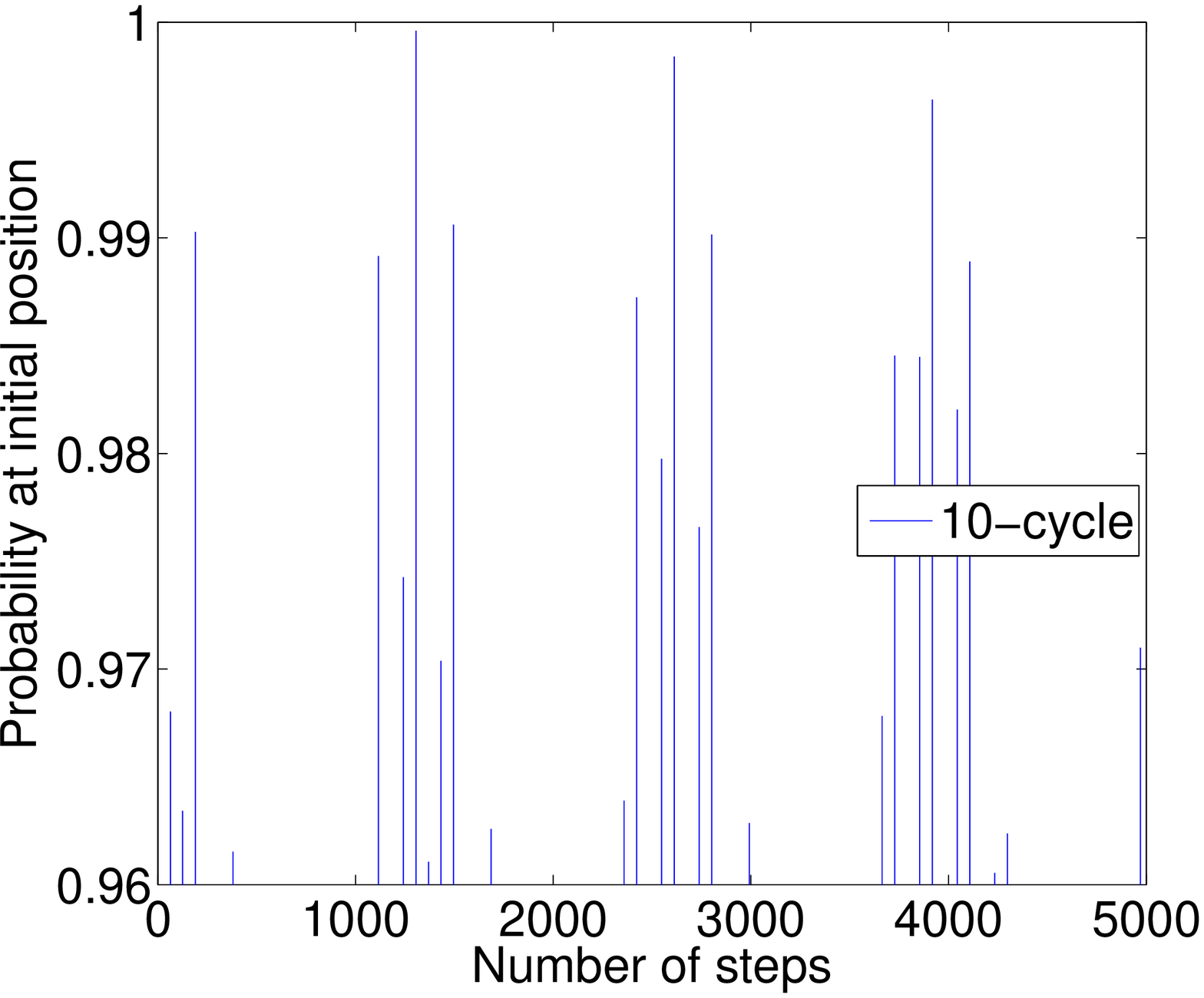, width=7.1cm}\\
(c)
\caption[Probability at the initial position on $10-$cycle. The distribution is obtained using Hadamard operation $B_{0, \pi/4, 0}$ for quantum walk up to 250 steps and 5000 steps respectively.]{Probability at the initial position on $10-$cycle. The distribution is obtained using Hadamard operation $B_{0, \pi/4, 0}$. A small deviation from complete recurrence can be seen. (a) and (b) are probability at initial position for quantum walk up to 250 steps and 5000 steps respectively and (c) is the close up of the probability and we note that the probability is not exact 1 at any time within 5000 steps.} 
\label{fig:rev10cycle}
\end{center}
\end{figure}
\begin{figure}
\begin{center}
\epsfig{figure=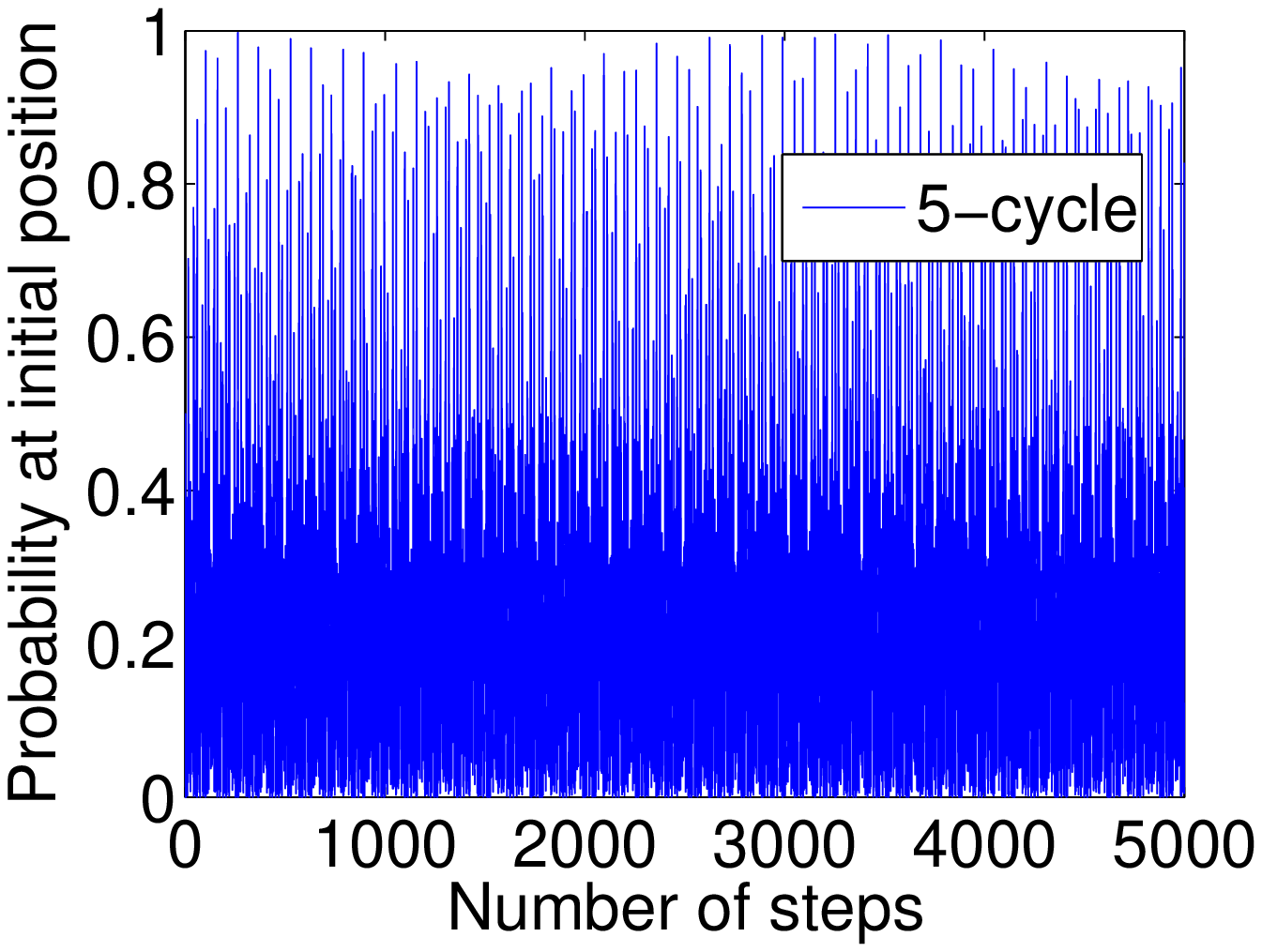, width=7.2cm}
\epsfig{figure=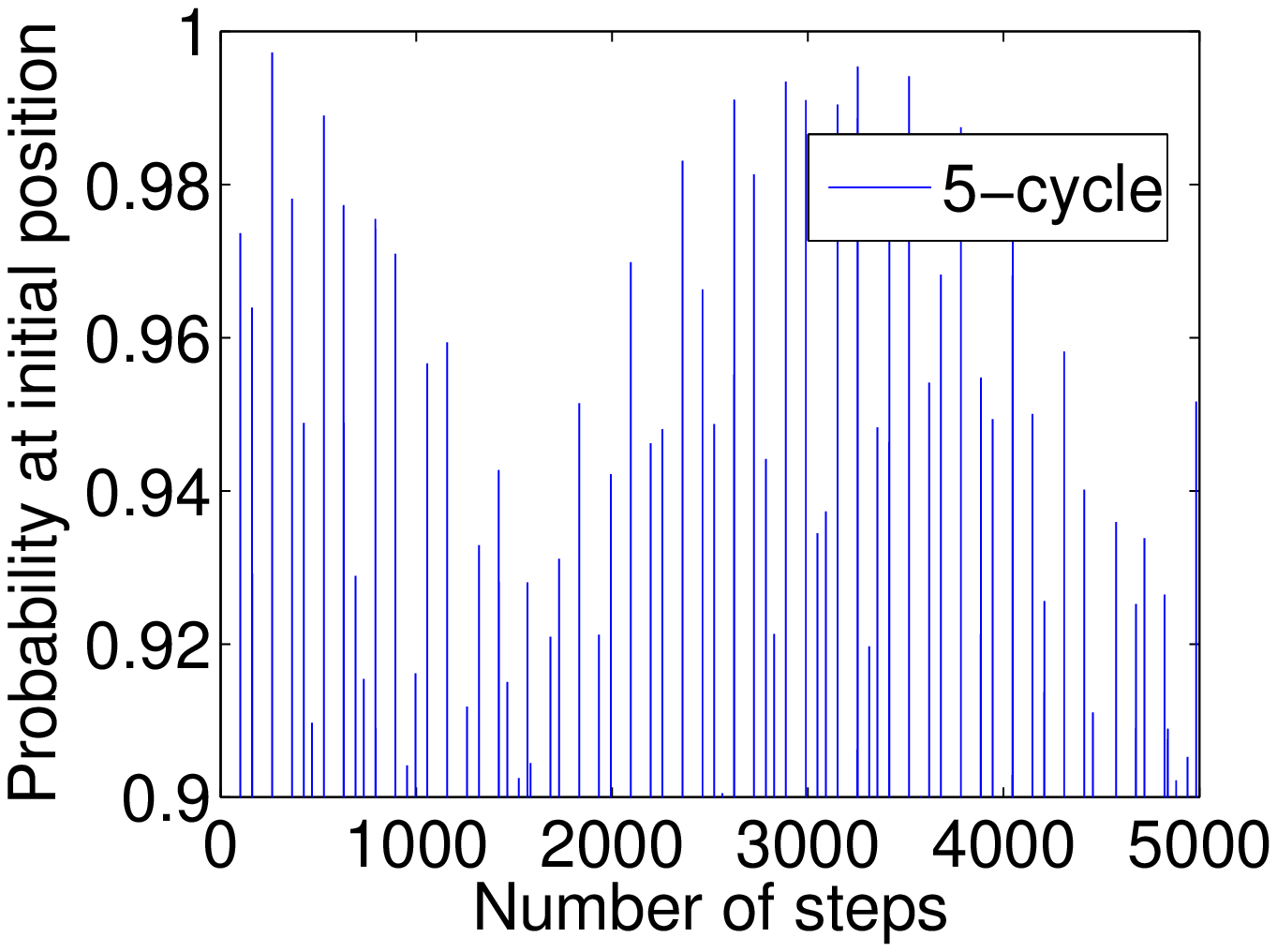, width=7.2cm}\\
~~~~(a)~~~~~~~~~~~~~~~~~~~~~~~~~~~~~~~~~~~~~~~~~~~~~~~~~(b)
\caption[Probability at the initial position after different number of steps of quantum walk on $5-$cycle.]{Probability at the initial position after different number of steps of quantum walk on $5-$cycle,  (a) is the complete plot whereas (b) is a close up of probability values between 0.9 and 1. At no step in the plot, probability is a unit value.  This numerically shows that the quantum walk on an $n-$ cycle does not recur for n=5.} 
\label{fig:rec5}
\end{center}
\end{figure}
From the numerically data for upto 5000 steps  of quantum walk we note that for small $n-$, especially when $n$ is even, the left and right propagating amplitude return back completely to the initial position (origin) before the mixing and repeated interference of the amplitudes takes over at non-initial position in the evolution process. Therefore, for a quantum walk on an $n-$cycle with $n$ being even up to $8$ a complete revival and recurrence of wave packet at initial position $j=0$ is seen, Figure \ref{fig:rev8cycle} \cite{TFM03}. For a quantum walk on particle initially in symmetric superposition state $|\Psi_{ins}\rangle$  with coin operation $B_{0, \theta, 0}$ and $n>8$, due to larger position Hilbert space the interference effect at the non-initial position dominates reducing the recurrence nature of the dynamics. In Figure \ref{fig:rev10cycle}, for quantum walk on $10-$cycle, a small deviation from complete recurrence is shown.

\par
For example, we will consider a small, odd $n=5$. If the positions on a cycle are marked as $j = {0, 1, 2, 3, 4}$  after the third step of the quantum walk, the left propagating amplitude move from position $2$ to $3$ and the right propagating amplitude move from $3$ to $2$. That is, after the second and third step of quantum walk using shift operator of the form $S^{c}$ for a walk on an $n-$cycle (\ref{eq:condshift1}) and Hadamard operation $H$ as coin operation on a particle initially in superposition state $|\Psi_{ins}\rangle$ (\ref{inista}) takes the form,
\be
S^{c}(H\otimes {\mathbbm
1})S^{c}|\Psi_{ins}\rangle = \frac{1}{2} \left \{ |0\rangle \otimes |\psi_{3}\rangle + \right ( |1\rangle + i |0\rangle \left ) \otimes |\psi_{0}\rangle - i |1\rangle \otimes |\psi_{2}\rangle \right \}
\ee
\begin{eqnarray}
S^{c}(H\otimes {\mathbbm
1})S^{c}(H\otimes {\mathbbm
1})S^{c}|\Psi_{ins}\rangle = \frac{1}{2 \sqrt{2}} \{ |0\rangle \otimes |\psi_{2}\rangle + \left ( |0\rangle + |1\rangle + i |0\rangle \right ) \otimes |\psi_{4}\rangle  \nonumber \\
- \left ( |1\rangle - i|0\rangle + i|1\rangle \right ) \otimes |\psi_{1}\rangle + i |1\rangle \otimes |\psi_{3}\rangle \}.
\end{eqnarray}
The left and right propagating amplitude crossover without suppressing the mixing, therefore the constructive interference effect continues to exist even at positions other than the origin during the  evolution. In Figure \ref{fig:rec5} the probability of finding the particle at the initial position is shown. Due to the small size of the Hilbert space, the probability is seen to be close to unity but the closer look reveals that its only a fractional revival.
\par
We also note that localization effect found in 2D \cite{IKK04}, in quantum walk using multi quantum coins to diminish the interference effect \cite{IKS05} or in walk using selective randomization of coin operations (Section \ref{rqw}) can result in increasing the fractional recurrence number ${\mathcal P}_{qw}$ on a line and higher dimension. If the particle wave packet is evolved in a {\it position} Hilbert space ${\mathcal H_{p}}$  with the edges that permits the wave packet to escape, the fractional recurrence nature of the quantum walk does not allow the quantum walk to be completely transient. Therefore, the fractional transient nature of the quantum walk is seen to complement the fractional recurrence nature.
\par
From this section we conclude that as long as the wave packet spread in position space interfering, forming mini wave packets during the evolution of the quantum walk process, it fails to satisfy the complete recurrence theorem. However, fractional recurrence can be seen.
\newpage

\section{Summary} 
\label{summary2}
\begin{itemize}
\item Expression used to understand the continuous- and discrete-time quantum walk which are mathematically identical to Schr\"odinger  and Dirac equations respectively are reviewed.  Our simple decoupling analysis of the evolution shows the similarity between the discrete-time quantum walk evolution expression and the Dirac equation. 
\item Our construction of discrete-time quantum walk model using three parameter quantum coin operation $B_{\xi, \theta, \zeta}$ from SU(2) group optimizes the control over the quantum walk evolution. For example, parameter $\theta$ to control the variance and maximize the variance for a walk on a line, control the mixing time on an $n-$cycle, and parameters $\xi$ and $\zeta$ to bias and control biasing in the walk. 
\item We have shown that the quantum walk evolution dominated by the interference of quantum amplitudes fails to satisfy complete recurrence theorem. However, fractional recurrence characterized by the quantum P\'olya number can be seen.
\end{itemize}

%% file: Chapters/Chapter3.tex
\chapter{Symmetries and noise on quantum walk} 
\label{Chapter3}
\lhead{Chapter 3. \emph{Symmetries and noise}} 

\section{Introduction}
\label{sec:intro}

The probability distribution of discrete-time quantum walk on a line remains invariant in position space when the operations to implement each step of the walk is augmented by certain operations. We refer to these discrete operations  as symmetries of the quantum walk. In this chapter we will study some of these symmetry operations. We further generalize the observations of these symmetries in the presence of environmental effects, modeled by  various  noise channels such as, bit flip, phase-flip (decoherence without net   dissipation), generalized amplitude damping (decoherence with dissipation) and squeezed generalized amplitude damping channels \cite{NC00, MKT00, TMK00, CSB07} on the coin space. We have found it convenient to explain the symmetries and effect of noise using quantum trajectories, and the numerical results are obtained by numerical integration by evolving the density operator of the system.
\par
We extend these studies to quantum walk on an $n-$cycle. Interestingly, we find that the symmetry operations is sensitive to the  walk topology, in the sense that the symmetry which holds for quantum walk on an one-dimensional line does not hold, in general, for a quantum  walk on an $n-$cycle but leads to other interesting behavior. 
The difference between the walk on the line and an $n-$cycle can be attributed to the different ways the interference occurs between the propagating wavefunction. Quantum walk on the line involves interference between the forward and backward propagating wavefunctions whereas, the walk on an $n-$cycle involves interference between forward (backward) propagating waves from both the sides of the loop along with the interference between the forward and backward propagating wavefunctions. Noise on an $n-$cycle tends  to  restore these symmetry both  by classicalizing the  walk and also desensitizing the
symmetry operation as a topology probe for the quantum walk.
\par
These observations can  have important implications  for a better insight into, and for simplifying certain implementations of, quantum walks and are also of relevance to  studies in  quantum
optics and condensed matter systems. Later in this chapter and in Chapter \ref{Chapter4} we show that the application  of these ideas can help simplify certain experimental implementations of  quantum  walk and can be used as an additional degree of freedom in applications of quantum walk in physical systems.
\par
This chapter is organized as follows. In Section \ref{sec:symmline}, we discuss symmetries and noise on quantum walk on a line with Section \ref{sec:symm} focusing on bit flip and phase flip symmetries and Section \ref{sec:env} focuses on the environmental effects. In Section \ref{qwpg} we extend the studies to walk on an $n-$cycle and discuss the breakdown in symmetry. 
In Section \ref{sec:qwbec} we discuss the experimental implications and conclude with summary in Section {\ref{summary3}.

\section{Symmetry and noise operations on a line}
\label{sec:symmline}
\subsection{Bit flip and phase flip symmetries}
\label{sec:symm}

As defined in Section \ref{dtqw} each step of the discrete-time quantum walk consists of the quantum coin operation $B_{\xi, \theta, \zeta}$ ({\ref{3paraU2}) followed by the shift operation
\begin{eqnarray}
\label{eq:condshift1}
S &=&|0\rangle \langle  0|\otimes \sum_{j \in \mathbb{Z}}
|\psi_{j-1}\rangle \langle \psi_{j} |+|1\rangle  \langle 1 |\otimes \sum_{j \in \mathbb{Z}} |\psi_{j+1}\rangle
\langle \psi_{j}| \nonumber \\
&\equiv & |0\rangle\langle0|\otimes \hat{a} + |1\rangle\langle1|
\otimes \hat{a}^{\dag}.
\end{eqnarray}
Here $\hat{a}$ and $\hat{a}^{\dag}$ are unitary operators that are
notationally reminiscent of annihilation and creation operations,
respectively. Lets consider the application of the modified conditional shift operator of the form
\begin{equation} 
\label{eq:modcondshift}
 S^{\prime}      =     (X      \otimes
\mathbb{I})\exp(-i(|0\rangle \langle 0| - |1\rangle \langle 1| )\otimes Pl), 
\end{equation}  
\begin{eqnarray}
S^{\prime} &=& |1\rangle  \langle 0|\otimes  \sum_{j  \in
\mathbb{Z}}|\psi_{j-1}\rangle  \langle \psi_{j} |+|0\rangle  \langle 1
|\otimes \sum_{j \in \mathbb{Z}} |\psi_{j+1}\rangle \langle \psi_{j}| \nonumber \\
&\equiv & |1\rangle\langle0|\otimes \hat{a} + |0\rangle\langle1|
\otimes \hat{a}^{\dag}
\end{eqnarray}
instead of $S$ (\ref{eq:condshift1}). Where $X = \sigma_{x}$ is  the Pauli $x$ operator.  Since  $S^{\prime} = (X   \otimes \mathbb{I})S$, i.e.,
it  is  equivalent  to  an  application of  bit  flip  following  $S$,
conditioned on the internal state being $|0\rangle$ ($|1\rangle$). That is, at any given position $j$, the particle will move to the  left (right) and changes its internal state
to       $|1\rangle$      ($|0\rangle$).  Thus,     
\be
S^{\prime}
(|0\rangle\otimes|\psi_{j}\rangle)=|1\rangle\otimes|\psi_{j-1}\rangle  {\rm ~~~~~ and~~~~}
S^{\prime}(|1\rangle\otimes|\psi_{j}\rangle)=|0\rangle\otimes|\psi_{j+1}\rangle.
\ee
A  relevant observation  in this  context is  that there  are physical
systems  where  the implementation of $S^{\prime}$ is  easier than that
of  $S$. We will discuss one such system in Section \ref{becqw} \cite{Cha06}.
In  that case,  applying a  compensatory bit flip on  the internal state,  after each application  of $S^{\prime}$, reduces the  modified quantum  walk to the  usual scheme. In  all, this would require  $(t-1)$ compensatory  bit  flip operation in addition for a $t$ step quantum walk, which adds to the complexity of the  experimental realization.  However, this additional complexity can be eliminated. For a quantum walk using $B_{\xi, \theta, \zeta}$ (\ref{3paraU2}) as coin operation, applying a bit flip in each  step can be shown to be equivalent to  a spatial inversion of the position probability distribution. A quick way to see why bit flips are harmless is to note that they are also equivalent  to relabeling the edges of the graph on which the quantum walk  takes place, so that each end of each edge has the same label 
\cite{KS05}\footnote{It  is worth noting  that in \cite{DRK02}, bit flips  are employed  to improve  the practical implementation of a quantum walk  on atoms in an optical lattice}.  We
may in  this sense call a  bit flip together with  spatial inversion a {\em  symmetry} of the quantum  walk  on  a line.   To  be specific, when any unitary operation augmented during each step of quantum walk which may leave the position probability distribution unaffected, then it is called a quantum  walk symmetry.

Experimentally, the symmetries are useful in identifying  variants of a given quantum  walk protocol that are equivalent to it.  This  motivates us to look for other (discrete)
symmetries of the  quantum walk, which we study  below.  We begin with
Theorem  \ref{thm:bias},  where  we  note  four  discrete  symmetries,
associated     with    the     matrices     $B^{\rm (f)}$    $({\rm f}=1,2,3,4)$
(\ref{eq:bias20}), of the  quantum walk. Thereafter  two of
these  symmetries,  $B^{(1)}$   and  $B^{(2)}$,  are  identified  with
operations  that  are  relevant   from  the  perspective  of  physical
implementation. It is an interesting open question with relevance to practical
implementation of quantum walks, whether other such symmetries of the
quantum walk exist.
\par
\begin{thm} If $B = B_{\xi, \theta, \zeta}$\footnote{ 
Three parameter quantum coin operation (\ref{U2}) is used instead of SU(2) operation so that $B$ reduces to Hadamard operation operation for $\xi =\zeta =0, \theta =\pi/4$} in (\ref{U2}) is replaced by any of $B^{(1)}$, $B^{(2)}$, 
$B^{(3)}$, or $B^{(4)}$, given by,
\begin{eqnarray}
B^{(1)} \equiv \left( \begin{array}{clcr}
 e^{i\xi}\cos(\theta)  & &   e^{i\zeta}\sin(\theta)   \\
 e^{i(\phi-\zeta)}\sin(\theta)  & & -e^{i(\phi-\xi)}\cos(\theta)
 \end{array} \right), \nonumber \\
B^{(2)} \equiv \left( \begin{array}{clcr}
 e^{i\xi}\cos(\theta)  & & e^{i(\phi+\zeta)}\sin(\theta)   \\
e^{-i\zeta}\sin(\theta)  & & -e^{i(\phi-\xi)}\cos(\theta)
 \end{array} \right),\nonumber \\
B^{(3)} \equiv \left( \begin{array}{clcr}
e^{i(\phi+\xi)}\cos(\theta)  & & e^{i(\phi+\zeta)}\sin(\theta)   \\
e^{-i\zeta}\sin(\theta)  & & -e^{-i\xi}\cos(\theta)
 \end{array} \right),\nonumber \\
B^{(4)} \equiv \left( \begin{array}{clcr}
e^{i(\phi+\xi)}\cos(\theta)  & & e^{i\zeta}\sin(\theta)   \\
e^{i(\phi-\zeta)}\sin(\theta)  & & -e^{-i\xi}\cos(\theta)
 \end{array} \right),
\label{eq:bias20}
\end{eqnarray}
the resulting position probability
distribution of the quantum walk remains invariant.
\label{thm:bias}
\end{thm}

\noindent {\bf  Proof.}  With the notation $B  \equiv \{b_{q,r}\}$ and $B^{({\rm f})} \equiv  \{b^{({\rm f})}_{q,r}\}$, we find  
\bea
b^{(1)}_{q,r}   = b_{q,r}e^{iq\phi} ~~;~~
b^{(2)}_{q,r}  = b_{q,r}e^{ir\phi} ~~;~~ 
b^{(3)}_{q,r} = b_{q,r}e^{i\bar{q}\phi}~~;~~ 
b^{(4)}_{q,r}   =  b_{q,r}e^{i\bar{r}\phi},~~
\eea
where the matrix indices $q,r$ take values 0 and 1,  $i  \equiv
+\sqrt{-1}$,   and   the  overbar   denotes   a   NOT  operation   ($0
\leftrightarrow  1$).  The  state  vector obtained,  after $t$  steps,
using  $B$  and  $B^{(1)}$  as  the  coin  rotation  operations,  are,
respectively, given by
\begin{eqnarray}
\label{eq:bitobit}
|\Psi_1\rangle &=& 
(SB)^t|\alpha,\beta\rangle = \sum_{q_1,q_2,\cdots,q_t}
b_{q_t,q_{t-1}}\cdots b_{q_2,q_1}b_{q_1,\alpha}
|q_t,\beta+2Q - t\rangle,\nonumber \\
|\Psi_2\rangle &=& 
(SB^{(1)})^t|\alpha,\beta\rangle \nonumber \\
&= & \sum_{q_1,q_2,\cdots,q_t}
b_{q_t,q_{t-1}}\cdots b_{q_2,q_1}b_{q_1,\alpha}
(e^{i\phi})^{q_{t-1}+ \cdots + q_1 + \alpha}
|q_t,\beta+2Q - t\rangle,
\end{eqnarray}
where $Q = q_1+\cdots+q_t$. Consider one of the element $|a,b\rangle$
in    the     computational-and-position    basis.    Now,    
\be
\langle a,b|\Psi_1\rangle  =  e^{i\eta\phi}\langle  a,b|\Psi_2\rangle,
\ee
where
$\eta  = q_{t-1}+  \cdots +  q_1 +\alpha$,  which is  fixed  for given
$\alpha$ and $b$, and determined  by $b = \beta+2Q-t$ and $q_t=a$.  As
a     result,  
\be
|\langle     a,b|\Psi_1\rangle|^2    +     |\langle
\bar{a},b|\Psi_1\rangle|^2 =  |\langle a,b|\Psi_2\rangle|^2 + |\langle
\bar{a},b|\Psi_2\rangle|^2.
\ee
A similar  proof of  invariance  of the position distribution can be demonstrated to 
hold when $B$ is replaced by  one of  the  other  $B^{({\rm f})}$'s $({\rm f}=2,3,4)$.   
On  account of  the linearity of quantum mechanics,  the invariance of the walk 
statistics under exchange of the $B^{({\rm f})}$'s  and $B$ holds even when the initial
state $|\alpha,\beta\rangle$ is replaced by a general superposition or
a mixed state.  \hfill $\blacksquare$

Interchanging $B$ and the $B^{({\rm f})}$'s may be considered as a discrete
symmetry operation $G: B \rightarrow B^{\star}$ (where $B^{\star}$
denotes any of the $B^{({\rm f})}$'s in (\ref{eq:bias20})), 
that leaves the positional probability 
distribution invariant. We express this by the statement that
\begin{equation}
\label{eq:symm}
\widehat{W} \simeq {\bf G}\widehat{W},
\end{equation}
where ${\bf G}$  refers to the application of $G$ at  each step of the
walk, and $\widehat{W}$  refers to the walk operation  of evolving the
initial state  through $t$  steps and then  measuring in  the position
basis.  Knowledge of this symmetry can help simplify practical quantum
walks. Below we identify two of these quantum walk symmetry operations
$B   \leftrightarrow  B^{(1)}$   and   $B  \leftrightarrow   B^{(2)}$,
associated with physical operations of interest.

We   first  consider   the   phase  shift   operation
\be
\label{eq:phi}
\Phi(\phi)   \equiv |0\rangle\langle0| +  e^{i\phi}|1\rangle\langle1|
\ee
as a  {\em symmetry operation} of a quantum walk.  In our model, the quantum operation for
each step is augmented by the insertion of $\Phi(\phi)$ just after the
operation $S(B \otimes {\mathbbm 1})$.  At each step, the walk evolves according to 
\begin{eqnarray}
\label{eq:uz}
W_{\Phi} &\equiv& \left(\begin{array}{ll}
1 & 0 \\
0 & e^{i\phi}
\end{array}\right)
\left[
\left(\begin{array}{lcl}
1  & 0 \\
0  & 0
\end{array}\right)\otimes \hat{a} +
\left(\begin{array}{lcl}
0 & 0 \\
0 &  1
\end{array}\right)\otimes \hat{a}^{\dag}\right]
\left( \begin{array}{clcr}
e^{i\xi}\cos(\theta)  &   e^{i\zeta}\sin(\theta)   \\
e^{-i\zeta}\sin(\theta)  &  -e^{-i\xi}\cos(\theta)
 \end{array} \right) \nonumber \\
&=&
\left[
\left(\begin{array}{lcl}
1 & 0 \\
0 &  0
\end{array}\right)\otimes \hat{a} +
\left(\begin{array}{lcl}
0 & 0 \\
0 &  1
\end{array}\right)\otimes \hat{a}^{\dag}\right]
\left( \begin{array}{clcr}
e^{i\xi}\cos(\theta)  &  e^{i\zeta}\sin(\theta)   \\
e^{i(\phi-\zeta)}\sin(\theta)  & -e^{i(\phi-\xi)}\cos(\theta)
 \end{array} \right).
\end{eqnarray}
This is equivalent to replacing  $B$ by $B^{(1)}$, which, according to
Theorem \ref{thm:bias}, leaves the walk distribution invariant. Thus
the operation $\Phi(\phi)$, applied at each step, is a symmetry of the
quantum walk.

As  a special case, the phase  flip operation $Z = \sigma_{z}$, a Pauli $z$ operation applied  at each step, obtained  by setting  $\phi=\pi$, is a  symmetry of  the quantum
walk.  Representing the inclusion of  operations $\Phi$ or $Z$ at each
step  of the  walk  by ${\bf  \Phi}$  or ${\bf  Z}$, respectively,  we
express this symmetry by the statements :
\begin{subequations}
\label{eq:sym1}
\begin{eqnarray}
\widehat{W} \simeq {\bf \Phi}\widehat{W}, \label{eq:sym1a} \\
\widehat{W} \simeq {\bf Z}\widehat{W}. \label{eq:sym1b} 
\end{eqnarray}
\end{subequations}
\par
Unlike the  phase flip operation,  bit flip is  not a symmetry  of the
quantum walk on a line.  However, the combined application of bit flip
along with angular  reflection $R$ ($\theta \rightarrow \pi/2-\theta$,
i.e.,    $\sin\theta    \leftrightarrow    \cos\theta$,    and    $\xi
\leftrightarrow  -\zeta$)  and  parity $P$  ($\hat{a}  \leftrightarrow
\hat{a}^{\dag}$) turns  out to be  a symmetry operation.   These three
operations commute  with each other.   By the inclusion of  $PRX$, the
walker evolves  by $([PRX]  SB)^t$. At each  step, the  walker evolves
according to
\begin{eqnarray}
W_P &\equiv& P R \left(\begin{array}{ll}
0 & 1 \\
1 & 0
\end{array}\right)
\left[
\left(\begin{array}{lcl}
1 & 0 \\
0 &  0
\end{array}\right)\otimes \hat{a} +
\left(\begin{array}{lcl}
0 & 0 \\
0 &  1
\end{array}\right)\otimes \hat{a}^{\dag}\right]
\left( \begin{array}{clcr}
e^{i\xi}\cos(\theta)  &  e^{i\zeta}\sin(\theta)   \\
e^{-i\zeta}\sin(\theta)  &  -e^{-i\xi}\cos(\theta)
 \end{array} \right) \nonumber \\
&=&
\left[
\left(\begin{array}{lcl}
1 & 0 \\
0 &  0
\end{array}\right)\otimes \hat{a} +
\left(\begin{array}{lcl}
0 & 0 \\
0 &  1
\end{array}\right)\otimes \hat{a}^{\dag}\right]
\left( \begin{array}{clcr}
e^{i\xi}\cos(\theta)  & -e^{i\zeta}\sin(\theta)   \\
e^{-i\zeta}\sin(\theta) &  e^{-i\xi}\cos(\theta)
 \end{array} \right).
\end{eqnarray}
This  is equivalent  to replacing  $B$ by  $B^{(2)}$  with $\phi=\pi$,
which, according  to Theorem  \ref{thm:bias}, should leave  the walk distribution invariant. Thus the operation $PRX$ applied at each step,
is a symmetry  of the quantum walk.  It  will be convenient henceforth
to choose  $\xi=\zeta=0$, so  that $R$ will  simply correspond  to the
replacement $\theta \rightarrow \pi/2-\theta$.

Representing the inclusion of operations  $P$, $R$ or $X$ at each step
of the  walk by ${\bf  P}$, ${\bf R}$  or ${\bf X}$,  respectively, we
express this symmetry by the statements:
\begin{subequations}
\label{eq:RX}
\begin{eqnarray}
\widehat{W} &\simeq& {\bf P R X}\widehat{W}, \label{eq:RX1} \\
{\bf X}\widehat{W} &\simeq& {\bf P R} \widehat{W}. \label{eq:RX2}
\end{eqnarray}
\end{subequations}
The above expression (\ref{eq:RX1}) was proved immediately above and (\ref{eq:RX2}) follows from (\ref{eq:RX1}), since the operations ${\bf P}$, ${\bf R}$  and ${\bf  X}$  mutually  commute, and  $X^2  = \mathbb{I}$.   It
expresses the fact that applying the  $X$ operation at  each step is equivalent  to  replacing  a  quantum  walk  by  its  angle-reflected,
spatially inverted counterpart. The  observation made at the beginning
of this  section pertains to the special  case of $\theta=45^{\circ}$.
By a similar technique the following symmetries may be proved,
\begin{equation}
\label{eq:sym3}
{\bf X}\widehat{W} ~\simeq~ {\bf X Z}\widehat{W}
~\simeq~ {\bf Z X}\widehat{W}.
\end{equation}
The first equivalence easily follows from (\ref{eq:sym1}).

\subsection{Environmental effects 
\label{sec:env}}

A  quantum  walk  implemented  on  a physical system is is  inevitably
affected by noise due to the environment. We consider
three physically relevant models of noise: a phase flip channel (which
is equivalent to  a phase damping or purely  dephasing channel), a bit
flip channel and a generalized  amplitude damping channel ($T \ge 0$).
In all  cases, our numerical implementation of  these channels evolves
the  density matrix  employing the  Kraus operator  representation for
them. However to explain symmetry effects, it is convenient to use the
{\em quantum trajectories} approach, discussed below.
\begin{figure}
\begin{center}
\subfigure[]{\includegraphics[width=7.2cm]{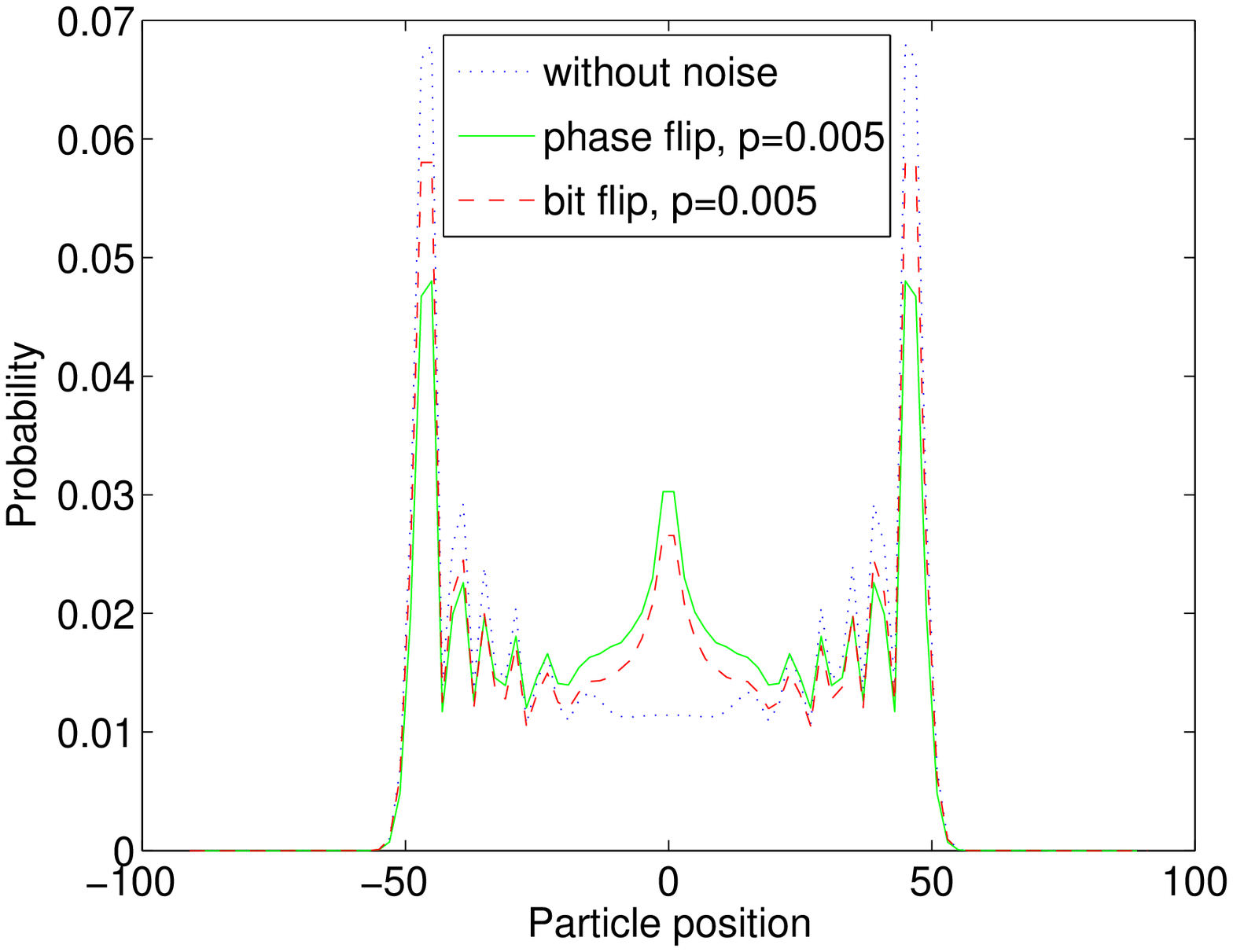}}
\hfill
\subfigure[]{\includegraphics[width=7.2cm]{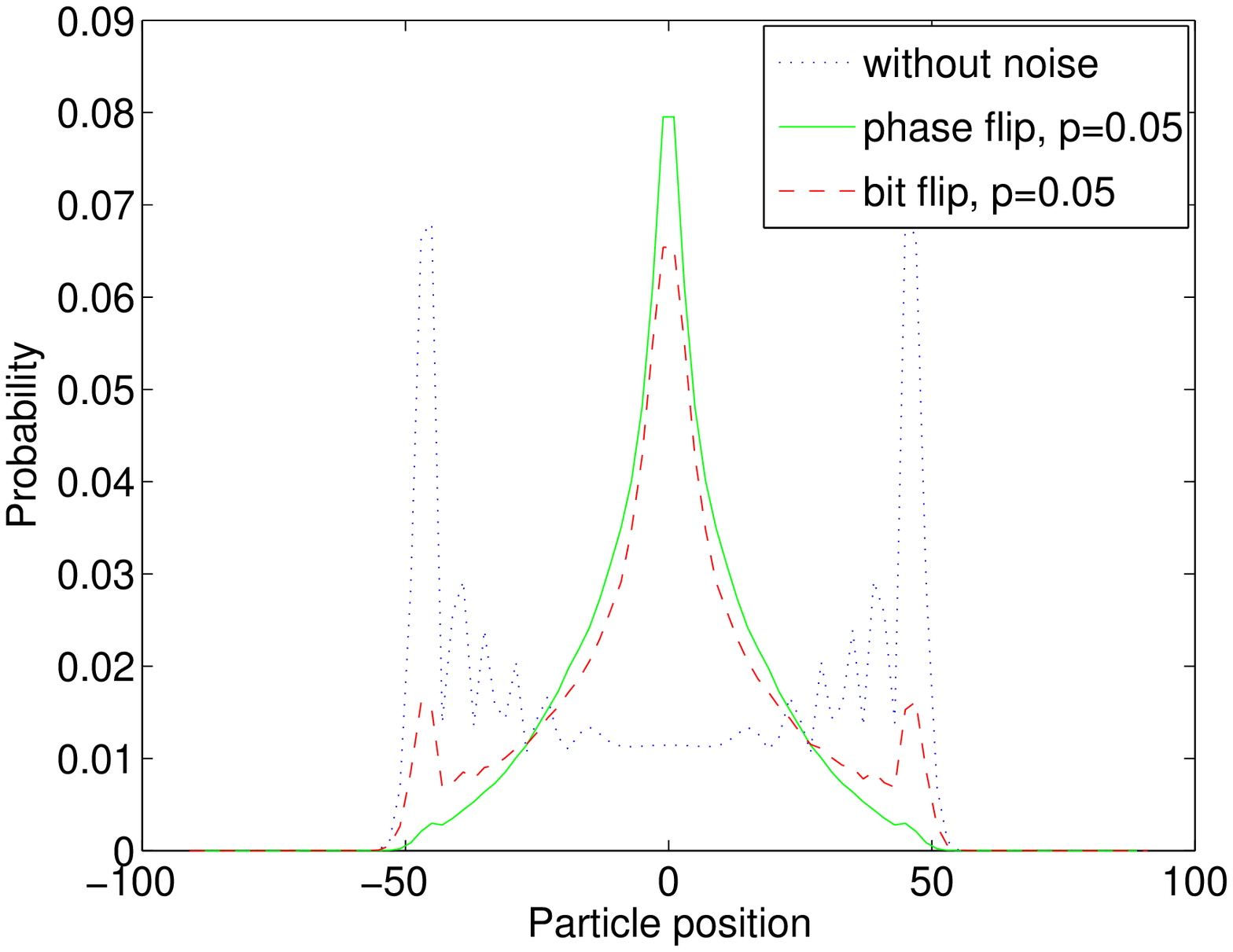}} \\
\subfigure[]{\includegraphics[width=7.2cm]{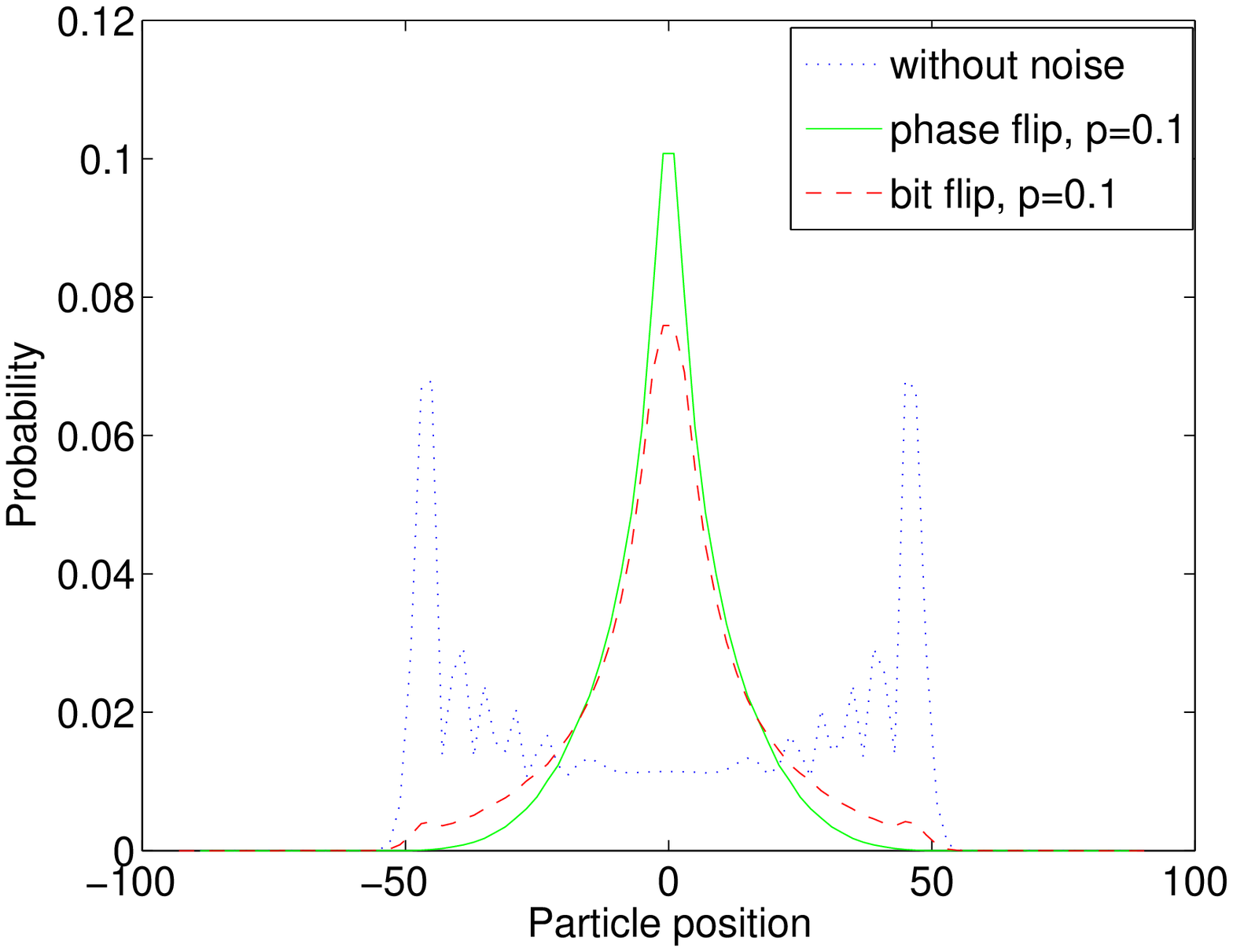}}
\hfill
\subfigure[]{\includegraphics[width=7.2cm]{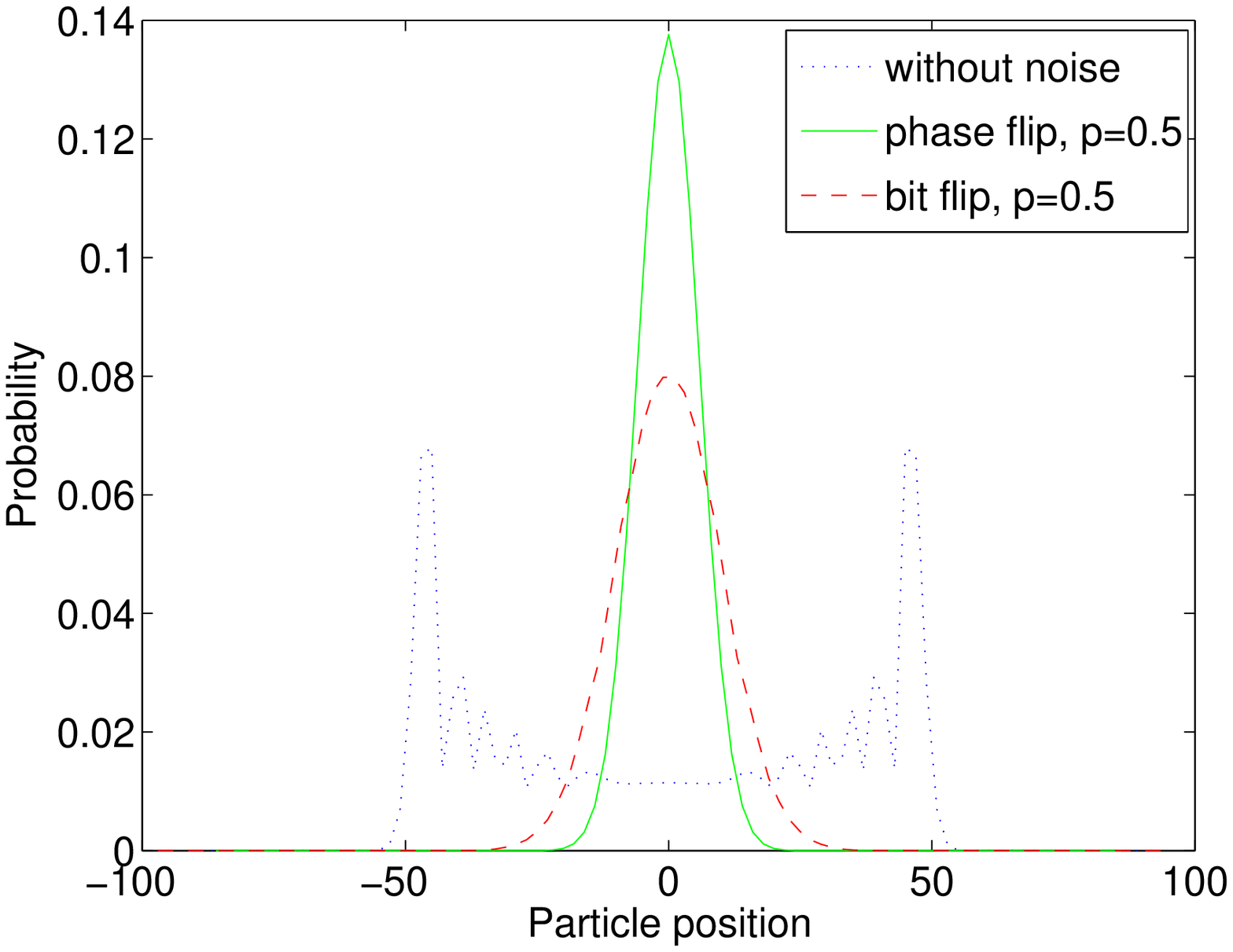}}
\caption[The  effect  of  environmental  decoherence on  the  position
probability  distribution of a quantum  walk with coin parameters  
$\theta = 60^{\circ}, \xi=\zeta=0$ subjected  to a
noisy  channel.]{The  effect  of  environmental  decoherence on  the  position
probability  distribution of a quantum  walk subjected  to a
noisy  channel. Coin  bias is  of the form (\ref{U2}) with 
$\theta = 60^{\circ}, \xi=\zeta=0$.   The noise is modeled as  a phase flip (solid
line)   and  bit   flip   (dashed  line)   channel,  characterized   by (\ref{eq:phaseflip})  and  (\ref{eq:bitflip}), respectively,  at
various noise levels  $p~$: (a) $p = 0.005$ ; (b) $p=0.05$; (c) $p=0.1$;
(d)   $p=0.5$,  which   corresponds  to   a  fully   classical  random
walk.  Comparing Figure  (d) with  Figure \ref{fig:env30}(d),  we note
that the  distribution in the  case of maximal  bit flip noise  is the
same. The distribution is for 100 steps.}
\label{fig:env60}
\end{center}
\end{figure}

\begin{figure}
\begin{center}
\subfigure[]{\includegraphics[width=7.2cm]{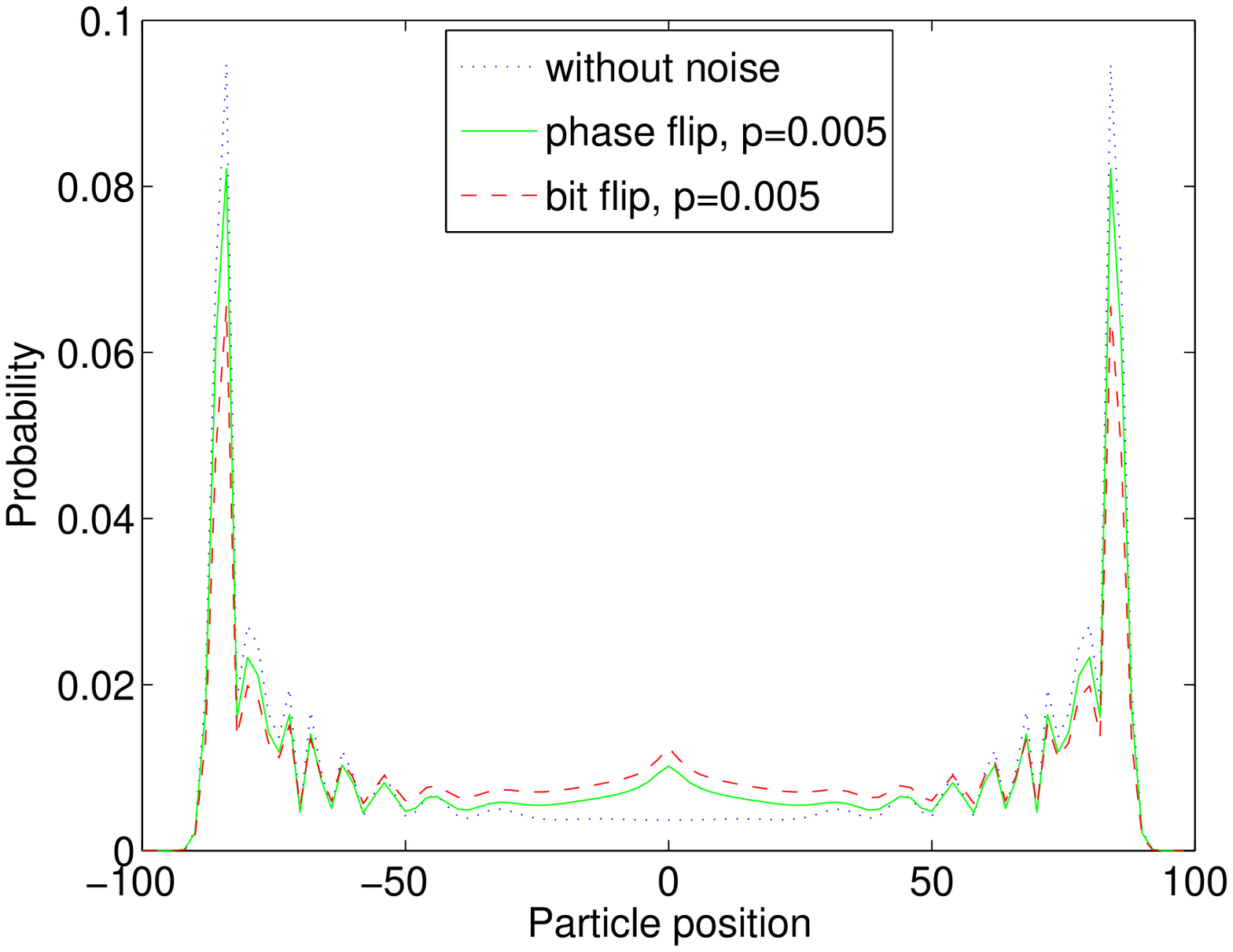}}
\hfill
\subfigure[]{\includegraphics[width=7.2cm]{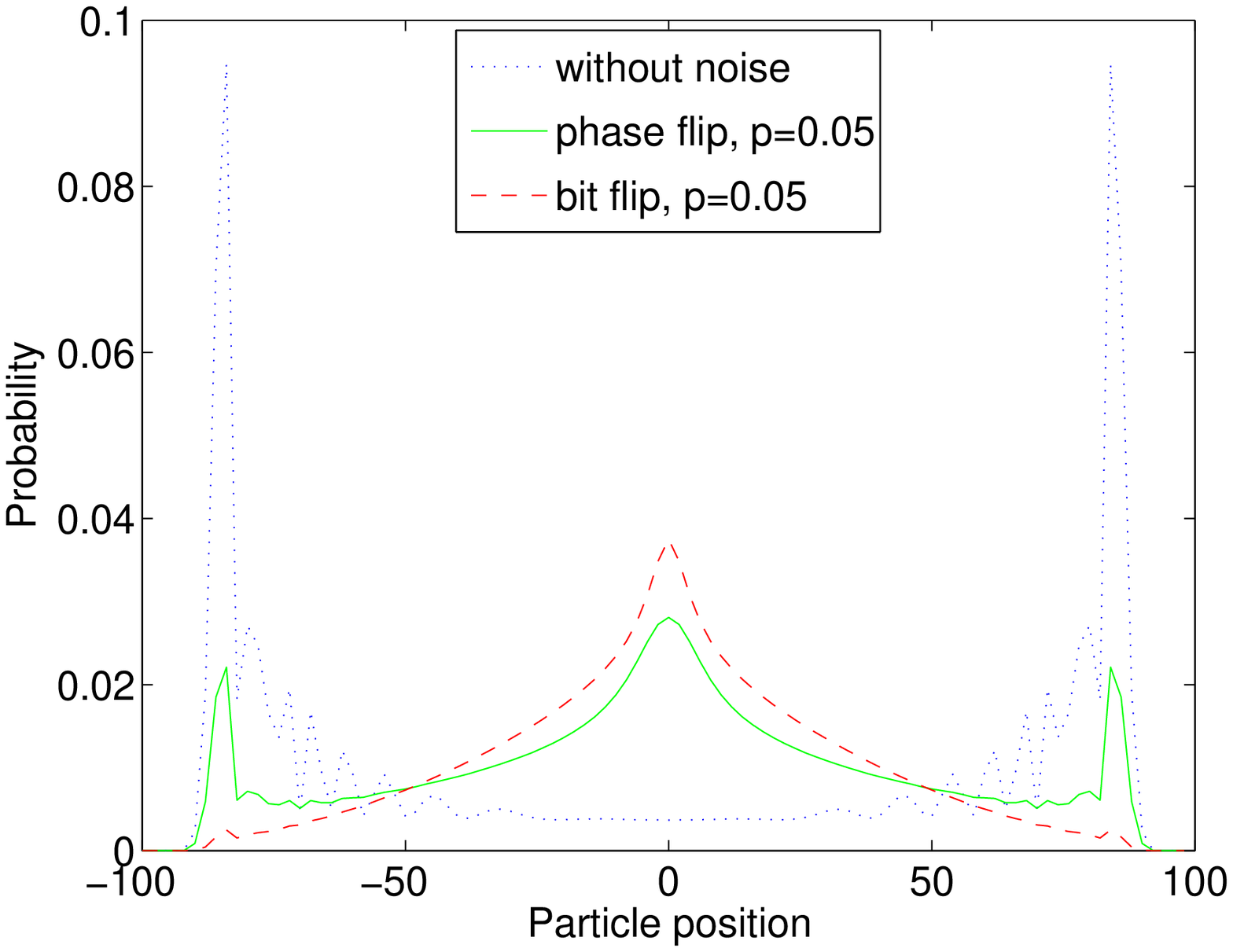}} \\
\subfigure[]{\includegraphics[width=7.2cm]{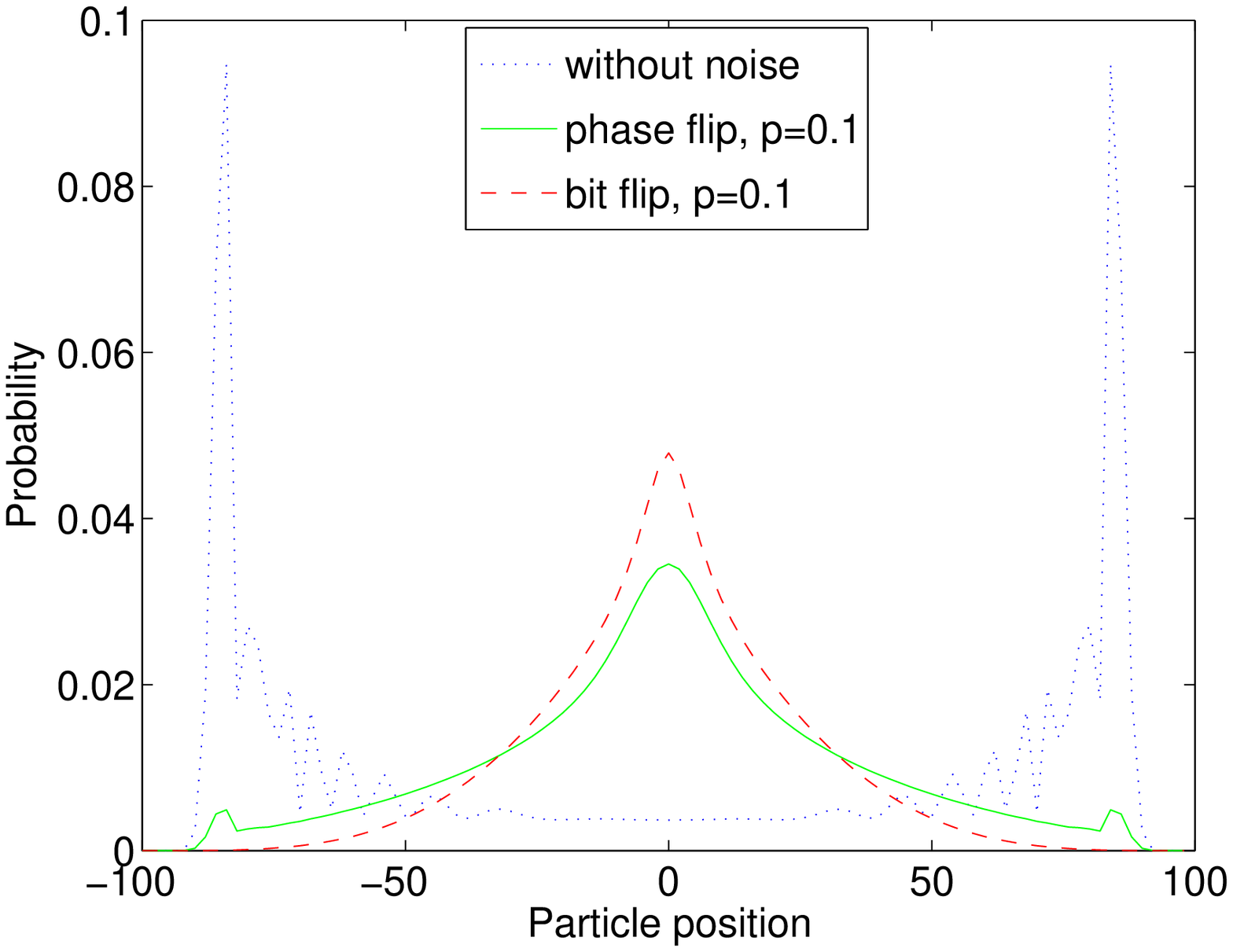}}
\hfill
\subfigure[]{\includegraphics[width=7.2cm]{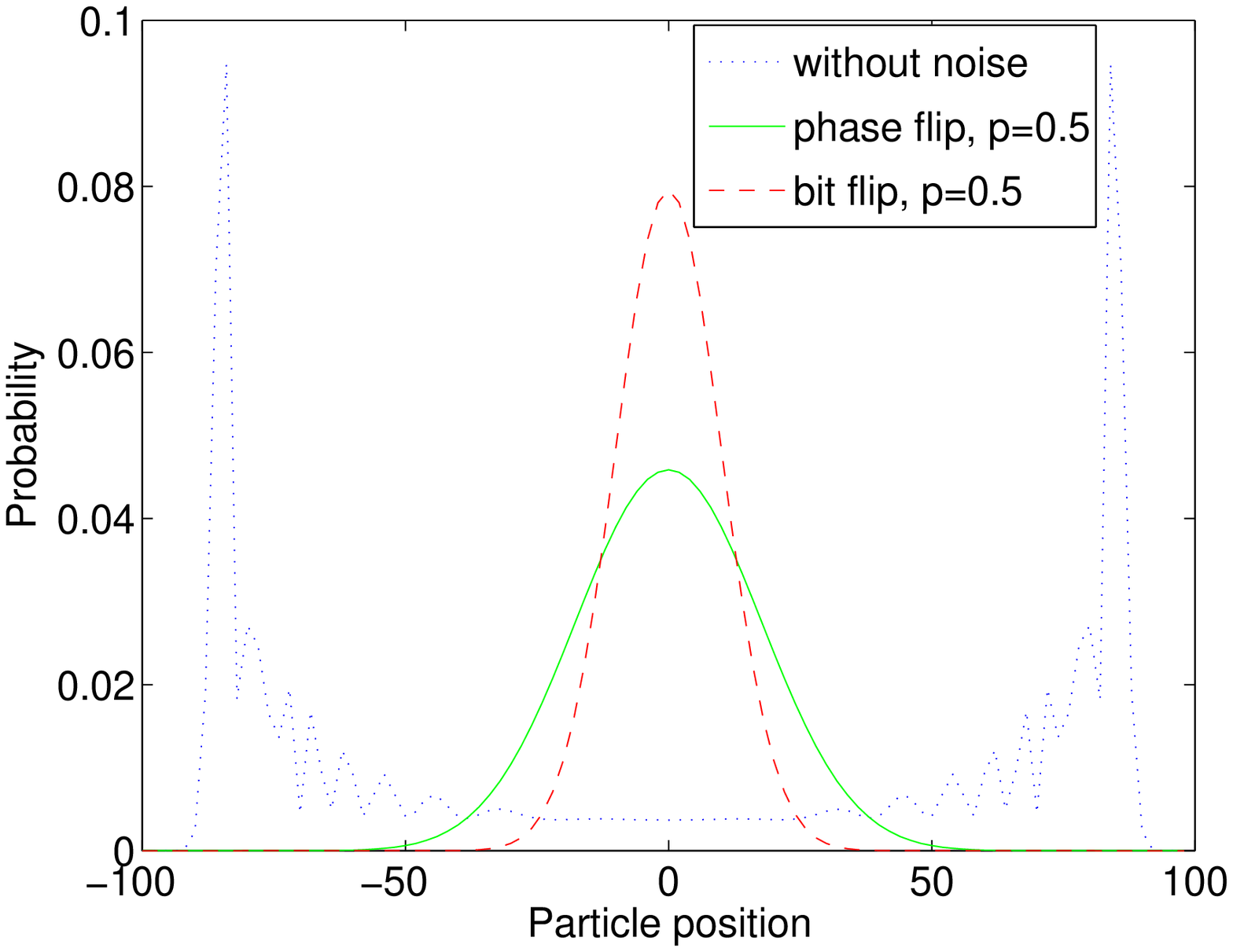}}
\caption[The  effect  of  environmental  decoherence on  the  position
probability  distribution of a quantum  walk with coin parameters  
$\theta = 30^{\circ}, \xi=\zeta=0$ subjected  to a
noisy  channel.] {The  effect  of  environmental  decoherence on  the  position
probability  distribution of a quantum  walk subjected  to a
noisy  channel. Coin  bias is  of the  form (\ref{U2}) with
$\theta = 30^{\circ}, \xi =\zeta=0$.   The noise is modeled as  a phase flip (solid line)   and  bit   flip   (dashed  line)   channel,  characterized   by (\ref{eq:phaseflip})  and  (\ref{eq:bitflip}), respectively,  at
various noise levels  $p$: (a) $p = 0.005$  (b) $p=0.05$; (c) $p=0.1$;
(d) $p=0.5$, which corresponds to a fully classical random walk. The distribution is for 100 steps.}
\label{fig:env30}
\end{center}
\end{figure}

\bc
\begin{figure}
\begin{center}
\includegraphics[width=8.6cm]{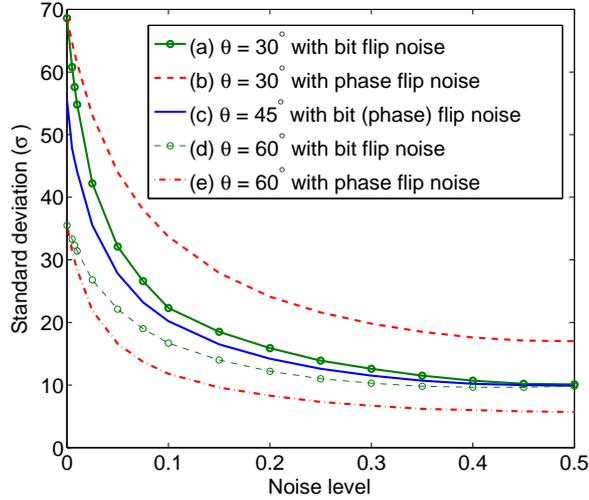}
\caption[Variation of  standard deviation  with noise level,  for both
phase  noise  and  bit flip  noise.]{Variation of  standard deviation  with noise level,  for both
phase  noise  and  bit flip  noise. (a) and (b) is for $\theta  =30^{\circ}$; (c) is for Hadamard walk ($\theta = 45^{\circ}$); (d) and (e) is for $\theta=60^{\circ}$ in the quantum coin operation $B_{0, \theta, 0}$.   In  the  classical  limit  of  $p=0.5$,  the  standard
deviation converges to a fixed  value for bit flip noise, irrespective
of  $\theta$, but  different  for phase  flip  noise. The  convergence
happens because, at maximum  bit flip noise ($p=0.5$), the measurement
outcome in  the computational basis is completely  randomized. 
On  the other  hand, the non-convergence in  the case of  phase flip noise  is due to  the fact
that the asymptotic  mixed state obtained via a  phase damping channel depends on the value of the $\theta$. The standard deviation was calculated for 100 steps of quantum walk.}
\label{fig:sd}
\end{center}
\end{figure}
\ec

\subsubsection{Phase damping and bit flip 
channels \label{sec:enva}}

In  studying the  status of  the walk  symmetries in  the  presence of
noise, it is advantageous  to employ the quantum trajectories approach
\cite{Bru02}.  This  simplifies the  description  of  an open  quantum
system in terms of a  stochastically evolving pure state, which allows
us to  adapt the symmetry  results for the  pure states, given  in the
preceding section, to mixed states.
\par
We  call the  sequence  of walk  step  operations,
\be
\widehat{X}  \equiv (SB_t)(SB_{t-1})\cdots(SB_{x})\cdots  (SB_1),
\ee
a  `quantum    trajectory'.    (More
precisely, a trajectory  refers to the sequence of  states produced by
these  operations,  for  which   the  above  serves  as  a  convenient
representation.) If all the  $B_{x}$'s with $x$ taking values from $1$ to $t$ are the same, then $\widehat{X}$ is the usual `homogeneous' quantum walk $\widehat{W}$. In general, the $B_{x}$'s may be different $U(2)$ operators, $B^{(\rm f)}$. More generally, each step of the walk  may include  generalized measurements  whose outcomes  are known (Section \ref{sec:envb}).   If each  walk  step  in $\widehat{X}$ is subjected to a fixed symmetry  operation $G$,  the result  is  a new
quantum trajectory        
\be
{\bf G}\widehat{X} \equiv (SB_t^{\star})(SB_{t-1}^{\star})\cdots  (SB_1^{\star}).
\ee
 We  have the following  generalization of  Theorem \ref{thm:bias}  to inhomogeneous
quantum walks on a line.

\begin{thm}
Given any quantum  walk trajectory $\widehat{X} = (SB_t)\cdots(SB_1)$,
the  symmetry  ${\cal  G}$   holds,  i.e.,  $\widehat{X}  \simeq  {\bf
G}\widehat{X}$. If  the operation ${\bf \Phi}$ (\ref{eq:phi}) is  restricted to ${\bf
Z}$, that is operation $Z$ at each step of the walk, then the symmetries hold even when 
some of the $S$'s are replaced
by $S^{\dag}$'s.
\label{thm:qutraj}
\end{thm}
{\bf Proof.}  In the proof of Theorem \ref{thm:bias}, we note that if,
in  each  step  of  the  walk,  we  alter  the  rotation  $B$  by  the
transformation  $G$,   the  proof   still  goes  through.    That  is,
$|\Psi_1\rangle                                                  \equiv
(SB_t)(SB_{t-1})\cdots(SB_1)|\alpha,\beta\rangle$  and $|\Psi_2\rangle
\equiv
(SB_t^{\star})(SB_{t-1}^{\star})\cdots(SB_1^{\star})|\alpha,\beta\rangle$
produce the same position distribution.
\par
Suppose  that  in  some  of  the  walk steps,  $S$  is  replaced  with $S^{\dag}$.  In place of (\ref{eq:bitobit}), we have
\begin{subequations}
\label{eq:bitobitb}
\begin{eqnarray}
\label{eq:bitobitb1}
|\Psi_1\rangle &=& 
(SB_{t})\cdots(S^{\dag}B_{x})\cdots(SB_1)|\alpha,\beta\rangle\nonumber\\
&=& \sum_{q_1,q_2,\cdots,q_t}
b_{q_t,q_{t-1}}\cdots b_{q_2,q_1}b_{q_1,\alpha}
|q_t,\beta+2(Q_1-Q_2) - (t_1- t_2)\rangle,\\
\label{eq:bitobitb2}
|\Psi_2\rangle &=& 
(SB_t^{(1)})\cdots(S^{\dag}B_{x}^{(1)})\cdots(SB_1^{(1)})|\alpha,\beta\rangle
\nonumber\\
&=& \sum_{q_1,q_2,\cdots,q_t}
b_{q_t,q_{t-1}}\cdots b_{q_2,q_1}b_{q_1,\alpha}
(-1)^{q_{t-1}+ \cdots + q_1 + \alpha}
|q_t,\beta+2(Q_1-Q_2) - (t_1-t_2)\rangle, \nonumber \\
\end{eqnarray}
\end{subequations}
where $Q_1 = \sum_k q_k$ for the $t_1$ steps $k$ where operator $S$ is
used, and  $Q_2 = \sum_l q_l$  for the $t_2$ steps  $l$ where operator
$S^{\dag}$ is  used.  Here $Q=Q_1+Q_2$ and  $t=t_1+t_2$.  Observe that
the  exponent   of  $(-1)$   is  effectively  evaluated   in  modulo-2
arithmetic. We  can thus  replace $Q$ by  $Q_1 -Q_2$ in  the exponent.
Following  the  argument  in  Theorem  \ref{thm:bias},  we  find  that
\be
\langle  a,b|\Psi_1\rangle = e^{i\Theta}  \langle a,b|\Psi_2\rangle,
\ee
where $\Theta = Q_1-Q_2-a+\alpha$.  \hfill $\blacksquare$
\bigskip

As a  corollary, the symmetries  ${\bf Z}$ and  ${\bf P R  X}$ hold
good because they reduce to special cases of ${\bf G}$.
A question of practical interest is whether
${\bf PRX}$ and ${\bf Z}$ are symmetries of a noisy quantum walk.
Suppose we are given a noise process ${\cal N}$ in the Kraus representation :
\begin{equation}
\label{eq:kraus}
\rho \longrightarrow {\cal N}(\rho) = \sum_{x=0}^{t-1} E_{x}\rho E^{\dag}_{x},
\hspace{0.5cm} \sum_{x} E^{\dag}_{x}E_{x}=\mathbb{I}.
\end{equation}
With the  inclusion of  noise, each step  of the quantum  walk becomes
augmented  to $(\Pi S  B_k)$, where  $\Pi$ is  a random  variable that
takes Kraus operator  values $E_{x}$. Thus, ${\cal N}$  corresponds to a
mixture  of upto  $t^t$ trajectories  or  `unravellings' 
\be
\widehat{X}_l
\equiv \left(\Pi(l_t)SB_t\right)\cdots\left(\Pi(l_1)SB_1\right),
\ee
each occurring with some probability  $p_l$, where $\sum_l p_l=1$.  If ${\bf
Z}$ and  ${\bf PRX}$ are symmetries of  an unraveling $\widehat{X}_l$,
then the  operations $\widehat{X}_l$ and  
\be
{\bf D}\widehat{X}_l \equiv
\left(\Pi(l_t)DUB_t\right)\cdots\left(\Pi(l_1)DUB_1\right),
\ee
 where ${\bf D}$ denotes  ${\bf  Z}$ or  ${\bf  PRX}$, must  yield the  same
position  probability  distribution.   In  the case  of  bit-flip  and
phase-flip channels,  there is a  representation in which  the $E_{x}$'s
are proportional to unitary operators. 
\par
\begin{thm}
If  trajectories  $\widehat{X}_l$  are  individually  symmetric  under
operation ${\bf G}$, then so  is any noisy quantum walk represented by
a collection $\{\widehat{X}_l, p_l\}$.
\label{thm:mix}
\end{thm}
{\bf Proof.}  The state  of the  system obtained via  ${\cal N}$  is a
linear  combination   (the  average)   of  states  obtained   via  the
$\widehat{X}_{x}$'s. Thus, the invariance of the $\widehat{X}_{x}$'s under
${\bf G}$ implies the invariance of the former.  \hfill $\blacksquare$
\bigskip

This result, together with those from the preceding Section, can
now be easily shown to imply that the symmetry ${\bf D}$ is preserved
in the case of phase-flip and bit-flip channels.

Decoherence via a purely dephasing channel, without any loss of
energy, can be modeled as a phase flip channel \cite{NC00}:
\begin{equation}
\label{eq:phaseflip}
{\cal E}(\rho) = (1-p)\rho + pZ\rho Z.
\end{equation}
An example of a physical process that realizes (\ref{eq:phaseflip}) 
is a two-level system interacting with its bath via
a quantum non-demolition (QND) interaction given by the Hamiltonian 
\begin{eqnarray}
{\bf H} &=& {\bf H}_S + {\bf H}_{R} + {\bf H}_{SR} \nonumber \\
{\bf H} &=&  {\bf H}_S + \sum\limits_k \hbar \omega_k \hat{b}^{\dagger}_k \hat{b}_k + {\bf H}_S 
\sum\limits_k g_k (\hat{b}_k+\hat{b}^{\dagger}_k) + {\bf H}^2_S \sum\limits_k 
{\frac {g^2_k}{\hbar \omega_k}}. \label{2a} 
\end{eqnarray} 
Here ${\bf H}_S$ stands for the Hamiltonian of the 
system (S), for our noise channels the system is only the particle (coin degree of freedom). ${\bf H}_R$ and ${\bf H}_{SR}$ are reservoir (R) and system-reservoir (SR) interaction, respectively.  Operators $\hat{b}$ and $\hat{b}^{\dag}$ are the annihilation and creation operators.
The last term on the RHS of (\ref{2a}) is a renormalization inducing `counter term'. Since $[{\bf H}_S, 
{\bf H}_{SR}]=0$, (\ref{2a}) is of QND type.

Following \cite{BS08} (apart from a change in notation
which switches $|0\rangle \longleftrightarrow |1\rangle$),
taking into account the effect of the
environment modeled as a thermal bath, 
the reduced dynamics of the system 
can be obtained, which can be described using Bloch vectors
as follows. Its action on an initial state
\begin{equation}
\rho_0 \equiv \begin{pmatrix}
\frac{1}{2}\left(1 + \langle \sigma_z(0) \rangle
\right) & \langle \sigma_-(0) \rangle \cr
\langle \sigma_+(0) \rangle & {\frac {1}{2}} \left(1 -
\langle \sigma_z(0) \rangle \right)
\end{pmatrix},
\label{eq:inista}
\end{equation}
is given in the interaction picture by
\begin{equation}
\label{eq:qnd2}
{\cal E}(\rho_0) = \begin{pmatrix}
\frac{1}{2}\left(1 + \langle \sigma_z(0) \rangle
\right) & \langle \sigma_-(0) \rangle e^{-(\hbar\omega)^2\gamma(t)} \cr
\langle \sigma_+(0) \rangle e^{-(\hbar\omega)^2\gamma(t)} & 
{\frac{1}{2}} \left(1 -
\langle \sigma_z(0) \rangle \right)
\end{pmatrix}.
\end{equation}
Here $\sigma_{+}$ and $\sigma_{-}$ are the standard raising and lowering operators given by
\be
\sigma_{+} = |1\rangle \langle 0| = \frac{1}{2} (\sigma_{x} + i \sigma_{y}) ~~~;~~~ \sigma_{-} = |0\rangle \langle 1| = \frac{1}{2} (\sigma_{x} - i \sigma_{y})
\ee 
where $\sigma_{x}$, $\sigma_{y}$ and $\sigma_{z}$ are the Pauli spin operator in x, y and z directions, respectively. The initial state (\ref{eq:inista}) may be mixed. (The derivation 
of the superoperator ${\cal E}$ in terms of environmental parameters for
the pure state case, given explicitly in \cite{BS08}, 
is directly generalized to the case of an arbitrary mixture of
pure states, since the environmental parameters are assumed to be
independent of the system's state.)

Comparing (\ref{eq:qnd2}) with (\ref{eq:phaseflip}) allows us
to relate the noise level $p$ in terms of physical parameters.
In particular,
\begin{equation}
p = \frac{1}{2}\left(1 - \exp\left[-(\hbar\omega)^2\gamma(t)\right]\right).
\end{equation}
When $\gamma(t)\approx0$ (either because the coupling with the environment is very weak or the interaction time is short or the temperature is low),
$p\approx0$, tending towards the noiseless case. On the other hand,
under strong coupling, $\gamma(t)$ is arbitrarily large, and $p\rightarrow1/2$,
the maximally noisy limit.
The result of implementing channel (\ref{eq:phaseflip}) is to
drive the position
probability distribution towards a classical Gaussian
pattern \cite{KS05}. The effect of increasing phase noise in the
presence on quantum walk is depicted
in Figure \ref{fig:env60}, for the case of $\theta=60^{\circ}$,
and in Figure \ref{fig:env30}, for the case of $\theta=30^{\circ}$.
The onset of classicality is observed in the Gaussianization
of the probability distribution. This is reflected also in the fall of
standard deviation, as shown in Figure \ref{fig:sd}.
\par
Decoherence can also be introduced by another noise model,
the bit flip channel \cite{NC00},
\begin{equation}
\label{eq:bitflip}
{\cal E}(\rho) = (1-p)\rho + pX\rho X.
\end{equation}
As with the phase damping channel,
the bit flip channel also
drives the probability distribution towards a classical, Gaussian
pattern, with increasing noise \cite{KS05}. 
The effect of increasing bit flip noise in the
presence of biased walk is depicted
in Figures \ref{fig:env60} and \ref{fig:env30}.
Here again, the onset of classicality is observed in the Gaussianization
of the probability distribution, as well as in the fall of
standard deviation, as shown in Figure \ref{fig:sd}.

A difference in the classical limit of these  two noise processes, as observed  in  Figure \ref{fig:sd},   is  that  whereas  the  standard deviation (in fact, the distribution) is unique in the case of the bit flip channel irrespective of $\theta$, in  the case of  phase flip noise, the classical  limit distribution is  $\theta$ dependent. This  is because phase flip  noise leads,  in the Bloch  sphere picture, to  a coplanar evolution of states towards the $\sigma_z$ axis. Thus all initial pure states corresponding to a  fixed $\theta$ evolve asymptotically to the same mixed state \cite{NC00, BS08}.  This   also   explains  the contrasting behavior of bit flip  and phase flip noise with respect to $\theta$,   as   seen   by   comparing Figures \ref{fig:env60}   and
\ref{fig:env30}.
\begin{figure}
\begin{center}
\includegraphics[width=8.6cm]{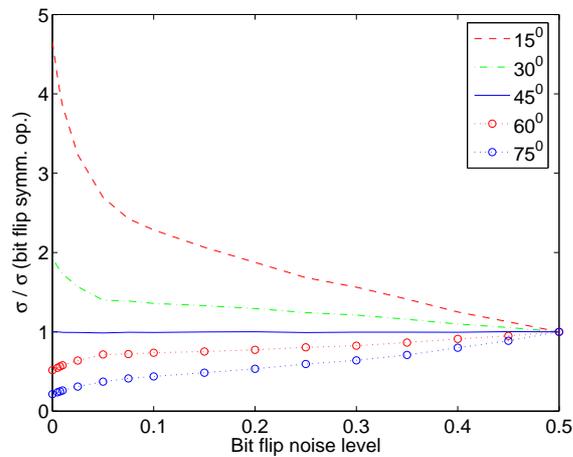}
\caption[Variation  of the  ratio  of standard  deviation without  any
symmetry operation to the  bit flip symmetry operation with increasing
bit flip noise level.]{Variation  of the  ratio  of standard  deviation without  any
symmetry operation to the  bit flip symmetry operation with increasing
bit flip noise level. The standard deviation was calculated for 100 steps of walk. }
\label{fig:sd2}
\end{center}
\end{figure}
\begin{figure}
\begin{center}
\includegraphics[width=8.6cm]{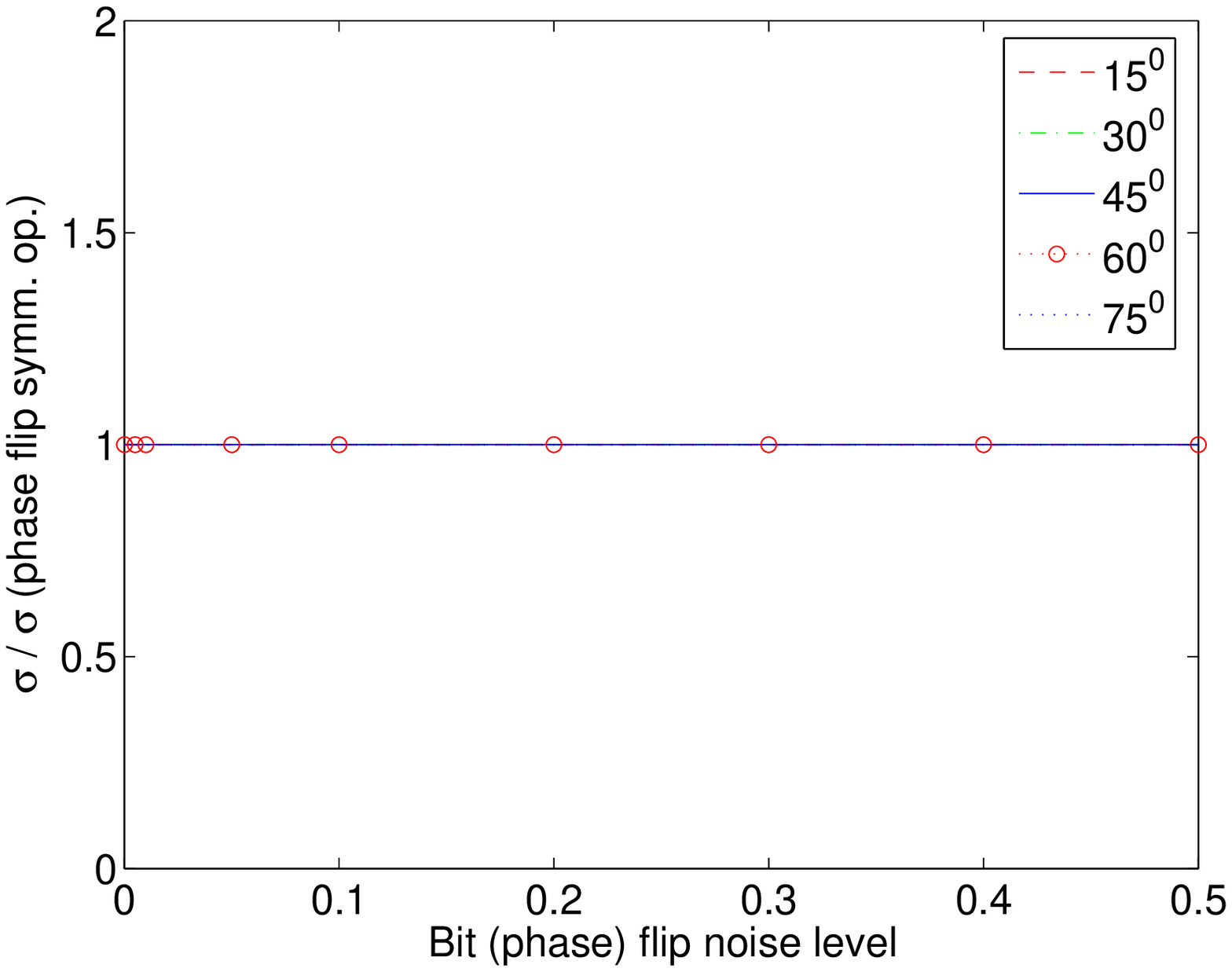}
\caption[Variation  of the  ratio  of standard  deviation without  any
symmetry  operation   to  the  phase  flip   symmetry  operation  with
increasing bit flip (phase flip) noise level.]{Variation  of the  ratio  of standard  deviation without  any
symmetry  operation   to  the  phase  flip   symmetry  operation  with
increasing bit flip (phase flip) noise level. The standard deviation was calculated for 100 steps of walk.}
\label{fig:sd3}
\end{center}
\end{figure}
\begin{figure}
\begin{center}
\includegraphics[width=8.6cm]{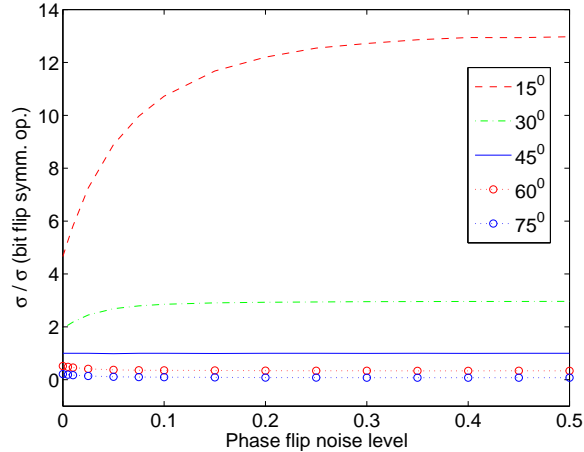}
\caption[Variation  of the  ratio  of standard  deviation without  any
symmetry operation to the bit flip symmetry operation with increasing phase
flip noise level.]{Variation  of the  ratio  of standard  deviation without  any
symmetry operation to the bit flip symmetry operation with increasing phase
flip noise level. The standard deviation was calculated for 100 steps of walk.}
\label{fig:sd5}
\end{center}
\end{figure}

Representing  the  walk   distribution  by  its  standard  deviation
$\sigma$, we  may describe symmetry  by the ratio of  $\sigma$ without
the  symmetry  operation  to  $\sigma$ with  the  symmetry  operation.
Figure \ref{fig:sd2} depicts  the  symmetry operation  ${\bf X}$  for
various  bit  flip  noise   levels.  The  convergence  of  the  curves
representing  various  $\theta$'s is  a  consequence  of the  complete
randomization of the measured  bit outcome in the computational basis.
This implies that  although {\bf X} is not a  symmetry of non-Hadamard walk ($\theta \neq \pi/4$), it does  become one in the  fully classical limit. On  the other hand, the  symmetries ${\bf P  R X}$  remain unaffected  by noise. We note
that, since the quantum walk  here is evolved from the symmetric state
$\frac{1}{\sqrt{2}} \left (|0\rangle +  i|1\rangle\right )$, and the bit  flip and phase  flip noise are
not  partial  to  the   state  $|0\rangle$  or  $|1\rangle$,  this  is
equivalent to setting  ${\bf P}$ to $1$, which  explains the fact that
the distributions  in Figures \ref{fig:env60}  and \ref{fig:env30} are
spatially symmetric.  Thus, ${\bf R  X}$ by itself becomes  a symmetry
operation,   which  is  manifested   in  the   fact  that   in  Figure \ref{fig:sd2} the values of the curve for complementary angles are the
inverse of each other. Figure \ref{fig:sd3} shows that for either of the two noises, ${\bf Z}$ is a walk symmetry.

Figure \ref{fig:sd5} depicts the symmetry of the ${\bf R X}$ operation
at all  phase flip noise  levels, as evident  from that fact  that the
values of the  curve for complementary angles are  the inverse of each
other.  From  Figures \ref{fig:sd2}, \ref{fig:sd3} and \ref{fig:sd5}, we  note that  for the  Hadamard walk,  all three  symmetry operations
${\bf Z}, {\bf  X}$ and ${\bf R}$ are  individually preserved. This is
expected because  here ${\bf P}=1$  as stated earlier, ${\bf  R}=1$ by
definition, so that the symmetry of ${\bf PRX}$ implies ${\bf X}=1$.
\par
With the non-Hadamard coin operation and an initial
arbitrary state, the full symmetries ${\bf Z}$ and ${\bf P R X}$ would be required, as proved by the
following theorem.
\begin{thm}
The operations ${\bf PRX}$ and ${\bf Z}$
are symmetries for the phase-flip and bit-flip
channels.
\label{thm:phasebit}
\end{thm}
{\bf Proof.} We may look upon the phase flip channel 
(\ref{eq:phaseflip}) as a probabilistic
mixture (in the discretized walk model) of $2^t$ quantum trajectories
with $\Pi \in \{\mathbb{I}, Z\}$. 
By virtue of Theorem \ref{thm:mix}, it suffices to show that any given
unraveling is invariant under ${\bf Z}$ and ${\bf PRX}$.
Consider an unraveling 
\be
\widehat{X}^1 \equiv \cdots(ISB)(ZSB)(ISB)\cdots
= \cdots(SB)(ZSB)(SB)\cdots.
\ee
This is the same as: $\cdots(SB)(SB^{\prime})(SB)\cdots$, where
$B^{\prime} = ZB$, noting that $Z$ commutes with $S$. Now,
\be
{\bf Z}\widehat{X}^1 = \cdots(IZSB)(ZZSB)(IZSB)\cdots 
= \cdots(ZSB)(ZSB^{\prime})(ZSB)\cdots,
\ee 
which, by Theorem \ref{thm:qutraj},
is equivalent to $\widehat{X}^1$.
Also, by Theorem \ref{thm:qutraj}
\begin{eqnarray}
{\bf PRX}\widehat{X}^1 
&=& \cdots(PRX~SB)(Z~PRX~SB)(PRX~SB)\cdots  \nonumber \\
&=& \cdots(PRX~SB)(PR~ZX~SB)(PRX~SB)\cdots  \nonumber \\
&=& \cdots(PRX~SB)(PR~ZXZ~ZSB)(PRX~SB)\cdots  \nonumber \\
&=& \cdots(PRX~SB)(PR~(-X)~SB^{\prime})(PRX~SB)\cdots, 
\end{eqnarray}
which, by Theorem \ref{thm:qutraj},
is equivalent to $\widehat{X}^1$, since an overall 
phase factor of $\pm 1$
is irrelevant. Thus, the phase-flip channel is symmetric
with respect to the operations ${\bf Z}$ and ${\bf PRX}$.

Regarding the bit-flip channel (\ref{eq:bitflip}):
as in the above case, consider an unraveling 
\be
\widehat{X}^2 \equiv \cdots(SB)(XSB)(SB)\cdots.
\ee
This is the same as: $\cdots(SB)(S^{\dag}B^{\prime\prime})(SB)\cdots$, where,
as may be seen by direct calculation,
$B^{\prime\prime} = XB$. Now,
\begin{eqnarray}
{\bf Z}\widehat{X}^2  
&=& \cdots(Z~SB)(X~Z~SB)(Z~SB)\cdots \nonumber \\
&=& \cdots(Z~SB)(X~ZX~XSB)(Z~SB)\cdots \nonumber \\
&=& \cdots(Z~SB)((-Z)~S^{\dag}B^{\prime\prime})(Z~SB)\cdots,
\end{eqnarray}
which, by Theorem \ref{thm:qutraj},
is equivalent to $\widehat{X}^2$, since an overall 
phase factor of $\pm 1$
is irrelevant. 
Further,
\begin{eqnarray}
{\bf PRX}\widehat{X}^2  
&=& \cdots(PRX~SB)(X~PRX~SB)(PRX~SB)\cdots \nonumber \\
&=& \cdots(PRX~SB)(PRX~X~SB)(PRX~SB)\cdots \nonumber \\
&=& \cdots(PRX~SB)(PRX~S^{\dag}B^{\prime\prime})(PRX~SB)\cdots,
\end{eqnarray}
by Theorem \ref{thm:qutraj},
is also equivalent to $\widehat{X}^2$.
\hfill $\blacksquare$
\bigskip

\subsubsection{Generalized
amplitude damping channel \label{sec:envb}}

Here we study the behavior of quantum walk subjected to a generalized
amplitude damping (with temperature
$T \ge 0$), which would reduce at $T=0$ to the
amplitude damping channel.
As an example of a physical process that realizes the generalized
amplitude damping channel, we consider a two-level system interacting
with a reservoir of harmonic oscillators, with the system-reservoir
interaction being dissipative
and of the weak Born-Markov type \cite{BP02,SB07} leading to a standard
Lindblad equation, which in the interaction
picture has the following form \cite{SB08} 
\begin{equation}
\frac{d}{dt}\rho^s(t) = \sum_{k=1}^2\left(
2 {\mathcal R}_k\rho^s {\mathcal R}^{\dag}_k - {\mathcal R}_k^{\dag} {\mathcal R}_k\rho^s - \rho^s {\mathcal R}_k^{\dag}{\mathcal R}_k\right),
\end{equation}
where 
\be
{\mathcal R}_1 = (\gamma_0(N_{\rm th}+1)/2)^{1/2}{\mathcal R}~~~~; ~~~~{\mathcal R}_2 = (\gamma_0N_{\rm th}/2)^{1/2}{\mathcal R}^{\dag}
\ee
and 
\be
N_{\rm th} = (\exp(\hbar\omega/k_B T) - 1)^{-1},
\ee
is the Planck distribution giving the number of thermal
photons at the frequency $\omega$, and $\gamma_0$ is the
system-environment coupling constant. Here
\be
{\mathcal R} = \sigma_-\cosh(r) + e^{i\delta}\sigma_+\sinh(r),
\ee

and the quantities $r$ and $\delta$ are the environmental squeezing parameters and
\be
\sigma_{\pm} = \frac{1}{2}\left(\sigma_x \pm i\sigma_y\right).
\ee
For the generalized amplitude damping channel, we set $r = \delta = 0$.
If $T=0$, so that $N_{\rm th}=0$, then ${\mathcal R}_2$ vanishes, and a single
Lindblad operator suffices.
\begin{figure}
\begin{center}
\subfigure[]{\includegraphics[width = 7.2cm]{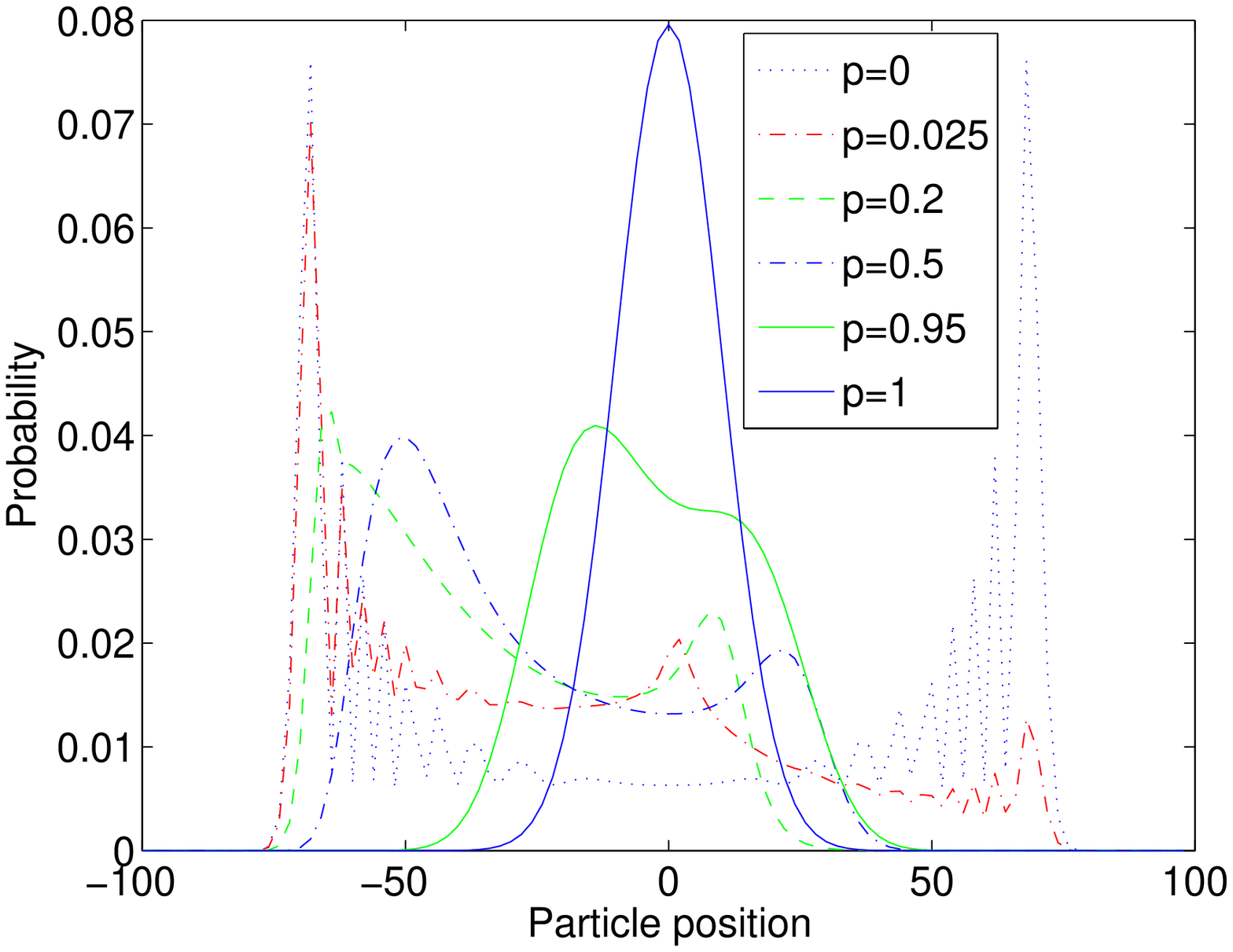}}
\subfigure[]{\includegraphics[width=7.2cm]{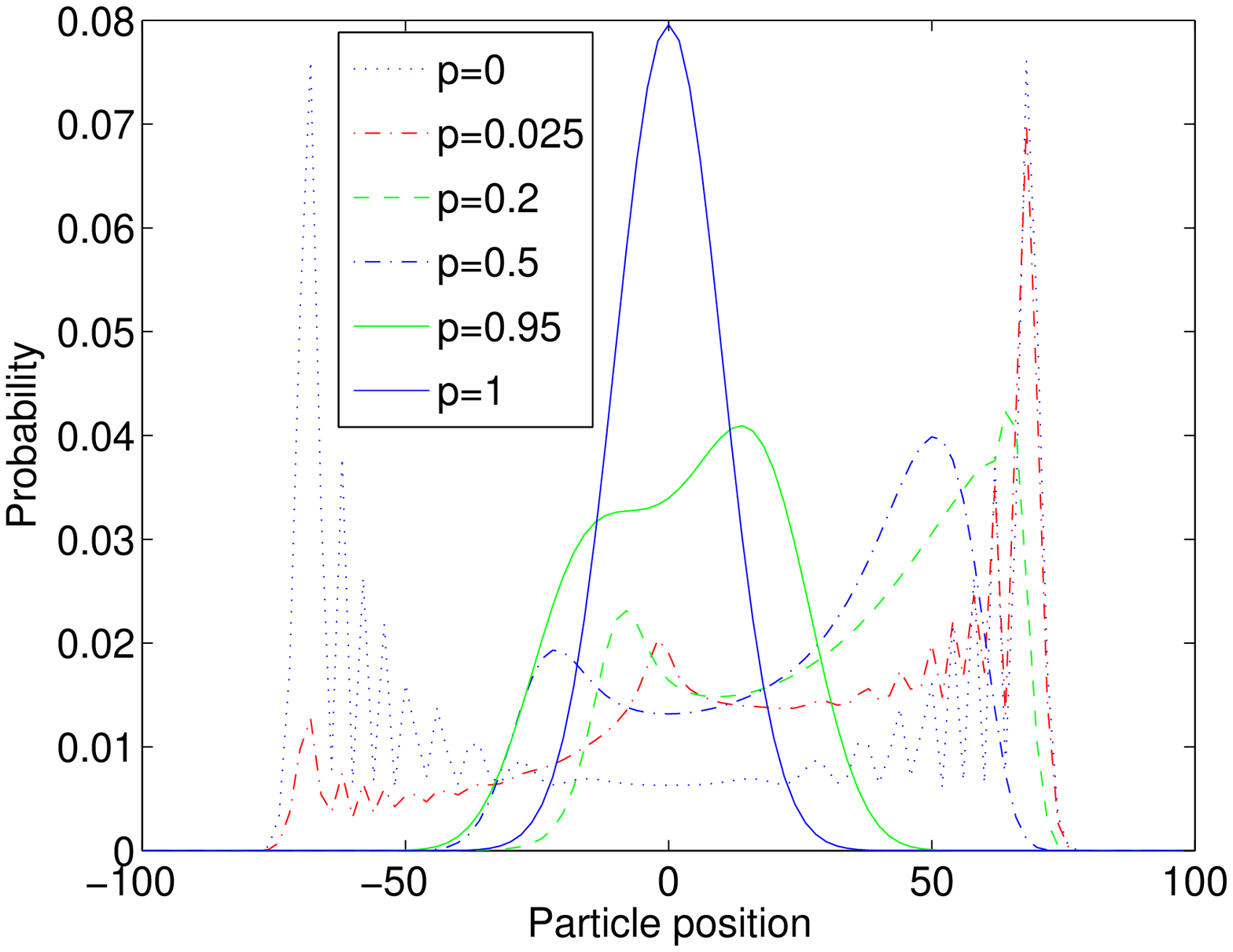}}
\caption[Amplitude damping channel acting on a Hadamard walk at temperature $T=0$. The distribution corresponding to intermediate values of $p$ clearly show the breakdown of the ${\bf R X}$ symmetry.
However, the extended symmetry, ${\bf P R X}$ (where ${\bf P}$ stands for parity operation (spatial inversion)) holds good.]{Amplitude damping channel acting on a Hadamard walk at temperature $T=0$. The distribution corresponding to intermediate 
values of $p$ clearly show the breakdown of the ${\bf R X}$ symmetry.
However, the extended symmetry, ${\bf P R X}$ (where ${\bf P}$ stands
for parity operation (spatial inversion)) holds good. This is observed
at all temperatures. (a) Probability distribution of finding the particle on which
amplitude damping  channel is acting.  This shows that
even at $T=0$, for sufficiently high coupling, the distribution turns
classical. (b) Amplitude damping channel with a bit flip symmetry.}
\label{fig:ampdamp}
\end{center}
\end{figure}
\begin{figure}
\begin{center}
\subfigure[]{\includegraphics[width=7.2cm]{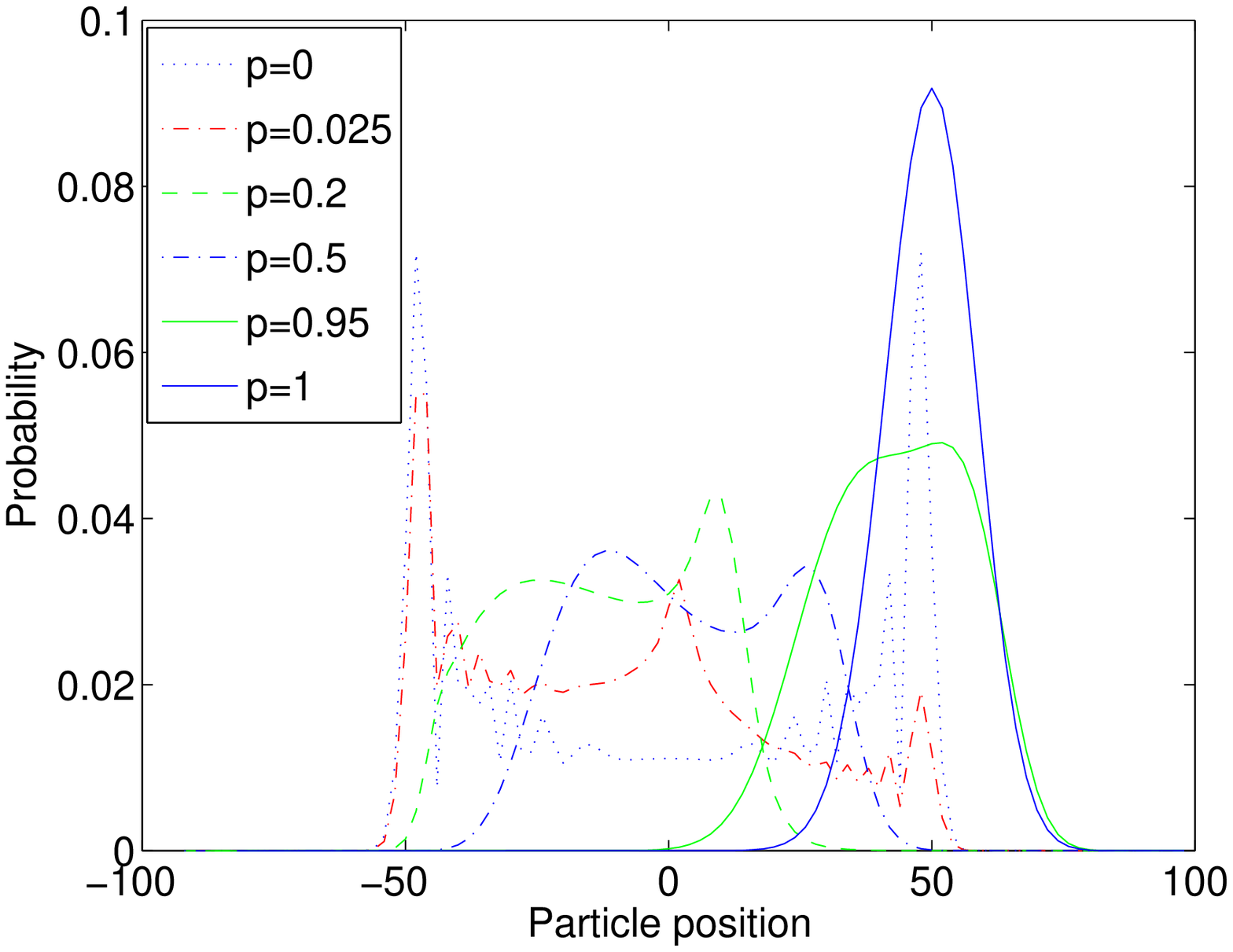}}
\hfill
\subfigure[]{\includegraphics[width=7.2cm]{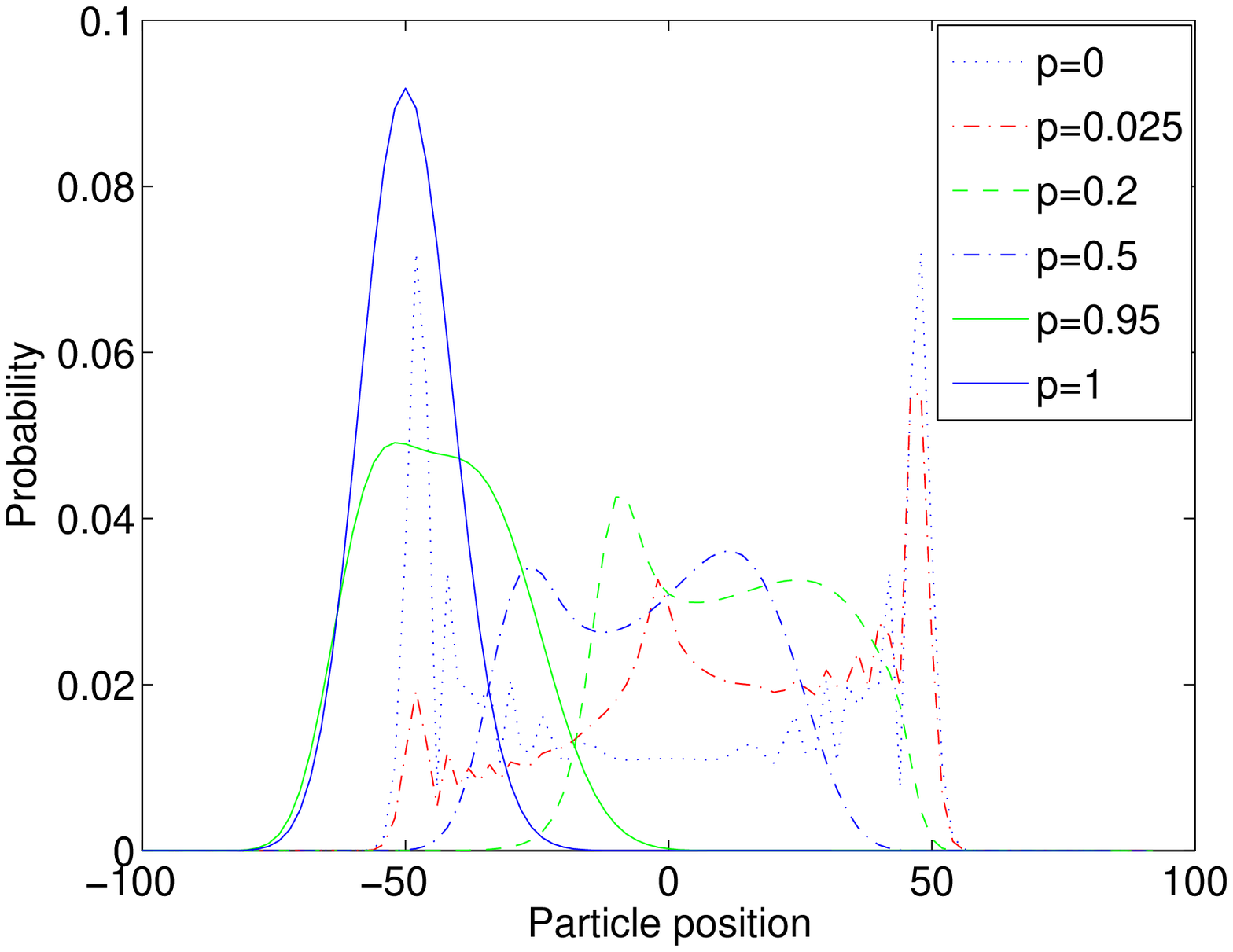}}
\caption[${\bf PRX}$ symmetry in quantum walk (with $\theta =60^{\circ}$ and $\theta =30^{\circ}$)
subjected to amplitude damping ($T=0$).]{${\bf PRX}$ symmetry seen to hold in quantum walk (with $\theta =60^{\circ}$ and $\theta =30^{\circ}$) subjected to amplitude damping ($T=0$). The two cases are spatial inversions of each other. This holds for a generalized amplitude damping
at any temperature. (a) walk with $\theta = 60^{\circ}$ ; (b) $\theta = 30^{\circ}$ and bit flip.}
\label{fig:ampdamp34}
\end{center}
\end{figure}
\begin{figure}
\begin{center}
\subfigure[]{\includegraphics[width=7.2cm]{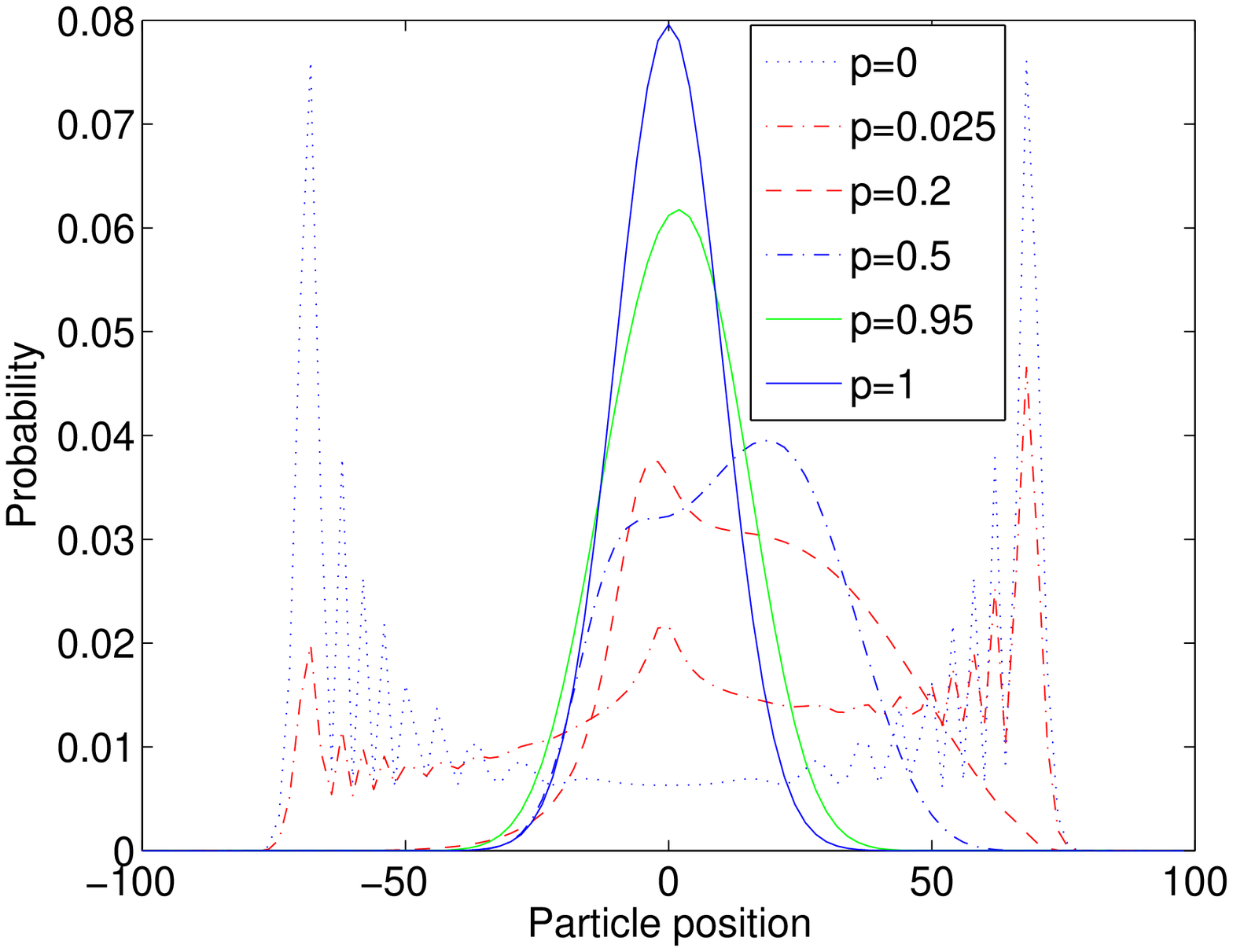}}
\subfigure[]{\includegraphics[width=7.2cm]{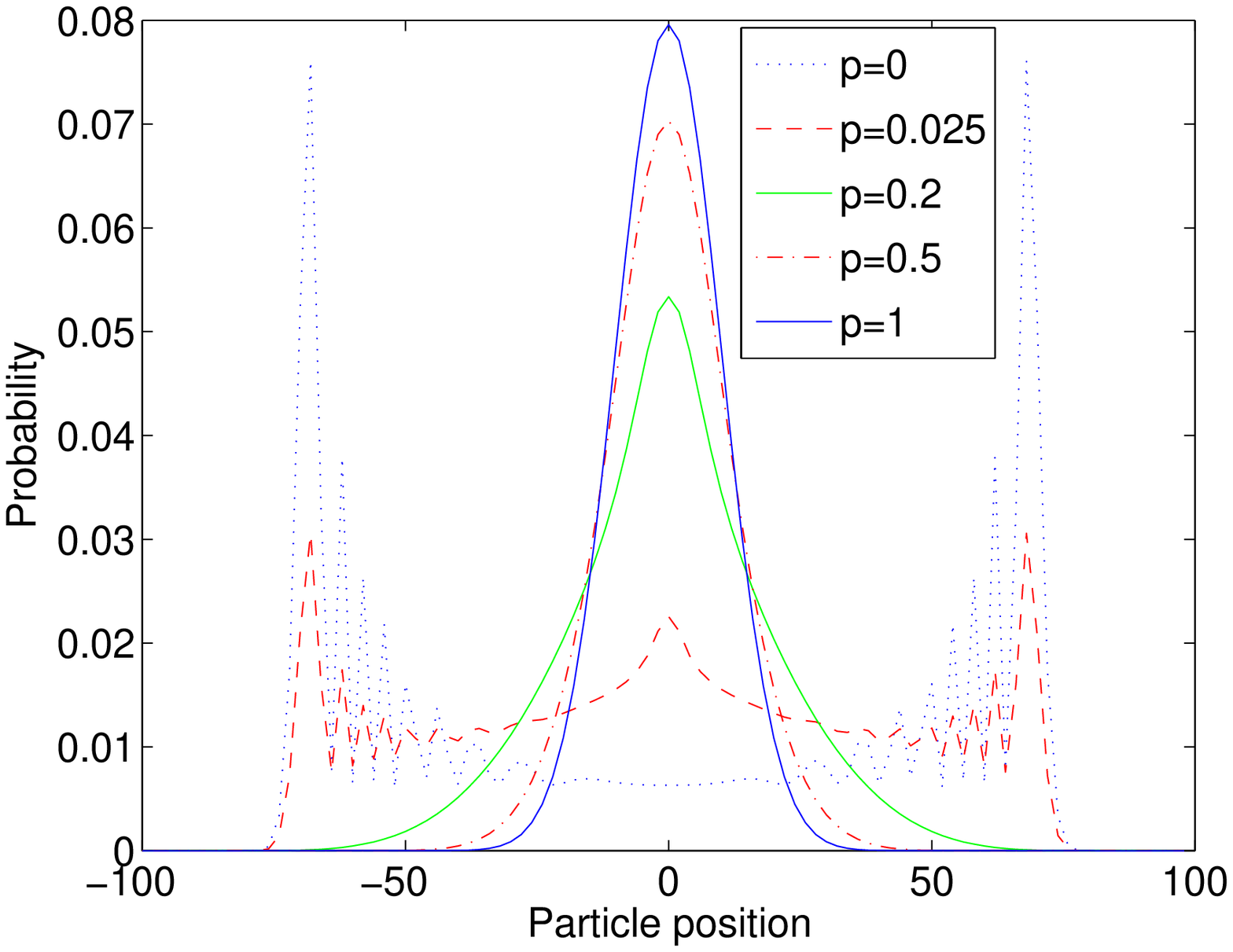}}
\caption[Onset of classicality accentuated in
a Hadamard walk subjected to generalized amplitude damping followed by bit flip with increasing temperatures.] {Onset of classicality is seen to be accentuated in
a Hadamard walk subjected
to generalized amplitude damping followed by bit flip with increasing temperatures (without bit flip the damping would have been to the right).
Figure \ref{fig:ampdamp}(a) depicts the $T=0$ case
($\chi=1$ in (\ref{eq:bma2_lambda})). 
(a) Finite temperature corresponding to $\chi=0.75$ 
(b) $T = \infty$, corresponding to $\chi=0.5$. 
It may be noted that even at $T=\infty$, for 
sufficiently small coupling the distribution remains non-classical.}
\label{fig:ampdamp12}
\end{center}
\end{figure}
\begin{figure}
\bc
\includegraphics[width=8.0cm]{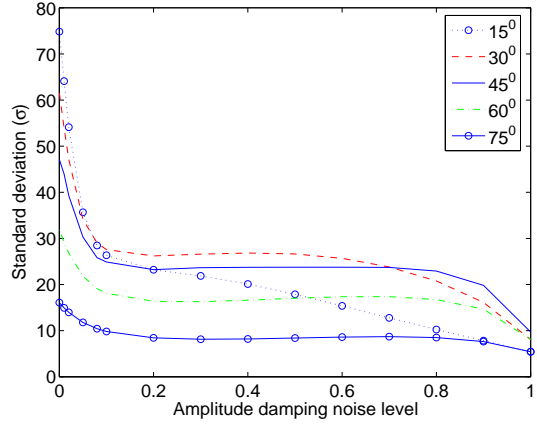}
\caption[Variation of standard  deviation with amplitude damping noise
level for various value of $\theta$.] {Variation of standard  deviation with amplitude damping noise
level for various value of $\theta$, $15^{\circ}$, $30^{\circ}$, 
$45^{\circ}$, $60^{\circ}$ and $75^{\circ}$. Note that the
standard deviation for complementary
angles converge to the same value.}
\label{fig:ampdamp5}
\ec
\end{figure}
The generalized amplitude damping channel is characterized by
the following Kraus operators,
\begin{eqnarray}
\label{eq:gbmakraus}
\begin{array}{ll}
E_0 \equiv \sqrt{\chi}\left[\begin{array}{ll} 1 & 0 \\ 0 & 
\sqrt{1-p(t)}
\end{array}\right]; ~~~~ &
E_1 \equiv \sqrt{\chi}\left[\begin{array}{ll} 0 & \sqrt{p(t)} \\ 0 & 0
\end{array}\right], \\
E_2 \equiv \sqrt{1-\chi}\left[\begin{array}{ll} 
\sqrt{1-p(t)} & 0 \\ 0 & 1
\end{array}\right]; ~~~~ &
E_3 \equiv \sqrt{1-\chi}\left[\begin{array}{ll} 0 & 0 \\ \sqrt{p(t)} & 0
\end{array}\right],  
\end{array}
\end{eqnarray} 
where 
\begin{equation}
\label{eq:bma2_lambda}
p(t) \equiv 1 - e^{-\gamma_0(2N_{\rm th} +1) t};\hspace{1.0cm} 
\chi \equiv \frac{1}{2}\left[1 + \frac{1}{2N_{\rm th} +1}\right].
\end{equation}
When $T = 0$, $\chi=1$, and for $T \rightarrow \infty$, $\chi=1/2$.

The density operator at a future time can be obtained as
\cite{SB08}
\begin{equation}
\label{eq:bmrhos}
\rho^s (t) = \begin{pmatrix} {\frac{1}{2}} (1 + A_1) & A_2
\cr A_2^*  & 
{\frac {1}{ 2}} (1 - A_1)
\end{pmatrix}, 
\end{equation}
where
\begin{equation}
A_1 \equiv \langle\sigma_z(t)\rangle
= e^{-\gamma_0 (2N_{\rm th} + 1)t} \langle 
\sigma_z (0) \rangle - {\frac{1}{(2N_{\rm th} + 1)}} 
\left(1 - e^{-\gamma_0 (2N_{\rm th} + 1)t} 
\right), \label{4m} 
\end{equation}
\begin{equation}
A_2 = e^{-{\frac{\gamma_0} {2}}(2N_{\rm th} + 1)t}
\langle \sigma_- (0) \rangle \rangle. \label{4n}
\end{equation}
Figures \ref{fig:ampdamp}, \ref{fig:ampdamp34} and \ref{fig:ampdamp12}
depict the onset of classicality with increasing coupling strength
(related to $p$) and temperature (coming from $\chi$).
Figure \ref{fig:ampdamp}, which shows the effect of an amplitude
damping channel on a Hadamard walk at zero temperature, illustrates
the breakdown of ${\bf R X}$ symmetry even though the initial
state is $\frac{1}{\sqrt 2}(|0\rangle + i|1\rangle)$. This is because,
in contrast to the phase-flip
and bit-flip channels, the generalized amplitude damping is not
symmetric towards the states $|0\rangle$ and $|1\rangle$.
However, the extended symmetry ${\bf PRX}$ is preserved both for
Hadamard as well as quantum walks with $\theta \neq \pi/2$, as seen
from Figures \ref{fig:ampdamp} and \ref{fig:ampdamp34}, respectively.

From  Figures \ref{fig:ampdamp}(a) and  \ref{fig:ampdamp12}(a,b), the
onset  of classicality  with increasing  temperature is  clearly seen.
Figure \ref{fig:ampdamp5} presents  the standard deviation for quantum
walks on  a line with  various biases, subjected to  amplitude damping
noise.  The  standard  deviation  for  complementary  angles  ($\theta
\leftrightarrow \pi/2-\theta$)  is seen to converge to  the same value
in   the  fully   classical  limit.    This  may   be   understood  as
follows. First,  we note that since  ${\bf PRX}$ is a  symmetry of the
quantum walk,  and the  effect of ${\bf  P}$ does  not show up  in the
standard  deviation  plots,  ${\bf  RX}$  by  itself  is  an  apparent
symmetry.   Further, in  the classical  limit the  measurement outcome
being  a  unique  asymptotic  state for  the  (generalized)  amplitude
damping channel, effectively ${\bf X} \simeq 1$, which makes ${\bf R}$
a symmetry operation.

The following theorem generalizes Theorem \ref{thm:qutraj} to an open
system subjected to a generalized amplitude damping channel.
\begin{thm}
The  operations ${\bf  Z}$  and  ${\bf PRX}$  are  symmetries for  the
generalized amplitude damping channel.
\label{thm:ampdamp}
\end{thm}
{\bf Proof.} By  virtue of Theorem \ref{thm:mix}, it  suffices to show
that  any given  unraveling is  invariant under  ${\bf Z}$  and ${\bf
PRX}$. Consider an unraveling
\begin{subequations}
\begin{eqnarray}
\label{eq:b} 
\widehat{X}^3 &\equiv& \cdots (E_0 SB)(E_1 SB)(E_2 SB)(E_3 SB)\cdots
 \\
&\equiv& \cdots (SB_{(0)})(S^{\dag}B_{(1)})(SB_{(2)})(S^{\dag}B_{(3)})
\cdots, \label{eq:b2}
\end{eqnarray}
\end{subequations}
where the non-unitary matrices are given by $B_{(x)} = E_{x}B$.
Now, 
\begin{eqnarray}
\label{eq:zb}
{\bf Z}\widehat{X}^3 
&=&  \cdots(E_0 Z SB)(E_1 ZSB)(E_2 ZSB)(E_3 ZSB)\cdots \nonumber \\
&=& \cdots (Z E_0 SB)((-Z) E_1 SB)(Z E_2 SB)((-Z)E_3 SB)\cdots \nonumber \\
&=& \cdots(Z SB_{(0)})((-Z) S^{\dag}B_{(1)})
(ZSB_{(2)})((-Z) S^{\dag}B_{(3)})
\cdots \nonumber \\
&=& \cdots(SB_{(0)}^{(1)})(-S^{\dag}B_{(1)}^{(1)})
(SB_{(2)}^{(1)})(-S^{\dag}B_{(3)}^{(1)}) \cdots .
\end{eqnarray}
Ignoring  the  overall  $\pm  1$  factor  in (\ref{eq:zb}),  and
comparing  it  with (\ref{eq:b}),  and  noting  that  that  the
derivation of  the proof of  Theorem \ref{thm:qutraj} did  not require
the matrices  $B_j$ to  be unitary, we  find along similar  lines that
${\bf Z}\widehat{X}^3$ is equivalent to $\widehat{X}^3$.

The following may be directly verified
\begin{subequations}
\label{eq:bprx}
\begin{eqnarray}
{\bf PRX}\widehat{X}^3 
&=&\cdots(E_0 PRX SB)(E_1 PRX SB)(E_2 PRX SB)(E_3 PRX SB) \cdots ~~~~~~~~\\
&=&\cdots(E_0 SB^{(2)})(E_1 SB^{(2)})(E_2 SB^{(2)})(E_3 SB^{(2)}) 
\cdots \label{eq:bprx2} \\
&=&\cdots(SB_{(0)}^{(2)})(S^{\dag}B_{(1)}^{(2)})(SB_{(2)}^{(2)})
 (S^{\dag}B_{(3)}^{(2)}) \cdots, \label{eq:bprx3}
\end{eqnarray}
\end{subequations}
which,    by    Theorem    \ref{thm:qutraj},    is    equivalent    to
$\widehat{X}^3$. For  proof of (\ref{eq:bprx2}), see  the proof of
Theorem \ref{thm:bias}. Equation (\ref{eq:bprx3}) is  obtained analogously
to (\ref{eq:b2}), except that the matrix $B^{(2)}$ is used instead
of $B$.  \hfill $\blacksquare$
\bigskip

This may be expressed by the statement 
\begin{subequations}
\begin{eqnarray}
\label{eq:symn}
{\cal N}\widehat{W} &\simeq& {\cal N}{\bf Z}\widehat{W}, \\
{\cal N}\widehat{W} &\simeq& {\cal N}{\bf PRX}\widehat{W}, 
\end{eqnarray}
\end{subequations}
which generalizes (\ref{eq:symm}).
These results show that the symmetries persist for dephasing (phase flip), bit flip and (generalized)
amplitude damping channels.

\section{Symmetry and noise operations on an $n-$cycle}
\label{qwpg}

In previous section we considered symmetries for a
quantum walk on a line, and the influence of noise on them. In this section we will extend the ideas to quantum walks on an $n-$cycle. We will show the breakdown of symmetry when the  quantum walk is implemented on  an $n-$cycle, the effect of noise and its influence on the restoration of symmetry. 
\subsection{Breakdown in symmetry}
\label{symbreak}

In contrast to the case of quantum walk on a line,
none of the four discrete symmetries of Theorem \ref{thm:bias}
hold in general for unitary quantum walk on an $n-$cycle or closed path. Thus, if $B$ (\ref{3paraU2}) is replaced by any of $B^{(1)}$, $B^{(2)}$, $B^{(3)}$, or $B^{(4)}$, given by (\ref{eq:bias20}), the spatial probability distribution is not guaranteed to be the same. 
\begin{thm} The operation $G: B \rightarrow B^{\star}$ 
is in general not a symmetry of the quantum walk on an $n-$cycle.
\end{thm}
\noindent {\bf Proof.} 
For the cyclic case, in place of (\ref{eq:bitobit}), we now have
\begin{subequations}
\begin{eqnarray}
\label{eq:cybitobit}
|\Psi_1\rangle &=& 
(SB)^t|\alpha,\beta\rangle = \sum_{q_1,q_2,\cdots,q_t}
b_{q_t,q_{t-1}}\cdots b_{q_2,q_1}b_{q_1,\alpha}
|q_t,\beta+2Q - t~({\rm mod}~ n)\rangle,  \nonumber \\
 \\
|\Psi_2\rangle &=& 
(SB^{(1)})^t|\alpha,\beta\rangle  \nonumber \\
&=& \sum_{q_1,q_2,\cdots,q_t}
b_{q_t,q_{t-1}}\cdots b_{q_2,q_1}b_{q_1,\alpha}
(e^{i\phi})^{q_{t-1}+ \cdots + q_1 + \alpha}
|q_t,\beta+2Q - t ~({\rm mod}~ n)\rangle, \nonumber \\
\end{eqnarray}
\end{subequations}
where $n$ is the number of sites in an $n-$cycle. 
For  an    arbitrary     state     $|a,b\rangle$    in     the
computational-and-position  basis, 
we have
\begin{subequations}
\begin{eqnarray}
\langle  a,b|\Psi_1\rangle &=& \sum_{q_1,q_2,\cdots,q_{t-1} \in {\cal Q}}
b_{a,q_{t-1}}\cdots b_{q_2,q_1}b_{q_1,\alpha} \label{eq:ny1} \\
\langle  a,b|\Psi_2\rangle &=& \sum_{q_1,q_2,\cdots,q_{t-1} \in {\cal Q}}
b_{a,q_{t-1}}\cdots b_{q_2,q_1}b_{q_1,\alpha}
(e^{i\phi})^{q_{t-1}+ \cdots + q_1 + \alpha} \nonumber \\
&\equiv& \sum_{q_1,q_2,\cdots,q_{t-1} \in {\cal Q}} b_{a,q_{t-1}}\cdots 
b_{q_2,q_1}b_{q_1,\alpha}(e^{i\phi\epsilon}), \label{eq:ny2}
\end{eqnarray}
\end{subequations}
where   ${\cal Q}$   is   the   set    of   binary   $(t-1)$-tuples
$q_1,q_2,\cdots,q_{t-1}$ such that 
\be
Q = q_1 + q_2 + \cdots + q_{t-1} + a
\ee
satisfies 
\be
b  =  \beta +  2Q-t  ~({\rm mod}~  n).
\ee
We find  that
\be
\epsilon = \alpha - a + Q {~~~\rm and~~~}  Q = (b + t - \beta)/2 + mR,
\ee
where $m =  0,1,2,\cdots,  \lfloor  t/n   \rfloor$.
Thus,  the  terms  in  the
superposition (\ref{eq:ny1})  are not in general  identical with those
in (\ref{eq:ny2}),  apart from a common factor,  unless $\phi=0, 2\pi,
4\pi,\cdots$.  A  similar argument  can be used  to show that the terms in 
$\langle  \overline{a},b|\Psi_1\rangle$ are not in general
the  same as those  in $\langle  \overline{a},b|\Psi_2\rangle$.  Given
the     independence    of     $\phi$     from    the     coefficients
$b_{q_2,q_1}b_{q_1,\alpha}$,  it  is   not  necessary  that 
\be
|\langle
a,b|\Psi_1\rangle|^2  +   |\langle  \overline{a},b|\Psi_1\rangle|^2  =
|\langle           a,b|\Psi_2\rangle|^2           +           |\langle
\overline{a},b|\Psi_2\rangle|^2.
\ee
The equality  holds in general (for arbitrary unitary  matrix $B$ and time $t$) if  and only if $\phi=0, 2\pi,  4\pi, \cdots$.  Repeating the  argument for $B^{(2)}$,
$B^{(3)}$ and $B^{(4)}$, we find that all the four discrete symmetries
of   Theorem   \ref{thm:bias}   break   down   in   general.    \hfill
$\blacksquare$
\bigskip

We  note  that for  phase  flip  symmetry,  where $e^{i\phi}=-1$,  the
superposition   terms   in (\ref{eq:ny2}) may differ  from   the
corresponding  terms  in (\ref{eq:ny1}) only  with  respect  to
sign. Given  that all the terms  like $b_{q_2,q_1}$, $b_{q_1,\alpha}$,
etc.  are built from  a small  set of trigonometric functions of the three parameters  $\theta$, $\zeta$ and  $\xi$, certain values  of $t$
may  render   the  right  hand  sides  of (\ref{eq:ny1})  and
(\ref{eq:ny2}) equal. However in  general, this equality will not hold
for arbitrary $t$.
\begin{figure}
\begin{center}
\subfigure[]{\includegraphics[width=7.2cm]{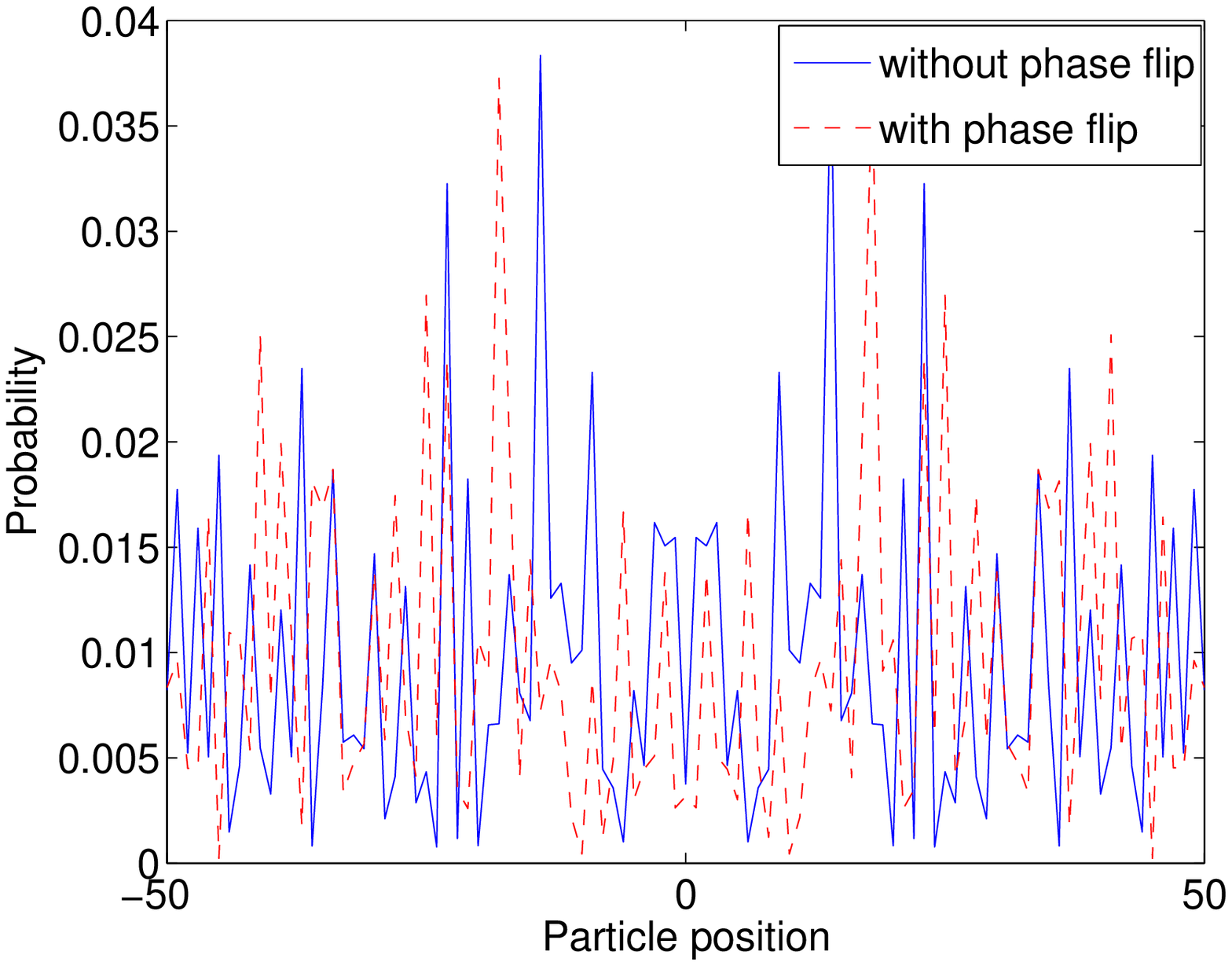}}
\hfill
\subfigure[]{\includegraphics[width=7.2cm]{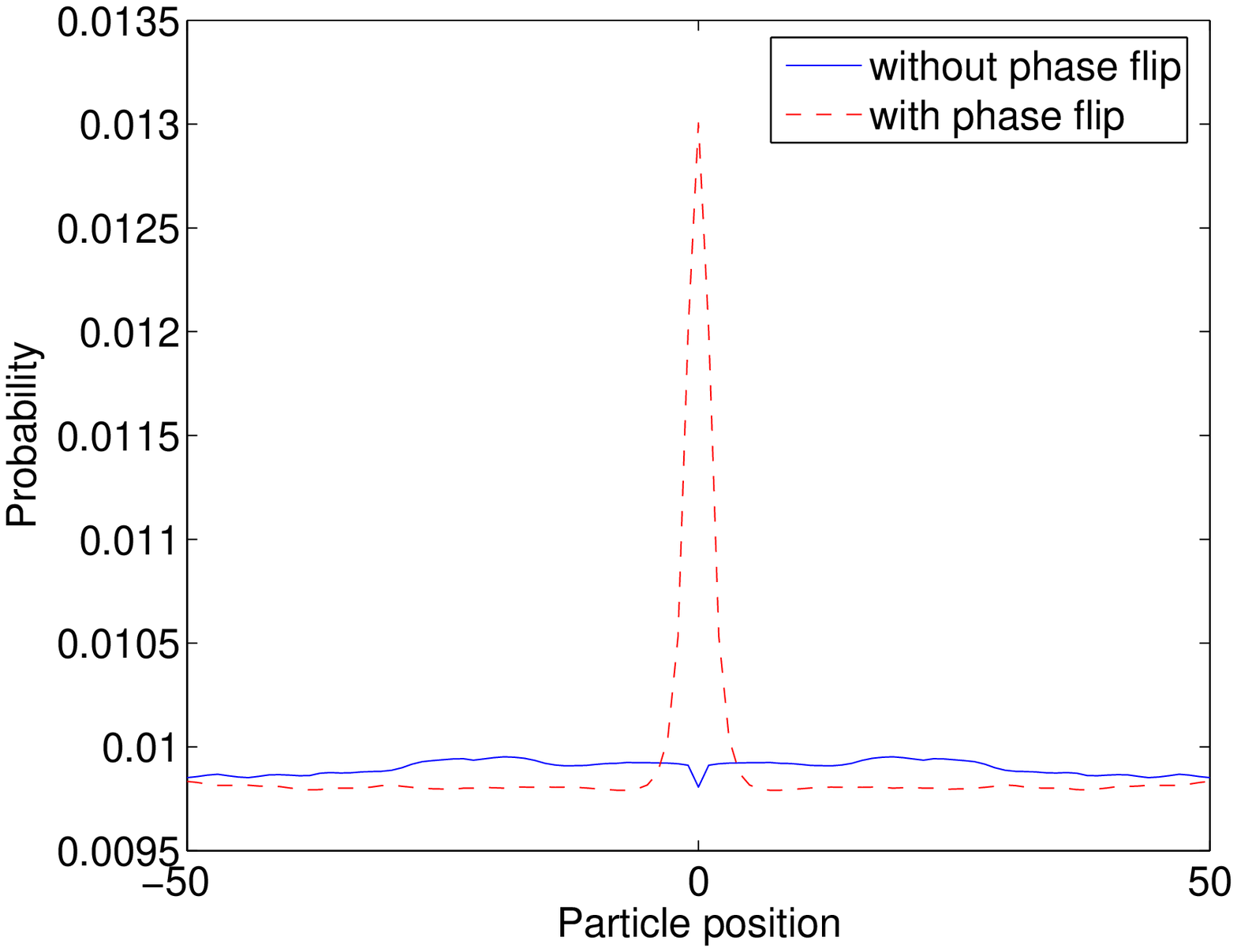}}
\caption[An instance of breakdown of phase flip symmetry in a 
unitary quantum walk on an $n-$cycle.] {An instance of breakdown of phase flip symmetry in a 
unitary quantum walk on a cycle, where the two extreme points 
(located at positions $\pm 50$) on
the plot are spatially adjacent. The number of sites is 101 and 
$t=5000$ (in units of discrete time-steps), 
with coin parameter $\theta=30^{\circ}$. 
(a) The solid curve represents the positional probability distribution
without any symmetry operation applied, while the dashed curve
represents that with a phase flip operation applied at each 
walk step. (b) The same as the above, but with time-averaging applied over every 50 steps, in order to more clearly bring out the breakdown in
symmetry.}
\label{fig:withoutnoise} 
\end{center}
\end{figure}

An instance of breakdown of phase flip 
symmetry in the unitary quantum  walk on a cycle is
demonstrated  in the example  of Figure \ref{fig:withoutnoise}.
The  profile of  the
position probability  distribution varies  depending on the  number of
sites and  the evolution time.  

\subsection{Breakdown using a generalized phase gate}
\label{breakgenpg}

We will introduce a generalized phase gate, an element from
the parameter group 
\be
\label{eq:pg}
G(\beta)=  \left(\begin{array}{ll}
1 & 0 \\ 0 & e^{i\beta}\end{array}\right),
\ee
to act on  the ${\cal H}_c$. We find that the operation $W  \longrightarrow GW$ leaves  the probability distribution
$p(j,t)$  of the particle  on the  line invariant;  hence the  walk is
symmetric  under the  operation 
\be
\label{eq:pgope}
G(\beta):  |q\rangle \mapsto
e^{i q\beta }|q\rangle
\ee
 for  $|q\rangle$ in the
computational basis (eigenstates of the Pauli operator $\sigma_z$) and
$q=0,1$.  The physical significance of $G$ is that it helps identify a
family  of quantum walks that are  equivalent  from the  viewpoint of  physical implementation,  which  can sometimes  allow  a significant  practical simplification \cite{CSB07}.   For example, suppose  the application of the  conditional  shift is  accompanied  by  a  phase gate.  The  walk symmetry implies  that this  gate need not  be corrected  for, thereby
resulting in  a saving of  experimental resources. The inclusion  of a
phase gate on the coin operator  is equivalent to a phase gate at each
lattice site  in the sense of  quantum lattice gas automata (QLGA) \cite{Mey96, Mey97}, with the  physical meaning  of a
constant potential. The
evolution rules for single-particle  QLGA can be classified into gauge
equivalent  classes, there  being a  difference between  the  class of
rules for periodic  ($n-$cycle) and non-periodic 1-dimensional lattice
and this  feature can  be used to  distinguish between  these two
spatial topologies \cite{Mey01}.

It turns  out that in  the case of quantum walk on an $n$-cycle,  this symmetry breaks down.   To see this, we  note that the  $t$-fold application of
the operation $GSB$ on  a particle with initial state $|\Psi_{in}\rangle$
on the line and on an $n-$cycle produces, respectively, the states
\begin{subequations}
\label{eq:gsup}
\begin{eqnarray}
(GSB)^t|\Psi_{in}\rangle =   
\sum_{q_1,q_2,\cdots,q_t}e^{iQ_t\beta} 
B_{q_t,q_{t-1}}\cdots B_{q_2,q_1}
(B_{q_1 0}a + B_{q_1 1}b)|q_t\rangle|2Q_t-t\rangle,~~\nonumber \\ 
\label{eq:gsupa} \\
(GSB)^t|\Psi_{in}\rangle = ~~~~~~~~~~~~~~~~~~~~~~~~~~~~~~~~~~~~~~~~~~~~
~~~~~~~~~~~~~~~~~~~~~~~~~~~~~~~~~~~~~~ \nonumber \\
\sum_{q_1,q_2,\cdots,q_t}e^{iQ_t\beta}  
B_{q_t,q_{t-1}}\cdots B_{q_2,q_1} 
(B_{q_1 0}a + B_{q_1 1}b)|q_t\rangle|2Q_t-t\mod n\rangle, \nonumber \\
\label{eq:gsupb} 
\end{eqnarray}
\end{subequations}
where  $Q_t =  q_1+q_2+\cdots+q_t$. All terms in superposition  (\ref{eq:gsupa}) contributing  to the  probability  to  detect the  walker  at a  given
position   $j  =   2Q_t-t$   have  the   {\it   same}  phase   factor, $e^{iQ_t\beta}$,  which is  fixed by  $Q_t = (j+t)/2$ (where,  it may be noted, $j$  and $t$ are both  even or both
odd).  Thus, this factor does not affect the probability to detect the walker at $j$, whence the symmetry.  In the case of quantum walk on an $n$-cycle the    breakdown    of     the    symmetry,    see    Figure \ref{fig:noisy_amp}, can  be attributed to the topology  of the cycle,
which introduces a periodicity in the walker position (determined by a
congruence relation  with modulus given  by the number of  sites), but
not in the phase of the  superposition terms.  As a result, fixing $j$
fixes $Q_t \mod n = (j+t)/2 \mod n$, but not $Q_t$ itself, so that the
phase terms  in the superposition  (\ref{eq:gsupb}) do  not factor
out globally. Thus if $\beta$ is non-vanishing, then  in general the symmetry $G$ is  absent in the cyclic
case.
\begin{figure}
\begin{center}
\subfigure[]{\includegraphics[width=7.2cm]{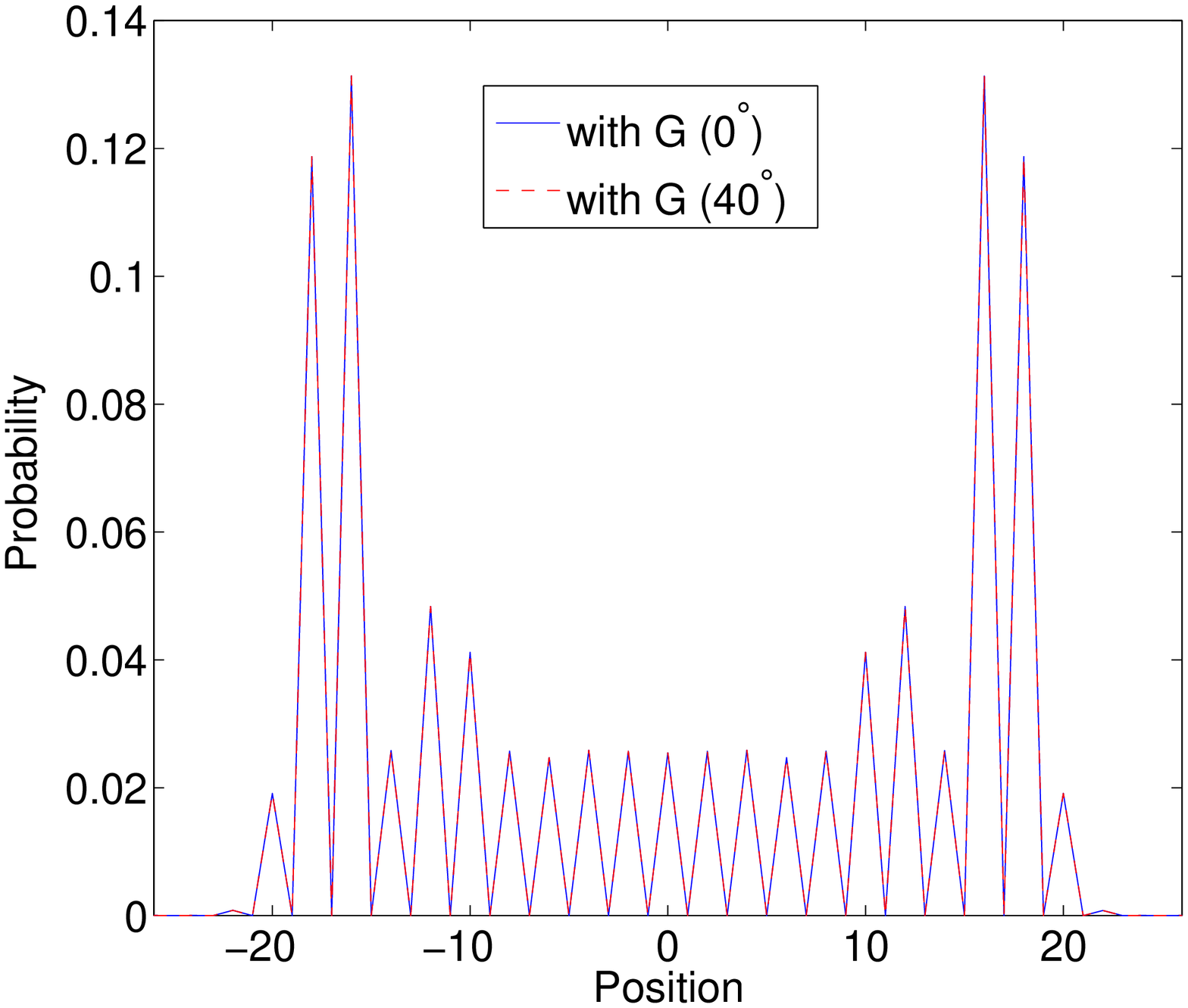}}
\hfill
\subfigure[]{\includegraphics[width=7.2cm]{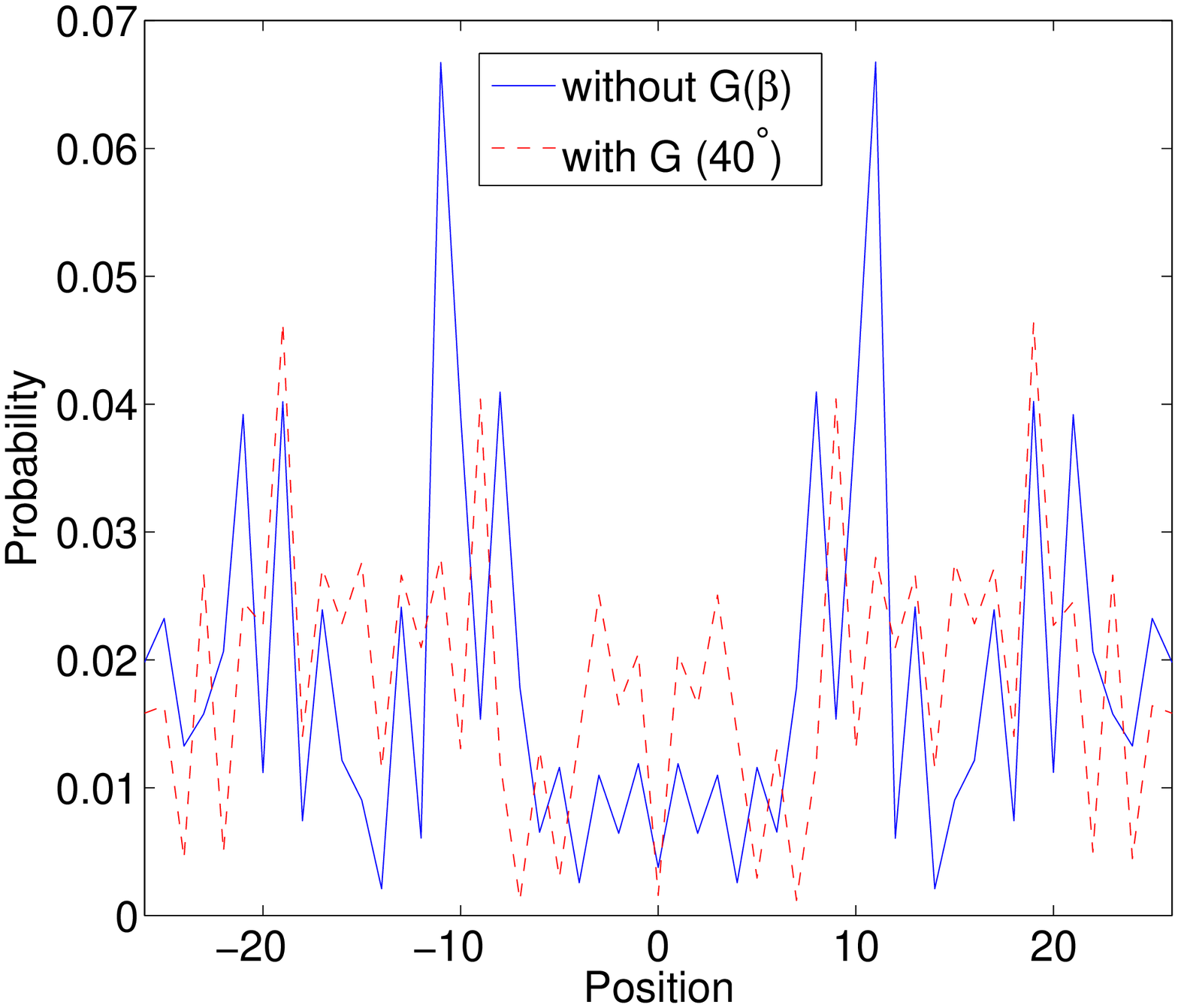}}
\caption[Position   probability  distribution  for  a Hadamard  walk on a line and an $n$-cycle for the unitary  case. Symmetry holds for walk on a line and breakdown  for a walk on an $n-$cycle.]
{Position   probability  distribution  for  a  Hadamard  walk, $B(0^{\circ},  45^\circ,0^{\circ})$ on (a) a line, and (b) an $n$-cycle, with  ($n=51$)  and  initial    state    $(1/\sqrt{2})(|0\rangle   +
i|1\rangle)$, for the unitary   case    ($\gamma_0=0$) and $\tau =50$. Each figure
presents the distribution with and without being subjected to
the phase  operation $G(40^\circ)$.
In (a), there is perfect symmetry, since both distributions coincide.
In (b) the two  plots  do  not   overlap,  indicating the  breakdown  of  the
  symmetry.}
\label{fig:noisy_amp} 
\end{center}
\end{figure}
\begin{figure}
\begin{center}
\includegraphics[width=8.0cm]{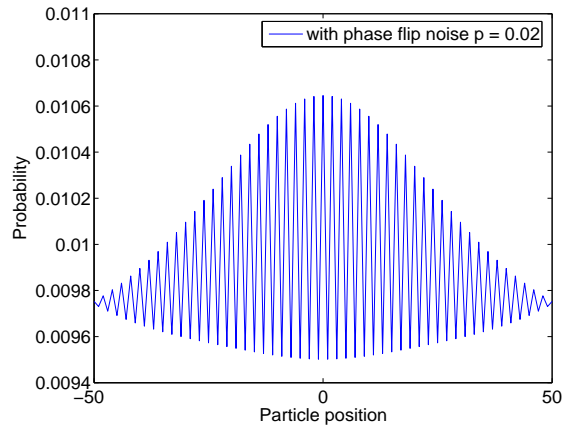}
\caption[Restoration of phase flip symmetry in a noisy quantum walk on
an $n-$cycle, where  the  two extreme  points  on the  plot are  spatially
adjacent.]{Restoration of phase flip symmetry in a noisy quantum walk on
a  cycle, where  the  two extreme  points  on the  plot are  spatially
adjacent. The number of sites is 101 and $t=5000$ (in units of discrete
time-steps), with coin parameter $\theta=30^{\circ}$.   The figure depicts the position
probability distribution with or without phase flip symmetry operation
applied  at  each  step,  with  phase  damping  noise  level  $p=0.02$ (\ref{eq:phaseflip}).  After sufficiently long time, the quantum walk  reaches the  uniform distribution,  typical of  classical random walk.}
\label{fig:withnoise} 
\end{center}
\end{figure}
We  quantify the  breakdown in  symmetry  by means  of the  Kolmogorov
distance  (or,   trace  distance   \cite{NC00}),  given  by 
\be
\label{eq:kd}
d(t)  =
(1/2)\sum_j    |P(j,t)-P^{\prime}(j,t)|,
\ee
between the particle position distributions obtained without and  with the symmetry operation, given
by $P(j,t)$ and $P^{\prime}(j,t)$, respectively.  The breakdown in symmetry for
a   noiseless  cyclic   quantum walk  is   depicted   by  the   bold  curve   in
Figure \ref{fig:kd_uni} as  a function  of the  number of  turns $\tau$
(where $t=\tau s$, with $n =2s+1$).

\subsection{Effect of noise and symmetry restoration}
\label{nqw}
Remarkably, this  symmetry is restored
above   a  threshold   value  of   noise.   The   pattern   in  Figure \ref{fig:withnoise} corresponds  to phase noise  with $p=0.02$ applied
to  a quantum  walk,  either with  or  without a  phase flip  symmetry
operation.   We note  that  the  introduction  of   noise  tends  to classicalize the random walk, hence causing it to asymptotically reach a  uniform  distribution \cite{Ken06, MK07}.  The  above-mentioned  symmetry
restoration happens  well before the uniformity sets  in.  The initial
lack of symmetry  gradually transit to full symmetry as the noise
level is increased. Thus, the role of symmetry  operations and noise
is quite different in the case of quantum walk  on an $n-$cycle as compared with that on a line.

We now  describe the $n$-cycle quantum walk  on the particle, a  two level
system, when subjected  to  noise. The situation is modeled as  an  interaction with a thermal  bath, characterized  by phase  damping or a
generalized  amplitude  damping  channel,  the  latter  process  being
represented by the Kraus operators (\ref{eq:gbmakraus}). 
The  density operator  $\rho_c$  of the  coin  evolves  according to 
\be
\rho_c
\rightarrow \sum_x  E_x \rho_c E_x^\dag.
\ee
The full  evolution of the
walker, described  by density operator $\rho(t)$, is  given by 
\be
\rho(t)v= \sum_j
E_x(W \rho(t-1) W^\dag)E_x^\dag,
\ee
where  the $E_x$'s are understood to
act only in the coin space.

The curves  in Figure \ref{fig:kd_uni}  plot $d(\tau)$ as a  function of
turns  in  the   case  of  unitary  and  noisy   quantum walk  (parametrized  by
$\gamma_0$), and demonstrate the  gradual restoration of symmetry with
time on account of the noise.  Although the figure employs generalized
amplitude damping  noise, qualitatively the same behavior  can be seen
for a phase damping noise.   Here, a general feature is that $d(\tau)$
is non-zero  when $\tau<2$, being equivalent to (noisy)  quantum walk on a
line.  Thereafter, $d(\tau)$ at first increases with increasing turns,
being dominated by unitary evolution, and eventually falls down, being
dominated by  noise.  It is  observed that for sufficiently  low noise
levels,  the time  at which  this  turnover in  slope happens  remains
constant,   for   given   $\theta$.    This  is   depicted   in   Figure \ref{fig:kd_uni}  for  the case  of  a  generalized amplitude  damping
channel corresponding  to a fixed temperature  and varying $\gamma_0$.
However, we  note that for  strong enough noise, the  turnover happens
earlier.
\begin{figure}
\begin{center}
\includegraphics[width=8.6cm]{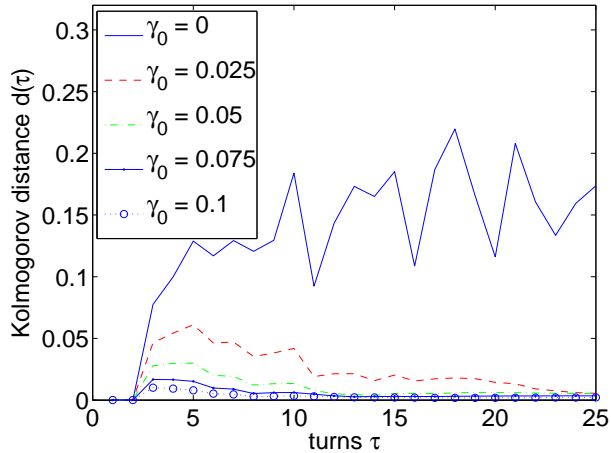}
\caption[Kolmogorov  distance $d(\tau)$  against  the
  number of turns ($\tau$) of the cyclic quantum walk in the noiseless and noisy
  case with $n=51$.]{Kolmogorov  distance $d(\tau)$  against  the
  number of turns ($\tau$) of the cyclic quantum walk in the noiseless and noisy
  case with $n=51$.  For the unitary case ($\lambda=0$, $\gamma_{0}$ ; bold line) the
  walk  becomes increasingly  asymmetric  as the  number  of turns  is
  increased, until about 7--10 turns, after which it fluctuates around
  $d\approx0.15$.  The  plots represent generalized  amplitude damping
  noise at different  noise levels at temperature $T  = 3.5$ (in units
  where   $\hbar    \equiv   k_B   \equiv    1$).  The  walk is evolved with  the initial state of the particle 
  parameters $\delta=30^\circ$, $\eta=40^\circ$ in (\ref{qw:ins}), with
  $B(20^\circ, 10^\circ,30^\circ)$ and $G(10^\circ)$.}
\label{fig:kd_uni}
\end{center}
\end{figure}
\begin{figure}
\begin{center}
\includegraphics[width=9.2cm]{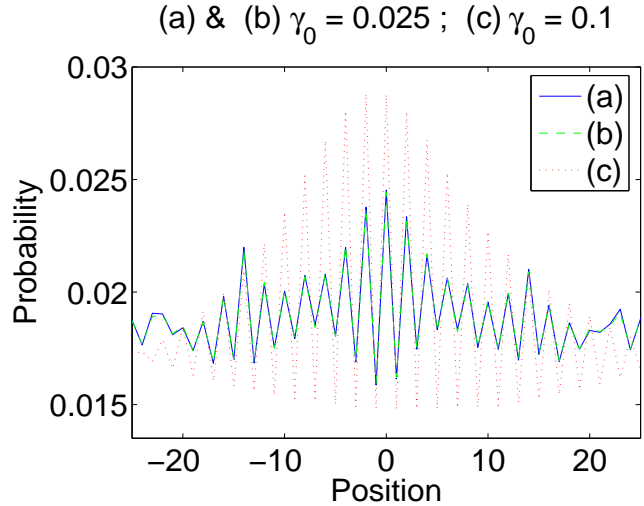}
\caption[Position   probability  distribution for  a
  Hadamard  walk on  an  $n$-cycle ($n=51$)  with
  initial   state when subjected to  generalized amplitude damping  noise.]
{Position   probability  distribution  for  a
  Hadamard  walk  $B(0, 45^\circ,0)$  on  an  $n$-cycle ($n=51$)  with
  initial   state  $(1/\sqrt{2})(|0\rangle   +  i |1\rangle)$  when
  subjected to  generalized amplitude damping  noise with $\Delta=0.1$
  and finite $T$  ($= 6.0$) for $\tau = 11$. For (a)  and (b) $\gamma_0 =
  0.025$. (b)  is the distribution  for the quantum walk augmented  by operation
  $G(20^\circ)$ and (a) is without any augmented operation;  note  the  overlap of (b) on (a), walk  remains quantum  yet  with symmetry  completely  restored ; (c)  classicalized  pattern
  (indicated by the regular envelope) obtained with larger noise level
  corresponding to $\gamma_0=0.1$.}
\label{fig:sym_restore}
\end{center}
\end{figure}

Typical    noisy   probability    distributions   are    depicted   in
Figure \ref{fig:sym_restore}(a),(b) at an instant where the symmetry has
been  almost fully  restored while the walk is  well  within the
quantum    regime.   Figure \ref{fig:sym_restore}(c)    represents a classicalized  distribution, indicated by  the regular  envelope (that will eventually turn into a uniform distribution).

We define coherence ${\bf C}$ as  the sum of the off-diagonal terms of
states  in ${\cal  H}_c \otimes  {\cal H}_p$,  where ${\cal  H}_c$ and
${\cal  H}_p$ are  the  coin  and position  Hilbert  space of  the
quantum walker, respectively.   If the state of the  quantum walker is
\be
\label{eq:rho}
\rho  =   \sum_{ab;jk}\alpha_{ab;jk}|a\rangle|j\rangle
\langle  b|\langle k|,
\ee
where $|a\rangle$,  $|b\rangle \in {\cal  H}_c$ and  
$|j\rangle$, $|k\rangle \in  {\cal H}_p$, then  
\be
{\bf C}  \equiv \left(\sum_{a  \ne b,j  =k} +
\sum_{a,b,j \ne  k} + \sum_{a \ne  b,j \ne k}\right)|\alpha_{ab;jk}|,
\ee
the sum of the absolute values  of all off-diagonal terms of $\rho$ in
the computational-position  basis.  The coherence  function is defined
as the  quantity $C(m)$, where  $m \in \{1,2,\cdots,M\}$,  obtained by
partitioning ${\bf C}$ into $M$ intervals of size $s/M$, such that for
the  $m$th  interval
\be
 (m-1)(s/M)  \le  |j-k|  < m(s/M).
 \ee
 Physically,
$C(m)$ is a measure of coherence  between two points on a (in general,
noisy) quantum walker, as a  function of their mutual separation.  Let
$C_0(m)$  represent  the   coherence  function  of  the  corresponding
noiseless walk. At any turn $\tau$, we define the normalized coherence
function  by
\be 
c(m) \equiv  C(m)/C_0(m),
\ee
and,  analogously, normalized Kolmogorov distance by 
\be
D(\tau) \equiv d(\tau)/d_0(\tau).
\ee

Since  noise tends  to destroy  superpositions, and  the  breakdown in
symmetry is  essentially a phenomenon of superposition  of the forward
and  backward waves,  noise tends  to restore  symmetry, as  seen from
Figure \ref{fig:sym_restore}.    This   is   brought   out   by   Figure \ref{fig:coherence} 
for two possible values of $G$. In the figure, in spite of its considerable spikiness, the
bold  curve,  representing $c(m=M)$,  shows  an  overall fall.
A similar trend as depicted in this figure, has been numerically 
checked for various other values of $G$. This
raises the question whether symmetry  restoration of the cyclic quantum walk can
be considered as a good indicator of classicalization.  
Here we note that from Figure \ref{fig:sym_restore}(a),(b) the
probability distribution  pattern is seen to be  clearly quantum, even
though symmetry has been almost fully restored.
This is suggestive of the notion that
that symmetry tends to be restored {\em even} in
the regime  where the walk  still possesses some quantum  features.

The reason $D(\tau)$  is not a faithful indicator  of classicalization of
the walk has to do with the  effect of noise on the sensitivity of the
symmetry  operation $G(\beta)$  to  the topology  of the  path.
Since  this operation  senses  the  closure of  the  path through  the
superposition of  the forward and  backward waves, the  suppression of
superposition through noise will also have the effect of desensitizing
the operation  to the  closure of the  path, thereby moving  the noisy
cyclic quantum walk  towards a noisy quantum walk on  a line from the  perspective of this
operation, before further classicalization transforms it into a cyclic classical random walk. And as shown in Section \ref{sec:symmline}, all the above symmetries are respected by  a (noisy) quantum  walk on a  line, both in the  case of phase damping noise and generalized amplitude damping\footnote{In NMR nomenclature, phase damping is called a
$T_2$ process, and generalized amplitude damping, which is a $T_1$ ,$T_2$ process \cite{Abr61}.}. This  brings out the point that decoherence ($T_2$  process)  is  the  principal  mechanism  responsible  for  the restoration of  symmetries.  It also highlights  the interplay between
topology  and noise  in a  quantum walk  on an  $n-$cycle.   A similar
interplay may be expected also in the case of quantum walk with other nontrivial
topologies.
\begin{figure}
\begin{center}
\includegraphics[width=8.5cm]{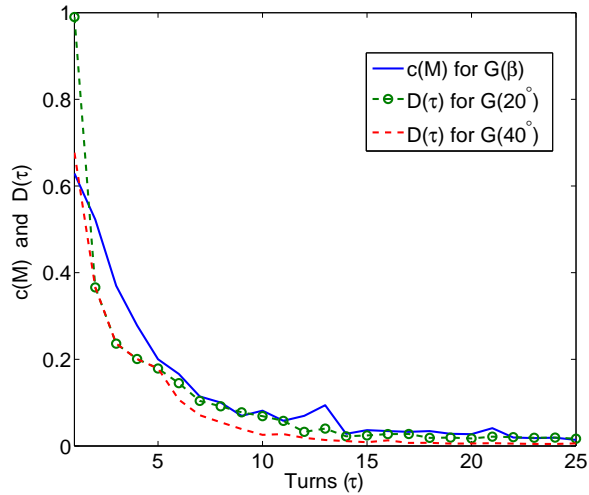}
\caption[Normalized  coherence function  $c(M)$
   and  the normalized  Kolmogorov  distance $D(\tau)$  for  an $n-$
   cyclic quantum walk  as a  function of turns  $\tau$.]{Normalized  coherence function  $c(M)$ (bold)
   and  the normalized  Kolmogorov  distance $D(\tau)$  (dashed) for  an $n-$
   cyclic quantum walk  as a  function of turns  $\tau$. An overall  reduction of
   both  $c(M)$ and  $D(\tau)$ with  time is  seen for  the generalized
   amplitude  damping noise  characterized by  temperature  $T=3.5$ and
   $\Delta=0.1$. The  Hadamard walk is evolved with  the initial state
   parameters (in degrees) $\delta=45^\circ$, $\eta=0^{\circ}$ (\ref{qw:ins}), and
   $G(\beta)$. The  line-with-circle plot  represents $D(\tau)$ with  $G(20^\circ)$ and dashed line represents for $G(40^\circ)$.
The  $c(M)$  (solid line) remains  roughly the same for both $\beta$.  For a clear depiction of the notion of symmetry 
restoration in the
walk even in the quantum regime, Figure (\ref{fig:sym_restore}).}
\label{fig:coherence}
\end{center}
\end{figure}

Further extensions would be quantum walks on a more
general  graph  \cite{Ken06, MK07}  or  in  higher dimensions  $d  >  2$
\cite{MBS02}. In the former, the  1D walk is generalized to an $n$-cycles
and to  hypercubes, including  the effect of  phase noise in  the coin
space, and decoherence in position space.  In the latter, the Hadamard
transformation is  generalized to  a non-entangling tensor  product of Hadamards,  or to  an  entangling discrete  Fourier  transform or  the
Grover  operator.  They  bring in  many novel  features absent  in the
quantum walk  on a line.
\par
The interplay  between geometry and decoherence has  been noted before
in  the  case  of  delocalized  bath modes  \cite{Wei98},  as  against
localized bath modes \cite{Wei98, AW95, Coh97}.   This is of relevance as the
noise processes considered here  \cite{SB08,CSB07} are described by the
interaction of the system with delocalized bath modes.

\section{Experimental implications}
\label{sec:qwbec}

Experimental  realization of quantum  walk using  any of  the proposed schemes is  not free  from noise due  to environmental  conditions and instrumental interference.  In particular,  noise can be a major issue
in the scaling  up of the number of steps  in already realized quantum
walk systems.  Understanding the symmetries of the noisy and noiseless
quantum walk could  greatly help in the improvement  of implementation technique (see for example Section \ref{becqw})  and in further  exploration of other possible  systems
where quantum walk can be realized  on a large scale.  
\par
The experimental study of the  decoherence and decay of quantum states
of  a trapped  atomic ion's  harmonic motion  subjected  to engineered
reservoirs, both of the phase damping and amplitude damping kind, have
been reported in \cite{MKT00, TMK00}.  The phase reservoir is simulated by
random variation  of the trap frequency $\omega$  without changing its
energy (non-dissipative),  while the amplitude  reservoir is simulated
by   random   electric   field    along   the   axis   of   the   trap
(dissipative).     Coupling     the     reservoirs     reported     in \cite{MKT00, TMK00}  to the  scheme  presenting  the combination  of
pulses required to implement  a quantum walk on a line and on  a cycle in an ion
trap \cite{TM02}  provides a  convenient set up  to demonstrate
the symmetry-noise interplay.

\subsection{NMR quantum-information processor}
\label{subsec:nmr}

Continuous time  \cite{DLX03} and discrete time  \cite{RLB05} quantum walk
have  been successfully  implemented in  a nuclear  magnetic resonance
(NMR) quantum-information  processor. 
In NMR spectroscopy  of the given system (molecule)  the extent of its
isolation from  the environment  is determined in  terms of  its phase
coherence time $T_{2}$ and its  energy relaxation time $T_{1}$. If the
pulse  sequence is  applied to  the NMR  quantum-information processor
within the  time $T  < {T_{2},  T_{1}}$, the environmental effects on the 
system are less significant. The pulse  sequence exceeding the time $T_{2}$
can be considered  to be affected by the dephasing  channel and the pulse
sequence exceeding the  time $T_{1}$ can be considered  to be affected
by  the  amplitude damping  channel.   In  experiments  of time  scale
greater than the time $T_{2}$  or $T_{1}$, a refocusing pulse sequence
is  applied to  compensate  for the  environmental  effects.  In the implementation 
of discrete-time quantum walk, the  pulse sequence was 
implemented within the time $T_{1}$ and $T_{2}$ and  by introducing dephasing,
 the transition from quantum walk to the classical random 
walk was shown \cite{RLB05}.

The environmental effect (noise)  on quantum walk symmetries presented
in this chapter can be verified in  the NMR system by  scaling up the
number of  steps of  quantum walk realized.  By applying  a controlled
amount  of the  refocusing  pulse sequence,  the  effect of  different
levels of noise can be experimentally verified.

\subsection{Ultracold atoms}
\label{subsec:atoms}

There have  been other proposals  for physical realization  of quantum
walk using  Bose-Einstein condensate (BEC)  \cite{Cha06} where the
unitary shift  operator induces a  bit flip.  A {\em  stimulated Raman
kick}  is  used   as  a  unitary  shift  operator   to  translate  the
Bose-Einstein  condensate   in  the  Schr\"odinger  cat   state  to  a
superposition in position  space. Two selected levels of  the atoms in
the  Bose-Einstein  condensate  are   coupled  to  the  two  modes  of
counter-propagating  laser  beams.   The  stimulated  Raman  kick,  in
imparting  a translation in  position space,  also flips  the internal
state of the  Bose-Einstein condensate.  An rf pulse  ($\pi$ pulse) is
suggested as  a compensatory mechanism  to flip the internal  state of
the condensate  back to  its initial value  after every  unitary shift
operator. From  the ${\bf PRX}$  symmetry pointed in this chapter, it
follows that  there is no need  for the compensatory  operation for an
unbiased quantum  walk for a particle initially in the state  $\frac{1}{\sqrt 2}(|0\rangle + i|1\rangle)$.
The availability of walk symmetries could also be useful for exploring
other  possible  physical  implementations  which  induce such
symmetry operations along with the translation.

In the  most widely  studied version of  quantum walk, a  quantum coin operation is  used after every displacement operation.
Continuous  external  operations on  a  particle  confined  in a  trap reduces the confinement  time of the particle. Reducing  the number of external operations will benefit the scaling up of the number of steps of the quantum walk. One can configure a system where a separate quantum coin  operation is eliminated by using a single coin-embedded conditional shift operator to shift the particle in superposition of position space retaining the superposition  of the coin state \cite{Cha08}. 
Further, systems of this kind are expected  to be affected  by amplitude  damping as  one of the basis states might be more stable than the other one in  the trap.  The present study of effect of noise could help to optimize and use quantum walk with in the system limitations.

\subsection{Other condensed matter systems}
\label{cms}

Breaking of  the symmetry  due to the  change in walk  topology causes long-range correlations to develop, in analogous to the hydrodynamics of ordered   systems   such   as   spin   waves   in:   ferromagnets, antiferromagnets   (where   it  is   the   spin   wave  of staggered
magnetization), second  sound in  He$^{3}$,  nematic liquid  crystals \cite{For95}.    Here  the   correlations  may   be   identified  with
symmetry-broken  terms  (whose   measurement  probability  depends  on $\beta$ in the walk augmented by $G(\beta)$) in the
superposition of  the quantum walker. One finds  that correlations are
set up  rapidly over large distances  with increase in  the winding of
the walker, until symmetry is broken throughout the cycle. However, as
noted  above,   the  randomization   produced  by  noise   causes  the
reappearance of  symmetries.  The  symmetry breaking and  the symmetry
restoring  agents  are  thus   different,  the  former  given  by  the
topological transition from  a line to an $n$-cycle,  the latter being
the noise-induced randomization.

Coherence is also widely used to understand quantum phase transitions,
the transition from  superfluid to Mott insulator state  in an optical
lattice   being   one  specific   example   \cite{JBC98, GME02}.   In  Chapter \ref{Chapter4} 
 the   quantum  phase  transition  using  quantum walk   in  a  one
dimensional optical lattice is discussed \cite{CL08}.  Using various lattice
techniques,  desired geometries to  trap and  manipulate atoms  can be
created.   In   most  physical   situations  one  deals   with  closed
geometries. The  characteristics of the $n-$cycle  walk, in particular
the   re-appearance   of   the   symmetry  (implying   a   family   of
implementationally  equivalent noisy  cyclic quantum walks)  while still  in the
quantum regime,  presented here could  be of direct relevance  to such
situations.   

The ubiquity of the ideas developed in this chapter can be seen from the
fact that the quantum dynamics of a particle on a ring (cycle) subject
to decoherence along  with dissipation finds its place  in the physics
of quantum dots.  The effective action of a quantum dot accounting for
the  joint   effect  of  charging  and  coupling   to  an  environment
 \cite{AES82, ESA84} mirrors the behavior of the quantum dynamics of a particle
on  a  ring  (cycle)   subject  to  a  dissipative  damping  mechanism
describing  the dissipation of  the energy  stored in  dynamic voltage
fluctuations  into   the  microscopic   degrees  of  freedom   of  the
quasi-particle continuum.   In the absence of  dissipation, the action
describes the  ballistic motion of a  quantum particle on  a ring. The
ring topology  reflects the  $2\pi$-periodicity of the  quantum phase,
which  is in  turn  related  to the  quantization  of charge,  thereby
highlighting  the point that  the main  source of  charge quantization
phenomena,  in   the  approach   developed  in  \cite{AES82, ESA84},   is  the
periodicity, of the relevant variable,  due to the ring topology. With
the  increase in  the effect  of dissipation,  the particle  begins to
forget  its   ring  topology  (full  traversal  of   the  ring  become
increasingly   unlikely),   leading  to   a   suppression  of   charge
quantization phenomena.   This behavior is  similar to that  seen here
for the case of quantum walk on a cycle, where with an increase in the
effect of the environment, i.e.,  with increasing noise, the walker in
unable to perceive  the cyclic structure of the  walk space.  That the
topology-noise  interplay studied  here has  an impact  on  a concrete
condensed  matter system,  viz.   the crossover  from  strong to  weak
charge  quantization  in a  dissipative  quantum  dot, highlights  the
generality and scope of these ideas.

\newpage

\section{Summary} 
\label{summary3}
\begin{itemize}

\item Our  work considers  variants of  quantum walks  on a  line  which are
equivalent  in  the  sense   that  the  final  positional  probability
distribution  remains the  same in  each variant.   In  particular, we
consider variants  obtained by the  experimentally relevant operations
of $Z$  or $X$ applied  at each quantum  walk step, with  the symmetry
operations  given  by  ${\bf  Z}$  and ${\bf  PRX}$.   This  could  be
experimentally advantageous since  practical constraints may mean that
one of the variants is preferred over the rest. What is especially interesting is
that these symmetries are preserved  even in the presence of noise, in
particular,  those  characterized by  the  phase  flip,  bit flip  and
generalized amplitude  damping channels. This is important because it
means that  the equivalence of these variants is not  affected by the
presence of noise, which would be inevitable in actual experiments.
The symmetry of the phase operation under phase noise is intuitive,
considering that this noise has a Kraus representation consisting of
operations that are symmetries of the noiseless quantum walk. However, for
the PRX symmetry under phase noise, and for any symmetry under other noisy
channels (especially in the case of generalized amplitude damping
channel), the connection was not obvious before the analysis was
completed.

\item An interesting  fact that comes out while extending the studies to walk on an $n-$cycle
is that the symmetry  breaks down in general but is restored above a certain noise level. 
This symmetry-topology-noise interplay presented would be of relevance to quantum information  processing systems, and have  wider implications  to the condensed matter systems.

\item Our results in this chapter to study noise model are supported by  several numerical examples  obtained by evolving  the density  operator in  the Kraus  representation. However, analytical proofs  of the effect  of noise on symmetries  are obtained using the quantum 
trajectories  approach, which we find convenient for this situation.
\end{itemize}

%% file: Chapters/Chapter4.tex
\chapter{Quantum phase transition using quantum walks} 
\label{Chapter4}
\lhead{Chapter 4. \emph{Quantum phase transition}}

\section{Introduction}

In Chapter \ref{Chapter2} we saw that the dynamics of the quantum walk can be controlled by varying the quantum coin parameters. From Chapter \ref{Chapter3} we concluded that the symmetries and addition of small amount of experimentally engineered noise or environmental effect can be used as an additional tool to control the dynamics and hence the probability distribution of the quantum walk. 
\par
These enhanced properties of the quantum walk purely due to quantum dynamics and small amount of engineered noise  \cite{KT03, CSB07, BCA03} can be used to enlarge the toolbox for controlling the dynamics in a physical system, to demonstrate coherent quantum control over, for example, atoms, photons or spin chain systems. Therefore, the use of the quantum walk to study the dynamics of particles in 1D magnetic systems \cite{OAF02}, atoms in optical lattice \cite{JBC98}, and photonic Mott insulators \cite {HP07} or to observe complex quantum phase transitions \cite{DZ02} and quantum annealing \cite{SBB08} would be of both theoretical and experimental relevance. In this chapter we will consider the use of a quantum walk to control and study the dynamics of ultracold bosonic atoms in an optical lattice.
\par
Traditionally, theoretical studies of the dynamics of atoms in an optical lattice are done using mean field approaches \cite{OSS01} and quantum Monte Carlo methods \cite{BRS02, KPS02, WAT04}. The quantum  correlation induced between atoms and position space by the quantum walk can serve as an alternate method for theoretical studies. In particular, we will consider the quantum phase transition from the Mott insulator (MI) - a regime where no phase coherence is prevalent - to the superfluid (SF) - a regime with long-range phase coherence \cite{JBC98, GME02} and vice versa by redistributing the density profile of atoms using quantum walk. The simulation of the quantum phase transition using the quantum walk occurs quadratically faster in one dimension (1D) compared to, varying the optical lattice depth and letting the atom-atom interaction follow the classical random walk behavior. Compared to ballistically moving the atoms, which just relocates the atoms to the new lattice site, quantum walk on atoms spread the wavepacket of each atom over the lattice sites (for example, SF regime). We will consider the discrete-time quantum walk with quantum coin operation $B_{0, \theta, 0}$ and small amounts of three physically relevant models of noise that can be experimentally induced: a bit-flip channel, a phase-flip channel and an amplitude-damping channel discussed in Chapter \ref{Chapter3} \cite{CSB07}, to act as an enhanced toolbox to control the redistribution of atoms in an optical lattice. We will also discuss the scheme for the experimental implementation of quantum walk on ultracold atoms in an optical lattice. 
\par
This chapter is organized as follows. In Section \ref{pt} we review phase transition in optical lattice using Bose-Hubbard model. The implementation of the quantum walk on atoms in 1D MI and SF regimes is discussed in Section \ref{impQW}, where we analyze the dynamics and present the density profile obtained by using some of the properties of the quantum walk. In Section \ref{qwnoise} we discuss the quantum walk with a noisy channel as a tool to control the redistribution of atoms. In Section \ref{impl} we propose a scheme for an experimental implementation of quantum walk on atoms (Bose-Einstein condensates) in optical lattice and conclude with summary in Section \ref{summary4}.
\section{Quantum phase transition in optical lattice}
\label{pt}

Bosons do not obey the exclusion principle, as temperature $T\rightarrow 0$ they all begin to get into the lowest energy eigenstate available, with energy $\epsilon =0$ and the effects of quantum degeneracy begin to emerge. That is, a large fraction of an atomic wave packets of bosonic atoms when cooled below a critical temperature $T_{c}$, begins to condense into the lowest quantum state called as Bose-Einstein condensate (BEC). A BEC at low enough temperature is a superfluid described by a wave function that exhibits long-range coherence \cite{Str01}. When the BEC is transferred to the lattice potential, the atoms move from one lattice site to the next by tunnel coupling. Bose-Hubbard model is the standard model used to discuss the dynamics of the boson and quantum phase transition in optical lattice. The simple possible approximation for the wavefunction of the many body system is a product of single particle state, usually known as Gross-Pitaevskii or mean-field approximation. Below we will briefly discuss the Bose-Hubbard model \cite{Sac00}.

\subsection{Bose-Hubbard model}

The elementary degree of freedom in this model are, as the name implies spinless bosons, this takes the place of the spin-1/2 fermionic electrons in the original Hubbard model introduced as a description of the motion of electrons in transition metals. The Hamiltonian of the Bose-Hubbard model is
\be
\label{bh}
{\bf H}_{B} =  -J \sum_{\langle jk \rangle}(\hat{b}_{j}^{\dagger}\hat{b}_{k} + \hat{b}_{k}^{\dagger}\hat{b}_{j}) - \mu_{j} \sum_{j}\hat{n}_{j} + \frac{U}{2} \sum_{j}\hat{n}_{j}(\hat{n}_{j}-1).
\ee
$J$ is the hopping element which allows hopping of bosons between the nearest neighbor pair of sites represented by $\langle jk \rangle$. $\hat{b}_{j}$ and their Hermitian conjugate $\hat{b}_{k}^{\dagger}$ are the boson operators, which annihilates and creates bosons on the sites $j$ and $k$ of a regular lattice in $D$ dimensions respectively. The two operators obey the commutation relation 
\be
[\hat{b}_{j}, \hat{b}_{k}^{\dagger}] = \delta_{jk}
\ee
while the two creation or annihilation operators always commute. 
\be
\hat{n}_{j}=\hat{b}_{j}^{\dagger}\hat{b}_{j}
\ee
is the boson number operator which counts the number of bosons on each site. $\mu_{j}$ is the chemical potential of the bosons, denotes the energy offset 
due to external harmonic confinement of the atoms in the $j$th lattice 
site, and $U$ is the repulsive interaction between two atoms in a single lattice site. $U>0$, represents the simplest possible on-site repulsive interaction between the bosons (on-site repulsion). ${\bf H}_{B}$ is invariant under global $U(1)\equiv O(2)$ phase transformation under which 
\be
b_{j}\rightarrow b_{j}e^{i\phi}.
\label{sym}
\ee
We can also notice that the $J$ term couples neighboring sites in a manner that prefers a state that breaks the global symmetry. However these terms compete with the $U$ term in ${\bf H}_{B}$, which are local and prefer states that are invariant under symmetry transformation. Therefore we can expect a quantum phase transition in ${\bf H}_{B}$ as a function of $J/U$ between a state in which the $U(1)$ symmetry (\ref{sym}) is unbroken to one in which it is broken. As the consequence of $U(1)$ symmetry we have the conservation on the total number of bosons
\be  
\hat{N}_{b} = \sum_{j} \hat{n}_{j},
\ee 
and this is easily verified by noting that the $N_{b}$ commutes with $\hat{H}$. We can notice that the chemical potential $\mu$ in ${\bf H}_{B}$ is coupled to the conserved total number of bosons $\hat{N}_{b}$. In contrast, it can be noted that ${\bf H}_{B}$ remains invariant under (\ref{sym}) for any value of $\mu$ and $\mu$ term does not break any symmetries. This gives no choice but to examine ${\bf H}_{B}$ for all $\mu$ and can be done by using a mean-field theory approach.
\par
Using the mean-field theory approach, the properties of ${\bf H}_{B}$ are modeled by the best possible sum, ${\bf H}_{MF}$, of single Hamiltonians
\be
{\bf H}_{MF} = \sum_{j} \left ( -\mu \sum_{j}\hat{n}_{j} + \frac{U}{2} \hat{n}_{j}(\hat{n}_{j} -1) - \Psi^{*}_{B}\hat{b}_{j}- \Psi_{B}\hat{b}^{\dagger}_{j} \right )
\label{mfbh}
\ee
where the complex number $\Psi_{B}$ is a variational parameter and a field to represent the influence of the neighboring sites; this field has to be self consistently determined. This breaks the $U(1)$ symmetry and does not conserve the total number of particles. As the mean-field Hamiltonian is the same on every site, the ground state does not spontaneously break a translational symmetry of the lattice. 
\par
To determine the optimal value of the parameter $\Psi_{B}$, the ground state wavefunciton of ${\bf H}_{MF}$ for an arbitrary $\Psi_{B}$ is determined. This wavefunction will simply be a product of single-site wavefunctions. Next the expectation value of ${\bf H}_{B}$ in the wavefunction is evaluated by adding and subtracting ${\bf H}_{MF}$ by ${\bf H}_{B}$. Then the mean-field value of the ground state energy is written in the form
\be
\frac{E_{0}}{M} = \frac{E_{MF}(\Psi_{B})}{M} - Z J\langle \hat{b}^{\dagger} \rangle \langle \hat{b}\rangle +\langle \hat{b} \rangle \Psi^{*}_{B}+\langle \hat{b}^{\dagger} \rangle \Psi_{B},
\label{mfwf}
\ee
where $M$ is the number of the lattice sites, $E_{MF}(\Psi_{B})$ is the ground state energy of ${\bf H}_{MF}$, $Z$ is the number of nearest neighbors around each lattice point (the coordination number), and the expectation values are evaluated in the ground state of ${\bf H}_{MF}$. Final step is to minimize (\ref{mfwf}) over variations in $\Psi_{B}$. By taking the derivative of (\ref{mfwf}) with respect to $\Psi_{B}$, it can be show that at the optimum the value of $\Psi_{B}$
\be
\Psi_{B} = Z J\langle \hat{b} \rangle;
\ee
this relation, however does not hold at a general point in parameter space.
\par
When $J=0$, the sites are decoupled, and the mean field theory is exact, $\Psi_{B} =0$, and we simply have to minimize the on-site interaction energy. The on-site Hamiltonian involves only the operator $\hat{n}$, and the solution involves finding the boson occupation number (which are integer-valued eigenvalues of $\hat{n}$) that minimizes ${\bf H}_{B}$. We get the ground state wavefunction 
\be
|m = n_{0} \left(\mu/U \right ) \rangle,
\label{m}
\ee
where the integer-valued function $n_{0} (\frac{\mu}{U})$ is given by
\be
n_{0}\left(\frac{\mu}{U}\right) = 
\begin{cases}
0, &   ~~{\rm for}~~~~ \mu/U < 0, \\
1, &   ~~{\rm for}~~~~ 0 < \mu/U < 1, \\
2, &   ~~{\rm for}~~~~ 1 < \mu/U < 2, \\
. &    ~~~~~~~~~~.~~~~~\\          
. &    ~~~~~~~~~~.~~~~~\\                    
n, &   ~~{\rm for}~~~~ n-1 < \mu/U < n
\end{cases}
\ee
\begin{figure}
\bc
\includegraphics[width=8.5cm]{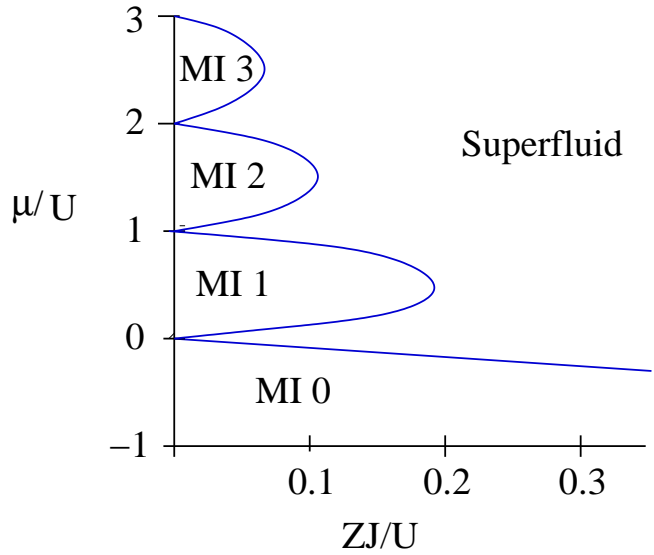}
\caption[Mean-field phase diagram of the ground state of the boson Hubbard model ${\bf H}_{B}$.]{Mean-field phase diagram of the ground state of the boson Hubbard model ${\bf H}_{B}$. The notation MI n refers to a Mott insulator with $n_{0}(\mu /U) = n$.}
\label{fig2}
\ec
\end{figure}
Each site has exactly the same number of bosons, which jumps discontinuously whenever $\mu/U$ goes through positive integer. When $\mu/U$ is exactly equal to a positive integer, there are two degenerate state on each site (with boson numbers differing by 1) and so the entire system has a degeneracy of $2^M$. This large degeneracy implies a macroscopic entropy which can be lifted once we turn on a nonzero $J$. Figure \ref{fig2} is the phase diagram of the ground state of the Bose-Hubbard model ${\bf H}_{B}$ (\ref{bh}). As shown in the Figure \ref{fig2}, the regions with $\Psi_{B} = 0$ survive in lobes around $J=0$, (\ref{m}) characterized by a given integer value of $n_{0}(\mu/U)$. 
Only at degenerate point with $\mu/U =$ integer does a nonzero $J$ immediately lead to a state with $\Psi_{B}\neq 0$. First we will consider the additional property of the lobes with $\Psi_{B}=0$. The expectation value of the number of each site is given by 
\be
\langle \hat{b}^{\dagger}_{j}\hat{b}_{j} \rangle = n_{0}(\mu /U).
\ee
 Existence of an energy gap and the fact that $\hat{N}_{b}$ commutes with $H_{B}$ are the important ingredients of the above expression.
First recall that for $J=0$, provided $\mu/U$ was not exactly equal to positive integer, there was a unique ground state, and there was a nonzero energy gap. As a result, when we turn on a small nonzero $J$ , the ground state will move adiabatically without undergoing any level crossings with other state. Now the $J=0$ state is exact eigenstate of $\hat{N}_{b}$ with eigenvalue $Mn_{0}(\mu/U)$, and the perturbation arising from a nonzero $J$ commutes with $\hat{N}_{b}$. Consequently the ground state will remain an eigenstate of $\hat{N}_{b}$, with precisely the same eigenvalue, $Mn_{0}(\mu/U)$, even for small nonzero $J$. These regions with a quantized value of the density and an energy gap to all excitations are known as Mott insulators. The Mott insulators are also known as incompressible because their density does not change under changes of the chemical potential $\mu$ or other parameter in ${\bf H}_{B}$:
\be
\frac {\partial \langle \hat{N}_{b}\rangle}{\partial \mu} = 0. 
\ee
The boundary of the Mott insulating phase is a second order quantum phase transition, i.e., a nonzero $\Psi_{B}$ turns on continuously.
\par
When we turn to the phase with $\Psi_{B} \neq 0$. The mean-field parameter $\Psi_{B}$ varies continuously as the parameters are varied. As a result all thermodynamic variables also change, and the density can be varied smoothly across any real positive value. So this is a compressible state in which 
\be
\frac {\partial \langle \hat{N}_{b}\rangle}{\partial \mu} \neq 0 
\ee
The presence of $\Psi_{B} \neq 0$ implies that the $U(1)$ symmetry is broken, and there is a nonzero stiffness to twists in the orientation of the order parameter. This state is a superfluid and that the stiffness is just a superfluid density.            
\par
From the above description we can conclude that when $J$ dominates the Hamiltonian, the ground-state energy is minimized if the single-particle wave functions of all $N$ atoms are spread out over the entire $M$-lattice site.  If $\mu_{j} =$ const. (homogeneous system) then the many-body ground state is called the superfluid state and is given by
\be
|\psi_{SF}\rangle_{U=0} \propto \left(\sum_{j=1}^{M} \hat{b}_{j}^{\dagger}\right)^{N}|V\rangle, 
\ee
where $|V\rangle$ is a vacuum state. In this state the probability distribution for the local occupation $n_{j}$ of atoms on a single lattice site is Poissonian. The state is well described by a macroscopic wave function with long-range phase coherence
throughout the lattice. With increase in the ratio $U/J$, the system 
reaches a quantum critical point, the fluctuations in atom number of a 
Poisson distribution become energetically very costly, and the ground state
of the system will instead undergo a quantum phase transition from
the superfluid state to the Mott insulator state, a product of local Fock states of $n$ atoms in each lattice site is given by \cite{JBC98, GME02}, 
\be
|\psi_{MI}\rangle_{J=0} \propto \prod_{j=1}^{M}
(\hat{b}_{j}^{\dagger})^{n}|V\rangle.
\ee

\section{Quantum walk on atoms in 1D Mott insulator and superfluid regime}
\label{impQW}

\begin{figure}[h]
\begin{center}
\epsfig{figure=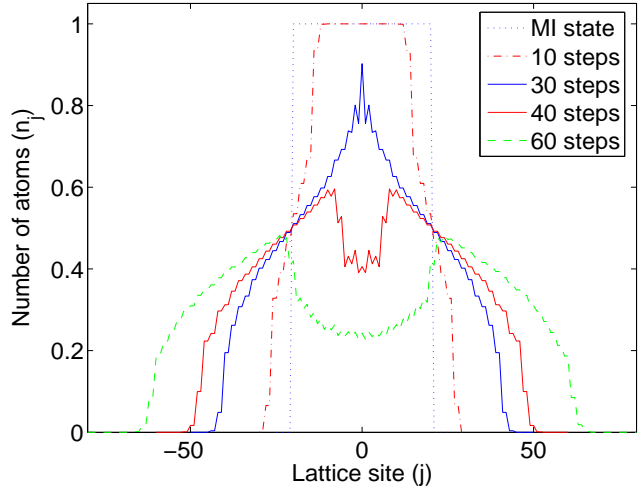, width=9.3cm}
\caption[Density profile of the evolution of 40
atoms starting from Mott insulator state in correlation with the position space when subjected to a quantum walk of different numbers of steps with the Hadamard operator as coin operation.]{\label{fig:multiMISF} Density profile of the evolution of 40
atoms starting from Mott insulator state in correlation with the position space when subjected to a quantum walk of different numbers of steps with the Hadamard operator $B_{0,45^{\circ},0}$  as the quantum coin. The distribution spreads with increase in the number of steps.}
\end{center}
\end{figure}

Lets first consider the localized atomic wave functions in the Mott insulator regime with one atom in each of the $M$ lattice sites and implement the quantum walk on it. Atoms are first initialized into a symmetric superposition of any of the two internal trappable state, hyperfine levels, $|0\rangle$ and $|1\rangle$\footnote{A $\pi/2$ radio frequency pulse can evolve the atoms into an equal superposition of the two internal trappable states} at position $j$, 
\be
|\Psi_{MI}\rangle_{J=0} \propto \prod^{\frac{M}{2}}_{j=-\frac{M}{2}}
\left( \frac {|0\rangle + i|1\rangle}{\sqrt{2}}\right)_{j}\otimes |j\rangle.
\label{mistate}
\ee 
The $\mathcal H_{c}$ of each atom is spanned by the two hyperfine levels and $\mathcal H_{p}$ is spanned by the lattice site. The total system is then in the Hilbert space $\mathcal H_{m} = (\Pi_{j}\mathcal H_{c_{j}}) \otimes \mathcal H_{p}$.
The unitary shift operation $S$ (\ref {eq:condshift}) written in the form (\ref{eq:condshift1}), on the above system will evolve each atom into superposition of the neighbor lattice site, establishing the quantum correlation between the states of the atom and the neighboring lattice site, 
\be
S|\Psi_{MI}\rangle_{J=0} \propto  \prod_{j =
-\frac{M}{2}}^{\frac{M}{2}} \left (
\frac{|0\rangle \otimes \hat{a}|j\rangle +
i|1\rangle \otimes \hat{a}^{\dagger}|j\rangle}{\sqrt 2}\right). 
\ee
For a system initially in state given by (\ref{mistate}), $t$ steps of quantum walk is implemented by iterating the process of $S$ followed by the quantum coin operation $B_{\xi, \theta, \zeta}$ (\ref{U2}), $W = (B_{\xi,\theta,\zeta} \otimes \mathbbm{1})S$, upto $t$ times. During the iteration, the atom and position correlations overlap resulting in,
\begin{eqnarray}
\label{eq:manyUMI}
(W)^{t}|\Psi_{MI}\rangle_{J=0}  \propto \prod_{j = -\frac{M}{2}}^{\frac{M}{2}} ( \beta_{j-t}|0\rangle \otimes |j-t\rangle +\nonumber \\
\beta_{j-(t+1)} |0\rangle \otimes |j-(t+1)\rangle +......+\beta_{j+t}|0\rangle \otimes |j+t\rangle \nonumber \\
+ \gamma_{j-t}|1\rangle \otimes |j-t\rangle + ......+ \gamma_{j+t})|1\rangle \otimes  |j+t\rangle ).
\end{eqnarray}
\noindent This can be written as
\begin{eqnarray}
\label{eq:manyUMIa}
(W)^{t}|\Psi_{MI}\rangle_{J=0}  \propto  
\prod_{j = -\frac{M}{2}}^{\frac{M}{2}}\left( \sum\limits_{x=j-t}^{j+t} [ \beta_x |0\rangle + \gamma_x |1\rangle ] \otimes |x\rangle \right ),  
\end{eqnarray}
$\beta_x$ and $\gamma_x$ are the probability amplitudes of state $|0\rangle$ and $|1\rangle$ at lattice site $x$, which range from $(j-t)$ to $(j+t)$. For a quantum walk of $t$ steps on a particle initially at position $j=0$ using $B_{0, \theta, 0}$ as the quantum coin, the probability distribution spread over the interval $(-t \cos(\theta), t \cos(\theta))$ in position space and ceases quickly outside this region \cite{NV01, CSL08}. Therefore, after $t$ steps the density profile of $M$ atoms initially between $\pm \frac{M}{2}$ will correlate with the position space and spread over the lattice site $\pm (\frac{M}{2} + t \cos(\theta))$.
 Figure \ref{fig:multiMISF} is the redistribution of atomic density of 40 atoms initially in the Mott insulator state when subjected to a quantum walk of different number of steps with Hadamard operator $H = B_{0,45^{\circ},0}$ as the quantum coin. In the lattice region range where all the  40 atoms are correlated can be seen as a superfluid region. 
\par
The redistributed atoms would be in either of their internal states and this can be retained to study the phase transition of the two-state bosonic atoms in an optical lattice \cite{AHD03} or all atoms can be transferred to one of the internal states. A technique based on adiabatic passage using crafted laser pulses can be used for a nearly complete transfer of population between two states \cite{VFS01}, and a popular example of one such technique is stimulated Raman adiabatic passage (STIRAP) \cite {BTS98}. 
\par
Along with the correlation between the atoms and the position space, (\ref{eq:manyUMI}) also reveals the overlap of the probability amplitude of different atoms in the position space. That is, during each step of the quantum walk, the amplitude of the states of each atom overlaps with the amplitude of the states of the atoms in the neighboring lattice site.  After the iteration of the $t$ steps $ \geq \frac{M/2}{\cos(\theta)}$, the overlap of all atoms could be seen at the central region of the lattice. 
\par
When the number of steps of the quantum walk is equal to the number of lattice site ($t=M$), the overlap of the fraction of amplitude of all $M$ atoms exists within the lattice sites $\pm\frac{M\cos(\theta)}{2}$ making the region identical to the superfluid state. Beyond $\pm\frac{M\cos(\theta)}{2}$ the number of atoms in correlation with that position space decreases and hence the number of overlapping atoms also decrease. To make all $M$ atoms to spread between lattice site $\pm\frac{M}{2}$ with the usual method of lowering the optical potential depth following the classical random walk protocol takes $M^{2}$ steps. Therefore, we can conclude that using quantum walk a long range correlation can be induced quadratically faster than any other technique using the classical random walk protocol.
\begin{figure}[h]
\begin{center}
\epsfig{figure=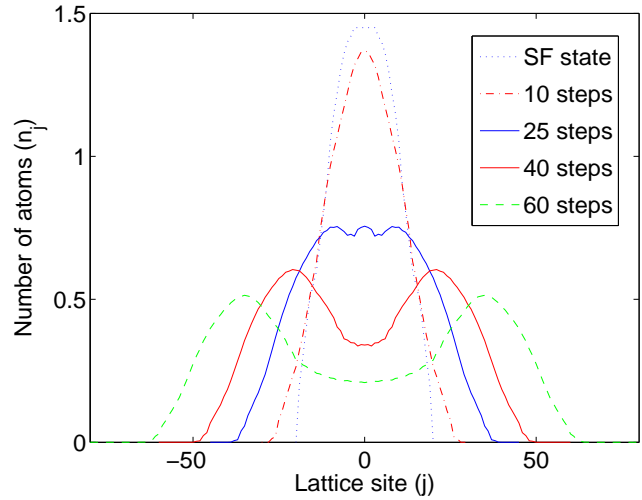, width=9.3cm}
\caption[Density profile of the evolution of 40
atoms starting from superfluid state in correlation with the position space when subjected to a quantum walk of different numbers of steps with the Hadamard operator as coin operation.]{\label{fig:multiSFMI} Density profile of the evolution of 40 atoms starting from superfluid state in correlation with the position space when subjected to a quantum walk of different numbers of steps with the Hadamard operator $B_{0,45^{\circ},0}$  as the quantum coin. The distribution spread with increase in the number of steps and is almost uniform between the lattice site $\pm 20$ when $t = 25$. At this stage the optical potential depth can be increased to cancel the correlation and obtain the Mott insulator state.}
\end{center}
\end{figure}
\par
Similarly, Figure \ref{fig:multiSFMI} is the redistribution of atomic density of 40 atoms initially in the superfluid states when subjected to the quantum walks of different number of steps using Hadamard operator $B_{0,45^{\circ},0}$ as the quantum coin. To reach Mott insulator state the distribution should be uniform over the lattice site without correlation. A distribution that is approximately uniform within the region $\pm \frac{M}{2}$ can be obtained using the quantum walk.  In Figure \ref{fig:multiSFMI}, when $t =25$ the distribution is almost uniform between the lattice site $\pm 20$ retaining the correlation with the position space.  
Once the distribution is uniform, the optical potential depth can be increased to cancel the correlation and obtain the Mott insulator state. 
The uniformity of the distribution can also be improved by introducing a noise channel, as we show later. 
\par
\begin{figure}[h]
\begin{center}
\epsfig{figure=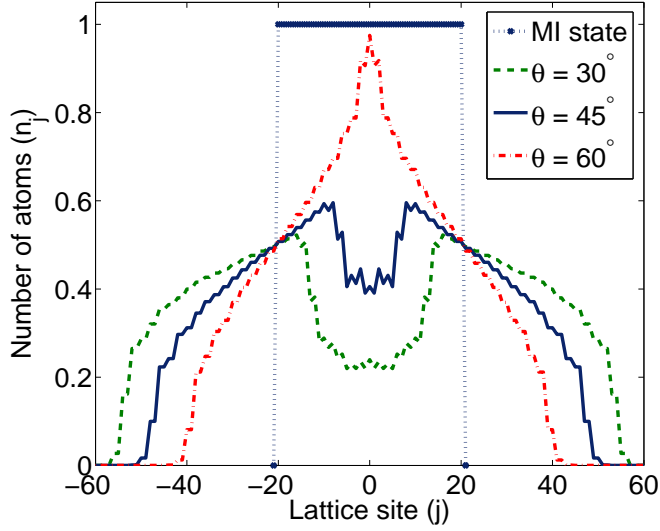, width=9.3cm} 
\caption[Distribution of atoms initially in Mott insulator state when subjected to quantum walk ($t=M=40)$ using different values for $\theta$ in the operator $B_{0, \theta, 0}$.]{\label{fig:multi1}Distribution of atoms initially in Mott insulator state when subjected to quantum walk ($t=M=40)$ using different values for $\theta$ in the operator $B_{0, \theta, 0}$. The spread is wider for $\theta = 30^{\circ}$ and decreases with increasing $\theta$.}
\end{center}
\end{figure}
\begin{figure}[h]
\begin{center}
\epsfig{figure=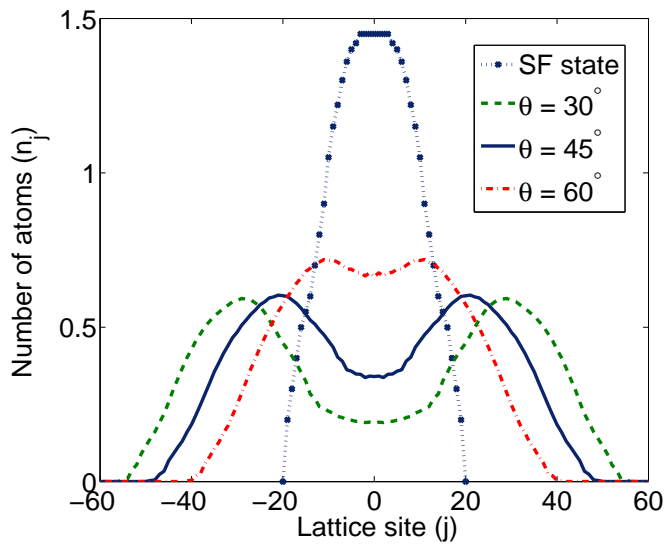, width=9.3cm}
\caption[Distribution of atoms initially in superfluid state when subjected to a quantum walk ($t=M=40)$ using different values for $\theta$ in the coin operator $B_{0, \theta, 0}$.]{\label{fig:multi2}Distribution of atoms initially in superfluid state when subjected to a quantum walk ($t=M=40)$ using different values for $\theta$ in the coin operator $B_{0, \theta, 0}$. The distribution is almost uniform for $\theta=60^{\circ}$ between the lattice site $\pm 20$. At this value the optical potential depth can be increased to cancel the correlation and obtain the Mott insulator state.}
\end{center}
\end{figure}

The variance and the probability distribution can be controlled using the parameters $\theta$, $\xi$ and $\zeta$ in $B_{\xi, \theta, \zeta}$. Figures \ref{fig:multi1} and \ref{fig:multi2} shows the density distribution obtained by implementing the quantum walk on atoms in Mott insulator and superfluid state $(t=M=40)$ respectively with different values of $\theta$ in the coin operator $B_{0,\theta,0}$.  The spread is wider for $\theta=30^{\circ}$ and decreases with increase in $\theta$. In Figure \ref{fig:multi2} the distribution is almost uniform for $\theta=60^{\circ}$ between the lattice site $\pm 20$. At this value the optical potential depth can be increased to cancel the correlation and obtain the Mott insulator state.
\par
\par
The Hamiltonian of the system in general can be described by the 1D Bose-Hubbard model for two-state atoms, 
\begin{eqnarray}
\label{eq:manyH}
{\bf H}_{BH_{2}} = -J_{\uparrow}\sum_{\langle j, k \rangle} \hat{b}_{j \uparrow}^{\dagger}\hat{b}_{k \uparrow} -J_{\downarrow}\sum_{\langle j, k \rangle} \hat{b}_{j \downarrow}^{\dagger}\hat{b}_{k \downarrow} +  \sum_{j, \alpha = \uparrow , \downarrow}\epsilon_{j, \alpha}\hat{n}_{j, \alpha} + \nonumber \\
U \sum_{j}(\hat{n}_{j \uparrow} -\frac{1}{2})(\hat{n}_{j \downarrow}-\frac{1}{2})+  \frac{1}{2}\sum_{j, \alpha = \uparrow , \downarrow}V_{\alpha} \hat{n}_{j \alpha}(\hat{n}_{j \alpha}-1) \nonumber \\
+ \sum_{j}\left ( d_{L} \hat{b}_{(j-1)\uparrow}^{\dagger}\hat{b}_{j\uparrow} + d_{R} \hat{b}_{(j+1)\downarrow}^{\dagger}\hat{b}_{j\downarrow } \right),
\end{eqnarray}
\noindent
here $\uparrow$ and $\downarrow$ represent the terms for atoms in state $|0\rangle$ and state $|1\rangle$ respectively. $U$ is the interaction between atoms in state $|0\rangle$ and $|1\rangle$; $V_{\uparrow (\downarrow)}$ is the interaction between atoms in same state. $d_L$ and $d_R$ are the left and right displacement terms in the Hamiltonian operation. The Hamiltonian is evolved with coin toss operation in a regular interval of time $\tau$, the time required to move the atom to the neighboring site. The optical lattice can be dynamically manipulated to evolve atoms in a superposition of lattice site without giving time for the atom-atom interaction, and the optical potential depth of the system can be configured just above the level where there is no direct tunneling. Then the dynamics of atoms in lattice will be dominated by the last term of the Hamiltonian in (\ref{eq:manyH}) ignoring the  atom-atom interaction. Therefore scaling up the scheme of using the quantum walk on systems with large number of atoms in each lattice site or to infinitely large numbers of lattice site is also straight forward; whereas, the atom-atom interaction play a prominent role during the usual method of varying the potential depth.  

\section{Quantum walk with a noisy channel as a toolbox}
\label{qwnoise}

We saw the effect of quantum walk with different number of steps and the coin parameters $\theta$ on atoms in optical lattice in the previous section. To demonstrate the effect of the quantum walk with a noisy channel on atoms in optical lattice we consider a bit-flip channel, a phase-flip channel and an amplitude-damping channel.  The bit-flip channel flips the state of the particle from $|0\rangle$ ($|1\rangle$) to $|1\rangle$ ($|0\rangle$) (Pauli $X$ operation) and a phase-flip channel flips the phase of $|1\rangle$ to $-|1\rangle$ (Pauli $Z$ operation). We use the notation $p$ for the noise level where, $0\le p \le 1$. Therefore, the bit-  (phase-)flip channel flips the state (phase) with probability $p$ during each step of the quantum walk. An amplitude-damping channel leaves state $|0\rangle$ unchanged but looses the amplitude of state $|1\rangle$ with probability $p$ resulting in an asymmetric distribution \cite{NC00}. From Chapter \ref{Chapter3} we know that the maximum decoherence effect using bit- and phase-flip channel is at $p=0.5$ due to symmetries induced by these two channels during the quantum walk; whereas, the amplitude-damping channel does not obey any symmetry and hence the maximum effect is for $p=1$ \cite{CSB07}. Our numerical implementation of these channels evolves the density matrix employing the Kraus operator representation. 
\par
\begin{figure}
\begin{center}
\epsfig{figure=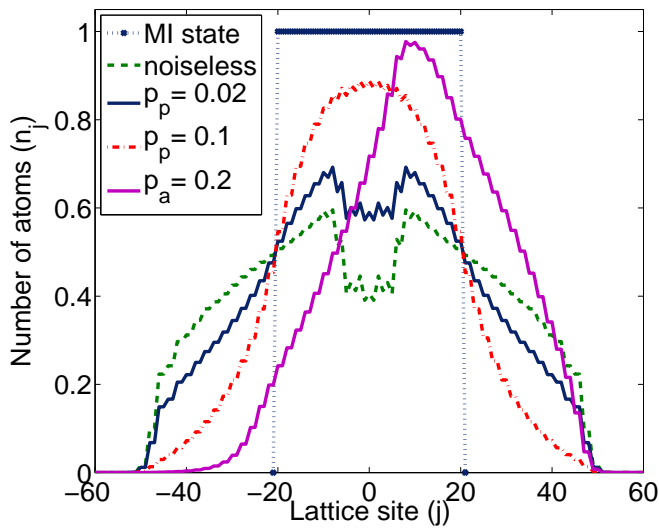, width=9.3cm}
\caption[ Atoms in Mott insulator state after implementing the quantum walk ($t = M = 40, \theta=45^{\circ})$ with noise channel.]{\label{fig:MISFPAdamp} Atoms in Mott insulator state after implementing the quantum walk ($t = M = 40, \theta=45^{\circ})$ with noise channel. With increased phase damping from $p_{p} = 0.02$ to $p_{p} = 0.1$ the distribution ($n_{j}$) at the central region gets closure to Gaussian. Amplitude damping of state $|1\rangle$ followed by a bit flip at each step introduces asymmetry to the distribution, $p_{a} = 0.2$ (without a bit flip the shift would have been to the left).}
\end{center}
\end{figure}
\begin{figure}
\begin{center}
\epsfig{figure=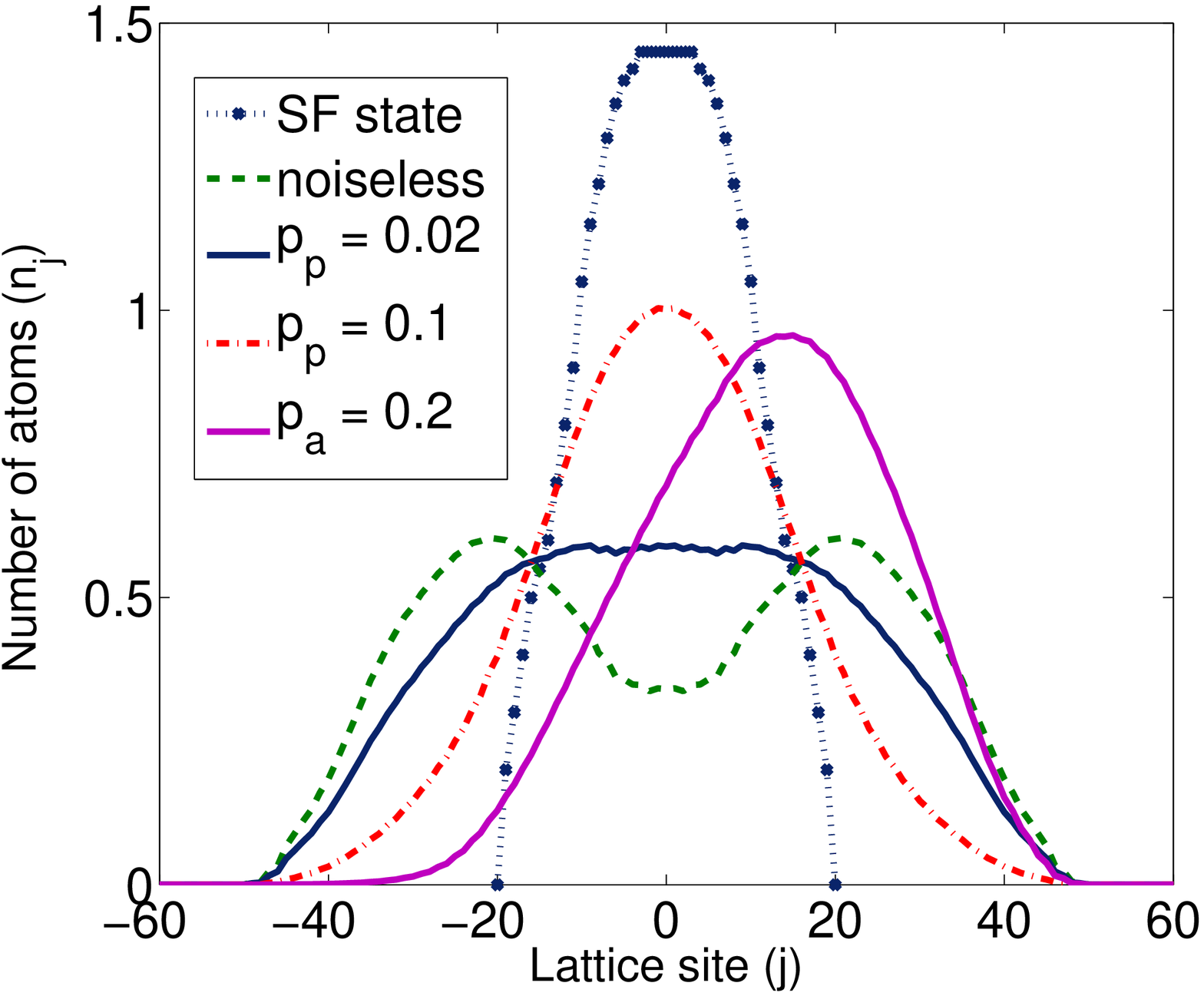, width=9.3cm}
\caption[Atoms in superfluid state after implementing a quantum walk $(t = M = 40, \theta=45^{\circ})$ with noise channel.]{\label{fig:multi5}
Atoms in superfluid state after implementing a quantum walk $(t = M = 40, \theta=45^{\circ})$ with noise channel. With noiseless quantum walk, the  atoms spread, getting close to a uniform distribution. For phase damping $p_{p} = 0.02$ the distribution is uniform between $\pm 20$ and gets closer to Gaussian at $p_{p} = 0.1$. Amplitude damping of state $|1\rangle$ followed by a bit flip introduces asymmetry to the distribution, $p_{a} = 0.2$  (without a bit flip the shift would have been to the left).}
\end{center}
\end{figure}

Figure \ref{fig:MISFPAdamp} is a comparison of the distribution obtained without noise channels to the distributions with phase-flip ($p = 0.02$ and $p=0.1$) and amplitude-damping channel ($p=0.2$) on atoms initially in the Mott insulator state. 
Similarly, the redistribution of atoms in superfluid state when subjected to quantum walk without and with noise channels is shown in Figure \ref{fig:multi5}. In Figure \ref{fig:multi5}, with the phase flip noise of $p=0.02$, a uniform distribution is obtained. Then the optical potential depth can be increased to cancel the correlation and obtain the Mott insulator state. Other plots shows the different redistribution of atoms that can be obtained using different noise channels.
\par
The effect of bit-flip on the distribution is close to the one obtained using phase-flip channel and hence the redistribution in the Figures \ref{fig:MISFPAdamp} and \ref{fig:multi5} are shown only for phase-flip and amplitude-damping channels. Increasing noise level affects the variance of the quantum walk \cite{CSB07}, which proportionally affect the atom-position correlation and the atom-atom overlap region. Therefore, higher noise level classicalize the dynamics and hence, it is important to restrict the noise level to a very low value ($p \lessapprox 0.1$) to use noisy channel as a useful tool.
\par
\begin{figure}
\begin{center}
\begin{tabular}{cc}
\epsfig{figure=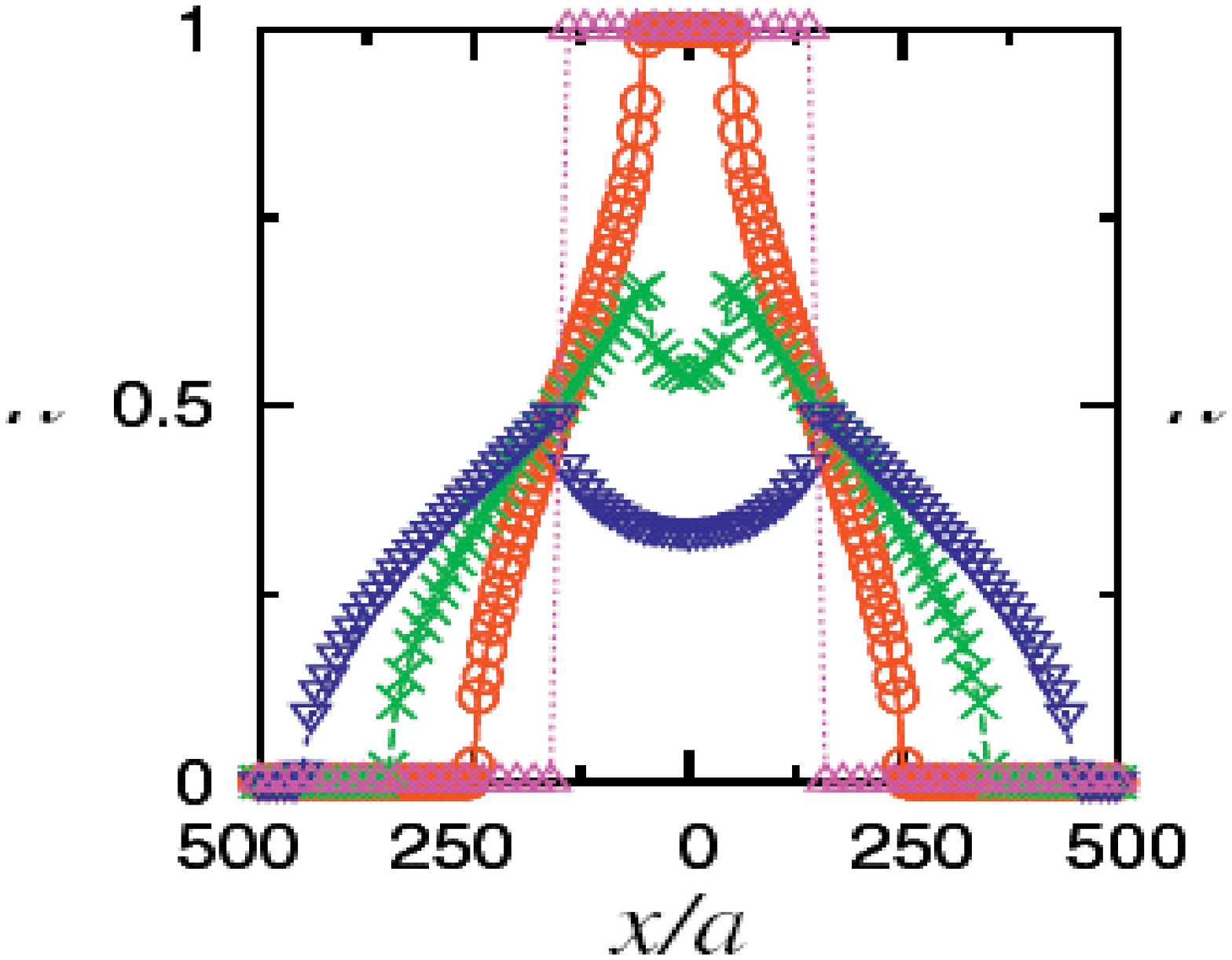, width=6.2cm}
\epsfig{figure=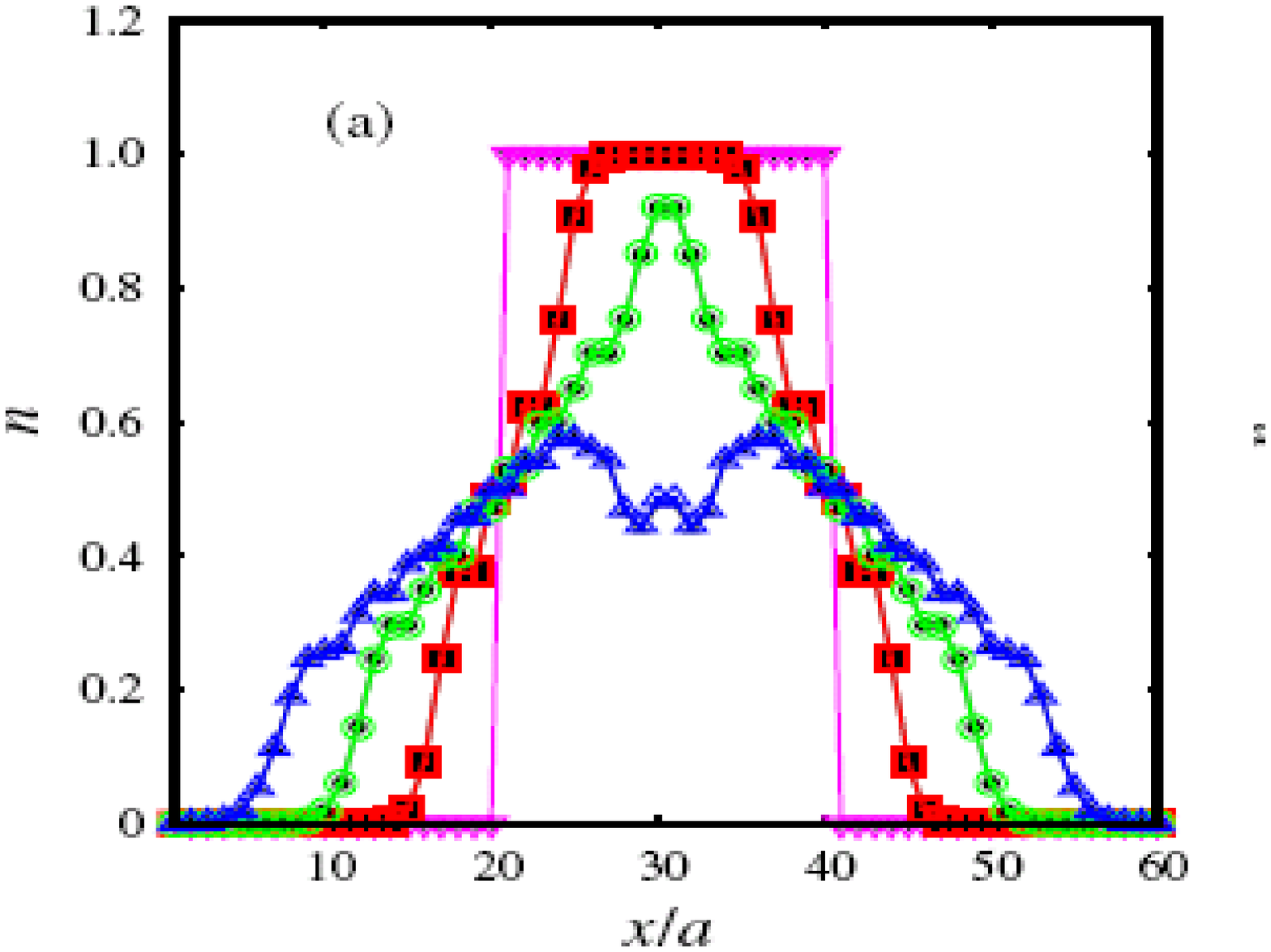, width=6.2cm}\\
(a) \hskip 5.3cm
(b)
\end{tabular}
\caption[Density profile of the evolution of atoms initially in MI state from other references. (a) Evolution of 100 atoms with time for atoms initially in MI state. (b) Evolution of  20 atoms with time (the dynamics is driven by the potential depth and tunneling of atoms between the lattice).]
{Density profile of the evolution of atoms initially in MI state from other references. (a) Evolution of 100 atoms with time for atoms initially in MI state \cite{RM05}. (b) Evolution of  20 atoms with time (the dynamics is driven by the potential depth and tunneling of atoms between the lattice)  \cite{RMR06}. These profile have visibly similar distribution compared to Figure \ref{fig:multiMISF}, Figure \ref{fig:multi1} and part of Figure \ref{fig:MISFPAdamp} in this chapter.}
\label{fig:ref1}
\end{center}
\end{figure}
\begin{figure}
\bc
\epsfig{figure=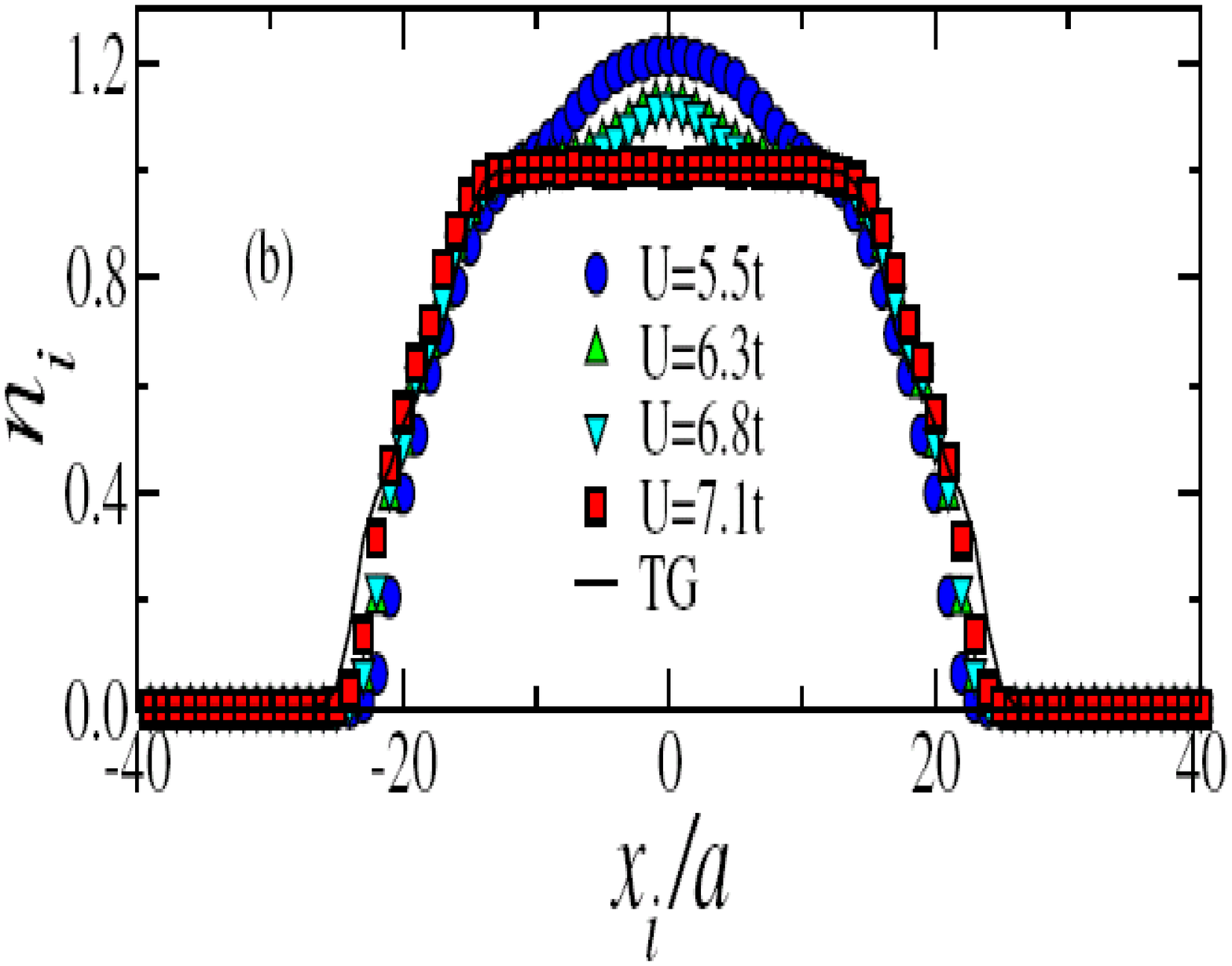, width=6.9cm}
\caption[Density profile of the evolution of 40 atoms initially in SF state from other reference.]{Density profile of the evolution of 40 atoms initially in SF state from \cite{SRB05}. The initial SF profile in our plots and in \cite{SRB05} are different and hence the similarity are not as close as we have for atoms in MI state). The profile in Figure \ref{fig:multiSFMI}
 after 25 steps of quantum walk is similar to the Profile at TG limit (Tonks-Girardeau limit - leading to MI state) in the above figure.}
\label{fig:ref2}
\ec
\end{figure}

Some of the distribution presented in Sections \ref{impQW} and \ref{qwnoise} using quantum walks are visibly similar to some of the distribution presented using different technique; in \cite{RMR06} using time evolution density matrix renormalization group (t-DMRG) and in \cite{SRB05, RM05} using quantum Monte Carlo simulation, Figure \ref{fig:ref1} and Figure \ref{fig:ref2}. These technique present the change in density profile distribution with time, whereas in this chapter we have discussed the change with number of steps of the quantum walk. 
\par
In this section, we have shown the use of the quantum walk to study the dynamics of atoms in an optical lattice and expedite the process of quantum phase transition. We have also used the quantum walk with experimentally realizable noisy channels to show the additional control one can have over the evolution and atomic density redistribution.  Theoretically, the evolution of the density profile with a quantum walk can be used in place of quantum Monte Carlo simulation to study the correlation and redistribution of atoms in optical lattices. We expect the quantum walk to play a wider role in simulating and expediting the dynamics in various physical systems.
\section{Implementation}
\label{impl}

Experiments on ultra cold atoms has been one of the active areas of research in the last two decades. Optical lattices ranging from a simple periodic, square and cubic, to a more exotic ones, such as hexagonal  and Kagome lattices using superlattice technique (see for example, \cite{GLV93, PCG94, DMK03, RDF03, SBC04, DFE05, IWM06}) have been created to trap and manipulate ultra cold atoms. Manipulation of cold atoms in time-varying optical lattice has also been reported \cite{BCF04, BGS06}. This has provided flexibility in designing and studying quantum phases and quantum phase transitions in coherent and strongly correlated ultracold atoms. 
\par
Various schemes have been proposed to implement quantum walk on neutral atoms in an optical lattice \cite{DRK02, EMB05}. In \cite{MGW03a}, the controlled coherent  transport and splitting of  atomic wave  packets in  spin dependent  optical lattice  has been experimentally  demonstrated  using  rubidium atoms.  A  BEC of up to $N = 3 \times  10^{5}$ atoms is initially created in a
harmonic  magnetic  trap.  A  three  dimensional  optical  lattice  is
superimposed  on the  BEC and  the intensity  is
raised  in order  to drive  the system  into a  Mott  insulating phase
\cite{GME02}.
\par
Two of the three orthogonal  standing wave light fields is operated at
one  wavelength, $\lambda_{y,z}  = 840$  nm and  the third  along the
horizontal direction is tuned to the wavelength $\lambda _{x} =785$ nm
between the  fine structure  splitting of the  rubidium $D1$  and $D2$
transitions.    Along  this  axis   a  quarter   wave  plate   and  an
electro-optical  modulator  (EOM)  is  placed  to  allow  the  dynamic
rotation of  the polarization  vector of the  retro-reflected laser
beam through an  angle $\theta$ by applying an  appropriate voltage to
the  EOM.  After  reaching  the Mott  insulating  phase, the  harmonic
magnetic  field is completely  turned off  but a  homogeneous magnetic
field  along the  $x$ direction  is  maintained to  preserve the  spin
polarization  of  the atoms.   The  light field  in  the  $y$ and  $z$
direction  is  adiabatically  turned  off to  reduce  the  interaction
energy, which  strongly depends on the  confinement of the  atoms at a
single lattice site.
\par
A  standing  wave  configuration  in  the $x$  direction  is  used  to
transport  the  atoms.  By  changing  the  linear polarization  vector
enclosing angle $\theta$, the separation between the two potentials is
controlled. By rotating the polarization angle $\theta$ by $\pi$, with
the  atom in  a superposition  of  internal states,  the spatial  wave
packets of the  atom in the $|0\rangle$ and  the $|1\rangle$ state are
transported  in opposite  directions.  The  final state  after  such a
movement   is   then    given   by   $1/\sqrt{2}(|0,x-1\rangle   +   i
\exp(i\beta_p)|1, x+1\rangle)$.  The phase $\beta_p$ between the separated
wave-packets depends  on the accumulated kinetic  and potential energy
phases in  the transport process and  in general will  be nonzero. The
coherence  between  the two  wave-packets  is  revealed by  absorption
imaging  of the momentum  distribution. A  $\pi/2$ microwave  pulse is
applied before  absorption imaging to erase  the which-way information
encoded in the hyperfine states.
\par
However,  to  increase the  separation  between  the two  wave-packets
further, one could increase the polarization angle $\theta$ to integer
multiples of $180^{\circ}$. To  overcome the limitation of the maximum
voltage  that can  be applied  to  the EOM,  a $\pi$  pulse after  the
polarization  is  applied,  thereby  swapping  the roles  of  the  two
hyperfine  states.   The   single  particle  phase  $\beta_p$  remains
constant throughout  the atomic cloud and is  reproducible.  After the
absorption imaging a Gaussian  envelope of the interference pattern is
obtained.
\par
One can build upon the  above technique to implement a quantum walk. 
The above setup can be modified by dividing the separations (splitting) into small 
steps and introducing a rotation ($\pi/2$ pulse for Hadamard rotation) after each 
separation without intermediate imaging. The absorption
imaging of the distribution of  the atomic cloud after $t$ steps would
give the interference pattern similar to the probability distribution of the 
quantum walk.
\par
This effect of  the addition of phase during  the quantum walk process
can be easily understood from  the phase damping channel and arbitrary
phase rotation discussed in Chapter ~\ref{Chapter2}. 
The addition of $\pi$ pulse to overcome the  limit of EOM is a bit flip 
operation in the quantum walk.  
\par
In the following section we propose a scheme to implement a discrete time quantum walk on ultra cold atoms in BEC state. BEC being a macroscopic wave packet, the scheme we propose can implement a quantum walk retaining the macroscopic features of the wave packet \cite{Cha06} or the scheme can be accordingly modified to implement the walk at an individual atom level.  For the transition from superfluid to the Mott insulator and vice versa using quantum walk, we need to consider the quantum walk on individual atom. 
\par
To implement a macroscopic quantum walk, atoms in BEC state is first evolved to macroscopic superposition state  and a {\em  stimulated Raman kick}, two selected levels of the atom are coupled to the two modes of counterpropagating laser beams to coherently impart a translation of atoms in the position space.  After each translation, the wave packet is again evolved into the macroscopic superposition state and the process is iterated to implement large number of steps of macroscopic quantum walk. With a certain modification to this scheme, that is, by evolving atoms to the superposition of the states at individual atom level and implementing the shift operation before the interatomic interaction takes over to form a macroscopic superposition, the quantum walk at individual atom level can be realized.

\subsection{Quantum walk using Bose-Einstein condensate}
\label{becqw}

As discussed in Section \ref{dtqw}, two degrees of freedom, the coin Hilbert space  $\mathcal H_{c}$ and  the position  Hilbert space $\mathcal H_{p}$ are required to implement the discrete-time quantum walk.  The state of the BEC formed from the atoms in one of the hyperfine states  $|0\rangle$  or $|1\rangle$ can be represented as $|0_{BEC}\rangle$ or $|1_{BEC}\rangle$ (macroscopic states of $N$ condensed atoms). The BEC formed is then transferred  to an optical  dipole   trap  with  long  Rayleigh   range $z_{R}$\footnote{Distance at which the diameter of the laser beam size increases by a factor of $\sqrt{2}$.}. With the appropriate choice     of      power     and      beam     waist, $\omega_{0}$\footnote{Minimum radius of the beam, at the focal point.} of the trapping  beam, the BEC can remain  trapped at any point within the  distance  $\pm z$  from  the focal  point  in the  axial
direction of the  beam (See Appendix~\ref{AppendixC} for calculations). 
\par
The position of the BEC formed in one of the Hyperfine state $|0_{BEC}\rangle$ or $|1_{BEC}\rangle$ and trapped at the center of  the optical trap is described by  a wave packet $|\Psi_{j_{0}}\rangle$  localized around a position $j_{0}$, i.e., the  function $\langle j | \Psi_{j_{0}}\rangle$
corresponds to a wave packet centered  around $j_{0}$. The BEC formed in one of the Hyperfine state is evolved  into the macroscopic superposition (Schr\"odinger cat) state 
\be
\label{wavefunction}
|\Psi_{in}\rangle= \left ( a |0_{BEC}\rangle+ b |1_{BEC}\rangle \right ) \otimes|\Psi_{j_{0}}\rangle
\ee
by applying the rotation (coin operation $B$), where $|a|^2 + |b|^2=1$. We will discuss this process in detail in Section~\ref{catstate}. The coin Hilbert space $\mathcal H_{c}$ can then be defined to be spanned by the two internal trappable macroscopic states $|0_{BEC}\rangle$  and  $|1_{BEC}\rangle$ of the BEC. 
The position  Hilbert space $\mathcal  H_{p}$ is spanned by the positions within the long Rayleigh range optical trap (trappable range). Once the BEC is in the superposition state, a unitary shift operator, {\it stimulated Raman kick}, Section \ref{ramankick}, corresponding  to
one step length $l$ in the form (\ref{eq:modcondshift}) is applied;
\be
S^{\prime}      =     (X      \otimes
\mathbb{I})\exp(-i(|0_{BEC}\rangle \langle 0_{BEC}| - |1_{BEC}\rangle \langle 1_{BEC}| )\otimes Pl),
\ee
\noindent
$P$ being the momentum operator. Step length $l$ is chosen to be less
than the spatial width of the condensate wave packet $|\Psi_{j_{0}}\rangle$.
\par
The application of the unitary shift operator ($S^{\prime}$) on the wave function of (\ref{wavefunction}) entangles the position and the
coin space and implements the quantum walk. Stimulated Raman kicks evolves the state to, 
\be
S^{\prime}|\Psi_{in}\rangle=\left ( a |1_{BEC}\rangle\otimes e^{-iPl}
+ b |0_{BEC}\rangle\otimes e^{iPl} \right ) \left (   \mathbb{I} \otimes |\Psi_{j_{0}}\rangle \right ),
\ee
\noindent
where the wave packet is  centered around $j_{0}\pm l$.  Note that the
values  in the  coin  space have  been  flipped. This can be corrected  by
applying a compensating  bit flip on the BEC, $S  \equiv     (X    \otimes
\mathbb{I})S^{\prime}$. From the symmetries of quantum walk discussed in Chapter \ref{Chapter3} we know that the distribution is invariant to bit flips at each step and therefore, augmenting $S^{\prime}$ with the $X$ operation is unnecessary. To realize a large number of steps  of the macroscopic quantum walk, the process of  shift operation followed by the quantum coin operation $(B_{\xi, \theta, \zeta}\otimes \mathbb{I} ) S^{\prime}$ is iterated  without resorting  to intermediate  measurement in  the long Rayleigh-range optical dipole trap. 
\par
However, the coin operation $B_{\xi, \theta, \zeta}$ can be performed such that, its action will evolve the individual atoms into the superposition of $|0\rangle$ and $|1\rangle$ and not the superposition of macroscopic state $|0_{BEC}\rangle$ and $|1_{BEC}\rangle$, Section \ref{catstate}.  This can be used to drive the transition between the superfluid and the Mott insulator state. To realize transition from the superfluid to the Mott insulator phase, atoms are first redistributed by implementing the quantum walk using the rotation operation (coin operation) $B_{\xi, \theta, \zeta}$ on individual atoms in a long-Rayleigh-range optical trap and the optical lattice is switched on at the end to confine atoms and cancel the correlation. For transition from Mott insulator to the superfluid, the atoms initially in the optical lattice are transferred to long-Rayleigh-range dipole trap and quantum walk is implemented.

\subsubsection{Macroscopic cat state of  Bose-Einstein condensate}
\label{catstate}
Various   schemes  have  been   proposed  for   producing  macroscopic
superposition      or       Schr\"odinger      cat      states      in BECs~\cite{CLM98, RCG98, GS99}.  
The scheme described  in \cite{CLM98},  shows  that  the  two species (two interacting Bose condensates) can be evolved to the macroscopic superposition of the internal atomic states.
In this scheme, atom-atom interactions are mediated through atom-atom collisions and a Josephson-like laser coupling that interchanges internal atomic states in a coherent manner.
In  certain parameter regimes the ground state of  the Hamiltonian  is  a  superposition of  two  states involving  a particle number imbalance between  the  two condensates.  Such a  state produced  by  the normal  dynamic  evolution of  the system represents a superposition of two states which are macroscopically (or mesoscopically)   distinguishable,   and  hence   can   be  called   a Schr\"odinger cat  state.  But, two interacting Bose condensates are needed for this scheme. In \cite{RCG98}, it is shown that  a macroscopic superposition 
state can  be  created  by a  mechanism involving the  coherent scattering of far-detuned 
light  fields and it neglects  the collisional  interactions between  particles.  The major
drawback of the  above schemes is that the time needed  to evolve to a
cat state  can be  rather long, and  thus problems due  to decoherence
would   be   greatly   increased. An other scheme proposed in \cite{GS99} involves the creation if macroscopic superposition by an adiabatic transfer of the ground
state  of  the Josephson-coupling  Hamiltonian,  that  is, after  the
initial state preparation the Josephson coupling is turned on for some
amount of time and turned off. The resulting modified quantum state is
a  Schr\"odinger  cat state.   But,  the  production  of such  a  state involves 
considerable experimental difficulty.  
\par 
The  ideal scheme to demonstrate the quantum  walk is to
confine the BEC that has an attractive interaction between  atoms in two
hyperfine  levels  $|0\rangle$ and  $|1\rangle$  in  a single  optical
potential well. The rotation operation is applied on the BEC in
the  potential well  to transfer  (or rotate)  the atoms  part  of way
between  states $|0\rangle$  and   $|1\rangle$  using  a  resonant  rf
pulse \footnote{Microwave pulses are also used.} of duration $\tau$ and
detuning $\Delta$ from the  rf resonance~\cite{BJT99}. The resonance rf
pulse couples the atomic hyperfine states $|0\rangle$ and $|1\rangle$ with
a coupling  matrix element $\hbar\omega_R/2$, where  $\omega_R$ is the
Rabi frequency and  the duration of pulse is much shorter than the
self-dynamics  of the BEC.   The amplitude  of  these states  evolves
according to the Schr\"odinger equation,
\begin{equation}
\label{rf}
i \hbar \frac{d}{d\tau}\left(\begin{array}{clr}
a\\
b
\end{array}\right)=\hbar  \left( \begin{array}{clcr}
 0  & &  \omega_{R}/2   \\
 \omega_{R}/2  & & \Delta
 \end{array} \right)\left(\begin{array}{clr}
a\\
b
\end{array}\right).
\end{equation}

\noindent
At  this  stage,  each  atom  evolves  into  the  superposition  state
$a(\tau)|0\rangle+b(\tau)|1\rangle$,                               with
$a(\tau)=\cos(\omega_R\tau/2)$  and $b(\tau)=\sin(\omega_R\tau/2)$ for
detuning $\Delta=0$. The $N$-particle wave function of the BEC is a  product 
of the single-particle superpositions of  $|0\rangle$ and  $|1\rangle$, that is,
it is still a microscopic superposition and is given by
\be
\label{super}
[a(\tau)|0\rangle+b(\tau)|1\rangle]^{N}= 
\sum_{n=0}^N  \sqrt \frac{N!}{n!(N-n)!} a(\tau)^{N-n}b(\tau)^{n}|N-n, n\rangle,
\ee
where $|N-n, n\rangle$ is the state with $N-n$ atoms in state $|0\rangle$ and $n$ atoms in state $|1\rangle$. The individual atoms in the superposition state interact among themselves.
Interatomic interactions, which provide nonlinear terms through binary collision as seen from the viewpoint of single-particle dynamics helps in generating highly entangled many body states~\cite{MGW03, DGP99}. That is, the Hamiltonian that governs the BEC with its attractive inter-atomic interactions, after time $\tau_{1}$, (\ref{super}) evolves into the macroscopic superposition state in which  all  atoms  are   simultaneously  in  level  $|0\rangle$ and level  $|1\rangle$~\cite{CLM98, DD02, MAK97},
\be
[a(\tau_{1})|0\rangle+b(\tau_{1})|1\rangle]^{N}= 
 a(\tau_{1})^{N}|N, 0\rangle + b(\tau_{1})^{N}|0,N\rangle
\ee
where $|N, 0\rangle=|0_{BEC}\rangle$  and $|0, N\rangle=|1_{BEC}\rangle$.
Symmetric probability distribution $a(\tau_{1})=b(\tau_{1})=\frac{1}{\sqrt  2}$  can  be  obtained  by  carefully choosing $\omega_R$ and $\tau$.
\vskip 2.2cm
{\bf Microscopic superposition state of atoms :} 
\par
Equation (\ref{super}) gives the microscopic superposition state of $N$-atoms. Implementing the unitary shift operator before the individual
atoms in the superposition states start interacting will prevent the evolution of atoms to macroscopic superposition state. Therefore we can choose between the macroscopic, Schr\"odinger cat state or the superposition of individual atoms, microscopic superposition state. 

\subsubsection{Unitary operator - Stimulated Raman kicks}
\label{ramankick}
A  unitary shift (controlled-shift) operation $S$ is applied on
the BEC  to spatially entangle the position and the coin space and implement 
the quantum walk. Various schemes have been
worked out to give momentum kick to ultracold atoms in a 
trap~\cite{MAK97, HKH99}. A technique was reported in \cite{MAK97} 
where coherent rf-induced transitions were used to change the internal state of the atoms in the magnetic trap from a trapped to an untrapped state and thus displacing
the untrapped atoms from the trap. This method, however, did not
allow the direction of the output-coupled atoms to be chosen.
Later, controlling the direction of the fraction of the outgoing BEC using a stimulated Raman process between magnetic sublevels was experimentally demonstrated~\cite{HKH99}. This technique was used to extract sodium atoms lasers from the trapped BEC.
\par
\begin{figure}
\begin{center}
\epsfig{figure=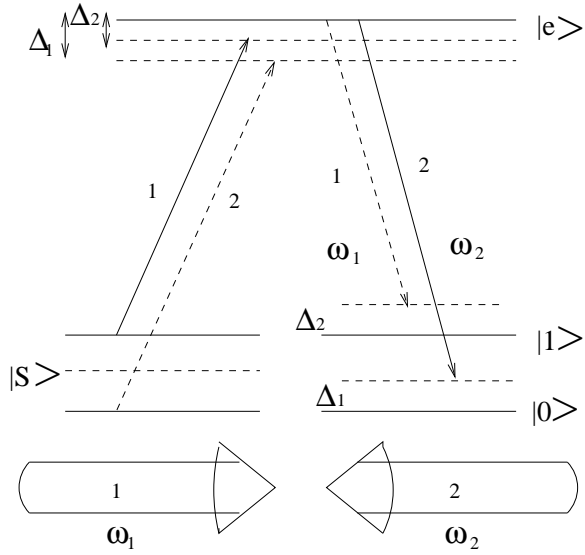, height= 7.3cm}
\caption[Light  field  configuration for a stimulated
Raman transition  process to give directional  momentum kick. ]
{\label{lightfield}Light  field  configuration for a stimulated
Raman transition  process to give directional  momentum kick. $\Delta$
is the detuning of the laser from its transition frequency.
$|S\rangle$ signifies $(1/\sqrt{2})(|0\rangle \pm |1\rangle)$.}
\end{center}
\end{figure}

A  stimulated Raman  process  can  also be  used  to drive the
transition between  two optically  trappable states  of the atom $|0\rangle$
and $|1\rangle$ using the virtual state $|e\rangle$ as an intermediary
state, and impart a well-defined momentum to spatially  translate
the atoms in the coherent state (BEC). A unitary shift operation 
$S^{\prime}$  thus  can  be applied  on  the atom in the BEC using 
a stimulated Raman process. A pair of counterpropagating laser beams 
1 and   2  (Figure~\ref{lightfield}) with  frequency   $\omega_{1}$  and 
$\omega_{2}$ and wave vectors $k_{1}$ and $k_{2}$ is  applied on 
the BEC for a $2N$-photon transition time ($N$ is the number  of atoms in the BEC)
to  implement one  unitary  shift  operation.   These beams  are
configured  to propagate  along  the axial  direction  of the
optical dipole trap. A stimulated Raman transition occurs when an
atom changes its  state by  coherently  exchanging photons
between  the two  laser fields,  absorption  of the photon  from
laser field  1  and  stimulated emission  into  laser  field 2  or
by  absorption  from field  2  and stimulated emission  into field
1.   The BEC initially in  eigenstates of the atom $|0\rangle$  ($|1\rangle$)
can  absorb  the photon from  field  1 (2)  and re-emit the photon into
field 2  (1).  This  inelastic  stimulated Raman scattering
process imparts well-defined momentum on the coherent atoms,
\be
{\bf P}=\hbar (k_{1}-k_{2})=\hbar \delta z \ee
to the left during $| 0\rangle \xrightarrow {\omega_{1}}|e\rangle\xrightarrow{\omega_{2}}| 1 \rangle $ and
\be
{\bf P}= \hbar
(k_{2}-k_{1})=-\hbar \delta z
\ee
 to the right during $| 1\rangle \xrightarrow{\omega_2} |e\rangle \xrightarrow{\omega_1}  | 0 \rangle $, where
 \bc
  $| k_{1}-k_{2}| = |k_{2}-k_{1}|=\delta$,
\ec
\noindent
Thus, conditioned to being  in coin state  $|0\rangle$ ($|1\rangle$),
the  atoms in the BEC receive  a  momentum kick (shift)  $\hbar  \delta z$  ($-\hbar
\delta  z$).   This  process  of  imparting momentum  is  called  
stimulated Raman kick and can  be analyzed as a photon absorption and
stimulated emission between  three bare states $|0\rangle$, $|1\rangle$,
and $|e\rangle$ driven by  two monochromatic light fields.  The matrix
element  for photon  absorption  and stimulated  emission during  each
stimulated  Raman kick  can  be written  as  \be \langle  n_{k_{1}}-1,
n_{k_{2}}+1, 1|\hat {H}_{I_{a}}|n_{k_{1}},n_{k_{2}}, 0 \rangle \ee
\noindent
during $| 0\rangle \xrightarrow {\omega_{1}}|e\rangle\xrightarrow{\omega_{2}}| 1 \rangle $ and
\be
\langle n_{k_{1}}+1, n_{k_{2}}-1, 0|\hat
{H}_{I_{b}}|n_{k_{1}},n_{k_{2}}, 1 \rangle
\ee
\noindent
during $| 1\rangle \xrightarrow{\omega_2} |e\rangle \xrightarrow{\omega_1}  | 0 \rangle $. ${H}_{I_{a}}$ and ${H}_{I_{b}}$ are the interaction Hamiltonians (electric dipole Hamiltonians). The field causing the transitions between bare states $|0\rangle$ and $|e\rangle$ is detuned from resonance by $\Delta_{1}$ and has a constant dipole matrix element  $V_{1}$ and phase $\varphi_{1}$. The field causing transitions between bare states $|e\rangle$ and $|1\rangle$ is detuned from resonance by $\Delta_{2}$ and has a constant dipole matrix element  $V_{2}$ and phase $\varphi_{2}$. The time-independent Hamiltonian in the rotating-wave approximation for this system can be written in terms of projection operations as
\[
\hat{H}_{I_{a}}=\hbar \Delta_1|e\rangle \langle e| + \hbar (\Delta_{1}+\Delta_{2})|1\rangle \langle 1|- \frac{\hbar V_{1}}{2} [|e\rangle\langle 0| \exp(-i\varphi_1)
\]
\be
+|0\rangle\langle e| \exp(i\varphi_1)]- \frac{\hbar V_{2}}{2} [|e\rangle\langle 1| \exp(i\varphi_2)+|1\rangle\langle e| \exp(-i\varphi_1)]
\ee
\noindent
In the above interaction picture, the energy of the bare  state $|0\rangle$ is chosen to
be zero.
\[
\hat{H}_{I_{b}}=\hbar \Delta_2|e\rangle \langle e| + \hbar (\Delta_{1}+\Delta_{2})|0\rangle \langle 0|- \frac{\hbar V_{2}}{2} [|e\rangle\langle 1| \exp(-i\varphi_2)
\]
\be  +|1\rangle\langle   e|  \exp(i\varphi_2)]-  \frac{\hbar  V_{1}}{2}
[|e\rangle\langle      0|      \exp(i\varphi_1)+|0\rangle\langle     e|
\exp(i\varphi_1)]. \ee  In the above  interaction picture the  energy of
the    bare    state  of the atom  $|1\rangle$    is    chosen    to   be    zero.
$V_{i}=\frac{\mu_{i}\cdot E_{0_{i}}}{\hbar}$, for $i={1,2}$. $\mu_{i}$
is  the  dipole  operator   associated  with  the  states  $|0\rangle$,$|e\rangle$ 
and $|e\rangle$,$|1\rangle$ and $E_{0}$  is the electric
field of the laser beam.  

With the appropriate choice of  $V$, $\Delta$, and  $\varphi$, the probability  of 
being  in bare  state  $|1\rangle$ (or $|0\rangle$  depending on the  starting state)  
after time  $t$ can be  maximized to be close to one, where $t$ is the time required for one  stimulated Raman kick. As discussed in Section~\ref{catstate}, after every stimulated Raman kick on atoms due to the attractive interaction between atoms, they evolve into the BEC.
\par
The coin state of the atoms in the BEC $|0\rangle$ ($|1\rangle$) after  stimulated 
Raman kick  flips  the states to $|1\rangle$  ($|0\rangle$) and symmetries due to bit flip discussed in Chapter \ref{Chapter3} will take care of these flip without affecting the distribution.

\subsubsection{Physical setup for the implementation}
\label{phy-setup}
Magnetic trap technique~\cite{KDS99} played a major role in the
first formation and early experiments on the BEC. But the magnetic
trap has a limitation, it can trap and manipulate atoms only of certain
sublevels. For example, If we consider $^{87}$Rb atom as an example, the $|F=2,{\it m_{f}}= 2 \rangle$
state can be confined in the magnetic trap whereas $|F=1,{\it
m_{f}}= 1 \rangle$ cannot be confined using the magnetic
trap~\cite{MVS02}. Under appropriate conditions, the dipole
trapping mechanism works independent of the particular sub-level of
the electronic ground state. Thus, internal ground state can thus be
fully exploited using an optical dipole trap technique and be widely
used for various experiments~\cite{GWO00}. Today the BEC from bosonic
atoms has been very consistently formed and manipulated using
various configurations of magnetic and optical  traps, an all
optical dipole trap technique has also been
developed~\cite{BSC01}.
\par
A BEC first formed in one  of their internal states (eigenstates) of atoms $|0\rangle$
or  $|1\rangle$,   using  any  of   the  techniques~\cite{KDS99,  BSC01}
mentioned above, is  transferred to a far detuned,  long Rayleigh range
($z_R$) optical  dipole trap.  A  sizable number of steps  of the quantum
walk  can  be implemented  within  the  axial  range $\pm  z$  without
decoherence in a long Rayleigh-range trap. Appendix \ref{AppendixC} has
calculated numerical values of potential depth, power required to trap
$^{87}$Rb  atoms at  distance $z=x_{n}$  from the  focal point  of the
trapping beam  (after compensating for gravity) using  light fields of
various frequency and  beam waists $\omega_{0}$. In the  same way one
can work out the required power and beam waists to trap ultracold atoms at
a distance $z$  from the focal  point for different  species of  atoms using
laser fields of different detuning from the resonance.
\par
Once the BEC is transferred into the optical dipole trap, a resonant rf
pulse (rotation) of duration $\tau$ and detuning $\Delta$ is
applied to make it evolve into the Schr\"odinger cat state (or a microscopic 
superposition state depending on the requirement) after time $\tau_{1}$. The Schr\"odinger
cat state for the $^{87}$Rb BEC trapped in one of the states
$|0\rangle$=$|F=1,{\it m_{f}}= 1 \rangle$ or $|1\rangle$=$|F=2,{\it m_{f}}= 2 \rangle$,
can be realized by applying a radio frequency (rf) pulse, fast laser pulse, a standard
Raman pulse or microwave techniques.
\par
\begin{figure}
\begin{center}
\epsfig{figure=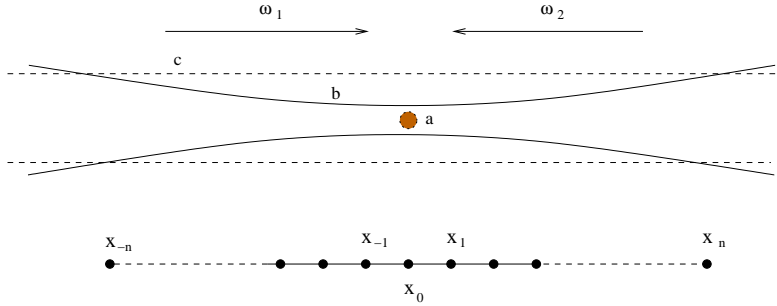, height= 4.0cm}
\caption[Physical setup  to implement the quantum walk
using the BEC.]{\label{physical-qrw}Physical setup  to implement the quantum walk
using the BEC. In the figure {\bf a} is a BEC, {\bf b} is a dipole trap with
a long Rayleigh range, and {\bf c} is the counterpropagating laser beam,
$\omega_1$  and $\omega_2$  used to  implement the stimulated  Raman kick,
unitary  operator.  A coin rotation and Raman  pulse are  applied
throughout the trap region. The stimulated Raman pulse and the rf pulse are
applied alternatively to realize the quantum walk.}
\end{center}
\end{figure}

The BEC in the superposition state is then subjected to a pair of
counterpropagating beams to implement the stimulated Raman kick for
duration  $t_{d}$,  $2N$  photon   transition  time  ($N$  being the number  of
photons). After implementing the stimulated Raman  kick, the BEC is left to
translate for a  duration $P/ml$, the time taken by the BEC to move  distance  $l$ during  or  after which  the  rf  $\pi$ pulse  is
applied.  This  process of applying  a rf pulse  (coin rotation) and a 
stimulated Raman  kick (shift), followed by a compensatory  rf pulse (bit
flip) is  iterated throughout  the trapping range  to realize  a large
number of steps (Figure~\ref{physical-qrw}).

\begin{figure}
\begin{center}
\epsfig{figure=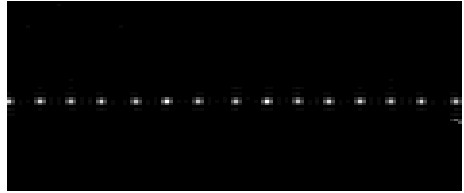, height= 2.5cm, , width= 6.00cm}
\caption[Multiple microtraps which can be used to confine the atom observe the distribution of atoms after implementing a considerable number of steps of the quantum walk on the BEC.]{\label{dipole}Multiple microtraps which can be used to confine the atom observe the distribution of atoms after implementing a considerable number of steps of the quantum walk on the BEC. }
\end{center}
\end{figure}
\vskip 1.0cm
\par
{\bf Measuring the probability distribution :}
\par

After  $t$ steps  of the quantum walk  the superposition  in position
space is made  to collapse by applying multiple  microtraps to confine
and redistribute atoms in the position space. The multiple microtraps 
with an equal spacing $l$ between each potential well can be created 
using the known techniques ~\cite{DVM02, BCF04} (Figure
~\ref{dipole}).  The microtraps are switched  on, turning off the long  Rayleigh  range
optical trap  simultaneously after  a time interval  of $T$.   
\be
T= tt_{d}+(t -1)\tau_{1} +t P/ml .
\ee
$T$ is the time taken by the BEC to travel a $t$ step quantum walk each of distance 
$l$ in a line. The  time $t_{d}$  is the $2N$-photon  transition time  required to
implement one  unitary shift operation, $\tau$  is the time duration to bring the BEC to the superposition state, $P/ml$ is the time taken by the BEC to move  one step distance $l$ with momentum $P$.
Fluorescence measurement is performed on the atoms in the microtrap to  
identify the  final position/distribution of atoms in the position space. 
\par
If the quantum walk was implemented on a BEC evolving the wave packet to the macroscopic superposition during each coin operation, the distribution of atoms will be similar to the distribution of single particle quantum walk.  We should note that by retaining the macroscopic behavior during the walk we get the quantum walk distribution in a single measurement at the end. Where as the experiment would have to be repeated a large number of times if we intend to get a probability distribution of the quantum walk using a single particle. 
\par
If the quantum walk is implemented only by evolving the atoms to the superposition at individual atom level during each coin operation we get the distributions similar to the ones presented in the earlier sections of this chapter. 

\par
{\bf Decoherence and physical limitations :}
\par
The decoherence of the BEC also leads to the decoherence
of the quantum walk. When the atoms in the trap are not coherent they no longer
follow the coherent absorption and stimulated emission of light. Some
atoms absorb light field 1 and emit light field 2 and bring atoms to
state $|1\rangle$ and some absorb light field 2 and emit light field 1
and bring atoms to state $|0\rangle$, displacing atoms in both directions
in space giving no signature of displacement in the superposition of position
space. This will contribute to collision and heating,  and finally, atoms
escape out of the trap. The number of implementable steps of quantum walk 
without decoherence depends mainly on (a) the Rayleigh range,
as the BEC moves away from the trap center the width of the wavepacket
increases and contributes to the internal heating of the atoms. Beyond a
certain distance $x_n$ from the dipole trap, center atoms in the trap
decohere resulting in the collapse of the quantum behavior. (b) The
stimulated Raman kick and rf pulse used to implement the quantum walk also
contribute to the internal heating and decoherence of the atoms after a few
iterations. With the careful selection of beam waist and laser power one can have
a trap with a long Rayleigh range $z_R$ (Appendix~\ref{AppendixC}) so that the one-dimensional
quantum walk can be implemented in a line of close to one centimeter length.

\newpage

\section{Summary} 
\label{summary4}
\begin{itemize}
\item  We proposed the use of the quantum walk to redistribute atoms, and studied its dynamics in an optical lattice and to expedite the process of quantum phase transition.  We have demonstrated the coherent control over the atoms using the coin degree of freedom during the evolution of the walk. We have also used an experimentally realizable noisy channels to show the additional control over the evolution and atomic density redistribution.  Theoretically, the evolution of the density profile with a quantum walk can be used in place of quantum Monte Carlo simulation or 
the time evolution density matrix renormalization group (t-DMRG) to study the correlation and redistribution of atoms in optical lattices. We expect the quantum walk to play a wider role in simulating and expediting the dynamics in various physical systems.

\item We proposed a scheme to implement a discrete-time quantum walk on Bose-Einstein condensate (BEC). BEC being a macroscopic wave packet, the scheme we propose can implement a quantum walk retaining the macroscopic features of the wave packet or the scheme can be accordingly modified to implement the walk at an individual atom level. To realize the quantum phase transition from superfluid to the Mott insulator state and vice versa, implementation of walk on individual atom level will be effective. 

\end{itemize}

%% file: Chapters/Chapter5.tex
\chapter{Spatial entanglement using quantum walk on many body system} 
\label{Chapter5}
\lhead{Chapter 5. \emph{Spatial entanglement}} 
\section{Introduction}
\label{intro}

Entanglement in many body systems has been more than
 just a computational resource, it has been used as a signature of quantum phase transition
\cite{ON02, OAF02, ORO06}.  Therefore, in the last few years, the study on entanglement in many body system has been one of important area of research interfacing between condensed matter system and quantum information sciences \cite{AFO08}.
\par
Measure of entanglement in a pure bipartite system is a function of the eigenvalues 
of the reduced density matrix. 
However, as the number of particles in the system increases, the complexity of
 finding the appropriate entanglement measures also increases making scalability of entanglement measure an enormous task.
 To address the scalability problems, a {\em scalable} entanglement measures which do not 
 diverge with the system size were proposed and global entanglement measure, a polynomial measure of multipartite entanglement is one such measure \cite{MW02}.
Using this global entanglement measure, namely the
Meyer-Wallach measure, we investigate the evolution of {\em spatial entanglement}-- particle-number entanglement between regions of space in 
a many distinguishable particle system subjected to  quantum walk process. 
Spatial entanglement has also been explored using different methods, for example, in 
an ideal bosonic gas it has been studied from off-diagonal long-range order  \cite{HAK07}. 
For our investigations of spatial entanglement using multi particle quantum walk, we consider
distinguishable particles whose dynamics can be controlled using the quantum coin parameters,
initial state of the particles, number of particles in the system and the number of steps of quantum walk.
 In particular we consider distinguishable particles in one dimensional open and closed chains. 
The spatial entanglement thus created can be used,
for example to create entanglement between distant atoms in optical
lattice \cite{SGC09} or as a channel for state transfer in spin chain
systems \cite{Bos03, CDE04, CDD05}.   At this stage, we find calculating spatial entanglement for indistinguishable particle case computationally enormous time consuming.
\par
This chapter is organized as follows. In Section \ref{entanglement} the entanglement 
between the particle and the position space has been discussed. In the same section we also introduce spatial entanglement and discuss spatial entanglement using single 
and many particle quantum walk. In Section \ref{mpent}, we present the measure
of the spatial entanglement on the system using Meyer-Wallach global
entanglement measure scheme for particles in one dimensional lattice
and in a closed chain ($n-$cycle). We also demonstrate the control
over the entanglement using the dynamical properties of quantum
walk. We conclude with a summary in Section \ref{summ}.


\section{Entanglement}
\label{entanglement}

\subsection{Position-particle entanglement}
\label{cpent}


Quantum walk entangles the particle Hilbert space $\mathcal H_{c}$ and the position Hilbert space $\mathcal H_{p}$. As discussed in Chapter \ref{Chapter2}, $H_{c}$  is spanned  by the basis state (internal state) of the  particle $|0\rangle$ and  $|1\rangle$ and $\mathcal H_{p}$  is spanned by the basis state of the position $|\psi_{j}\rangle$, where $j  \in \mathbb{Z}$. 
Lets consider a discrete-time quantum walk on a particle initially in state given by symmetric superposition state (\ref{inista})  at position $j$
with coin operation
\be 
\label{coin1}
B_{0,\theta, 0}
\equiv    \left(   \begin{array}{clcr}   \cos(\theta)    &   &
\sin(\theta)    \\        \sin(\theta)    &    &
-\cos(\theta)
\end{array} \right).
\ee
After the first step, $W_{\theta} = S(B_{0, \theta, 0} \otimes   {\mathbbm 1})$,  the state takes the form
\begin{eqnarray}
| \Psi_{1} \rangle &=& W_{\theta} | \Psi_{j}\rangle   \nonumber \\
                   &=& \gamma \left( |0 \rangle \otimes |\psi_{j-1} \rangle
                   \right) + \delta \left( |1 \rangle \otimes |\psi_{j+1}
                     \rangle \right) 
  \label{evolution01}
\end{eqnarray}
where $\gamma = \left( \frac{\cos(\theta) + i\sin(\theta)}{\sqrt{2}} \right)$ and
$\delta = \left( \frac{\sin(\theta) - i\cos(\theta)}{\sqrt{2}} \right)$. 
Schmidt rank of   $|\Psi_1\rangle$ is
$2$ which implies entanglement in the system. 
The value of entanglement can be further quantified with increase in
number of steps by  
computing the Von Neumann  entropy of the reduced density matrix of
the position subspace.  Position-particle entanglement 
in quantum walk on regular graph has been studied in detail in \cite{CLX05}.

\begin{figure}
\begin{center}
\includegraphics[width=9.0cm]{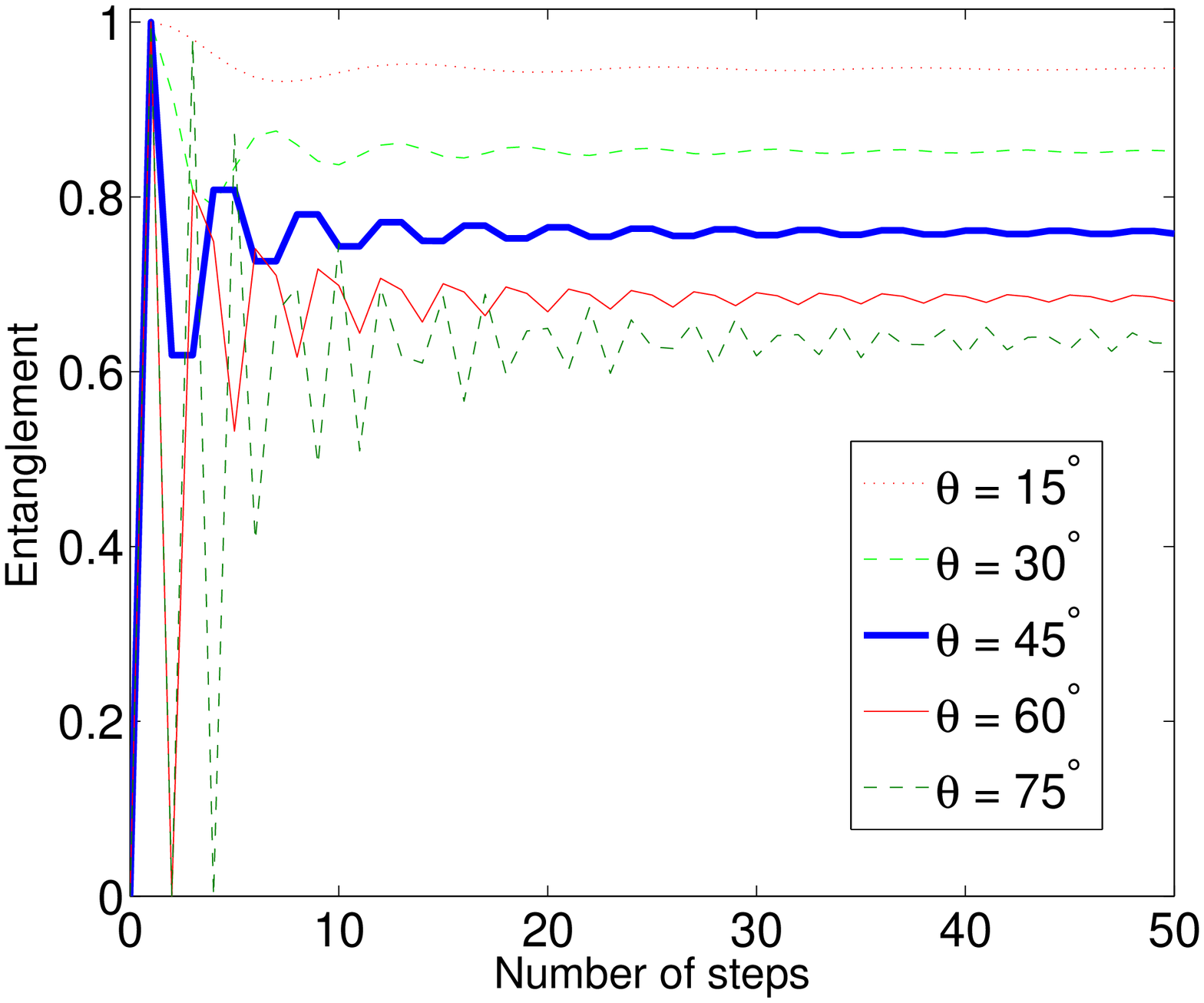}
\caption[Entanglement of single particle with position
  space when subjected to quantum walk. The initial state of the particle is
  $\frac{1}{\sqrt 2}(|0\ra + i |1\ra)$ and is evolved in position
  space using different values for $\theta$ in the quantum coin
  operation $B_{0, \theta, 0}$.]{Entanglement of single particle with position
  space when subjected to quantum walk. The initial state of the particle is
  $\frac{1}{\sqrt 2}(|0\ra + i |1\ra)$ and is evolved in position
  space using different values for $\theta$ in the quantum coin
  operation $B_{0, \theta, 0}$. The entanglement initially oscillates
  and approaches asymptotic value with increase in number of
  steps. For smaller values of $\theta$ the entanglement is higher and
  decreases with increase in $\theta$. Initial oscillation is also
  larger for higher $\theta$.} 
\label{enta}
\end{center}
\end{figure}
\begin{figure}
\begin{center}
\includegraphics[width=9.0cm]{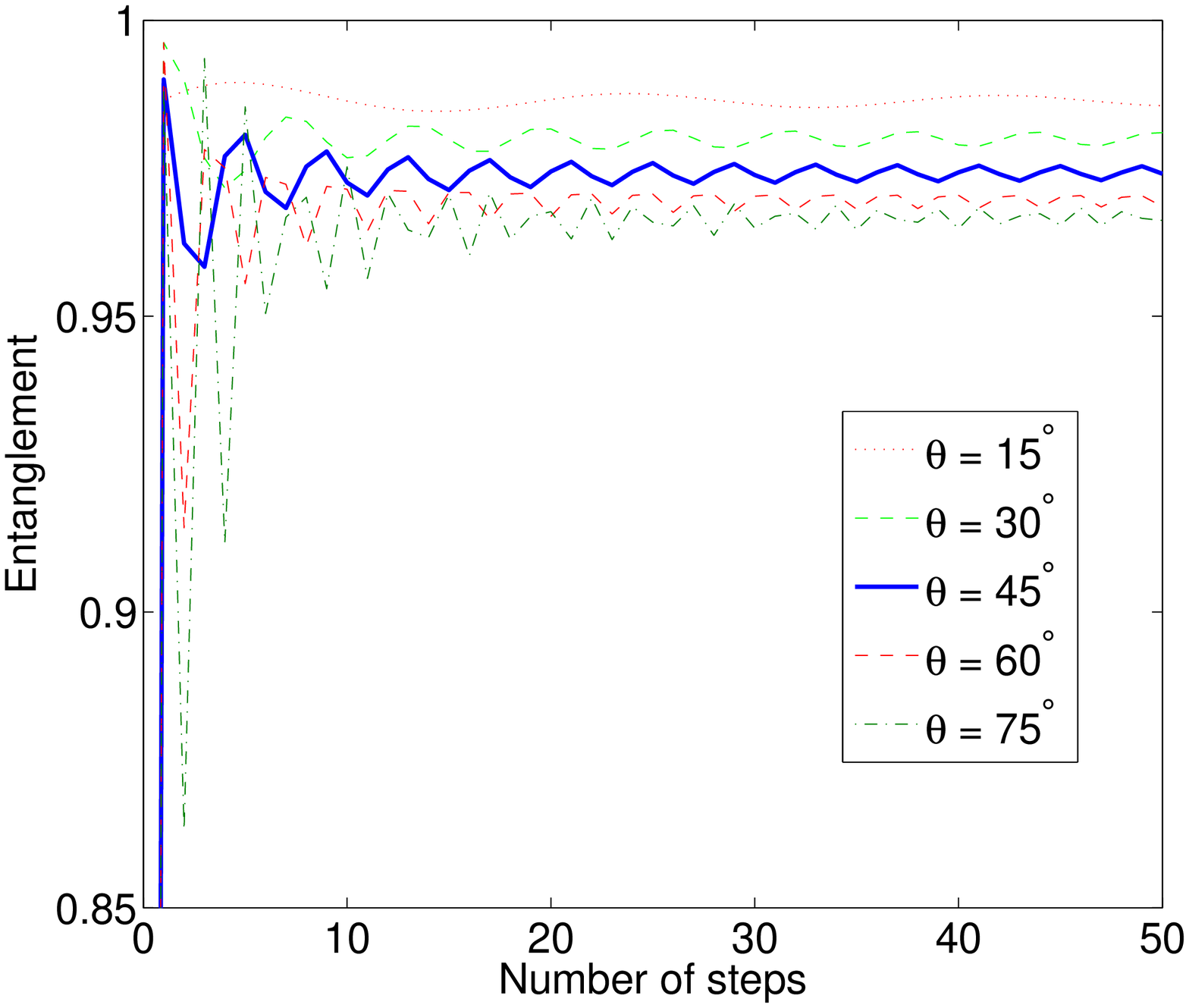}
\caption[Entanglement of single particle with position
  space when subjected to quantum walk. The initial state of the particle is
  given by parameter $\delta = 40^{\circ}$ and $\eta
  = 30^{\circ}$ and is evolved in position space using different
  values for $\theta$ in the quantum coin operation $B_{0, \theta,
    0}$.]{Entanglement of single particle with position
  space when subjected to quantum walk. The initial state of the particle is
  given by form (\ref{qw:ins}) with $\delta = 40^{\circ}$ and $\eta
  = 30^{\circ}$ and is evolved in position space using different
  values for $\theta$ in the quantum coin operation $B_{0, \theta,
    0}$. The entanglement initially oscillates and approaches
  asymptotic value with increase in number of steps. For smaller
  values of $\theta$ the entanglement is higher and decreases with
  increase in $\theta$. Initial oscillation is also larger for higher
  $\theta$.} 
\label{enta1}
\end{center}
\end{figure}

Figure \ref{enta} is the plot of the entanglement against the number of
steps of the quantum walk on particle initially in symmetric
superposition state using different values for $\theta$ in the
operation $W_{\theta}$. Von Neumann entropy of the reduced density matrix of the coin is used to quantify the entanglement between the coin and the particle position,
\be
E_{c}(t) = - \sum_{j} \lambda_{j} \rm{log}_{2}(\lambda_{j})
\ee
where $\lambda_{j}$ are the eigenvalues of the reduced density matrix of the coin after $t$ steps (time). The entanglement initially oscillates and
reaches an asymptotic value 
with increasing number of steps.  In the asymptotic limit,
entanglement value decreases with increase in $\theta$ and this
dependence can be attributed to spread of the amplitude distribution
in position space.  That is, with increase in $\theta$, constructive
interference of quantum amplitudes towards the origin gets prominent
narrowing the distribution in the position space. 
In Figure  \ref{enta1}, the process is repeated for a particle initially
in a non symmetric superposition state $|\Psi_{in}\rangle =
[\cos(40^{\circ}) |0\rangle + e^{i30^{\circ}}
\sin(40^{\circ})|1\rangle ] \otimes |\psi_{0}\rangle$.

\subsection{Spatial Entanglement}
\label{spentqw}
{\em Spatial entanglement} is the entanglement between the lattice
points. This entanglement takes the form of non-local
particle number correlations between spatial modes.  
To observe spatial entanglement we need to first associate the lattice
with state of the particle.

\subsubsection{Using single particle quantum walk}
\label{spentqw}

In a single particle quantum walk, each lattice point is associated
with a Hilbert space spanned by two subspaces. The first is the
zero-particle subspace which does not involve any coin (particle)
states. The other is the one-particle subspace spanned by the two
possible states 
of the coin, $|0\rangle$ and $|1\rangle$.  To get the spatial
entanglement we will write the state of
  the particle in the form of the state of the
lattice. Following from (\ref{evolution01}) the state of the
  particles after 
first two steps of quantum walk takes the form : 
  \begin{eqnarray}
    | \Psi_{2} \rangle &=& W_{\theta} | \Psi_{1} \rangle \nonumber \\  
    & = & \gamma \left [
      \cos (\theta) |0\rangle |\psi_{j-2} \rangle  + \sin (\theta)
      |1\rangle|\psi_{j} \rangle 
    \right ]  \nonumber \\
&&    + \delta \left [  \sin(\theta) |0\rangle |\psi_{j} \rangle -
  \cos(\theta) |1\rangle|\psi_{j+2} 
      \rangle \right ].
    \label{evolution02}
  \end{eqnarray}
In order to get the state of the lattice we can redefine the position state in the 
following way: the occupied position state $|\psi_j\rangle$ as  $|1_{j}\rangle$, which means 
that the $j$-th position is occupied and rest of the lattice is empty. Therefore, we can rewrite (\ref{evolution02}) as, 
\begin{eqnarray}
    | \Psi_{2} \rangle    & = & \gamma \left [
      \cos (\theta) |0\rangle | 1_{j-2} \rangle  + \sin (\theta)
      |1\rangle| 1_{j} \rangle 
    \right ]  \nonumber \\
&&    + \delta \left [  \sin(\theta) |0\rangle | 1_{j} \rangle -
  \cos(\theta) |1\rangle| 1_{j+2} 
      \rangle \right ].
    \label{evolution03}
  \end{eqnarray}
When $j-$th position is unoccupied (empty) the lattice state is written as $|0_{j}\rangle$.
Since we are interested in the spatial entanglement we project this state 
into one of the coin state so that we can ignore the entanglement between the coin and the position state and consider only the lattice states.
Here we will choose the coin state to be $|0\rangle$ and take the projection to obtain the state of the lattice in the form :
 \begin{equation}
  |\Psi_{lat}\rangle = |0\rangle \left(\gamma \cos(\theta)|1_{j-2}\rangle + \delta \sin(\theta)
    |1_{j}\rangle
  \right). 
  \label{latticestate}
\end{equation}
Each lattice site $j$ can be considered as a Hilbert space with the basis state $|1_{j}\rangle$ (occupied state) and $|0_{j}\rangle$ (unoccupied state). Then, the above expression (\ref{latticestate}) in the extended Hilbert space of each lattice can be rewritten in terms of occupied and unoccupied lattice states as
 \begin{equation}
  |\Psi_{lat}^{\prime}\rangle =  \gamma \cos(\theta) |1_{j-2} \,0_{j}\rangle+\delta \sin(\theta) |0_{j-2} \,1_{j}  \rangle. 
  \label{latticestate1}
\end{equation}
We can see that after first two steps of quantum walk the lattice
states $|1_{j}\rangle$ and $|1_{j-2}\rangle$ are entangled. One
  can check that the lattice states $|1_{j}\rangle$ and $|1_{j+2}\rangle$ are entangled if we choose the coin state to be $|1\rangle$. With increase in number of steps the state of the particle spread in position space and the projection over one of the coin state reduces the state considered to measure spatial entanglement. Therefore, with increase in number of steps the spatial entanglement from single
particle quantum walk decreases.

\subsubsection{Using many particle quantum walk}
\label{mbqw}

We will extend the  study of the evolution of spatial entanglement as
the quantum walk progresses on a many particle system.   
\par
\begin{figure}
\begin{center}
\includegraphics[width=8.5cm]{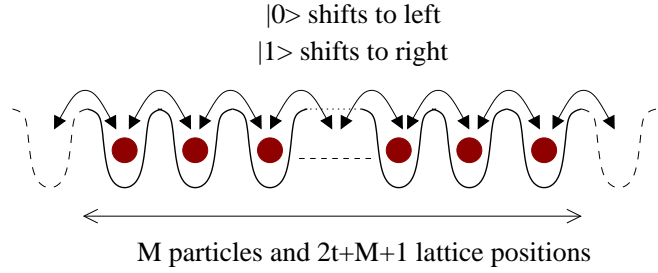}
\caption{Many particle state with one non-interacting particle at each
  position space.} 
\label{mi}
\end{center}
\end{figure}

To define a many particle quantum walk we will consider a one
dimensional lattice with one particle at each position as initial
state, Figure \ref{mi}.  $M$ independent identical particles in $M$
lattice with each particle having its own coin and position Hilbert
space will have a  total Hilbert space $\mathcal{H}=\left(
  \mathcal{H}_c \otimes \mathcal{H}_p 
\right)^{\otimes M }$. We consider the particles to be distinguishable for the
time being.
\par
The evolution of each particle is independent as they are
not interacting, and hence the evolution operator is simply
$W_{\theta}^{\otimes M}$. 
The initial state that we will consider for many particle system in one dimension will be    
\begin{equation}
|\Psi_{0}^{M}\rangle = \bigotimes_{j=-\frac{M-1}{2}}^{j=\frac{M-1}{2}}
  \left( \frac{|0\rangle + i|1\rangle}{\sqrt{2}} \right) \otimes
  |\psi_{j}\rangle.
  \label{initialMBQWstate}
\end{equation}
\par
For an $M$ particle system after $t$ steps of quantum walk, the
  Hilbert space consists of the tensor products of single lattice
  position Hilbert space which are $(2t+M+1)$ in number. That
  is, after $t$ steps of quantum walk, each of $M$ particles spread  
  between $(j-t)$ to $(j+t)$. In principle, each lattice point can be empty without any 
  particle or can have a particle in state $|0\rangle$ or $|1\rangle$. That is, at each lattice point, each of $M$ distinguishable particles are spanned by the state $|0\rangle$, $|1\rangle$ and no particle state.  Therefore, after redefining the particle-position Hilbert space in the form of state of the lattice, the dimension of each lattice point will be $3^{M}$ and the dimension of total Hilbert space is $(3^{M})^{\otimes M}$.
\par
Let us first consider the analysis of the first two steps of Hadamard
walk ($\theta = \pi/4$ in (\ref{coin1})) on a three particle
system with initial state 
\begin{equation}
 |\Psi_{0}^{3p}\rangle = \bigotimes_{j=-1}^{+1}
  \left( \frac{|0\rangle + i|1\rangle}{\sqrt{2}} \right) \otimes
  |\psi_{j}\rangle. 
  \label{initialMBQWstate3p}
\end{equation}
We will label the three particles at positions $-1$, $0$ and
$1$ as ${\rm A}$, ${\rm B}$ and ${\rm C}$. Since the evolution of these particles are
independent, we write down the state after the first step as a tensor product of each of the three particle,
\begin{align}
 |\Psi^{3p}_{1}\rangle = W_{\theta}^{\otimes 3}|\Psi_{0}^{3p}\rangle
 &= \left[ \gamma  |0\rangle |-2  \rangle + \delta |1\rangle| 0 \rangle  \right]_{\rm A}  \nonumber \\ 
  &\otimes \left[ \gamma  |0\rangle |-1
        \rangle + \delta |1\rangle| +1 \rangle \right]_{\rm B}  \nonumber \\ 
 &\otimes \left[ \gamma  |0\rangle |0
        \rangle + \delta |1\rangle| +2 \rangle  \right]_{\rm C} 
  \label{threeparticle1step} 
\end{align}
where $\gamma = (1+i)/2$ and $\delta = (1-i)/2$.  After  two step the 
tensor product of each of the three particle is given by
\begin{align}
  |\Psi^{3p}_{2}\rangle &= \left[ \gamma \left( \frac{|0\rangle |-3
        \rangle+|1\rangle|-1 \rangle}{\sqrt{2}} \right) + \delta
    \left( \frac{|0\rangle |-1 \rangle - |1\rangle|+1
        \rangle}{\sqrt{2}} \right) \right]_{\rm A} \nonumber \\ 
  &\otimes  \left[ \gamma \left( \frac{|0\rangle |-2
        \rangle+|1\rangle|0 \rangle}{\sqrt{2}} \right) + \delta \left(
      \frac{|0\rangle |0 \rangle - |1\rangle|+2 \rangle}{\sqrt{2}}
    \right) \right]_{\rm B} \nonumber \\ 
  &\otimes \left[ \gamma \left( \frac{|0\rangle |-1
        \rangle+|1\rangle|+1 \rangle}{\sqrt{2}} \right) + \delta \left(
      \frac{|0\rangle |+1 \rangle - |1\rangle|+3 \rangle}{\sqrt{2}}
    \right) \right]_{\rm C}. 
  \label{threeparticlestate} 
\end{align}
 By projecting this
state onto a 1-D coin state  
we can get the state of the lattice to calculate spatial
entanglement. We choose this state to be 
$|0\rangle \otimes |0\rangle \otimes \dots |0\rangle$. Then the
state of
the lattice after projection and normalization is 
 \begin{align}
    |\Psi_{lat}\rangle =&  \gamma^3 \,
    |{\rm A}\rangle_{-3}|{\rm B}\rangle_{-2}|{\rm C}\rangle_{-1}  \nonumber \\
  &  + \gamma^2 \delta \left( \,
      |{\rm A}\rangle_{-3}|{\rm B}\rangle_{-2}|{\rm C}\rangle_{1}
      + |{\rm A}\rangle_{-3}|{\rm B}\rangle_{0}|{\rm C}\rangle_{-1}  
     +|{\rm AC}\rangle_{-1}|{\rm B}\rangle_{-2} \right) \nonumber \\
    & + \gamma \delta^2  \left( \,
      |{\rm A}\rangle_{-3}|{\rm B}\rangle_{0}|{\rm C}\rangle_{1} +
      |{\rm AC}\rangle_{-1}|{\rm B}\rangle_{0} 
     + |{\rm A}\rangle_{-1}|{\rm B}\rangle_{-2}|{\rm C}\rangle_{1} \right) \nonumber \\
    &  + \delta^3 \, |{\rm A}\rangle_{-1}|{\rm B}\rangle_{0}|{\rm C}\rangle_{1}.
    \label{mblatticestate}
  \end{align}
From the above expression the spatial entanglement due to
quantum walk on a three particle system can be calculated.  But with increase in the 
number of particles and number of steps of quantum walk, the measure of
entanglement gets complicated. For which a scalable entanglement
measure scheme we discuss in next section would be useful. 

\section{Calculating spatial entanglement in a multipartite system} 
\label{mpent}

In a system with two particles, the state is separable if we can write
it as a tensor product of the individual particle states, and
entangled if not. However, it is somewhat more involved to define
entanglement for a system with more than two particles. In general a
state is said to be \emph{partially} entangled if it can be written as  

\begin{equation}
  | \psi \rangle = | \phi_1 \rangle \otimes | \phi_2 \rangle \otimes
  \cdots | \phi_k \rangle 
  \label{partialentangled}
\end{equation}
where $k < M$. In case $k=M$ the state is said to be fully separable
and $|\phi_i\rangle$ will then denote the state of the $i-$th
particle. If on the other hand $k=1$ then the state will be fully
entangled.  
\par
\begin{figure}
\begin{center}
\includegraphics[width=12cm]{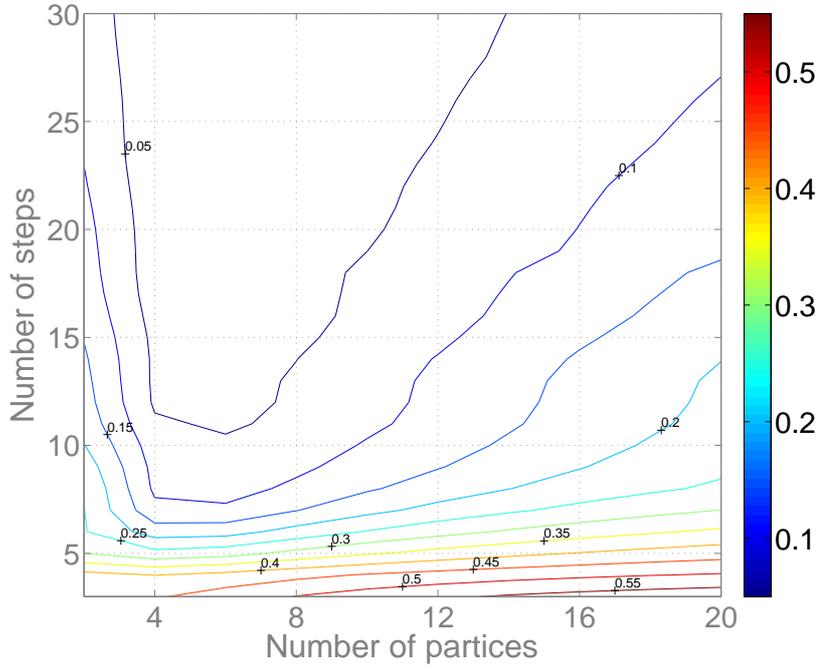}
\caption[Evolution of spatial entanglement with increase in the number
  of steps of the quantum walk for different number of particles in an
  open one dimensional lattice chain.]
  {Evolution of spatial entanglement with increase in the number
  of steps of the quantum walk for different number of particles in an
  open one dimensional lattice chain.  The entanglement first
  increases and with the further increase in the the number of steps,
  the number of lattice positions exceeds the number of particles in
  the system resulting in the decrease of the spatial entanglement.  
The distribution is obtained by implementing quantum walk on particles
in the initial state $\frac{1}{\sqrt 2}(|0\ra + i |1\ra)$ and Hadamard
operation $B_{0, \pi/4, 0}$ as quantum coin operation.} 
\label{eeqw}
\end{center}
\end{figure}
\begin{figure}
\begin{center}
\includegraphics[width=8.5cm]{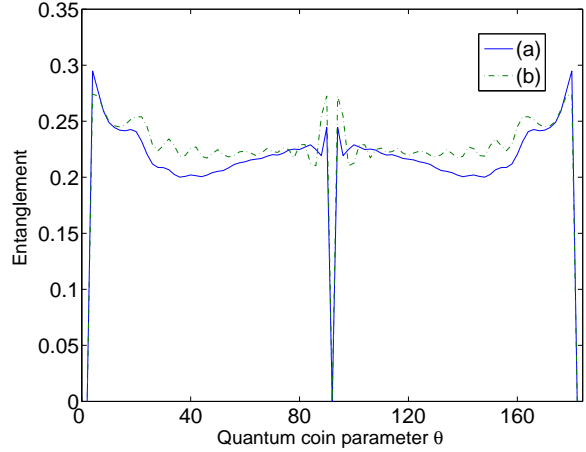}
\caption[Value of spatial entanglement for 20 particles on a one
  dimensional lattice after 20 steps of quantum walk using different
  values of $\theta$ in the quantum coin operation $B_{0, \theta, 0}$.]
  {Value of spatial entanglement for 20 particles on a one
  dimensional lattice after 20 steps of quantum walk using different
  values of $\theta$ in the quantum coin operation $B_{0, \theta, 0}$.  
(a) and (b) are the distributions for particles initially in state
$\frac{1}{\sqrt{2}}(|0\rangle + i|1\rangle)$ and state $|0\rangle
~(|1\rangle)$ respectively. In (a), with increase in $\theta$ from $0$
to $\pi/2$ the spread of the distribution in lattice position
decreases first up to $\theta = \pi/4$ and again increase. Due to
asymmetric distributions for particles in case of (b) the above effect
is not very prominent. For both (a) and (b) when $\theta = \pi/2$, for
every even number of steps of quantum walk, the system returns to the
initial state where entanglement is $0$. Entanglement is 0 for
$\theta= 0$.} 
\label{entqwtheta}
\end{center}
\end{figure}

There are quite a few good entanglement measures for the multipartite
state \cite{CKW00, BL01,  EB01, MW02, VDM03, Miy08, HJ08}. We will be
using the Meyer-Wallach (M-W) entanglement measure in the lattice as
it does not diverge with increasing system size and it is relatively easy
to calculate \cite{MW02}. The M-W measure is the entanglement of a
single particle to the rest of the system, averaged over the whole of
the system. For pure states, linear entropy can serve as a good
measure for the 
entanglement of a single particle with the rest of the system. It is
given by 
\begin{equation}
 E = \frac{d}{d-1} \left[ 1 - {\rm Tr} \rho^2 \right]
  \label{linentropy}
\end{equation}
for a $d$-dimensional particle Hilbert space. The M-W measure follows : 

\begin{equation}
  E = \frac{d}{d-1} \left[ 1 - \frac{1}{L} \sum_{i=1}^{L} {\rm Tr} \rho_i^2 \right]
  \label{W-M}
\end{equation}
where $L$ is the system size.
In a multipartite quantum walk the dimension of each lattice point,
after projection over 
and one of the coin (particle) state, is $2^M$ where $M$ is the number of
particles. So the expression for entanglement will be

\begin{align}
E(|\Psi_{lat}\rangle) & =
\frac{2^M}{2^M-1}\left(1-\frac{1}{2t+M+1}\sum_{j=-(t+\frac{M}{2})}^{t+\frac{M}{2}}{\rm
    tr}\rho_j^2\right) 
\end{align}
where $t$ is the number of steps and $\rho_j$ is the reduced
density matrix of $j^{th}$ lattice point. $\rho_j$ can be written as
\begin{align}
\rho_j &= \sum p^j_k|k\rangle\langle k|
\end{align}
 where $|k\rangle$ is one of the  $2^M$ possible states available for
 a lattice point and $p^j_k$ can be calculated once we have the
 probability distribution of individual particle on that lattice.

Since we have $M$ distinguishable particles, we have $2^M$
configuration depending upon whether a given particle is present in
the lattice point or not after freezing the state of the particle. That forms the
basis for a single lattice 
point Hilbert space. Now we can calculate $p^j_k$, the probability of
one of the possible $|k\rangle$ state of particle in the $j-$th lattice point as:
let us say $a_j^{(l_i)}$ is the probability of $i-$th particle to be
or not to be in the in the $j-$th lattice point depending on
  $l_i$. If $l_i$ is $1$ 
then it gives us the probability of the particle to be in the lattice
point. If $l_i$ is $-1$ then $a^{(l_i)}_j$ is the probability of the
particle not to be in the lattice point so $a_j^{(-1)} =
  1-a_j^{(1)}$. And hence we can write 
\begin{align}
p^j_k &= \prod_{k}a^{l_k}_j.
\end{align}
Thus the spatial entanglement can be conveniently calculated once  
we have the probability distribution of individual particle on that
lattice.  
Since quantum walk is a controlled evolution, one can obtain a
probability distribution of each particle over the entire lattice
positions. In fact one can easily control the probability distribution 
by varying the quantum coin parameters during the quantum walk process
and hence the entanglement.
\par
Figure \ref{eeqw} is
  the phase diagram of the spatial entanglement using the many
  particle quantum walk. This figure gives us the information about
the entanglement in the space for different number of particles with
increasing number of steps of the quantum walk. 

Here we have chosen Hadamard operation $B_{0, \pi/4, 0}$ and
$\frac{1}{\sqrt 2}(|0\rangle  +  i |1\rangle)$ as quantum coin
operation and initial state of the particles respectively for the walk
evolution.  To see the variation of entanglement for a fixed number of
particles with increase in steps, we can pick a line parallel to $y$
axis i.e, for fixed number of particles and see the entanglement
varying with number of steps.
\par
We see that the entanglement for an instance, after first iteration
goes to maximum from zero 
and with the further increase in the number of steps, the number of
lattice positions exceeds the number of particles in the system
resulting in the decrease of the spatial entanglement. The decrease in
entanglement before the number of steps is equal to number of
particles should be noted. This is because, for Hadamard walk the
spread of the probability distribution after $t$ steps is between
$\frac{-t}{\sqrt 2}$  and $\frac{t}{\sqrt 2}$ and not between $-t$ and
$t$ \cite{CSL08}. To see the variation of entanglement with parameter
$\theta$ more clearly, in Figure  \ref{entqwtheta} the spatial
entanglement for 20 particles after 20 steps of quantum walk using
different values of $\theta$ in the quantum coin operation $B_{0,
  \theta, 0}$ is presented. Variation of entanglement with $\theta$
for two case, all particles in equal superposition state
$\frac{1}{\sqrt 2}(|0\ra + i|1\ra)$ (unbiased quantum walk) and in one
of the basis state $|0\ra$ or $|1\ra$ (biased quantum walk) as initial
state is presented. The effect of biasing the quantum walk by
different magnitude can also be reproduced using the other two
parameters $\xi, \zeta$ in the coin operation $B_{\xi, \theta,
  \zeta}$.  
\par
In Figure  \ref{fig:qw1a},  we can see that for a single particle quantum walk, with increase in $\theta$ from $0$ to $\pi/4$ in the coin operation, along with decrease in the spread, the maximum point (two peaks) in the probability distribution in the position space also decrease. Whereas, with increase in $\theta$ from $\pi/4$ to $\pi/2$, the maximum point in the probability distribution in position space increases along with decrease in the spread.  That is, we can see that the minima of the maximum point in the probability distribution is for $\theta = \pi/4$.  Therefore, this attributes for the minimal in the spatial entanglement at $\theta = \pi/4$ in Figure  \ref{entqwtheta}. Note that we have ignored the extreme values of $\theta = 0$ and $\pi/2$ which evolves without any quantum interference effect. 
\par
For fixed number of steps of quantum walk, the spatial entanglement first decreases and then it starts reviving as we increase the number of particles.
\par
{\bf Closed chain :} 
\par
Since most of the physical system that will be considered for implementation will be of definite dimension we extend our calculations to one of the simple example of closed geometry, $n-$ cycle.
\begin{figure}
\begin{center}
\includegraphics[width=8.5cm]{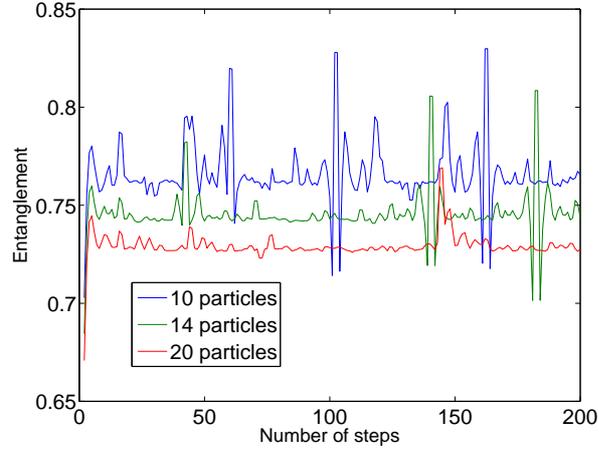}
\caption[Evolution of spatial entanglement for a system with
  different number of particles in a closed chain.]
  {Evolution of spatial entanglement for a system with
  different number of particles in a closed chain.  With increase in
  the number of steps, the entanglement value remains close to  
asymptotic value with some peaks in between. The peaks can be
accounted for the cross over of the left and right propagating
amplitudes of the internal state of the particle during quantum
walk. The peaks are more for chain with smaller number of
particles. Increase in the number of particles in the system results
in the decrease in the entanglement value. The distribution is
obtained by using $\frac{1}{\sqrt{2}}(|0\ra + i|1\ra)$ as the initial
states of all the particles and Hadamard operation $B_{0, 45^{\circ},
  0}$ as quantum coin operation.} 
\label{Ering}
\end{center}
\end{figure}
\begin{figure}
\begin{center}
\includegraphics[width=8.5cm]{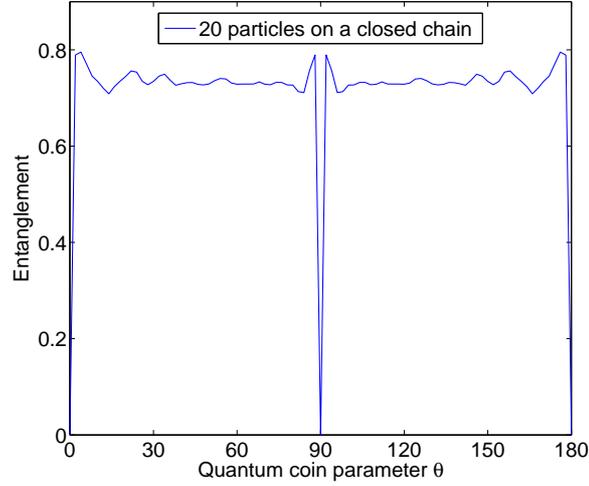}
\caption[Value of spatial entanglement for 20 particles on a closed
  chain after 20 steps of quantum walk using different values of
  $\theta$ in the quantum coin operation $B_{0, \theta, 0}$.]
  {Value of spatial entanglement for 20 particles on a closed
  chain after 20 steps of quantum walk using different values of
  $\theta$ in the quantum coin operation $B_{0, \theta, 0}$. (a) and
  (b) are the distributions for particles initially in state
  $\frac{1}{\sqrt{2}}(|0\rangle + i|1\rangle)$ and state $|0\rangle
  ~(|1\rangle)$ respectively. Since the system is in the closed chain,
  the quantum walk does not expand the position Hilbert space
  therefore for all values of $\theta$ from $0^{\circ}$ to $\pi/2$ the
  entanglement value remains close to the asymptotic value except for
  a small peak at smaller values of $\theta$.  For $\theta=0^{\circ}$
  when number of steps equal to number of particles, the amplitudes
  goes round the chain and returns to its initial state making the
  entanglement $0$ and for $\theta = \pi/2$, for every even number of
  steps of quantum walk, the system returns to the initial state where
  entanglement is again $0$.} 
\label{Eringtheta}
\end{center}
\end{figure}
\par
When we consider a many particle system in a closed chain, with number of lattice position equal to number of particles $M$, the quantum walk process does not expand the position Hilbert space like it does on an open chain. Therefore the spatial entanglement does not decrease, remains close to the asymptotic value with increase in the number of steps of quantum walk. Figure  \ref{Ering} shows the evolution of entanglement for a system with different number of particles in a closed chain. The peaks seen in the plot can be accounted for the cross over of the left and right propagating amplitudes of the internal state of the particle during the quantum walk process. Therefore the frequency of the peaks are more for smaller number of particles (smaller closed chain). In general, one can also notice that the increase in the number of particles and number of lattice point in the closed cycle results in the decrease in the spatial entanglement of the system. 
\par
In Figure  \ref{Eringtheta}, the value of spatial entanglement for 20 particles on a closed chain after 20 steps of quantum walk using different values of $\theta$ in the quantum coin operation $B_{0, \theta, 0}$ is presented. For all values of $\theta$ from $0$ to $\pi/2$ the entanglement value remains close to the asymptotic value except for the extreme values of $\theta$.  For $\theta=0$ when number of steps equal to number of particles the amplitudes goes round the ring and returns to its initial state making the spatial entanglement value $= 0$. For $\theta = \pi/2$, for every even number of steps of quantum walk, the system returns to the initial state where the spatial entanglement is again $0$. 
\par
Therefore, spatial entanglement on a large lattice space can be created, controlled and optimized for maximum entanglement value by varying the quantum coin parameters and number of particles in the multi particle quantum walk.  In \cite{SGC09} a scheme to implement a scalable quantum information processing using ultra cold molecules is proposed.  Two different species of atoms held in independent optical lattices are entangled  by translating the separate lattices to overlap. Implementing quantum walk on one of the species and creating spatial entanglement can possibly make the above scheme more robust.

\newpage

\section{Summary} 
\label{summ}
\begin{itemize}

\item We have presented the evolution of spatial entanglement in many particle system subjected to the quantum walk process. By considering many particles in the one dimensional open and closed chain we have shown that the spatial entanglement of the system can be controlled by controlling the dynamics of the quantum walk, the quantum coin parameter, initial state and number of steps. Developing on this approach, many particle entanglement can be generated and controlled in a many body system using quantum walk as a tool.
\end{itemize}

%% file: Chapters/Chapter6.tex

\chapter{Conclusion and future perspective} 
\label{Chapter6}
\lhead{Chapter 6.  \emph{Conclusion and future perspective}} 

\section{Conclusion}

In this thesis we briefly reviewed  the continuous- and discrete-time quantum walk which are structurally identical to Schr\"odinger and Dirac equations, respectively.  By simple decoupling analysis of the evolution, structural similarity of the discrete-time quantum walk and the Dirac equation was shown.  To optimize the discrete-time quantum walk, the use of three parameter quantum coin operation $B_{\xi, \theta, \zeta}$ from SU(2) group. Numerical data with supporting mathematical analysis, it was shown that parameter $\theta$ can be used to control the variance 
and parameters $\xi$ and $\zeta$ to bias and control the biasing in the walk.  The use of quantum coin $B_{\xi, \theta, \zeta}$ for walk on an $n-$cycle and its effect on optimizing mixing time was shown. 
\par
Recurrence is an important phenomenon in the study of dynamics. Recurrence we considered was 
the return of unit probability amplitude at the origin during the dynamics. Based on the numerical data and analysis we have shown that the quantum walk evolution dominated by the interference of quantum amplitudes fails to satisfy complete recurrence theorem. However, fractional recurrence characterized by the quantum P\'olya number can be seen. 
\par
Symmetries and effect of noise on quantum walk on a line and $n-$cycle was studied.  We considered the noise on the coin space in our study. Variants of quantum walks  on a  line was considered and showed that they are equivalent  in  the  sense   that  the  final  positional  probability
distribution  remains the  same in  each variant (symmetry in the probability distribution).   In  particular, we considered variants  obtained by the  experimentally relevant operations
of $Z$  or $X$ applied  at each quantum  walk step, with  the symmetry
operations  given  by  ${\bf  Z}$  and ${\bf  PRX}$.  Interestingly we observed that these symmetries were preserved even in the presence of noise, in particular,  those  characterized by  the  phase  flip,  bit flip  and generalized amplitude  damping channels. 
This is  important because it
means that  the equivalence of these  variants is not  affected by the
presence of noise, which would be inevitable in actual experiments. However, the symmetry of the phase operation under phase noise was intuitive, considering that this noise has a Kraus representation consisting of operations that are symmetries of the noiseless quantum walk. But, for
the PRX symmetry under phase noise, and for any symmetry under other noisy
channels (especially in the case of generalized amplitude damping
channel), the connection was not obvious before the analysis was
completed. When the studies were extended to the walk on an $n-$cycle, an interesting fact that we observed was the breakdown of the symmetries found in the walk on a line. Further, when the noise was introduced, above a certain noise level these symmetries were restored. 
Results to study noise model were supported by  several numerical examples  obtained by evolving  the density  operator in  the Kraus representation and analytical proofs  of the effect  of noise on symmetries were obtained using the quantum trajectories  approach, which we found convenient for this situation.
\par
In this thesis two of the applications of quantum walk was considered. Use of quantum walk to redistribute atoms in optical lattice and evolution of spatial entanglement in many body system. 
We proposed the use of the quantum walk to redistribute atoms, and studied its dynamics in an optical lattice and to expedite the process of quantum phase transition.  The coherent control over the atoms using the coin degree of freedom during the evolution of the walk was demonstrated. 
Experimentally realizable noisy channels studied in the dynamics part of this thesis was used 
to show the additional control over the evolution and atomic density redistribution. 
Making use of the stimulated Raman transition, we have proposed stimulated Raman kicks to act as shift operator along with rf-pulse as coin operation to implement  discrete-time quantum walk on Bose-Einstein condensate (BEC). Our scheme  can retain the macroscopic features of the wave packet or can be accordingly modified to implement the walk at an individual atom level which will be effective for transition from superfluid to the Mott insulator state and vice versa.
\par
Entanglement is many body system is one of the important topic of interest. As a step towards 
using quantum walk to generate and control entanglement in many body system, both distinguishable and indistinguishable,  we have presented the evolution of spatial entanglement in many distinguishable particle system. We considered Meyer-Wallach measure for our study.
By considering many particles in one dimensional open and closed chain we have shown that the spatial entanglement of the system can be controlled by controlling the dynamics of the quantum walk, the quantum coin parameter, initial state and number of steps. 

\section{Future perspective}

In this thesis we considered quantum walk on a line and on an $n-$cycle. Extension of these studies presented to general graph and higher dimension could reveal other interesting features of the quantum walk dynamics. For example, structural similarity of discrete-time quantum walk with Dirac equation was presented. Discrete-time quantum walk in higher dimension and its analysis following the decoupling approach used to walk on a line could reveal interesting structural similarity.  
\par
Derivation of dependency of variance on the parameter $\theta$ for a walk on a line ($1- \sin(\theta))t^2$ from first principles (Fourier analysis) which has not be done in this thesis would be a step in a direction of deriving the dependency of variance in two- and higher dimensions. 
\par
The results from symmetries and effects of noise on quantum walk can be extended and generalized to walk on most of the symmetric closed graphs. This symmetry-topology-noise interplay presented would be of relevance to quantum information  processing systems, and have  wider implications  to the condensed matter systems.
\par
Theoretically, the evolution of the density profile with a quantum walk can be used in place of quantum Monte Carlo simulation \cite{SRB05, RM05}  or the time evolution density matrix renormalization group (t-DMRG) to study the correlation and redistribution of atoms in optical lattices \cite{RMR06}. We expect the quantum walk to play a wider role in simulating and expediting the dynamics in various physical systems.
Similarly, entanglement in many body system using quantum walk could be an interesting topic to explore quantum phase transition and quantum annealing problems in various physical systems. Presently, the spatial entanglement studied in this thesis seems to have very little practical interest. Developing on this, further studies can be considered to extend the evolution of spatial entanglement using indistinguishable particle quantum walk. However, many particle-position entanglement using quantum walk will be an interesting and bigger problem to explore and use it to study phase transitions in quantum systems.  
\par
In general quantum walk seems to serve as a tool to understand dynamics in various quantum systems which can further motivate to be used for various applications in quantum systems.

%% file: Appendices/Appendix0.tex

\chapter{Quantum walk and Klein-Gordon equation}
\label{Appendix0}
\lhead{Appendix A. \emph{Quantum walk and Klein-Gordon equation}}

\section{Decoupling the coupled expression} 
Getting Equation (\ref{eq:dec}) from Equations (\ref{eq:comp}) and (\ref{eq:compb})} \\
\\
From Equation (\ref{eq:compb}), solving for $\Psi_L$ we get
$$
\Psi_L(j+1, t) = \frac{i}{\sin(\theta)}\left [ \Psi_R(j, t+1)
- \cos(\theta) \Psi_R(j-1,t) \right ].
$$
Therefore
$$
\Psi_L(j, t+1) = \frac{i}{\sin(\theta)} \left [ \Psi_R(j-1, t+2)
- \cos(\theta) \Psi_R(j-2, t+1) \right ].
$$
By substituting the above for $\Psi_L(j+1, t)$  and $\Psi_L(j ,t+1)$ in (\ref{eq:comp}), we get (\ref{eq:dec}).

\section{Getting the difference operator that corresponds to the differential operators}
Getting Equation (\ref{eq:k-g}) from Equation (\ref{eq:dec})}

The difference $ \nabla_t$ operator that corresponds to the differential
operator $\partial/\partial t$ is
$$
 \nabla_t = \frac{\Psi(j, t+ \frac{h}{2}) - \Psi(j,t-\frac{h}{2} )}{h}.
$$
By setting the small incremental time to 1 ($h=1$) difference operator
$$
\nabla_t = \Psi(j, t+0.5) - \Psi (j, t-0.5),
$$
corresponds to the difference operator $\partial/ \partial t$.
Therefore, the operator $\partial^2/\partial t^2$ will correspond to
applying the difference operator in each of the above two terms, which
yields
$$
 \nabla^2_t = \frac{1}{h} \times
\frac{[\Psi(j, t+1) - \Psi(j,t)] - [\Psi(j,t)-\Psi(j,t-1)]}{h} 
$$
$$
 = \frac{(\Psi(j,t+1) - 2\psi(j,t) +\Psi(j,t-1)}{h^2},
$$
when the small incremental time step $h=1$, it corresponds to $\partial^2 / \partial t^2$.
The difference operators $\nabla_j$ and $\nabla^2_j$ corresponding to corresponds to $\partial / \partial j$
and corresponds to $\partial^2 / \partial j^2$ are also defined
analogously for $j$ keeping $t$ constant.

%% file: Appendices/AppendixA.tex

\chapter{Variation of the variance as a function of $\theta$}
\label{AppendixA}
\lhead{Appendix B. \emph{Variation of the variance}}

For  a quantum walk using $B_{0, \theta,  0}$ as  quantum  coin, after $t$ 
steps the probability distribution is spread over the interval $(-t\cos(\theta),
t\cos(\theta))$ and shrink  quickly outside this  region. The height of the distribution with $\theta$ is proportional to $\sin(\theta)$. 
Therefore the integral of the probability distribution with in the interval can be approximated to, 
\begin{eqnarray}
\int_{-t\cos(\theta)}^{t\cos(\theta)} P(j) d j \approx 1
\label{eq:proba}
\end{eqnarray}
By approximating the probability distribution to fit the envelop of the quantum walk distribution,
\begin{eqnarray}
\int_{-t\cos(\theta)}^{t\cos(\theta)} \frac{ [1+ \cos^{2}(2 \theta)]}{\sqrt t} \exp \left [ K(\theta)\left ( \frac{j^2}{t^{2}\cos^{2}(\theta)} - 1 \right ) \right ] d j \approx 1,
\label{eq:proba1}
\end{eqnarray}
\noindent where $K(\theta) =\frac{\sqrt t}{2} \cos(\theta)[1+\cos^{2}(2 \theta)][1 + \sin(\theta)]$.
\par
The position $j$ in the interval $(-t\cos(\theta), t\cos(\theta))$ in position space can be represented as a function of $\phi$,
\be
j \approx f(\phi) = t \cos(\theta)\sin(\phi)
\ee
where $\phi$ range from $-\frac{\pi}{2}$ to $\frac{\pi}{2}$. For a walk with coin $B_{0, \theta, 0}$, the mean of the distribution is zero and hence the variance can be analytically obtained by extrapolating, 
\begin{eqnarray}
\sigma^{2}  = \int P(j)j^2dj  \approx \int_{-t\cos(\theta)}^{t\cos(\theta)} P(j) j^2 d j = \int_{-\frac{\pi}{2}}^{\frac{\pi}{2}} P(f(\phi))(f(\phi))^2 f^{\prime}(\phi) d\phi.
\label{eq:vari}
\end{eqnarray}
\begin{eqnarray}
\sigma^{2} \approx  \int_{-\frac{\pi}{2}}^{\frac{\pi}{2}} \frac{ [1+ \cos^{2}(2 \theta)]}{\sqrt t} e^{K(\theta)\left ( \frac{t^2\cos^2(\theta)\sin^2(\phi)}{t^2\cos^2(\theta)} - 1 \right )} 
\left( t\cos(\theta)\sin(\phi)\right )^2 \times \left(t\cos(\theta)\cos(\phi) \right) d \phi.
\label{eq:vari1}
\end{eqnarray}

\begin{eqnarray}
\sigma^2 \approx t^{\frac{5}{2}}  [1+ \cos^{2}(2 \theta)]\cos^3(\theta) \int_{-\frac{\pi}{2}}^{\frac{\pi}{2}} e^{K(\theta)(\sin^2(\phi)-1)} \ \sin^2(\phi) \cos(\phi) \ d\phi.
\label{eq:vari2}
\end{eqnarray}

\begin{eqnarray}
\sigma^2 \approx t^{\frac{5}{2}}  [1+ \cos^{2}(2 \theta)]\cos^3(\theta) 
\int_{-1}^{1} e^{K(\theta)(r^2 -1)} \ r^2 \ d r  \nonumber \\
= t^{\frac{5}{2}} \sin(\theta)\cos^3(\theta) \int_{-1}^{1} r \ e^{K(\theta)(r^2 -1)} \ r \ d r  \nonumber \\ 
= t^{\frac{5}{2}}  [1+ \cos^{2}(2 \theta)]\cos^3(\theta) \frac{1}{e^{K(\theta)}} \left[ \frac{r e^{K(\theta)\ r^{2}}}{2 K(\theta)}\vline_{-1}^{1} - \int_{-1}^{1} \frac{e^{K(\theta)r^2}}{2 K(\theta)} \ dr \right ]  \nonumber \\ 
= t^{\frac{5}{2}}  [1+ \cos^{2}(2 \theta)]\cos^3(\theta) \frac{1}{e^{K(\theta)}2 K(\theta)} \left[2  e^{K(\theta)} - \int_{-1}^{1} e^{K(\theta)r^2} \  dr \right ].
\label{eq:vari3}
\end{eqnarray}

The integration of $\int_{-1}^{1} e^{K(\theta)r^2} dr$ can be done using Taylor expansion,
\begin{eqnarray}
I &= &\int_{-1}^{1} e^{K(\theta)r^2} dr \nonumber \\ 
&= &\int_{-1}^{1} \left [ 1 + K(\theta)r^2 + \frac{K(\theta)^2r^4}{2!} + \frac{K(\theta)^3r^6}{3!}+ \frac{K(\theta)^4r^8}{4!} + .....+ \frac{K(\theta)^nr^{2n}}{n!}\right] dr \nonumber \\
&=& 2 \left[ 1 + \frac{K(\theta)}{3} + \frac{K(\theta)^2}{5*2!} + \frac{K(\theta)^3}{7*3!}+ \frac{K(\theta)^4}{9*4!} + ..........+ \frac{K(\theta)^n}{(2n+1)*n!}\right ] 
\label{eq:vari4}
\end{eqnarray}

\begin{eqnarray}
I =  2 \left [ \sum_{0}^{n} \frac{K(\theta)^n}{(2n+1)*n!}\right ]  =  \sum_{0}^{n} \left [ \frac{2K(\theta)^n}{n!} -  \frac{ 4 \ n \ K(\theta)^n}{(2n+1)n!} \right ].
\label{eq:vari5}
\end{eqnarray}
For large $n$, $\frac{4n}{2n+1} \approx 2$ in the above expression. When $n$ is small  $I \approx e^{K(\theta)}$. Substituting in (\ref{eq:vari3}), the variance is simplified to, 
\begin{eqnarray}
\sigma^2  \approx \frac{t^{\frac{5}{2}}  [1+ \cos^{2}(2 \theta)]\cos^3(\theta)}{ 2K(\theta)} = \frac{t^{\frac {5}{2}}  [1+ \cos^{2}(2 \theta)]\cos^3(\theta)}{\sqrt t \cos(\theta) [1+ \cos^{2}(2 \theta)][1+\sin(\theta)]} \nonumber \\ 
= \frac{t^2 (1-\sin(\theta))(1+\sin(\theta))}{1+\sin(\theta)} = (1 - \sin(\theta))t^2.
\label{eq:vari8}
\end{eqnarray}
From the solution obtained empirically through numerical integration we find that the variation of $B_{\theta}$ with $\theta$ also fits $(1  - \sin(\theta))$. That is,
\begin{eqnarray}
\sigma^{2} = C_{\theta}t^{2} \approx (1  - \sin(\theta))t^{2}.
\label{eq:varib}
\end{eqnarray}

%% file: Appendices/AppendixC.tex

\chapter{Dipole trap for $^{87}$Rb atoms using light of different wavelengths}
\label{AppendixC}
\lhead{Appendix C. \emph{Dipole trap}}

Effective laser detuning $\Delta$ for alkalis can be calculated
using,

\begin{equation}
\frac{1}{\Delta}=\left(\frac{1}{\Delta_{1}}+\frac{2}{\Delta_{2}}\right)
\end{equation}

where $ \Delta_{i} $ is detuning from $D_{i}$ line.\\
Maximum potential depth is calculated using

\begin{equation}
U_{0}=\frac{\hbar\Gamma}{2}\frac{P\Gamma}{\pi W_{0}^{2}I_{0}\Delta},
\end{equation}

where $\Gamma$ is natural linewidth, $P$ is laser power, and $I_{0}$ is the saturation intensity given by,

\begin{equation}
I_{0}=\frac{\pi^{2}h c
\Gamma}{3\lambda^{3}}=\frac{\pi}{3}\frac{hc}{\lambda^{3}\tau},
\end{equation}
where $\Gamma=1/\tau$. The numerical value of saturation intensity
$I_{0}$ for rubidium atoms is calculated to be 1.67
mW/cm$^{2}$=16.7 W/m$^{2}$.

\section{Frequency-detuning and potential depth for different wavelength}

Below is the table with numerical values of detuning and potential
depth for dipole traps using laser lights of different
wavelengths.\\
\begin{center}
\begin{tabular}{|c|c|c|}
\hline
wavelength (nm) & Detuning $\Delta$ & Potential Depth $U_{0}$ W/m$^{2}$\\
\hline
$1064$ & $-25\times10^{5}\Gamma$ & $1.53\times10^{-35}\frac{P}{W_{0}^{2}}$\\
$850$ & $-6.93\times10^{5}\Gamma$ & $5.5\times10^{-35}\frac{P}{W_{0}^{2}}$\\
$820$ & $-3.46\times10^{5}\Gamma$ & $11.04\times10^{-35}\frac{P}{W_{0}^{2}}$\\
$800$ & $-0.6\times10^{5}\Gamma$ & $63.66\times10^{-35}\frac{P}{W_{0}^{2}}$\\
\hline
\end{tabular}\\
\end{center}
The ac stark shift creates a potential, $U$ proportional to the
light intensity

\begin{equation}
U(r,z)=U_{0}\left[\frac{e^{\frac{-2r^{2}}{W(z)^{2}}}}
{1+(\frac{z}{z_{R}})^{2}}\right]
\end{equation}

$r$ and $z$ being the radial and axial co-ordinates,
$z_{R}=\frac{\pi W_{0}^{2}}{\lambda}$ is the Rayleigh
range\footnote{Distance at which the diameter of the spot size
increases by a factor of $\sqrt{2}$.} at wavelength $\lambda$ and
beam waist $W_{0}$. $W(z)$ is the beam radius as a function of
axial position $z$ and is given by,

\begin{equation}
W(z)=W_{0}\sqrt{1+\left(\frac{z}{z_{R}}\right)^{2}}.
\end{equation}

\noindent The overall potential due to effect of gravity is given by,

\begin{equation}
U_{g}(r,z)= - mgr -
U_{0}\left[\frac{e^{\frac{-2r^{2}}{W(z)^{2}}}}{\left(1 +
\frac{z}{z_{R}}\right)^{2}}\right].
\end{equation}
To calculate the minimum power required to trap $^{87}$Rb atoms at
distance $z$ from the focal point of the beam in the axial
direction of the beam, the above equation is differentiated,

\begin{equation}
\frac{d^{2}U}{dr^{2}}= -mg+\frac{4U_{0}r}{W(z)^{2}}\left[\frac{e^
{\frac{-2r^{2}}{W(z)^{2}}}}{1 +
(\frac{z}{z_{R}})^{2}}\right]=0
\end{equation}

\begin{equation}
\Longrightarrow r =
\frac{mg}{4U_{0}}\frac{W(z)^{2}}{e^{\frac{-2r^{2}}{W(z)^{2}}}}
\left[1+\left(\frac{z}
{z_{R}}\right)^{2}\right].
\end{equation}

By substituting the appropriate values for the above equation one
can calculate the minimum power required to trap atoms at distance
$z$ from the beam focus in the axial direction. Below is the table
with the calculated power required to trap  $^{87}$Rb atoms using laser
light with different wavelengths and beam waist.\\

For light with beam waist, $W_{0}=50 \mu$m \\

\begin{center}
\begin{tabular}{|c|c|c|c|}
\hline
$W_{0}=50 \mu$m  & 1064 nm & 850 nm & 820 nm\\
\hline
$z=$ 5 cm & 3.33 W & 450 mW & 210 mW\\
$z=$ 2 cm & 30 mW & 37 mW & 26.5 mW\\
$z=$1 cm & 16 mW & 8.4 mW & 4 mW\\
$z=$ 0.5 cm & 11.6 mW & 3.8 mW & 1.5 mW\\
$z=$ 0 cm & 9 mW  & 2.7 mW & 1.4 mW\\
\hline
\end{tabular}\\
\end{center}
For $W_{0}=100 \mu$m \\
\begin{center}
\begin{tabular}{|c|c|c|c|}
\hline
$W_{0}=100 \mu$m  & 1064 nm & 850 nm & 820 nm\\
\hline
$z=$ 5cm & 610 mW & 110 mW & 47 mW\\
$z=$ 2cm & 140 mW & 31 mW & 15 mW\\
$z=$ 1cm & 91 mW & 23 mW & 12 mW\\
$z=$ 0.5cm & 78 mW & 22 mW & 11 mW\\
$z=$ 0cm & 77 mW  & 22 mW & 11 mW\\
\hline
\end{tabular}\\
\end{center}
For $W_{0}=200 \mu$m \\
\begin{center}
\begin{tabular}{|c|c|c|c|}
\hline
$W_{0}=200\mu$m  & 1064nm & 850nm & 820nm\\
\hline
$z=$ 5cm & 1.53 W & 200 mW & 1.1 W\\
$z=$ 2cm & 650 mW & 180 mW & 900 mW\\
$z=$ 1cm & 610 mW & 180 mW & 900 mW\\
$z=$ 0.5cm & 600 mW & 180 mW & 900 mW\\
$z=$ 0cm & 600 mW  & 180 mW & 900 mW\\
\hline
\end{tabular}\\
\end{center}
The maximum photon scattering rate $\Gamma_{sc}$ is given by
\begin{equation}
\Gamma_{sc}=\frac{\Gamma}{\Delta}\frac{U_{0}}{\hbar}.
\end{equation}

For the laser light of 850nm the photon scattering rate at beam waist $3.8\mu m$ 
and power 0.12 mW will be, 

\begin{equation}
\Gamma_{sc}=\frac{\Gamma}{6.93\times10^{5}\Gamma}
\frac{5.5\times10^{-35}}{1.054\times10^{-34}}\left(\frac{P}
{W_{0}^{2}}\right) = 6.25 \times 10^{6}\rm{photons/sec}.
\end{equation}
